\documentclass[pdflatex,sn-mathphys-num]{sn-jnl}

\usepackage{graphicx}%
\usepackage{multirow}%
\usepackage{amsmath,amssymb,amsfonts}%
\usepackage{amsthm}%
\usepackage{mathrsfs}%
\usepackage[title]{appendix}%
\usepackage{xcolor}%
\usepackage{textcomp}%
\usepackage{manyfoot}%
\usepackage{booktabs}%
\usepackage{algorithm}%
\usepackage{algorithmicx}%
\usepackage{algpseudocode}%
\usepackage{listings}%
\usepackage{relsize}%
\usepackage{lmodern}%
\usepackage{anyfontsize}%
\usepackage{bookmark}%
\usepackage[utf8]{inputenc}%

\begin{document}

\title[Article Title]{Electrostatic force between two charged spherical conductors in an electric field: Matched asymptotic expansion approach}


\author*[1]{\fnm{Natarajan} \sur{Thiruvenkadam}}\email{natarajanvenkadam@gmail.com}

\author[1]{\fnm{Vishwanath} \sur{ Kadaba Puttanna}}\email{vishy@nitk.edu.in}
\equalcont{These authors contributed equally to this work.}

\affil*[1]{\orgdiv{Department of Mathematical and Computational Sciences}, \orgname{National Institute of Technology Karnataka}, \orgaddress{\street{Surathkal}, \city{Mangaluru}, \postcode{575025}, \state{Karnataka}, \country{India}}}
%
%


\abstract{In this study, we derive the asymptotic expressions for the electrostatic force between two charged spherical conductors in an electric field. Davis \cite{davis1964two} initially provided an expression for these forces, which are split into components along the $X$ and $Z$ directions. These forces are influenced by several factors, including the separation distance between the spheres, their sizes, their charges, the applied external electric field, and 10 specific force coefficients. These coefficients involve complex series expressions, and computing them requires summing up the 24 infinite series. To tackle the challenge of series convergence, previous researchers (\cite{lekner2012electrostatics}, \cite{lekner2013forces}, \cite{arp1977particle}) have used various methods to derive asymptotic expressions. In this work, the method of matched asymptotic expansions has been implemented to derive the asymptotic forms of these 24 series, specifically for the scenario involving two equal-sized conducting spheres in an external electric field. Accurate calculation of these series is crucial, especially when the distance between the spheres is very small, making the series calculations computationally intensive. We utilize the method proposed by Cox and Brenner \cite{Cox1967} to derive the asymptotic expressions and hence compute the 10 force coefficients and finally apply them to compute the electrostatic force for small separations. Additionally, we assess the accuracy of these asymptotic expressions by comparing them to the actual series and analyzing their deviations through percentage errors.}




\maketitle
\section{Introduction}
\label{sec:sample1}
The interaction between two charged spherical conductors in an electric field is a fascinating and significant problem of interest with numerous practical applications, such as rain dissipation and cloud physics. Extensive research has been conducted on the electrostatic interactions between two conducting spheres, depending on the impact of an external electric field on the interaction. The findings of these studies shed light on the complex nature of these phenomena and their potential applications in various fields. Davis \cite{davis1964two} has solved the Laplace equation for the electric potential field using the separation of variables in bispherical coordinates. The outcome of this analysis has helped to determine the variation of the applied electric field in the regions that exclude the two spheres. Furthermore, this approach has also enabled the determination of the induced surface charges on the spheres, which has significant implications in several applications. The force expressions given in \cite{davis1964two} involve ten force coefficients, namely, $F_1, F_2, . . ., F_{10}$ which are functions of the size of the spheres, charges on the spheres, the separation between the spheres and the external electrical field. Love \cite{love1975dielectric} derived the analytical expressions for the dipole moment of two identical dielectric spheres with no charge in an external electric field.  O'Meara Jr and Saville \cite{o1981electrical} solved the boundary value problem pertaining to the electrostatic interactions between two conducting spheres in contact in the presence of an external electric field. The studies conducted by Stoy(\cite{stoy1989solution1,stoy1989solution2}) provides a solution for the potential field, both inside and outside of two spheres with varying sizes and permittivities, under an imposed electric field directed along and normal to the line joining the centers.
These studies( \cite{davis1964two},  \cite{love1975dielectric}, \cite{o1981electrical},\cite{stoy1989solution1,stoy1989solution2} ) utilize a series expression to analyze the electrostatic force between two spherical conductors, whether charged or uncharged. However, as the gap between the two spheres approaches zero, the convergence of these series becomes computationally expensive, as it requires numerous terms. In such cases, it becomes necessary to have analytical expressions that can substitute these series to address the convergence issue for close separations. In their study, Friesen and Levine \cite{friesen1992electrostatic} evaluated the electric-field-induced forces between two uncharged spheres by computing the electrostatic energy of the system based on a uniform electric field. They observed that the dipole moment components along and perpendicular to the line joining the centers were the only contributors to the system energy. Subsequently, Lekner \cite{lekner2013forces} extended the formulation of Friesen and Levine \cite{friesen1992electrostatic} to a system of uncharged spherical conductors by generalizing the Landau and Lifshitz's \cite{landauelectrodynamics} theorem for a single uncharged spherical conductor. This work proposes a method for expressing the energy of a system in terms of the polarizability tensor hence when calculating the electrostatic energy and forces in a two-sphere system, only the longitudinal and transverse components of the polarizability tensor are necessary. To determine the electric-field-induced forces at small interparticle distances, Lekner \cite{lekner2011polarizability} derived exact analytical expressions for the longitudinal and transverse polarizabilities in close separations of the two spheres of any size ratio. The forces that act along the line joining the centers of the two spheres are equal and opposite and are related to the derivatives of the polarizabilities with respect to the separation. The force that is perpendicular to the line of centers generates torque on the two-sphere system that is proportionate to the difference between the longitudinal and transverse polarizabilities. This torque always acts to align the line of centers with the direction of the external electric field. Lekner \cite{lekner2012electrostatics} also derived an analytical expression for the electrostatic force between two charged spherical conductors without an electric field. Lekner derived asymptotic expressions for the force coefficients $F_5$, $F_6$ and $F_7$ for arbitrary charged spheres \cite{lekner2012electrostatics} and the force coefficients $F_1$, $F_2$ and $F_8$ for equal size uncharged spheres in the presence of electric field \cite{lekner2013forces}.

When two spheres are positioned in close proximity to each other, they cause disturbances in the velocity fields around each other. This, in turn, results in an increase in hydrodynamic resistance on each individual sphere. It follows that, apart from electrostatic interactions, spheres also interact with each other through hydrodynamic interactions. The influence of hydrodynamic interactions on the relative motion between a pair of spheres in Stokes flow conditions has been thoroughly investigated(S. Kim, S. J. Karrila \cite{kim2013microhydrodynamics}). In close separations, the hydrodynamic resistance considerably diminishes the relative velocity, and this effect depends on the two-sphere relative geometry, which includes the non-dimensional separation and size ratio.  If $\xi$ is used to denote the dimensionless separation, the hydrodynamic resistance present in the lubrication region is $O(1/f(\xi))$. When two rigid spheres interact via continuum hydrodynamics, the function $f(\xi) = \xi$ is utilized as per Batchelor's findings \cite{batchelor1972hydrodynamic}. On the other hand, when two spherical viscous drops interact via continuum hydrodynamics, the function $f(\xi) = \sqrt{\xi}$ is applied in accordance to Davis \cite{davis1989lubrication}.

Thiruvenkadam et al. \cite{thiruvenkadam2023pair} analysed the interaction of two uncharged spherical conductors in the presence of an external electric field for the arbitrary size ratio and discussed the behaviour of continuum hydrodynamic interactions between the spheres as they move relative to one another. They calculated the asymptotic variation of interparticle separation using the near-field asymptotic expressions for the electric-field-induced forces using Lekner's \cite{lekner2011polarizability} polarization formula for arbitrary spheres, exploring the role of hydrodynamic interactions in interparticle motion parallel and perpendicular to the electric field. They got the asymptotic expression for the force coefficients $F_1$,$F_2$ and $F_8$  for arbitrary uncharged spherical conductors in an electric field by using Lekner(\cite{lekner2011polarizability,lekner2012electrostatics}).The force coefficient $F_1$ have $O(\xi^{-1}[\ln\xi]^{-2})$ singularity near contact. Hence, the electrostatic force between two charged/uncharged spherical conductors in an electric field has singularity near contact, and it is a function of multiple infinite series. 

As mentioned earlier, convergence issues occur in small separations. To address the issue at hand, we may employ perturbation techniques, specifically regular and singular perturbation. However, the regular perturbation technique is not viable in this case due to the absence of uniform convergence of the series expressions for different separation values ($\xi \to 0$). Therefore, the method of matched asymptotic expansion has been utilized. This technique involves dividing the original infinite series into two parts, referred to as "inner" and "outer" parts. These two parts are then matched within their common area of validity using similar techniques to those utilized in singular perturbation theory (refer to \cite{Van},\cite{orszag1978advanced} for more information). This method was used by Cox and Brenner  \cite{Cox1967} to derive an asymptotic expression for the hydrodynamic force on a moving sphere in a viscous fluid approaching a plane. Arp and Mason \cite{arp1977particle} analyzed the behaviour of an uncharged particle with shear flow in the presence of an external electric field. They derived the asymptotic expressions for the three force coefficients ($F_1$, $F_2$, and $F_8$) using the method of matched asymptotic expansions developed by Cox and Brenner \cite{Cox1967} based on a simplified form of the force components which was presented by Davis \cite{davis1964two}. 

In prior research (\cite{lekner2012electrostatics},\cite{lekner2013forces},\cite{arp1977particle}), to the best of our knowledge, the asymptotic expressions for the  24 infinite series, which are crucial in the computation of 10 force coefficients, have not been obtained till the date. The objective of this study is to develop an asymptotic expression for, specifically, charged conducting spheres of equal size in the presence of an electric field. Notably, the force components encompass all 10 force coefficients denoted as $F_1$, $F_2$, ..., $F_{10}$. Hence, it becomes imperative to establish the asymptotic expression for these force coefficients. This study involves the utilization of Cox and Brenner's method of matched asymptotic expansions to derive the asymptotic expressions for 24 series, hence compute the 10 force coefficients and finally apply them to compute the electrostatic force under the small separation limit for two charged conducting spheres of equal size in the presence of an electric field. We thoroughly assess the accuracy of the asymptotic expressions by comparing them with the original series and analyzing the deviations using percentage errors.
 In Section(\ref{sec:2}), we give a brief description of the problem. In Section(\ref{sec:3}), we provide an overview of the matched asymptotic expansion method. We derive the asymptotic expression for two series, namely ($T_0(\eta_1)$,$U_0(\eta_1)$) in Section(\ref{sec:4}) and Section(\ref{sec:5}) respectively. In Section(\ref{sec:6}), we discuss the asymptotic expressions about all 24 series, force coefficients, and force components in detail. In Section(\ref{sec:7}), we summarize our results and their consequences.

\section{Problem formulation} \label{sec:2}

\begin{figure}
	\centerline{\includegraphics[width=1\textwidth]{ 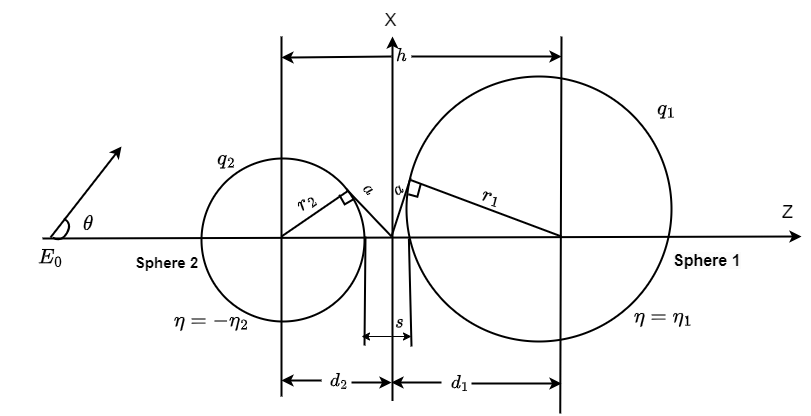}}
\caption{Schematic representation of the problem adopted from Davis(\cite{davis1964two}) }
	\label{fig 1}
\end{figure}

We consider a system of two charged conducting spheres subjected to an external electric field $\boldsymbol{E_0}$ making an angle $\theta$ with $Z$-axis. The spheres 1 and 2 of radii $r_1$ and $r_2$ with charges $q_1$ and $q_2$ respectively, are separated by a distance $s$. The distance between centers of the spheres is denoted by $h=r_1+r_2+s=d_1+d_2$(Figure \ref{fig 1}). In \cite{davis1964two} derived the electric potential between the conducting spheres, denoted as $U$, through the solution of Laplace's equations using bispherical coordinates($\eta,\psi,\phi$). In this context, $\eta=\eta_1 $ and $\eta=-\eta_2$ are denoted as spheres 1 and 2, respectively. Subsequently, this potential was utilized to calculate the electrostatic force ($\boldsymbol{F}$) acting on either sphere as follows:
\begin{equation}
	\boldsymbol{F}\cdot \hat{\boldsymbol{p}}=\frac{\epsilon}{8\pi}\int_{\displaystyle \scriptstyle _{S}}\Big(\frac{\partial U}{\partial n}\Big)^2 \hat{\boldsymbol{p}}\cdot \hat{\boldsymbol{n}}\, dS  
	\label{Force_formula}  
\end{equation}
where \\ $\hat{\boldsymbol{p}}$ - arbitrary unit vector \\  $\hat {\boldsymbol{n}}$ - the unit vector normal to the surface element $dS$ \\ $\epsilon$ - the permittivity of the medium \\ $\displaystyle \frac{\partial U}{\partial n}$ - the derivative of the  electric potential along the normal ($\hat {\boldsymbol{n}}$) direction. 
\vspace{0.5 cm}

Let $\boldsymbol{F(1)}$ and $\boldsymbol{F(2)}$ are the forces on the spheres 1 and 2 respectively. Computing the force acting on one of the spheres,
\begin{equation}
	\boldsymbol{F(1)}=\boldsymbol{E_0}(q_1+q_2)-\boldsymbol{F(2)}  
\end{equation}

Integrating the square of the normal derivative of the potential(Equation  \ref{Force_formula}), we get the explicit form of forces on sphere 2 in $Z$ and $X$ directions  as follows (\cite{davis1964two}) :
\begin{equation}\label{eq:davisforce_Z}
	\begin{split}
		F_Z(2)=\bigg[\epsilon 
		{r_2}^2{E_0}^2(F_1 \cos^2\theta+F_2 \sin^2\theta)+E_0\cos\theta(F_3  q_1+F_4 q_2)\\
		+\frac{1}{\epsilon {r_2}^2}(F_5 {q_1}^2+F_6 q_1q_2+F_7 {q_2}^2) \bigg]+E_0 q_2 \cos\theta  
	\end{split} 
\end{equation} 
and
\begin{equation}\label{eq:davisforce_X}
	F_X(2)=\bigg[\epsilon {r_2}^2{E_0}^2F_8\sin2\theta+E_0\sin\theta(F_9q_1+F_{10}q_2)\bigg]+E_0q_2\sin\theta  
\end{equation}
where $F_Z(2)$ and $F_X(2)$ are force components on sphere 2 in $Z$ and $X$ directions respectively. \par

The nondimensional force components obtained by multiplying $\mathlarger{\frac{\epsilon {r_2}^2}{{q_2}^2}}$ on both sides of the equations (\ref{eq:davisforce_Z}) and (\ref{eq:davisforce_X}) is given by:
\begin{equation}
	\begin{split}
		F_Z(2)= \bigg[ {\beta}^2(F_1\cos^2\theta+F_2\sin^2\theta)+\beta\cos\theta(F_3\alpha+F_4)\\+(F_5{\alpha}^2+F_6\alpha+F_7) \bigg]+\beta\cos\theta 
	\end{split}
	\label{nondimensionalized_davisforce_Z}
\end{equation}
and
\begin{equation}
	F_X(2)=\bigg[ {\beta}^2F_8\sin2\theta+\beta\sin\theta(F_9\alpha+F_{10}) \bigg]+\beta\sin\theta
	\label{nondimensionalized_davisforce_X}
\end{equation}
Note that $F_Z(2)$ and $F_X(2)$  have been retained after nondimensionalization. Here, $\mathlarger{\alpha = \frac{q_1}{q_2}}$ is the ratio of the charges and $\mathlarger{\beta=\frac{E_0\epsilon{r_2}^2}{q_2}}$ is the ratio of strength of the impressed electric field to that of the electric field induced by charge $q_2$.

\par The coefficients $F_1, F_2, ..., F_{10}$ (hereafter called as the force coefficients) are the complicated series expressions of the size ratio ($\kappa=r_2/r_1$) and separation($\xi=2s/(r_1+r_2)$). These force coefficients depend on the following infinite series:
\begin{eqnarray}
	\displaystyle T_m(p)&=&\sum_{n=0}^{\infty} \dfrac{ (2n+1)^m e^{(2n+1)p}}{(e^{(2n+1)(\eta_1+\eta_2)}-1)^2}
	\label{E1}\\
	U_m(p)&=&\sum_{n=0}^{\infty} \dfrac{(2n+1)^m e^{(2n+1)p}}{(e^{(2n+1)(\eta_1+\eta_2)}-1) (e^{(2n+3)(\eta_1+\eta_2)}-1)}
	\label{E2}
\end{eqnarray}
where, $m$=0, 1, 2, 3 and $p$=$\eta_2$, $\eta_1+\eta_2$,  $2\eta_1+\eta_2$, which are 24 in number.

\section{Outline of the method of matched asymptotic expansion for infinite series}\label{sec:3}
In this section, we discuss the outline of the matched asymptotic expansion method for infinite series(\cite{Cox1967}). The following are important points to note before using this method :
\begin{enumerate}
	\item For small values of $\eta_1$, the expansion of these 24 series becomes highly singular. 
	\item For sufficiently large values of the summation index $n$, the product $n\eta_1$ in the series will not remain small, no matter how small $\eta_1$ is.
	\item Thus, the usual approach for expanding infinite series for small values of $\eta_1$ involves truncating the series at a sufficiently high value of $n$. This approach becomes impractical when $n\eta_1 = O(1)$.
\end{enumerate}

In light of this, \cite{Cox1967} employed a valuable technique to effectively solve infinite series of this nature. The technique involved partitioning the series into two components, which were referred to as the ``inner" and ``outer" sums. The ``inner" sum is the sum of the first $N+1$  terms, and the ``outer" sum is the sum of the remaining infinite number of terms. Then, the asymptotic expression for the inner and outer sums are individually calculated and added together to get the asymptotic expression for the entire infinite series. The forthcoming sections will provide a detailed discussion of the selection process for $N$ and obtaining asymptotic expressions for ``inner" and ``outer" sums. In the next sections (\ref{sec:4}, \ref{sec:5}, \ref{sec:6}), we derive the asymptotic expression of these series as $\eta_1\to0$ using this method. In these cases, $\kappa=\frac{r_2}{r_1}\approx\frac{\eta_1}{\eta_2}$ and we derive the asymptotic expressions for 24 series in the small separation limit for two equally sized charged conducting spheres ($\kappa=1$) in the presence of an electric field. Additionally, $\eta_1$ depends on the separation ($\xi$) between the two charged conducting spheres, which implies that $\eta_1$ and $\eta_2$ are approximately equal to $\sqrt{\xi}$(i.e., $\eta_1\approx\eta_2\approx\sqrt{\xi}$). Hence, our $p$ values change as follows: $p=\eta_1$, $2\eta_1$, $3\eta_1$. We now derive the asymptotic expressions of the series $T_0(\eta_1)$ and $U_0(\eta_1)$ in sections (\ref{sec:4} and \ref{sec:5}) respectively. 
\section{Asymptotic expression for the first T-series} \label{sec:4}
The series $T_0(\eta_1)$ can be obtained by substituting $m=0$ and $p=\eta_1$ in Equation (\ref{E1}),and it is given by,
\begin{eqnarray}
	T_0(\eta_1)&=&\sum_{n=0}^{\infty} \dfrac{ e^{(2n+1)\eta_1}}{(e^{(4n+2)\eta_1}-1)^2}
	\label{1_E1}
\end{eqnarray}
As outlined in the previous section, the series $T_0(\eta_1)$ is decomposed to an ``inner expansion'' ($f_{i}$) and an ``outer expansion'' ($f_{o}$) as follows:  
\begin{eqnarray}
	T_0(\eta_1)&=& f_{i}(\eta_1,N)+f_{o}(\eta_1,N)
	\label{1_E2}
\end{eqnarray}

where
\begin{eqnarray}
	f_{i}(\eta_1,N)&=&\sum_{n=0}^{N} \dfrac{ e^{(2n+1)\eta_1}}{(e^{(4n+2)\eta_1}-1)^2}
	\label{1_E3}
	\\
	\text{and}\hspace{0.2cm}	f_{o}(\eta_1,N)&=&\sum_{n=N+1}^{\infty} \dfrac{ e^{(2n+1)\eta_1}}{(e^{(4n+2)\eta_1}-1)^2} 
	\label{1_E4}
\end{eqnarray}
Here, the inner and outer expansions are valid, where $n=O(\eta_1^{-1})$ and $n=O(\eta_1^{0})$ respectively and $N$ is an arbitrary natural number lying in the region of validity of both the inner and outer expansions. Thus, $\eta_1\to 0$ implies that $N \to \infty$ and $N$ is a function of $\eta_1$. For example, one can choose
\begin{equation}
	N=a \eta_1^{-(1-b)}
	\label{1_E5}
\end{equation}
where $a$ and $b$ are positive constants and independent of $\eta_1$. A constant $b$ is chosen such that $b\in(0,1)$. It is evident that $N$ is within the validity region of the inner and outer expansions.
\par For the following analysis, we introduce the parameter $X$ that will replace $N$ in such way that $X$ is a small($<<1$) positive parameter defined such that $N$ is the integer immediately less than $\displaystyle \frac{X}{\eta_1}$, i.e.
\begin{equation}
	(N+1)\eta_1>X>=N\eta_1
	\label{1_E6}
\end{equation}
Note that $X\to0$ as $\eta_1\to0$. Then, one can write Equation (\ref{1_E2}) in terms of $X$ as follows,
\begin{eqnarray}
	T_0(\eta_1)&=& f_{i}(\eta_1,X)+f_{0}(\eta_1,X)
	\label{1_E7}
\end{eqnarray}
where $f_{i}$ and $f_{o}$ are Equation (\ref{1_E3}) and  Equation (\ref{1_E4}) respectively. Combining Equations (\ref{1_E5} and \ref{1_E6}), then we get
\begin{equation}
	X\approx a \eta_1^{b} 
	\label{1_E8}
\end{equation}
In view of the definition of $X$, which is arbitrary in nature, the final expression of  $T_0(\eta_1)$ cannot include $X$. However, it may appear in the individual inner ( $f_i$) and outer ($f_o$) expansions of $T_0(\eta_1)$. 

\subsection{Inner expansion}\label{subsec_1}
In this subsection, we derive the asymptotic expression for the inner expansion of $T_0(\eta_1)$.
The inner expansion can be written as follows,
\begin{eqnarray}
	f_{i}(\eta_1,N)&\approx&\frac{1}{4}(I_1 {\eta_1}^{-2}-I_2 {\eta_1}^{-1}+\frac{1}{6}I_3)
	\label{1_E9}
\end{eqnarray}
where
\begin{eqnarray}
	I_1&=&\sum_{n=0}^{N}\frac{1}{(2n+1)^2},
	\label{1_E10}\\
	I_2&=&\sum_{n=0}^{N}\frac{1}{2n+1}
	\label{1_E11}\\
	\text{and}\hspace{0.2cm}	I_3&=&\sum_{n=0}^{N}1
	\label{1_E12}
\end{eqnarray}
Hence, as $N\to\infty$,
\begin{eqnarray}
	I_1&\sim&\frac{\pi^2}{8}-\frac{1}{4N}+\frac{1}{4N^2},
	\label{1_E13}\\
	I_2&\sim&\frac{1}{2}\big(\gamma+\log{4N}+\frac{1}{N}\big)
	\label{1_E14}\\
	\text{and}\hspace{0.2cm}	I_3&=&N+1
	\label{1_E15}
\end{eqnarray}
Here $\gamma$ is Euler's constant, and it is approximately equal to 0.57721.  \\
Substituting Equations (\ref{1_E13}-\ref{1_E15}) into Equation (\ref{1_E9}), the asymptotic expression for $f_{i}$ can be obtained in terms of $N$ and $\eta_1$  as follows:
\begin{eqnarray}
	f_{i}(\eta_1,N)&\sim&\bigg(\frac{\pi^2}{32}-\frac{1}{16N}+\frac{1}{16N^2}\bigg){\eta_1}^{-2}\nonumber\\ 
	&&   -\bigg(\frac{ 1}{8}\big(\gamma+\log{4N}+\frac{1}{N}\big)\bigg){\eta_1}^{-1}+\frac{N+1}{24}
	\label{1_E16}
\end{eqnarray}

After introducing intermediate vanish $\displaystyle N=\frac{X}{\eta_1}$ and simplifying Equation (\ref{1_E16}), we get,
\begin{eqnarray}
	f_{i}(\eta_1,X)&\sim&\frac{\pi^2}{32{\eta_1}^2}-\frac{\gamma+\log{\frac{4}{\eta_1}}}{8\eta_1}+\frac{1}{24}\nonumber\\
	&& +\frac{1}{16X^2}-\frac{1}{8X}-\frac{1}{16X \eta_1}-\frac{\log{X}}{8\eta_1}+\frac{X\eta_1^{-1}}{24}
	\label{1_E17}
\end{eqnarray}
Equation (\ref{1_E17}) is the asymptotic inner expansion which involves  $\eta_1$ and $X$ (which should vanish after summing with the outer expansion). 
In the next subsection (\ref{subsec:4.2}), we derive the outer expansion for $T_0(\eta_1)$.
\subsection{Outer expansion}\label{subsec:4.2}
To obtain the outer expansion $f_{o}$, let $m=n\eta_1$ in Equation (\ref{1_E4}) and consider the case where $\eta_1\to0$ for a fixed value of $m$. After substituting the $n$ value in Equation (\ref{1_E4}), we get outer expansion in terms of $m$ and $\eta_1$ as follows,  
\begin{eqnarray}
	f_{o}(\eta_1,N)&=&\sum_{\substack{m=n\eta_1 \\ n=N+1}}^{\infty} \dfrac{ e^{2m+\eta_1}}{(e^{4m+2\eta_1}-1)^2}  
	\label{1_E18}
\end{eqnarray}
For fixed $m$,  the function expansion for $\eta_1\to 0$ is as follows:
\begin{eqnarray}
	f_{o}(\eta_1,N)&=& \sum_{\substack{m=n\eta_1 \\ n=N+1}}^{\infty}\Bigg(\frac{e^{2m}}{(-1+e^{4m})^2}-\frac{e^{2m}(1+3e^{4m})}{(-1+e^{4m})^3}\eta_1 \Bigg)
	\label{1_E19}
\end{eqnarray}
To apply the Euler-Maclaurin summation formula, we express Equation (\ref{1_E19}) as follows:
\begin{eqnarray}
	f_{o}(\eta_1,N)&=& \frac{1}{\eta_1}\sum_{\substack{m=n\eta_1 \\ n=N+1}}^{\infty}f_{1}(m)\Delta m-\sum_{\substack{m=n\eta_1 \\ n=N+1}}^{\infty}f_{2}(m)\Delta m
	\label{1_E20}
\end{eqnarray}

where, 
\begin{eqnarray}
	f_{1}(m)&=&\frac{e^{2m}}{(-1+e^{4m})^2} 
	\label{1_E21}\\
	\text{and}\hspace{0.2cm}	f_{2}(m)&=&\frac{e^{2m}(1+3e^{4m})}{(-1+e^{4m})^3}
	\label{1_E22}
\end{eqnarray}
Define $\Delta m =m_{n+1}-m_n $. This makes $\Delta m=(n+1)\eta_1-n\eta_1=\eta_1$.
Hence, the outer expansion may be written in the form 
\begin{eqnarray}
	f_{o}(\eta_1,X)&=&\frac{1}{\eta_1}\sum_{m=X}^{\infty}f_{1}(m)\Delta m -\sum_{m=X}^{\infty}f_{2}(m)\Delta m
	\label{1_E23}
\end{eqnarray}
\begin{figure}
	\includegraphics[width=1\textwidth]{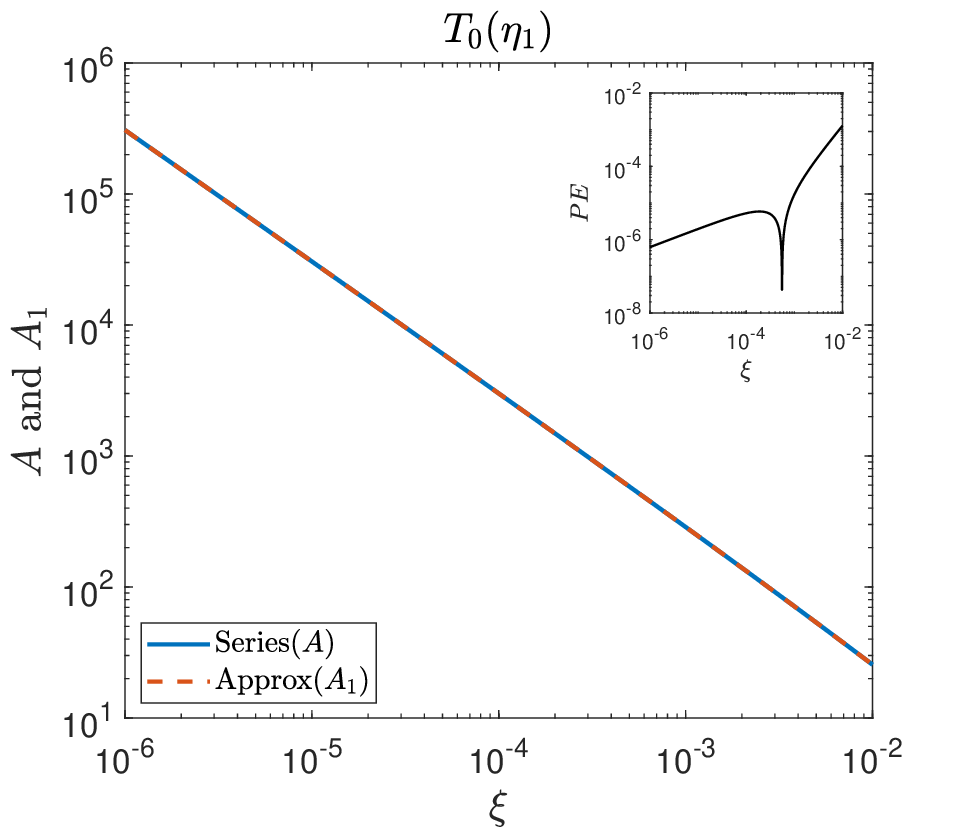}
	\caption{$T_{0}$-series and approximation for $\eta_1$}
	\label{T_0_series_1}
\end{figure}
where we have noted from the definition of X that 
$m = X$ at the lower summation limit. Now, according to the Euler-Maclaurin summation formula 
\begin{eqnarray}
	\sum_{m=X}^{\infty}f(m)\Delta m &=&\int_{X}^{\infty}f(m)dm+\frac{\Delta m }{2}(f(\infty)-f(X))
	\label{1_E24}
\end{eqnarray}
This expression is asymptotically valid as $\Delta m\to 0$. Now, from Equations (\ref{1_E21} and \ref{1_E22}) we get $f_1(\infty)=0$ and $f_2(\infty)=0$. Furthermore, as $X\to 0$, 
\begin{eqnarray}
	f_{1}(X)&\sim&\dfrac{1}{16 X^2}-\dfrac{1}{8 X}+\dfrac{1}{24},
	\label{1_E25}
	\\
	\text{and}\hspace{0.2cm}f_{2}(X)&\sim&\dfrac{1}{16 X^3}-\dfrac{1}{16 X^2}
	\label{1_E26}
\end{eqnarray}

Since $\Delta m=\eta_1$, from Equations (\ref{1_E23}-\ref{1_E26}) as $\eta_1\to0$, we have
\begin{eqnarray} 
	f_{o}(\eta_1,X) &=& \frac{1}{\eta_1}\int_{X}^{\infty}f_{1}(m)dm-\frac{1}{32X^2}+\frac{1}{16X}-\frac{1}{48}\nonumber\\
	&& -\int_{X}^{\infty}f_{2}(m)dm+\textit{o(1)}
	\label{1_E27}
\end{eqnarray}
The integral appearing above is convergent for the large value of $m$. As $\Delta m \to 0$ Equation(\ref{1_E27}), can be written  as , 
\begin{eqnarray}
	f_{o}(\eta_1,X)&\sim&-\frac{1}{16X^2}+\frac{1}{8X}+\frac{1}{16X \eta_1}+\frac{\log{X}}{8\eta_1}-\frac{X\eta_1^{-1}}{24}\nonumber\\
	&&+\frac{\eta_1^{-1}}{24}-\frac{1}{48}+\frac{K_{11}}{\eta_1}-C_{13},
	\label{1_E28}
\end{eqnarray}
\\ where
\begin{eqnarray*}
	K_{11}&=&C_{11}+C_{12}=-0.0416667,  \\
	C_{11}&=&\int_{1}^{\infty} \Bigg({f_1(m)}-\frac{1}{16m^2}\Bigg)dm = -0.0620776, \nonumber\\
	C_{12}&=&\int_{0}^{1} \Bigg({f_1(m)}-\frac{1}{16m^2}+\frac{1}{8m}-\frac{1}{24}\Bigg)dm =0.0204109,\nonumber\\
	\text{and}\hspace{0.2cm}	C_{13}&=&\int_{0}^{\infty} \Bigg({f_2(m)}-\frac{1}{16m^3}+\frac{1}{16m^2}\Bigg)dm =0.0208333
\end{eqnarray*}

Here, Equation(\ref{1_E28}) is the asymptotic outer expansion for $T_0(\eta_1)$, which also involves  $\eta_1$ and $X$.

Adding Equation (\ref{1_E17}) and Equation (\ref{1_E28}), we obtain the asymptotic expression for $T_0(\eta_1)$ in the the small separation region as follows:
\begin{eqnarray}
	T_0(\eta_1)&\sim&\frac{\pi^2}{32{\eta_1}^2}-\frac{\gamma+\log{\frac{4}{\eta_1}}}{8\eta_1}+\frac{1}{48}
	+\frac{\eta_1^{-1}}{24}+\frac{K_{11}}{\eta_1}-C_{13}
	\label{1_E29}
\end{eqnarray}

At this stage, it can be observed that the final asymptotic expression is independent of $X$.

As discussed earlier, the calculation of force coefficients requires the computation of asymptotic expressions of 12 T-series and 12 U-series. In section (\ref{sec:4}), we have shown the computation of the asymptotic expression of $T_0(\eta_1)$. Similarly, in section (\ref{sec:5}), an asymptotic expression for $U_0(\eta_1)$ has been derived.

\section{Asymptotic expression for the first U-series }\label{sec:5}
The series $U_0(\eta_1)$ can be obtained by substituting $m=0$ and $p=\eta_1$ in Equation (\ref{E2}),and it is given by,
\begin{eqnarray}
	U_0(\eta_1)&=&\sum_{n=0}^{\infty} \dfrac{ e^{(2n+1)\eta_1}}{(e^{(4n+2)\eta_1}-1) (e^{(4n+6)\eta_1}-1)}
	\label{13_E1}
\end{eqnarray}
In this case, $U_0(\eta_1)$ is decomposed into an ``inner expansion'' ($g_{i}$) and an ``outer expansion'' ($g_{o}$) as follows:
\begin{eqnarray}
	U_0(\eta_1)&=& g_{i}(\eta_1,N)+g_{o}(\eta_1,N),
	\label{13_E2}
\end{eqnarray}
where
\begin{eqnarray}
	g_{i}(\eta_1,N)&=&\sum_{n=0}^{N} \dfrac{ e^{(2n+1)\eta_1}}{(e^{(4n+2)\eta_1}-1) (e^{(4n+6)\eta_1}-1)},
	\label{13_E3}
	\\
	\text{and}\hspace{0.2cm}	g_{o}(\eta_1,N)&=&\sum_{n=N+1}^{\infty} \dfrac{ e^{(2n+1)\eta_1}}{(e^{(4n+2)\eta_1}-1) (e^{(4n+6)\eta_1}-1)} 
	\label{13_E4}
\end{eqnarray}

Following similar steps as shown in  Section (\ref{sec:4}), we find the inner ($g_i$) and outer ($g_o$) expansions in the next two subsections.

\subsection{Inner expansion}\label{subsec:5.1}
In this subsection, we derive the inner expansion $g_{i}$ for $U_0(\eta_1)$.

\begin{eqnarray}
	g_{i}(\eta_1,N)&\approx&\frac{1}{4}\Bigg(I_1 ({\eta_1}^{-2}-\frac{2}{3})+I_2 (1-{\eta_1}^{-1})+\frac{1}{6}I_3\Bigg),
	\label{13_E9}
\end{eqnarray}
\\
where
\begin{eqnarray}
	I_1&=&\sum_{n=0}^{N}\frac{1}{(2n+1)(2n+3)},
	\label{13_E10}\\
	I_2&=&\sum_{n=0}^{N}\frac{1}{2n+1}
	\label{13_E11}\\
	\text{and}\hspace{0.2cm}	I_3&=&\sum_{n=0}^{N}1
	\label{13_E12}
\end{eqnarray}
As $N\to\infty$, $I_1$, $I_2$ and $I_3$ is given by,
\begin{eqnarray}
	I_1&\sim&\frac{1}{2}-\frac{1}{4N}+\frac{3}{8N^2},
	\label{13_E13}\\
	I_2&\sim&\frac{1}{2}\big(\gamma+\log{4N}+\frac{1}{N}\big)
	\label{13_E14}\\
	\text{and}\hspace{0.2cm}	I_3&=&N+1
	\label{13_E15}
\end{eqnarray}
Substituting Equations (\ref{13_E13}-\ref{13_E15}) in Equation (\ref{13_E9}), the asymptotic expression for $g_{i}$ can be obtained in terms of $N$ and $\eta_1$ as follows:
\begin{eqnarray}
	g_{i}(\eta_1,N)&\sim&\frac{1}{4}\Bigg(\bigg(\frac{1}{2}-\frac{1}{4N}+\frac{3}{8N^2}\bigg)\bigg({\eta_1}^{-2}-\frac{2}{3}\bigg) \nonumber\\&&+\bigg(\frac{\big(\gamma+\log{4N}+\frac{1}{N}\big)}{2}(1-{\eta_1})+\frac{N+1}{6})\bigg) \Bigg) 
	\label{13_E16}
\end{eqnarray}

After introducing intermediate vanish $N=\displaystyle\frac{X}{\eta_1}$ and simplifying Equation (\ref{13_E16}), we get,
\begin{eqnarray}
	g_{i}(\eta_1,X)&\sim&\frac{1}{8{\eta_1}^2}+\frac{(1-\eta_1^{-1})(\gamma+\log{\frac{4}{\eta_1}})}{8}-\frac{1}{24}\nonumber\\
	&& +\frac{3}{32X^2}-\frac{1}{8X}-\frac{1}{16X \eta_1}+\frac{(1-\eta_1^{-1})\log{X}}{8}+\frac{X\eta_1^{-1}}{24}
	\label{13_E17}
\end{eqnarray}
Equation (\ref{13_E17}) can be used to calculate the inner expansion $g_{i}$ for $U_0(\eta_1)$. In the next subsection (\ref{subsec:5.2}), we derive the outer expansion for $U_0(\eta_1)$.
\subsection{Outer expansion}\label{subsec:5.2}
To obtain the outer expansion, let $m=n\eta_1$ in Equation (\ref{13_E4}) and consider the case where $\eta_1\to0$ for a fixed value of $m$. After substituting the $n$ value in Equation (\ref{13_E4}), we get outer expansion in terms of $m$ and $\eta_1$ as follows,  
\begin{eqnarray}
	g_{o}(\eta_1,N)&=&\sum_{\substack{m=n\eta_1 \\ n=N+1}}^{\infty} \dfrac{ e^{(2m+\eta_1)}}{(e^{(4m+2\eta_1)}-1) (e^{(4m+6\eta_1)}-1)}  
	\label{13_E18}
\end{eqnarray}
For fixed $m$,  the function expansion for $\eta_1\to 0$ is as follows:
\begin{eqnarray}
	g_{o}(\eta_1,N)&=&  \sum_{\substack{m=n\eta_1 \\ n=N+1}}^{\infty}\Bigg(\frac{e^{2m}}{(-1+e^{4m})^2}-\frac{e^{2m}(1+7e^{4m})}{(-1+e^{4m})^3}\eta_1 \Bigg)
	\label{13_E19}
\end{eqnarray}
To apply the Euler-Maclaurin summation formula, we express Equation (\ref{13_E19}) as follows:
\begin{eqnarray}
	g_{o}(\eta_1,N)&=& \frac{1}{\eta_1}\sum_{\substack{m=n\eta_1 \\ n=N+1}}^{\infty}g_{1}(m)\Delta m-\sum_{\substack{m=n\eta_1 \\ n=N+1}}^{\infty}g_{2}(m)\Delta m
	\label{13_E20}
\end{eqnarray}
where,
\begin{eqnarray}
	g_{1}(m)&=&\frac{e^{2m}}{(-1+e^{4m})^2},
	\label{13_E21}\\
	\text{and}\hspace{0.2cm}	g_{2}(m)&=&\frac{e^{2m}(1+7e^{4m})}{(-1+e^{4m})^3}
	\label{13_E22}
\end{eqnarray}
Define $\Delta m =m_{n+1}-m_n $. This makes 
$\Delta m=(n+1)\eta_1-n\eta_1=\eta_1$
Hence, the outer force may be written in the form 
\begin{eqnarray}
	g_{o}(\eta_1,X)&=&\frac{1}{\eta_1}\sum_{m=X}^{\infty}g_{1}(m)\Delta m -\sum_{m=X}^{\infty}g_{2}(m)\Delta m
	\label{13_E23}
\end{eqnarray}
where we have noted from the definition of X that 
$m = X$ at the lower summation limit. 
This expression is asymptotically valid as $\Delta m\to 0$. Now, from Equations (\ref{13_E21} and \ref{13_E22}) we get $g_1(\infty)=0$ and $g_2(\infty)=0$. Furthermore, as $X\to 0$, 
\begin{eqnarray}
	g_{1}(X)&\sim&\dfrac{1}{16 X^2}-\dfrac{1}{8 X}+\dfrac{1}{24},
	\label{13_E25}
	\\
	\text{and}\hspace{0.2cm}	g_{2}(X)&\sim&\dfrac{1}{8 X^3}-\dfrac{1}{16 X^2}-\dfrac{1}{8 X}
	\label{13_E26}
\end{eqnarray}

Since $\Delta m=\eta_1$, From Equations (\ref{1_E24},\ref{13_E23},\ref{13_E25} and \ref{13_E26}) as $\eta_1\to0$, we have
\begin{eqnarray} 
	g_{o}(\eta_1,X) &=& \frac{1}{\eta_1}\int_{X}^{\infty}g_{1}(m)dm-\frac{1}{32X^2}+\frac{1}{16X}-\frac{1}{48}\nonumber\\
	&& -\int_{X}^{\infty}g_{2}(m)dm+\textit{o(1)}
	\label{13_E27}
\end{eqnarray}
\begin{figure}
	\includegraphics[width=1\textwidth]{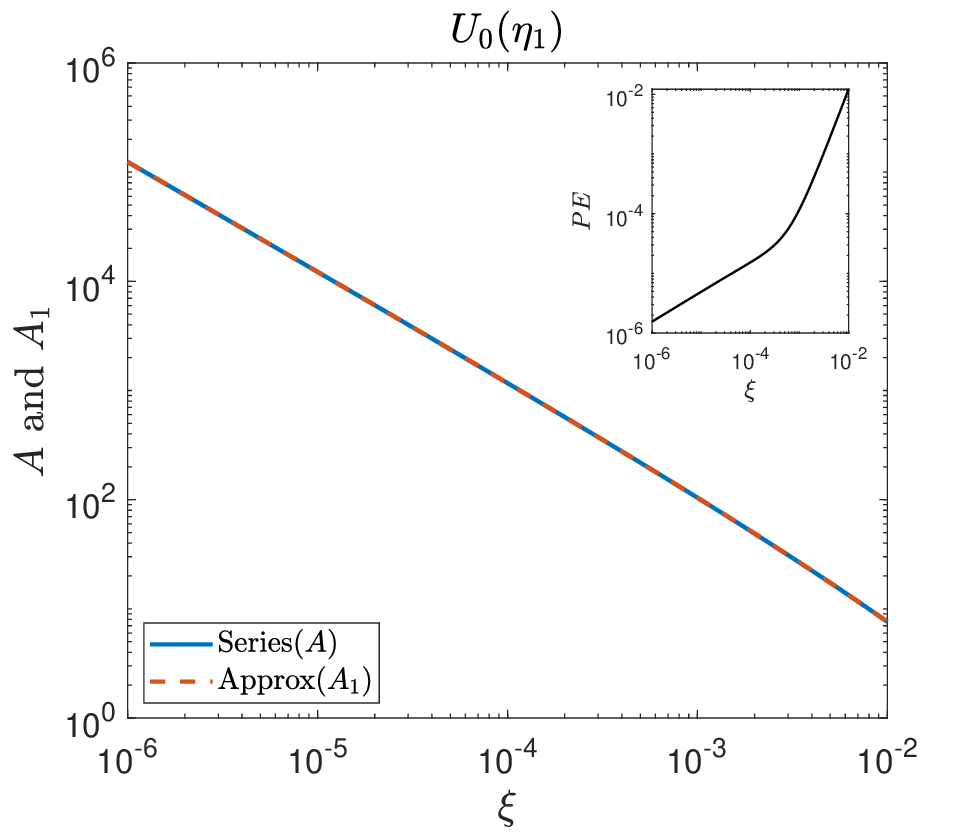}
	\caption{$U_0$-series and approximation for  $\eta_1$}
	\label{U_0_series_1}
\end{figure}
The integral appearing above is convergent for the large value of m. As, $\Delta m \to 0$, Equation(\ref{13_E27}) can be written as follows:
\begin{eqnarray}
	g_{o}(\eta_1,X)&\sim&-\frac{3}{32X^2}+\frac{1}{8X}+\frac{1}{16X \eta_1}-\frac{(1-\eta_1^{-1})\log{X}}{8}-\frac{X\eta_1^{-1}}{24}\nonumber\\
	&&+\frac{\eta_1^{-1}}{24}-\frac{1}{48}+\frac{K_{131}}{\eta_1}-K_{132}
	\label{13_E28}
\end{eqnarray}
\\ where,
\begin{eqnarray*}
	K_{131} &=& C_{131}+C_{132}=-0.0416667 , \nonumber \\ 
	K_{132} &=& C_{133}+C_{134}=0.0416667,
	\nonumber \\
	C_{131} &=& \int_{1}^{\infty} \Bigg({g_1(m)}-\frac{1}{16m^2}\Bigg)dm = -0.0620776,\nonumber \\
	C_{132} &=&\int_{0}^{1} \Bigg({g_1(m)}-\frac{1}{16m^2}+\frac{1}{8m}-\frac{1}{24}\Bigg)dm =0.0204109,\nonumber \\
	C_{133} &=&\int_{1}^{\infty} \Bigg({g_2(m)}-\frac{1}{8m^3}+\frac{1}{16m^2}\Bigg)dm= 0.0029945,\nonumber \\
	\text{and}\hspace{0.2cm}	C_{134} &=&\int_{0}^{1} \Bigg({g_2(m)}-\frac{1}{8m^3}+\frac{1}{16m^2}+\frac{1}{8m}\Bigg)dm=0.0386722 \nonumber \\
\end{eqnarray*}
Here, Equation (\ref{13_E28}) is the asymptotic outer expansion for $U_0(\eta_1)$ , which also involves  $\eta_1$ and $X$.

Adding Equation (\ref{13_E17}) and Equation (\ref{13_E28}), the asymptotic expression for $U_0(\eta_1)$ in the the small separation region is given by,
\begin{eqnarray}
	U_0(\eta_1)&\sim&\frac{1}{8{\eta_1}^2}+\frac{(1-\eta_1^{-1})(\gamma+\log{\frac{4}{\eta_1}})}{8}-\frac{1}{16}+\frac{\eta_1^{-1}}{24}+\frac{K_{131}}{\eta_1}-K_{132}
	\label{13_E29}
\end{eqnarray}
In sections (\ref{sec:4}) and (\ref{sec:5}) we have derived the asymptotic expression for $T_0(\eta_1)$ and $U_0(\eta_1)$. The asymptotic expressions for the rest of the $22$ infinite series can be obtained in a similar manner.
In the section (\ref{sec:6}), we provide the asymptotic expressions for all 24 infinite series, and discuss their preliminary characteristics.

\section{Results and discussion}\label{sec:6}
In this section, we discuss asymptotic expressions for 24 infinite series that aid in finding the $10$ force coefficients, which in turn contribute to obtaining the force components of charged spherical conductors with an electric field.\\ 

\subsection{Asymptotic expressions for first T-series and first U-series}
\label{subsec:6.1}
The plots of the series $T_0(\eta_1)$ and $U_0(\eta_1)$ as a function of separation $\xi$ maintaining size ratio $\kappa = 1$ are as shown in the figures (\ref{T_0_series_1}) and (\ref{U_0_series_1}) respectively. In these figures, the actual series is represented by $A$, and its asymptotic expression is represented by $A_1$ for both $T_0(\eta_1)$ and $U_0(\eta_1)$ series. The Inset/ subplot in the figures shows the percentage errors versus separation ($\xi$) of the spheres. The actual series are represented by continuous blue lines and their asymptotic approximations by dashed red lines. The percentage error plot is represented by continuous black lines. Since we consider the gap between the spheres to be very small, we consider our separation range from $10^{-6}$ to $10^{-2}$ on a $\log-\log$ scale. The following characteristics can be observed:
\begin{itemize}
	\item Both the curves show a decreasing trend in the values of $T_0(\eta_1)$ and $U_0(\eta_1)$ as a function of separation.
	\item Both the series and its asymptotic expressions match in the range of $10^{-6}$ to $10^{-2}$.
	\item The percentage error between the full series computations and the asymptotic expressions increases as the separation increases between the spheres.
	\item This infers that the expression for $T_0(\eta_1)$ and $U_0(\eta_1)$ can be used for very small separations.
	\item $T_0(\eta_1)$ has the highest percentage error of $1.28\times10^{-3}$ at $10^{-2}$
	\item $U_0(\eta_1)$ has the highest percentage error of $1.22\times10^{-2}$ at $10^{-2}$
	
\end{itemize}

\subsection{Asymptotic expressions for 24 infinite series:}
\label{subsec:6.2}
\begin{figure}
	\includegraphics[width=1\textwidth]{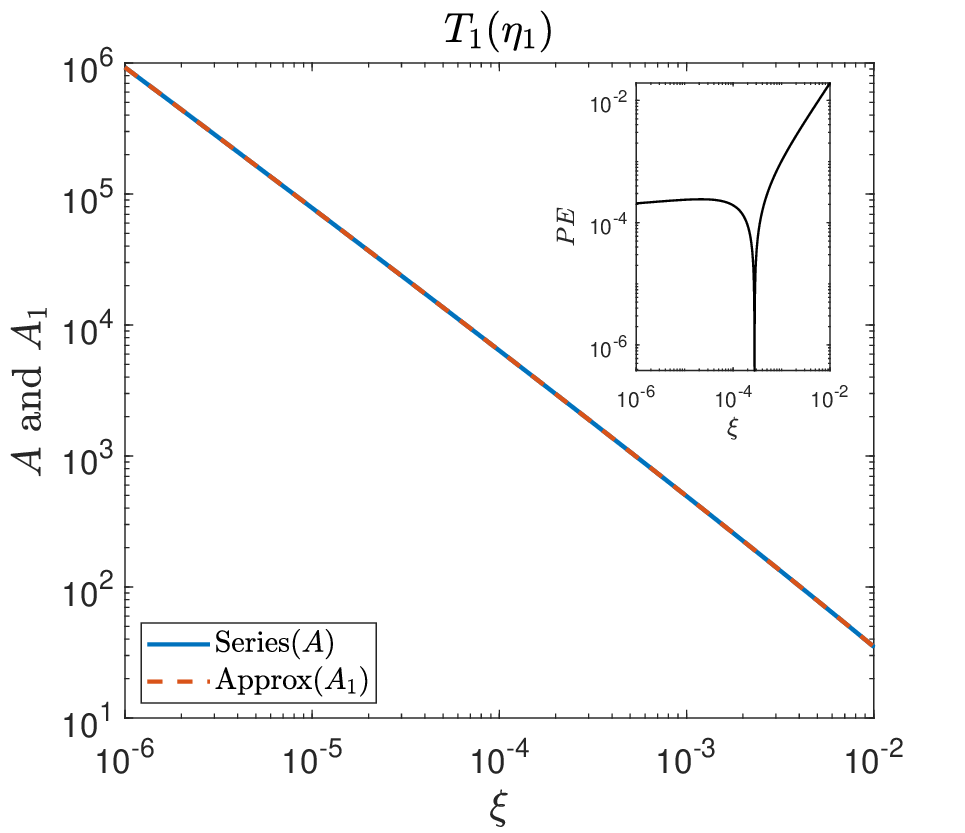}
	\caption{$T_{1}$-series and approximation for $\eta_1$}
	\label{T_1_series_1}
\end{figure}

\begin{figure}
\includegraphics[width=1\textwidth]{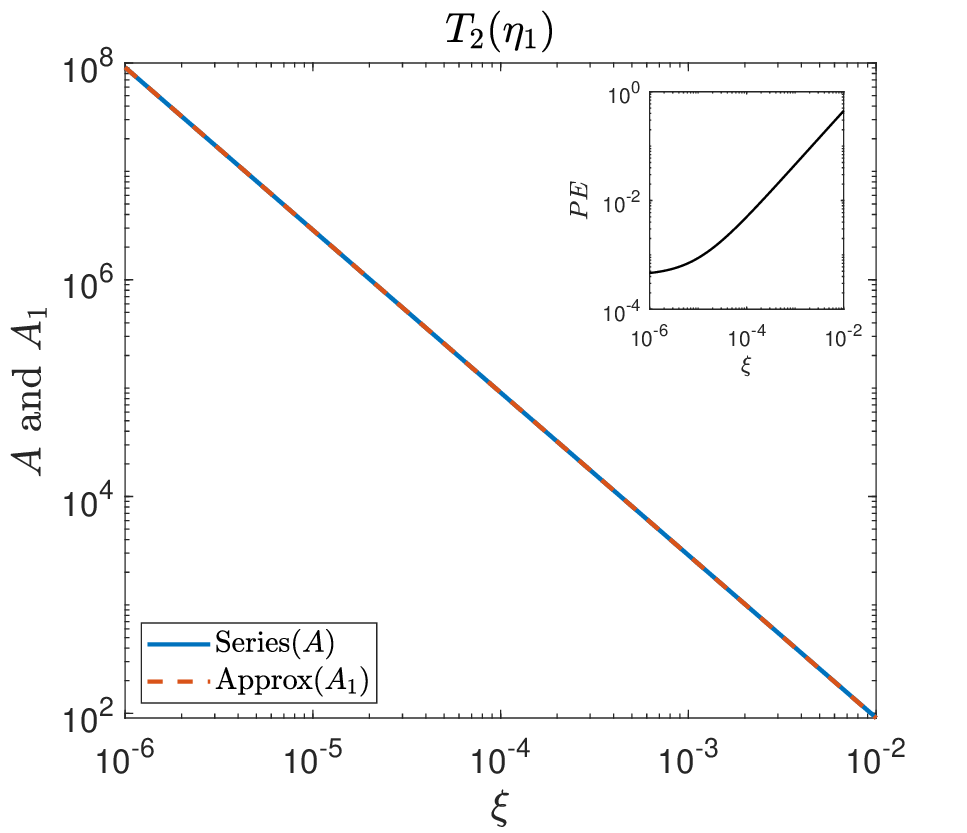}
\caption{$T_2$-series and approximation for $\eta_1$}
\label{T_2_series_1}
\end{figure}

\begin{figure}
		\includegraphics[width=1\textwidth]{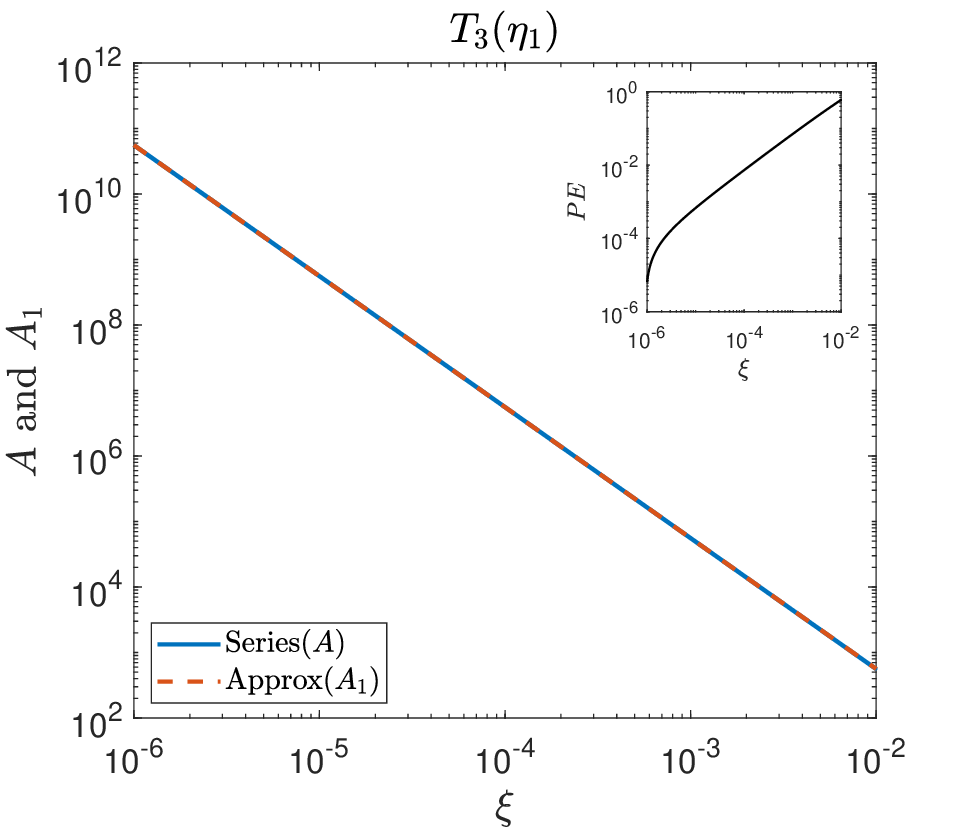}
		\caption{$T_3$-series and approximation for $\eta_1$}
		\label{T_3_series_1}
 \end{figure}

\begin{figure}
	\includegraphics[width=1\textwidth]{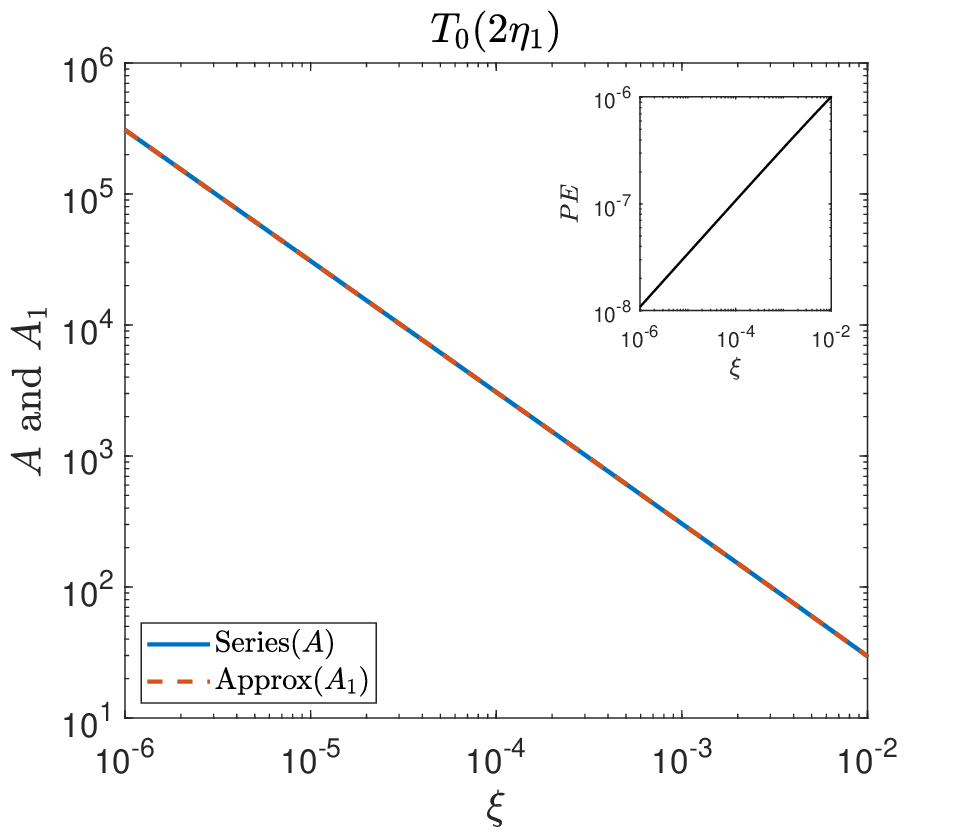}
		\caption{$T_0$-series and approximation for $2 \eta_1$}
		\label{T_0_series_2}
\end{figure}

\begin{figure}
		\includegraphics[width=1\textwidth]{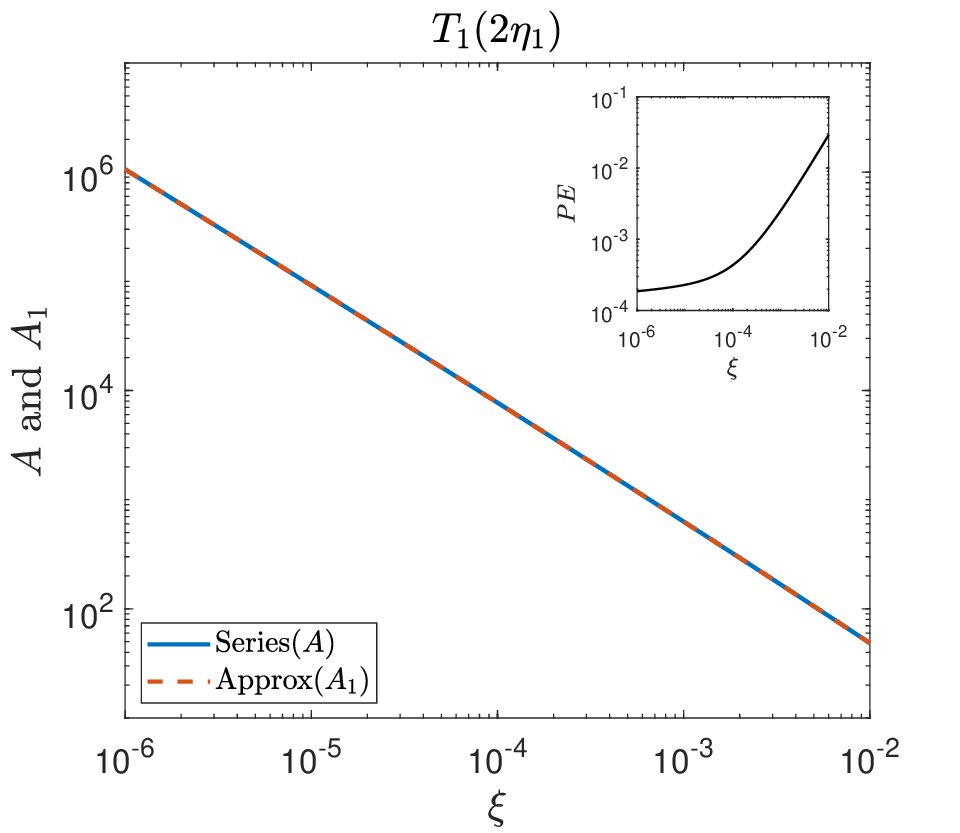}
		\caption{$T_1$-series and approximation for $2 \eta_1$}
		\label{T_1_series_2}
\end{figure}

\begin{figure}
		\includegraphics[width=1\textwidth]{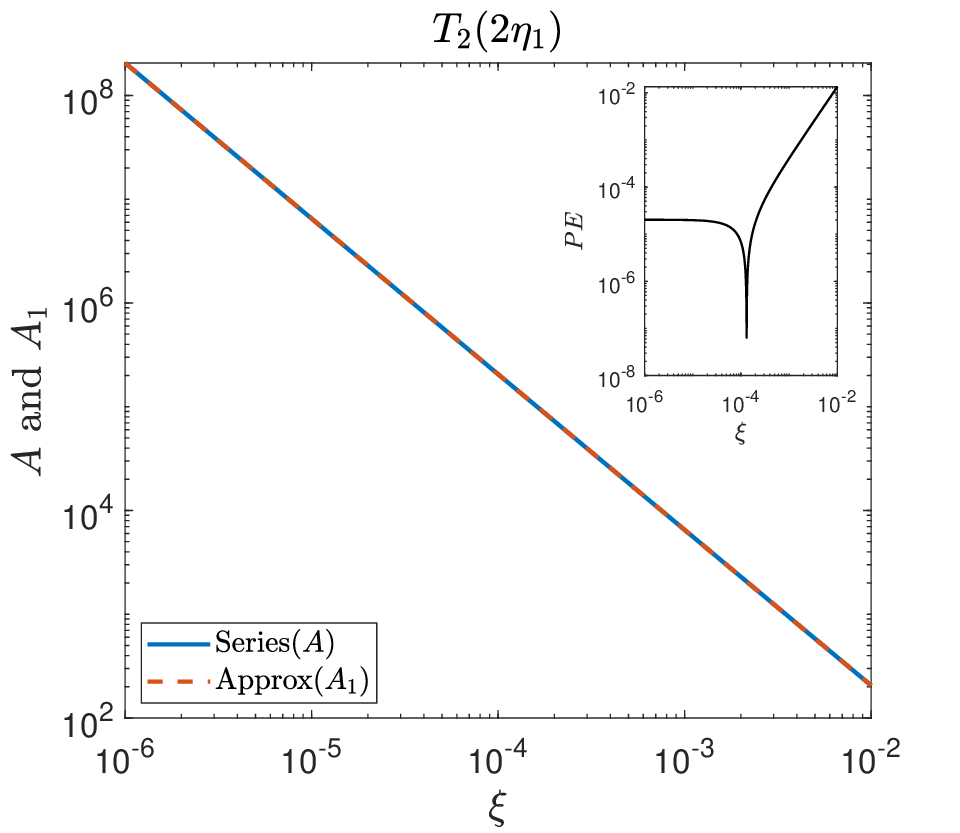}
		\caption{$T_2$-series and approximation for $2 \eta_1$}
		\label{T_2_series_2}
\end{figure}

\begin{figure}
		\includegraphics[width=1\textwidth]{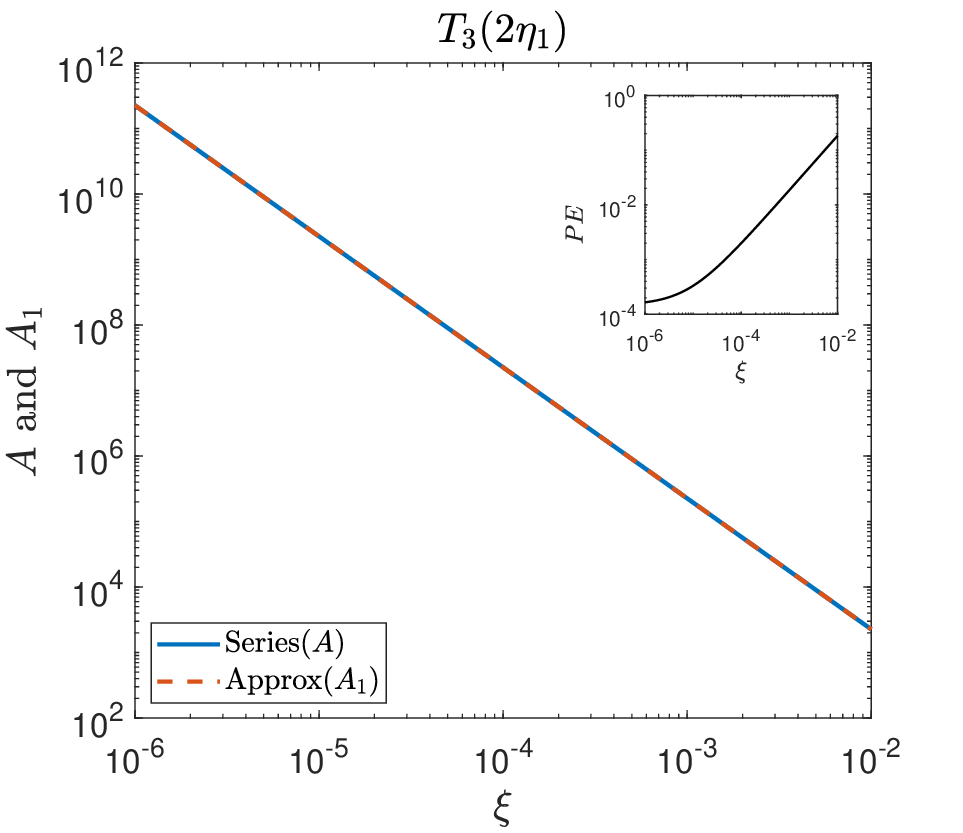}
		\caption{$T_3$-series and approximation for $2 \eta_1$}
		\label{T_3_series_2}
\end{figure}

\begin{figure}
 	\includegraphics[width=1\textwidth]{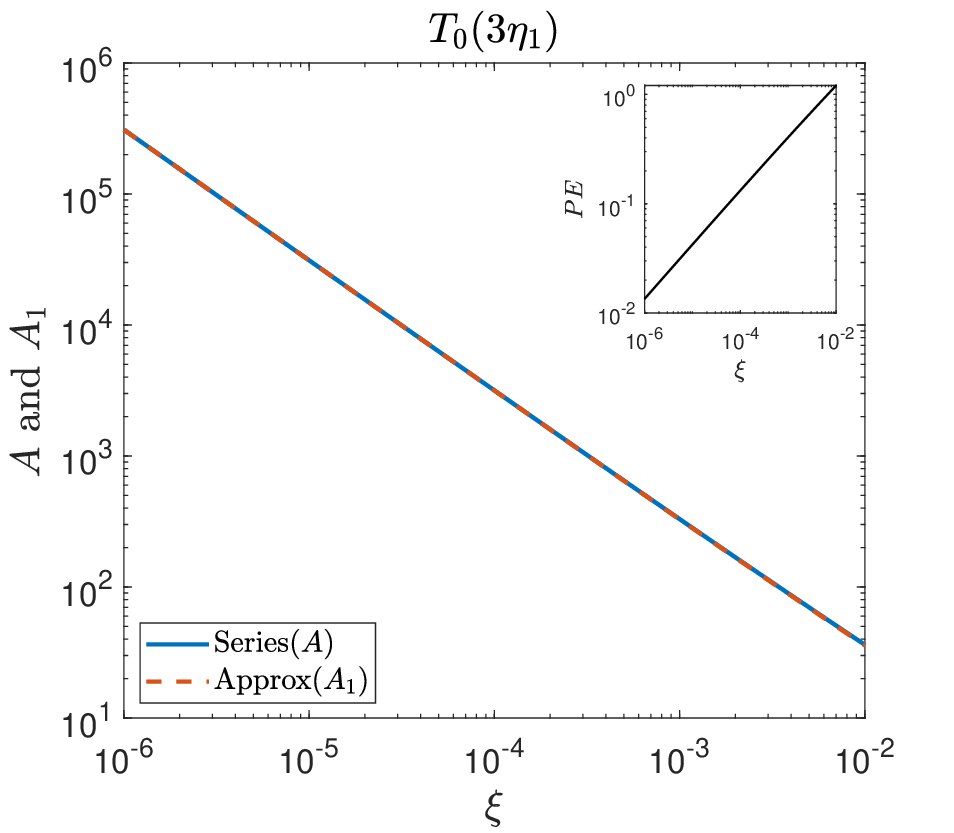}
		\caption{$T_0$-series and approximation for $2\eta_1$}
		\label{T_0_series_3}
\end{figure}

\begin{figure}
		\includegraphics[width=1\textwidth]{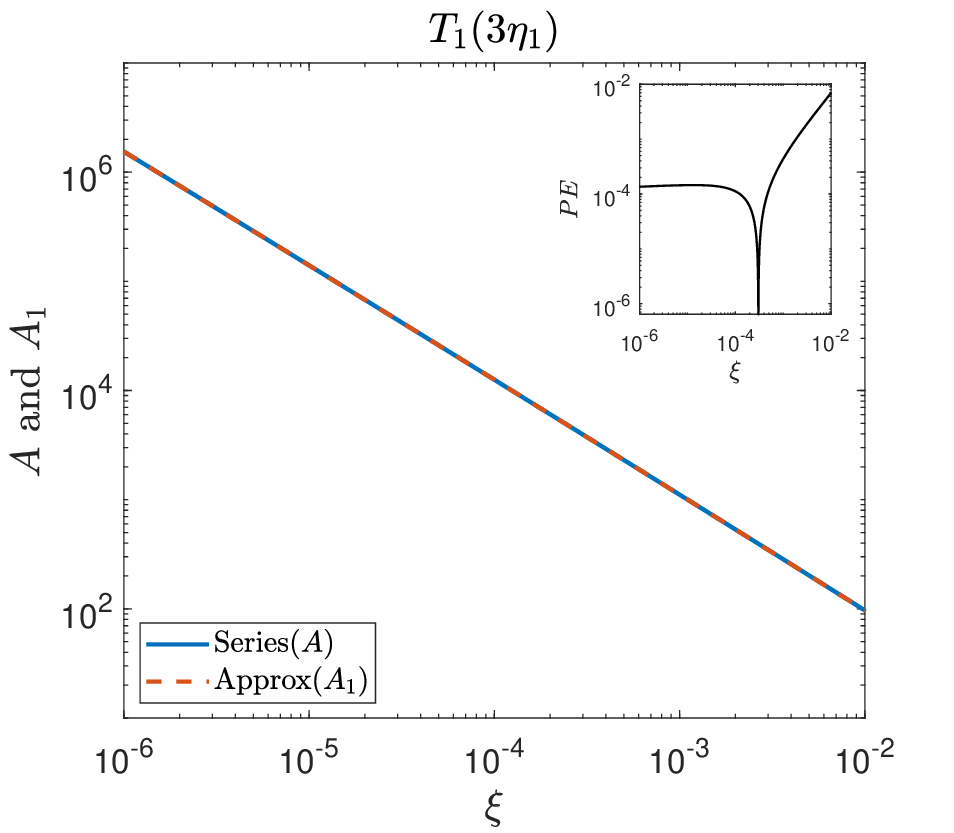}
		\caption{$T_1$-series and approximation for $3\eta_1$}
		\label{T_1_series_3}
\end{figure}

\begin{figure}
		\includegraphics[width=1\textwidth]{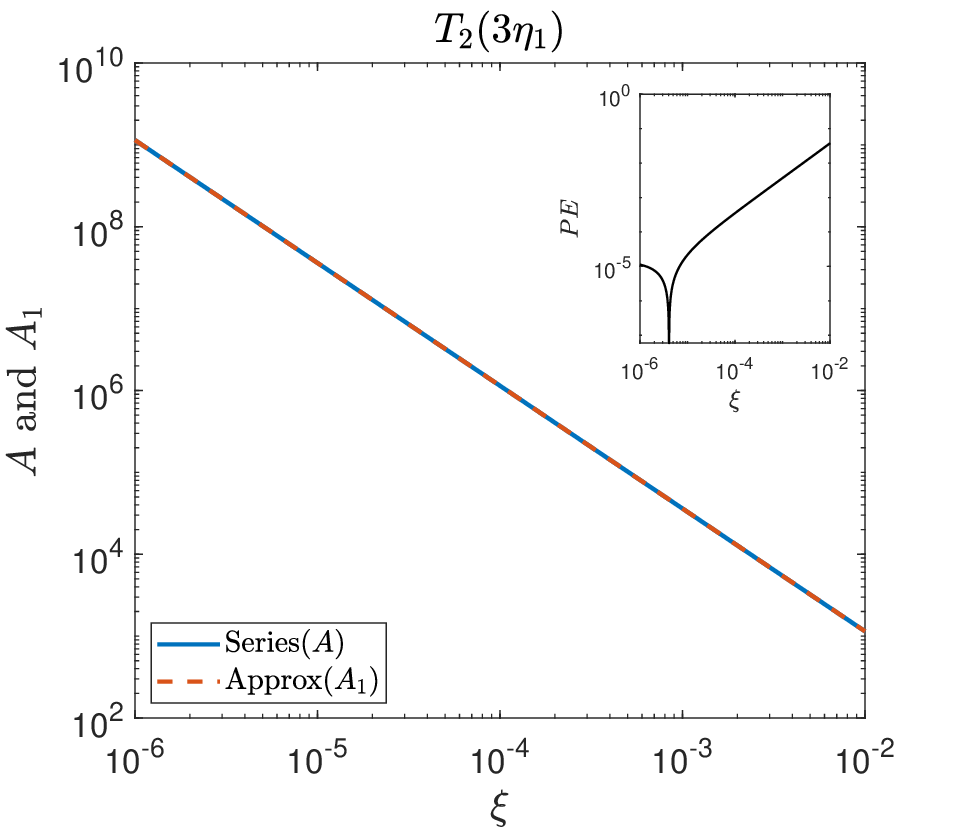}
		\caption{$T_2$-series and approximation for $3\eta_1$}
		\label{T_2_series_3}
\end{figure}

\begin{figure}
		\includegraphics[width=1\textwidth]{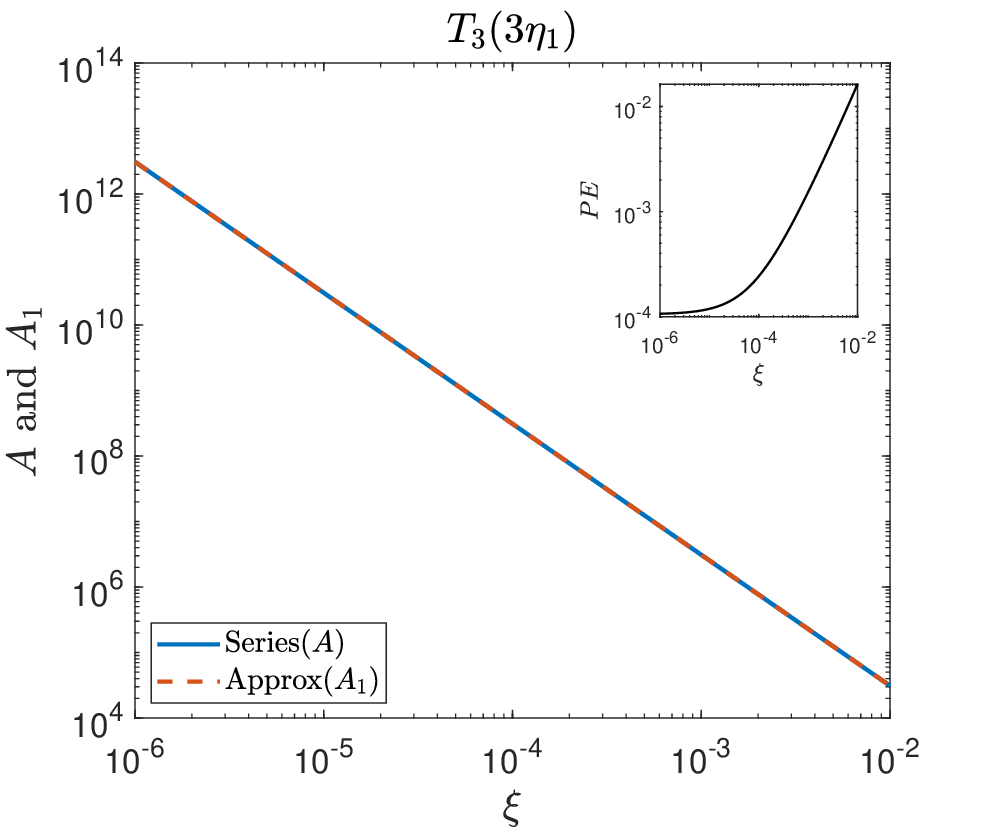}
		\caption{$T_3$-series and approximation for $3\eta_1$}
		\label{T_3_series_3}
\end{figure}

\begin{figure}
	\includegraphics[width=1\textwidth]{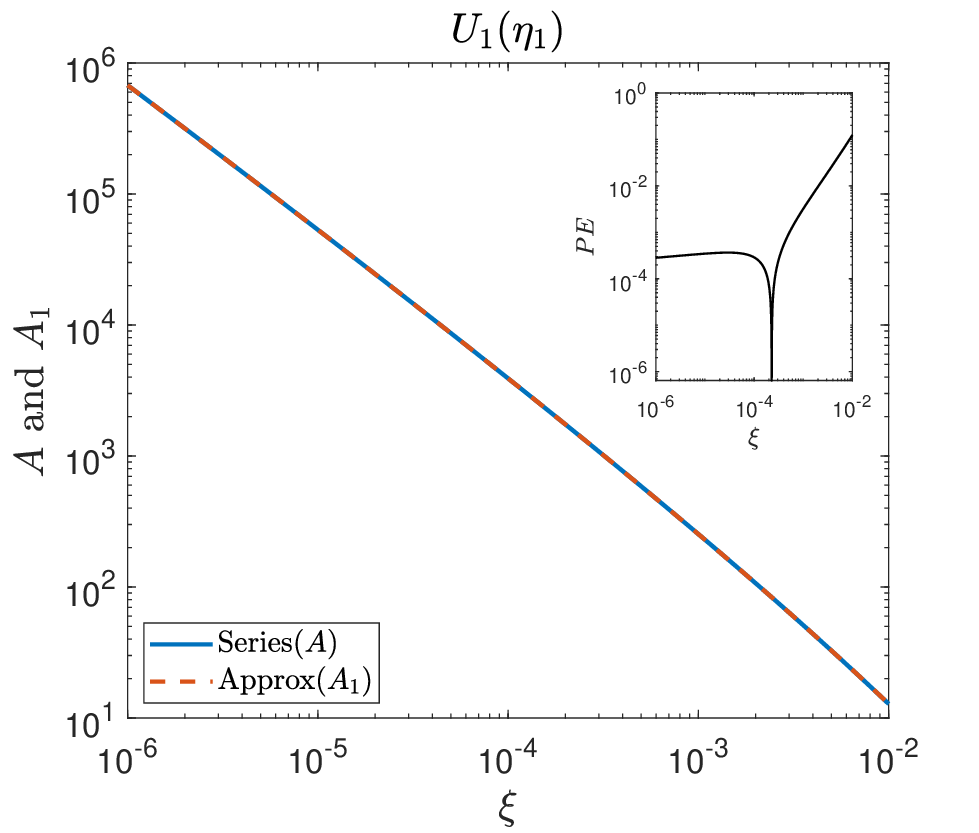}
		\caption{$U_1$-series and approximation for $\eta_1$}
		\label{U_1_series_1}
 \end{figure}

\begin{figure}
		\includegraphics[width=1\textwidth]{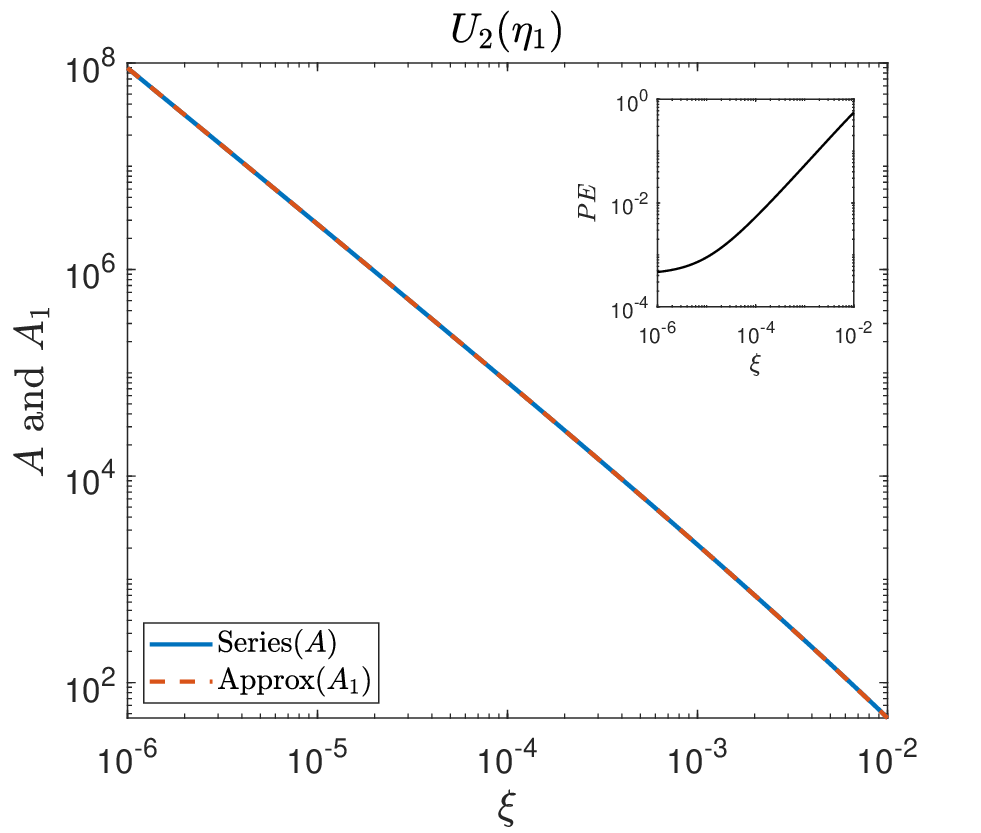}
		\caption{$U_2$-series and approximation for $\eta_1$}
		\label{U_2_series_1}
 \end{figure}

\begin{figure}
		\includegraphics[width=1\textwidth]{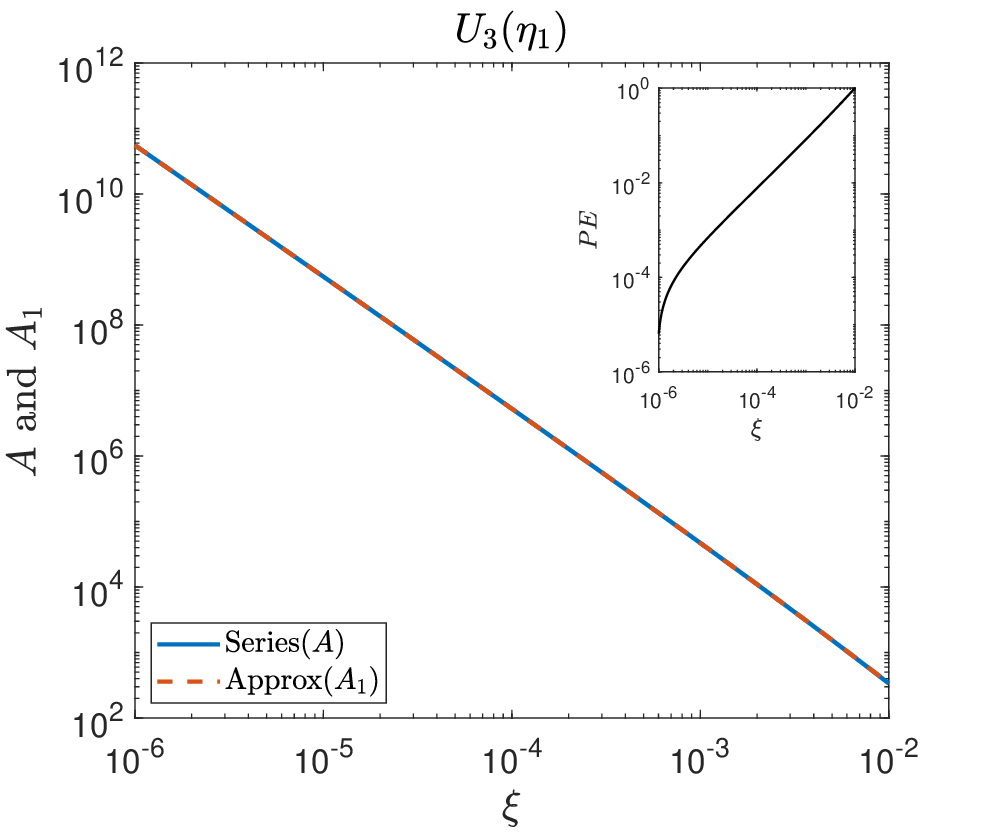}
		\caption{$U_3$-series and approximation for $\eta_1$}
		\label{U_3_series_1}
\end{figure}

\begin{figure}	\includegraphics[width=1\textwidth]{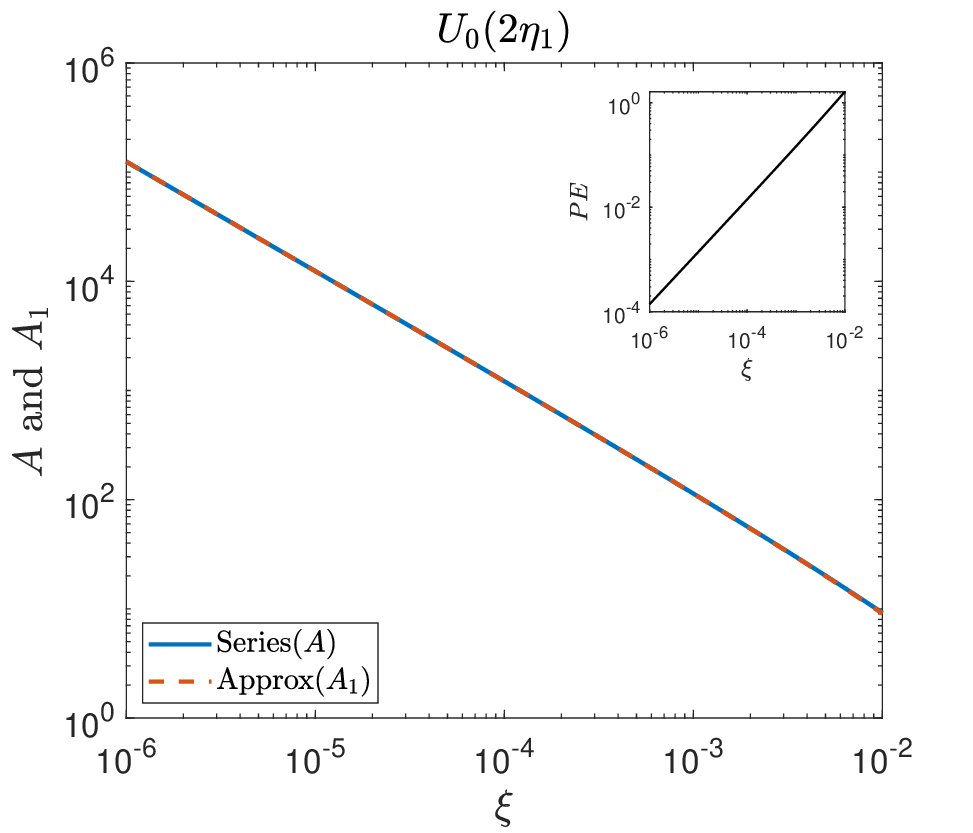}
		\caption{$U_0$-series and approximation for $2 \eta_1$}
		\label{U_0_series_2}
 \end{figure}

\begin{figure}
		\includegraphics[width=1\textwidth]{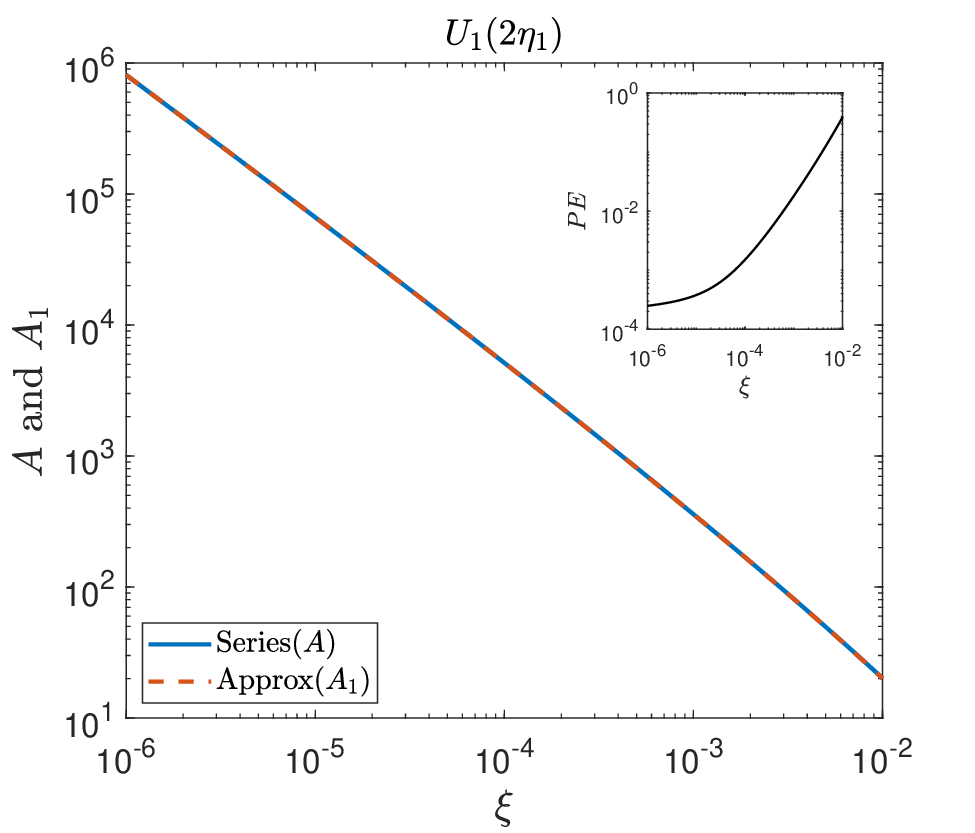}
		\caption{$U_1$-series and approximation for $2 \eta_1$}
		\label{U_1_series_2}
 \end{figure}

\begin{figure}
		\includegraphics[width=1\textwidth]{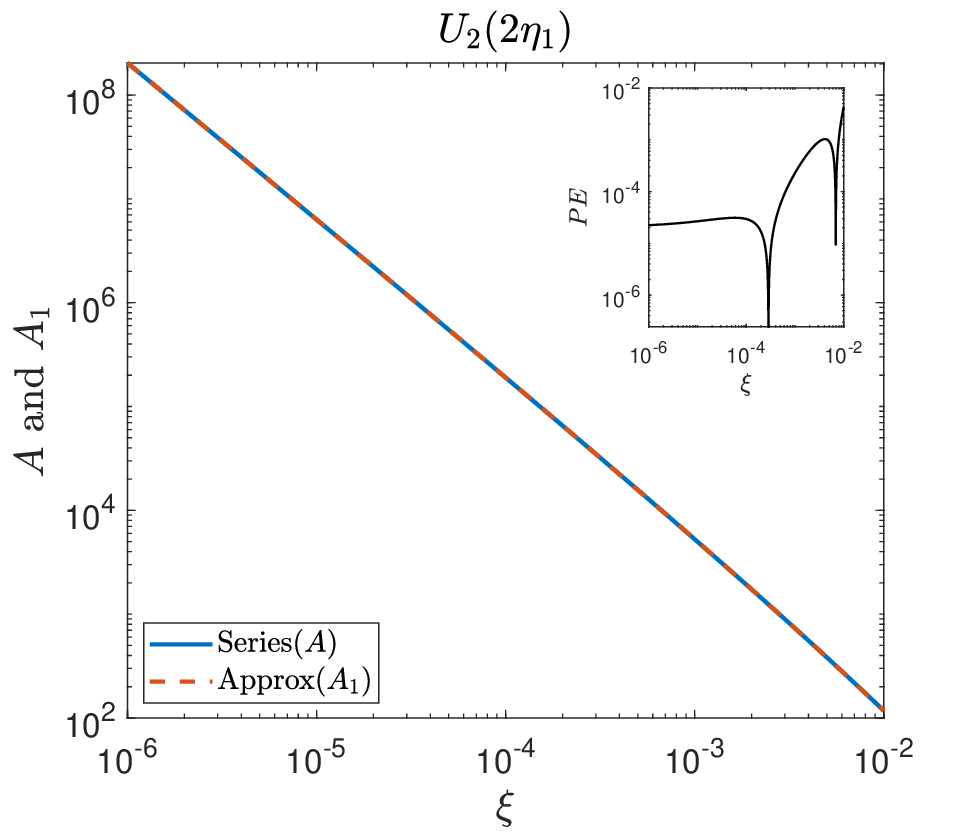}
		\caption{$U_2$-series and approximation for $2 \eta_1$}
		\label{U_2_series_2}
\end{figure}

\begin{figure}

		\includegraphics[width=1\textwidth]{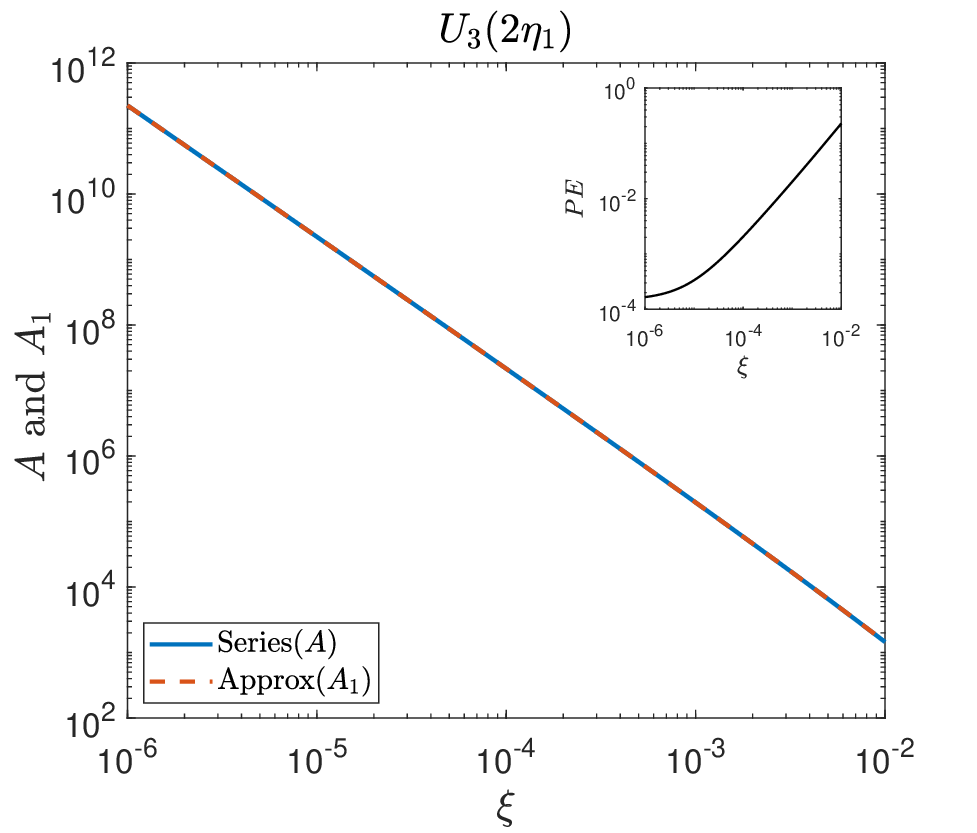}
		\caption{$U_3$-series and approximation for $2 \eta_1$}
		\label{U_3_series_2}
\end{figure}

\begin{figure}	\includegraphics[width=1\textwidth]{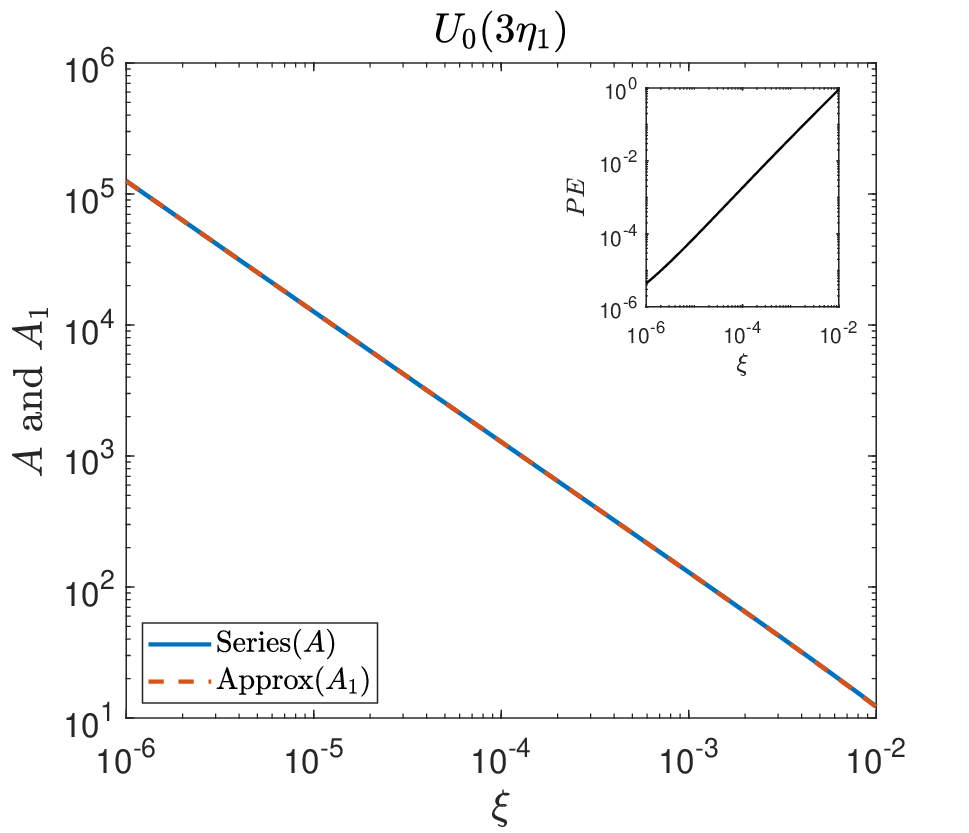}
		\caption{$U_0$-series and approximation for $3 \eta_1$}
		\label{U_0_series_3}
\end{figure}

\begin{figure}
	\includegraphics[width=1\textwidth]{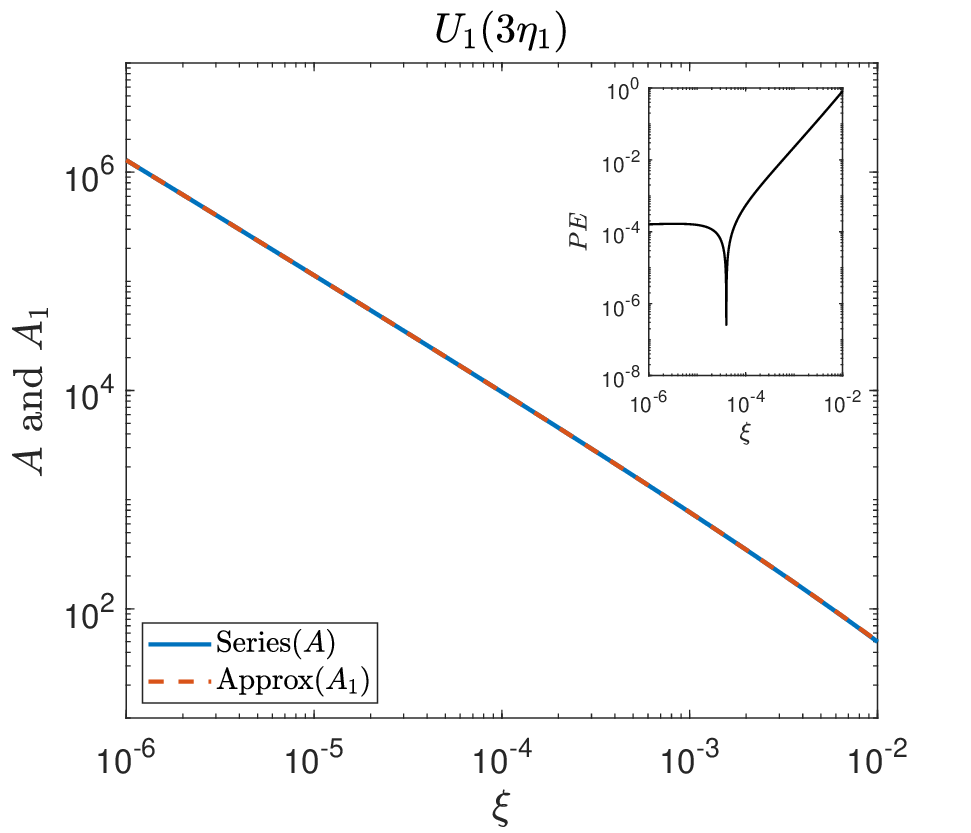}
		\caption{$U_1$-series and approximation for $3 \eta_1$}
		\label{U_1_series_3}
 \end{figure}

\begin{figure}
		\includegraphics[width=1\textwidth]{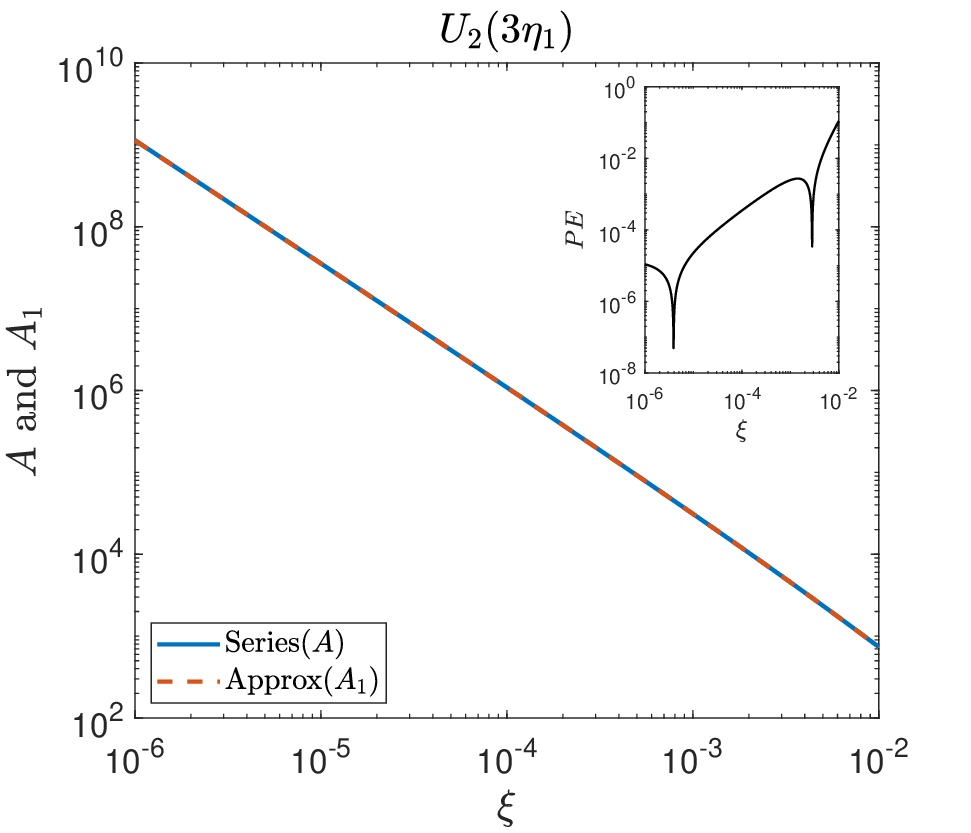}
		\caption{$U_2$-series and approximation for $3 \eta_1$}
		\label{U_2_series_3}
\end{figure}

\begin{figure}
		\includegraphics[width=1\textwidth]{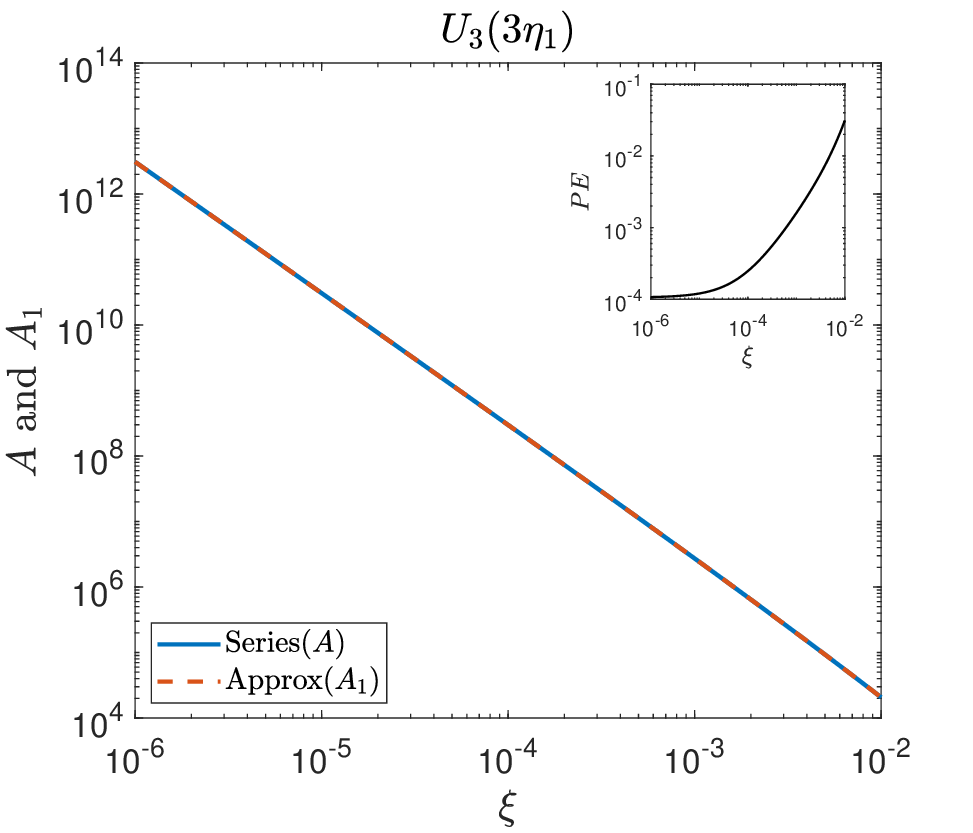}
		\caption{$U_3$-series and approximation for $3 \eta_1$}
		\label{U_3_series_3}
\end{figure}

In the previous sections \ref{sec:4} and \ref{sec:5} we obtained the asymptotic expression for $T_0(\eta_1)$ and $U_0(\eta_1)$, respectively. The asymptotic expressions for the rest of the 22 series are obtained using the method outlined in section \ref{sec:3} and described in sections
 \ref{sec:4} and \ref{sec:5} for the $T$-series and $U$-series, respectively. The following tables (\ref{table_1} and \ref{table_2}) consolidate the asymptotic expressions of 12 $T$-series and 12 $U$-series, respectively as a function of separation ($\xi$) and size ratio ($\kappa$). Further,   $\Gamma=\gamma+\log[\frac{4}{\eta_1}]$ and $\kappa=1$. The plots for remaining 22 series are as shown in the figures(\ref{T_1_series_1}-\ref{U_3_series_3}). These 22 figures are also plotted as mentioned in Subsection(\ref{subsec:6.1}). In all the above cases separation range is maintained from $10^{-6}$ to $10^{-2}$ and plotted on a $\log-\log$ scale.

\begin{table}
	\makebox[1 \textwidth][c]{       
		\resizebox{7in}{!}{
			\begin{tabular}{|p{2.2cm}|p{16cm}|}
				\hline 
				Series Name & Approximated Expression \\
				\hline
				$T_0(\eta_1)$&$\displaystyle [(\frac{\pi^2}{32})\eta_1^{-2}-\frac{\eta_1^{-1}}{8}(\Gamma)+\frac{1}{48}+\frac{\eta_1^{-1}}{24}-\eta_1^{-1}(0.0416667)-0.0208333]$ \vspace{.1cm}\\
				$T_1(\eta_1) $&$\displaystyle [\frac{\eta_1^{-2}}{8}(\Gamma)-\frac{\eta_1^{-1}}{12}-\frac{\eta_1^{-2}}{4}+\frac{1}{48}+\eta_1^{-2}(0.0665749)+\eta_1^{-1}(0.0833333)-0.0104167]$\vspace{.1cm}\\
				$T_2(\eta_1) $&$\displaystyle[\frac{\eta_1^{-3}}{4}-\frac{\eta_1^{-2}}{8}-\frac{\eta_1^{-1}}{12}+\frac{1}{48}-\eta_1^{-3}(0.15905)+\eta_1^{-2}(0.125)+\eta_1^{-1}(0.0208333)-0.00698606]$\vspace{
					.2cm}\\
				$T_3(\eta_1) $&$\displaystyle [\frac{\eta_1^{-3}}{4}-\frac{\eta_1^{-2}}{8}-\frac{\eta_1^{-1}}{12}+\frac{1}{48}+\eta_1^{-4}(0.0556826)-\eta_1^{-3}(0.25)+\eta_1^{-2}(0.1875)+\eta_1^{-1}(1.9893\times10^{-9})-0.005243]$\vspace{.1cm}\\
				\hline
				$T_0(2 \eta_1)$&$\displaystyle [(\frac{\pi^2}{32})\eta_1^{-2}-\frac{\eta_1^{-1}}{12}-\frac{1}{24}-\eta_1^{-1}(0.0416667)+0.0416667]$\vspace{.1cm}\\
				$T_1(2 \eta_1) $&$\displaystyle [\frac{\eta_1^{-2}}{8}(\Gamma)-\frac{\eta_1^{-1}}{12}-\frac{1}{24}-\eta_1^{-2}(0.0482868)+\eta_1^{-1}(0.0833333)+0.0208333]$\vspace{.1cm}\\
				$T_2(2 \eta_1) $&$\displaystyle[\frac{\eta_1^{-3}}{4}+\frac{\eta_1^{-2}}{8}-\frac{\eta_1^{-1}}{12}-\frac{1}{24}-\eta_1^{-3}(0.0443832)-\eta_1^{-2}(0.125)+\eta_1^{-1}(0.0833333)+0.0138888]$\vspace{.1cm}\\
				$T_3(2 \eta_1)$&$\displaystyle [\frac{\eta_1^{-3}}{4}+\frac{\eta_1^{-2}}{8}-\frac{\eta_1^{-1}}{12}-\frac{1}{24}+\eta_1^{-4}(0.225386)-\eta_1^{-3}(0.25)-\eta_1^{-2}(0.0625)+\eta_1^{-1}(0.0833333)+0.010415]$\vspace{.1cm}\\
				\hline
				$ T_0(3 \eta_1)$&$\displaystyle [(\frac{\pi^2}{32}) \eta_1^{-2}+\frac{\eta_1^{-1}}{8}(\Gamma)-\eta_1^{-1}(0.0416667)-0.0208333]$\vspace{.1cm}\\
				$T_1(3 \eta_1) $&$\displaystyle [\frac{\eta_1^{-2}}{8}(\Gamma)+\frac{\eta_1^{-1}}{6}+\frac{\eta_1^{-2}}{4}+\frac{1}{48}+\eta_1^{-2}(0.183425)-\eta_1^{-1}(0.166667)-0.0104167]$\vspace{.1cm}\\
				$T_2(3 \eta_1) $&$\displaystyle[\frac{\eta_1^{-3}}{4}+\frac{3\eta_1^{-2}}{8}+\frac{\eta_1^{-1}}{6}+\frac{1}{48}+\eta_1^{-3}(0.89275)-\eta_1^{-2}(0.375)-\eta_1^{-1}(0.104167)-0.00690272]$\vspace{.1cm}\\
				$T_3(3 \eta_1) $&$\displaystyle [\frac{\eta_1^{-3}}{4}+\frac{3\eta_1^{-2}}{8}+\frac{\eta_1^{-1}}{6}+\frac{1}{48}+\eta_1^{-4}(3.09972)-\eta_1^{-3}(0.25)-\eta_1^{-2}(0.3125)-\eta_1^{-1}(0.0833333)-0.00517646]$\vspace{.1cm}\\ 
    \hline
\end{tabular}
	} }
	\caption{Asymptotic expressions for T-series}
	\label{table_1}
\end{table}

    \begin{table}
	\makebox[1 \textwidth][c]{       
		\resizebox{7in}{!}{
			\begin{tabular}{|p{2.2cm}|p{16cm}|}
				\hline 
				Series Name & Approximated Expression \\
				\hline
				$ U_0(\eta_1) $&$\displaystyle [\frac{\eta_1^{-2}}{8}+\frac{(\Gamma)(1-\eta_1^{-1})}{8}+\frac{\eta_1^{1}}{24}-\frac{1}{16}-\eta_1^{-1}(0.0416667)-0.0416667]$\vspace{.1cm}\\
				$U_1(\eta_1) $&$\displaystyle [(\frac{\eta_1^{-2}-\frac{2}{3}}{8}(\Gamma-2)-\frac{\eta_1^{-2}}{4}+\frac{\eta_1^{-1}}{6}+\frac{7}{48}+\eta_1^{-2}(0.0665749)+\eta_1^{-1}(0.141758)-0.204861]$\vspace{.1cm}\\
				$U_2(\eta_1) $&$\displaystyle [(\frac{-\eta_1^{-2}}{4}+\frac{1}{6})(\Gamma-2)+\frac{\eta_1^{-3}}{4}-\frac{\eta_1^{-2}}{8}+\frac{1}{16}-\eta_1^{-3}(0.15905)+\eta_1^{-2}(0.2759)-\eta_1^{-1}(0.473734)+0.0747]$\vspace{.1cm}\\
				$U_3(\eta_1) $&$\displaystyle [(\frac{\eta_1^{-2}}{2}-\frac{1}{3})(\Gamma-2)-\frac{\eta_1^{-3}}{4}-\frac{3\eta_1^{-2}}{8}+\frac{\eta_1^{-1}}{3}+\frac{11}{48}+\eta_1^{-4}(0.0556826)-\eta_1^{-3}(0.0785338)+\eta_1^{-2}(0.302367)+\eta_1^{-1}(0.309695)-0.275837]$\vspace{.1cm}\\
				\hline
				$U_0(2 \eta_1)$&$\displaystyle [(1/8)(\eta_1^{-2}-(1/3)\eta_1^{-1}+\frac{1}{8}-\eta_1^{-1}(0.0416667)+0.159166]$\vspace{.1cm}\\
				$U_1(2 \eta_1) $&$\displaystyle  [(1/8)(\eta_1^{-2}-2\eta_1^{-1}+\frac{4}{3})(\Gamma-2)-\frac{\eta_1^{-1}}{12}-\frac{1}{24}-\eta_1^{-2}(0.0482868)+\eta_1^{-1}(0.304907)-0.321327]$\vspace{.1cm}\\
				$U_2(2 \eta_1) $&$\displaystyle [(\frac{-\eta_1^{-2}}{4}+\frac{\eta_1^{-1}}{2}-\frac{1}{3})(\Gamma-2)+\frac{\eta_1^{-3}}{4}-\frac{3\eta_1^{-2}}{8}+\frac{1}{8}-\eta_1^{-3}(0.0443832)-\eta_1^{-2}(0.0646599)+\eta_1^{-1}(0.164342)+0.170431]$\vspace{.1cm}\\
				$U_3(2 \eta_1) $&$\displaystyle [(\frac{\eta_1^{-2}}{2}-\eta_1^{-1}+\frac{2}{3})(\Gamma-2)-\frac{\eta_1^{-3}}{4}+\frac{3\eta_1^{-2}}{8}-\frac{\eta_1^{-1}}{6}-\frac{5}{24}+\eta_1^{-4}(0.225386)-\eta_1^{-3}(0.817622)+\eta_1^{-2}(1.44523)-\eta_1^{-1}(0.977839)-0.0566794]$\vspace{.1cm}\\
				\hline
				$U_0(3 \eta_1)$&$\displaystyle [(1/8)(\eta_1^{-2}+(-3+\eta_1^{-1})(\Gamma-2))+\frac{3}{16}-\frac{5\eta_1^{-1}}{24}-\eta_1^{-1}(0.0416667)-0.0416667]$\vspace{.1cm}\\
				$U_1(3 \eta_1) $&$\displaystyle [(\frac{\eta_1^{-2}}{8}-\frac{4\eta_1^{-1}}{8}+\frac{11}{12})(\Gamma-2)-\frac{7\eta_1^{-1}}{12}+\frac{\eta_1^{-2}}{4}-\frac{17}{48}+\eta_1^{-2}(0.183425)-\eta_1^{-1}(0.841942)+2.52884]$\vspace{.1cm}\\
				$U_2(3 \eta_1) $&$\displaystyle [(\frac{-\eta_1^{-2}}{4}+\eta_1^{-1}-\frac{11}{6})(\Gamma-2)+\frac{\eta_1^{-3}}{4}-\frac{5\eta_1^{-2}}{8}+\frac{3\eta_1^{-1}}{4}+\frac{9}{16}+\eta_1^{-3}(0.89275)-\eta_1^{-2}(3.7951)+\eta_1^{-1}(7.43967)-10.0125]$\vspace{.1cm}\\
				$U_3(3 \eta_1)$&$\displaystyle [(\frac{\eta_1^{-2}}{2}-2\eta_1^{-1}+\frac{11}{3})(\Gamma-2)-\frac{\eta_1^{-3}}{4}+\frac{9 \eta_1^{-2}}{8}-\frac{23\eta_1^{-1}}{12}-\frac{61}{48}+\eta_1^{-4}(3.09972)-\eta_1^{-3}(12.4774)+\eta_1^{-2}(24.9142)-\eta_1^{-1}(32.675)+33.9727]$\vspace{.1cm}\\ 
				\hline
			\end{tabular}
	} }
	\caption{Asymptotic expressions for U-series}
	\label{table_2}
\end{table}
	




\newpage
We have further classified the 24 series into three categories depending on the increase in the percentage error and when the separation between the conducting spheres lies in the range of $10^{-3}$ to $10^{-2}$  for all the series. 
\begin{enumerate}
\item In the first category, the percentage errors for the series, namely, $T_0(\eta_1)$,  $T_1(\eta_1)$,$T_0(2\eta_1)$, $T_1(2\eta_1)$, $T_2(2\eta_1)$,$T_1(3\eta_1)$,$T_2(3\eta_1)$, $T_3(3\eta_1)$, $U_0(\eta_1)$,$U_2(2\eta_1)$, $U_3(3\eta_1)$ increases at an extremely low rate. In this case, the percentage error is less than 0.1( figures $\ref{T_0_series_1}$,$\ref{T_1_series_1}$ $\ref{T_0_series_2}, $\ref{T_1_series_2}, $\ref{T_2_series_2}$ ,$\ref{T_1_series_3}$, $\ref{T_2_series_3}$, $\ref{T_3_series_3}$, \ref{U_0_series_1}, $\ref{U_2_series_2}$, $\ref{U_3_series_3}$). In these series, $T_2(3 \eta_1)$ has the highest percentage error of 0.0376 at $10^{-2}$( table \ref{table_3}).\\
	
	\item In the second category, the percentage errors for the series,  namely, $T_2(\eta_1)$, $T_3(2 \eta_1)$,   $U_1(\eta_1)$, $U_1(2 \eta_1)$,  $U_3(2\eta_1)$, $U_2(3 \eta_1)$  increases gradually. In this case, the percentage error is between 0.1 and 0.5( figures \ref{T_2_series_1}, \ref{T_3_series_2}, \ref{U_1_series_1}, \ref{U_1_series_2}, \ref{U_3_series_2}, \ref{U_2_series_3}). In these series, $T_2(\eta_1)$ has the highest percentage error of 0.444 at $10^{-2}$ ( table \ref{table_4}).\\
	
	\item Finally, the percentage errors for the series, namely,   $T_3(\eta_1)$, $T_0(3 \eta_1)$, $U_2(\eta_1)$, $U_3(\eta_1)$, $U_0(2 \eta_1)$, $U_0(3 \eta_1)$, $U_1(3 \eta_1)$ increases significantly. In this case, the percentage error is more than 0.5 ( figures  \ref{T_3_series_1},  \ref{T_0_series_3}, \ref{U_2_series_1},\ref{U_3_series_1}, \ref{U_0_series_2}, \ref{U_0_series_3}, \ref{U_1_series_3}). Among these series, $U_0(2 \eta_1)$ has the highest percentage error of 1.62 at $10^{-2}$( table \ref{table_5}).
\end{enumerate}

\begin{table}
	\makebox[1 \textwidth][c]{       
		\resizebox{5in}{!}{
			\begin{tabular}{|p{2.2cm}|p{5cm}|p{5cm}|}
				
				\hline 
				Series Name & Percentage Error at $10^{-3}$& Percentage Error at $10^{-2}$ \\
				\hline
				$T_0(\eta_1) $  & $1.70\times10^{-5}$ & $1.28\times10^{-3}$ \vspace{.1cm}\\
				$T_1(\eta_1)$&$1.02\times10^{-3}$ & $1.93\times10^{-2}$\vspace{.1cm}\\
				$T_0(2 \eta_1)  $ & $3.35\times10^{-7}$ & $1.01\times10^{-6}$ \vspace{.1cm}\\
				$T_1(2\eta_1)  $ & $2.52\times10^{-3}$ & $2.9\times10^{-2}$ \vspace{.1cm}\\
				$T_2(2 \eta_1) $ &$4.07\times10^{-4}$ &  $1.35\times10^{-2}$\vspace{.1cm}\\
				$T_1(3 \eta_1)$& $4.36\times10^{-4}$ & $6.95\times10^{-3}$\vspace{.1cm}\\
				$T_2(3 \eta_1)$& $3.67\times10^{-3}$ &  $3.76\times10^{-2}$\vspace{.1cm}\\
				$T_3(3 \eta_1) $ & $1.54\times10^{-3}$ & $1.63\times10^{-2}$ \vspace{.1cm}\\
				$U_0(\eta_1) $ &1.$18\times10^{-4}$& $1.22\times10^{-2}$\vspace{.1cm}\\
				$U_2(2 \eta_1) $& $2.28\times10^{-4}$& $4.36\times10^{-3}$\vspace{.1cm}\\
				$U_3(3 \eta_1) $ & $1.6 \times 10^{-3}$  & $3.12 \times 10^{-2}$\vspace{.1cm}\\
				
				\hline
			\end{tabular}
	} }
	\caption{Category 1}
	\label{table_3}
\end{table}

\begin{table}
	\makebox[1 \textwidth][c]{       
		\resizebox{5in}{!}{
			\begin{tabular}{|p{2.2cm}|p{5cm}|p{5cm}|}
				
				\hline 
				Series Name & Percentage Error at $10^{-3}$ & Percentage Error at $10^{-2}$ \\
				\hline
				$T_2(\eta_1) $& $4.58\times10^{-2}$ & $4.44\times10^{-1}$ \vspace{.1cm}\\
				$T_3(2 \eta_1) $ & $1.86\times10^{-2}$ & $1.83\times10^{-1}$ \vspace{.1cm}\\
				$U_1(\eta_1) $  & $3.33\times10^{-2}$ & $1.24\times10^{-1}$\vspace{.1cm}\\
				$U_1(2 \eta_1) $ & $1.8\times10^{-2}$ & $3.88\times10^{-1}$\vspace{.1cm}\\
				$U_3(2 \eta_1) $ & $2.02\times10^{-2}$ & $2.28\times10^{-1}$ \vspace{.1cm}\\
				$U_2(3 \eta_1)$& $2.46\times10^{-3}$& $1.1\times10^{-1}$\vspace{.1cm}\\
				\hline
			\end{tabular}
	} }
	\caption{Category 2}
	\label{table_4}
\end{table}

\begin{table}
	\makebox[1 \textwidth][c]{       
		\resizebox{5in}{!}{
			\begin{tabular}{|p{2.2cm}|p{5cm}|p{5cm}|}
				
				\hline 
				Series Name & Percentage Error at $10^{-3}$ & Percentage Error at $10^{-2}$ \\
				\hline
				$T_3(\eta_1)$& $7.0\times10^{-2}$ & $6.0\times10^{-1}$\vspace{.1cm}\\
				$T_0(3\eta_1) $ & $4.057\times10^{-1}$& $1.208$  \vspace{.1cm}\\
				$U_2(\eta_1) $ & $5.33\times10^{-2}$& $5.55\times10^{-1}$\vspace{.1cm}\\
				$U_3(\eta_1)$& $8.33\times10^{-2}$ &  $1.02$ \vspace{.1cm}\\
				$U_0(2 \eta_1)$& $1.46\times10^{-1}$& $1.62$\vspace{.1cm}\\
				$U_0(3 \eta_1)$& $4.26\times10^{-2}$& $9.1\times10^{-1}$\vspace{.1cm}\\
				$U_1(3 \eta_1) $ & $2.25\times10^{-2}$ & $8.12\times10^{-1}$\vspace{.1cm}\\
				\hline
			\end{tabular}
	} }
	\caption{Category 3}
	\label{table_5}
\end{table}

\subsection{Force coefficients}\label{subsec:6.3}
\begin{figure}[!htbp]
	\includegraphics[scale=0.6,angle=0]{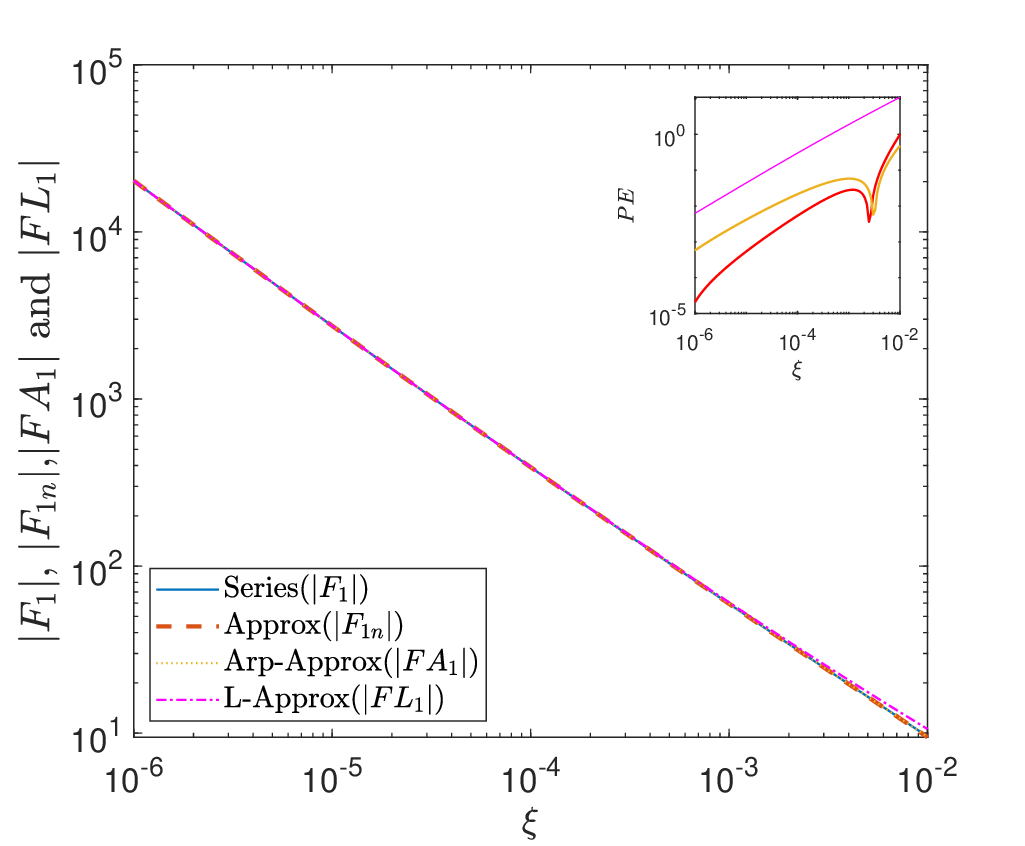}
	\caption{$F_1$-series and approximations}
	\label{F_1}
\end{figure}

We know that the force on the spheres depends on the force coefficients $F_1$, $F_2$,...$F_{10}$, which are a function of separation ($\xi$) between the spheres. Based on the 24 series (see Tables \ref{table_1} and \ref{table_2}) the plots of the absolute value of the force coefficients $F_1$, $F_2$,...,$F_{10}$ versus separation($\xi$) between the spheres and its approximations are as given in figures (\ref{F_1} to \ref{F_{10}}). We consider absolute value of all the force coefficients because plots are on a $\log-\log$ scale and our separation range from $10^{-6}$ to $10^{-2}$. In the figure, the blue continuous line represents the force coefficients $F_1$, $F_2$,...,$F_{10}$ computed using infinite series as given by \cite{davis1964two}, and the red dashed line represents the force coefficients $F_1$, $F_2$,...$F_{10}$ from our asymptotic expressions. The Inset/subplot in the figures shows the percentage errors versus separation ($\xi$) of the spheres. The plot illustrates the percentage error using a continuous black line for the force coefficients $F_3$, $F_4$, $F_9$, and $F_{10}$. Percentage error plots for the other force coefficients are also indicated by the same colour as the series approximations. The trend indicates that all ten force coefficients decrease as the separation($\xi$) increase. The percentage error value is less than  or equal to 1 when the separation range is from $10^{-6}$ to $10^{-3}$. However, for the remaining range of separation, the value of the percentage error is increasing significantly. Among these, we have categorized the force coefficients into three categories based on charged/uncharged spheres and the presence/absence of an external electric field. It should be noted that, we are dealing with spheres of equal size.

\begin{figure}[!htbp]
	\includegraphics[scale=0.6,angle=0]{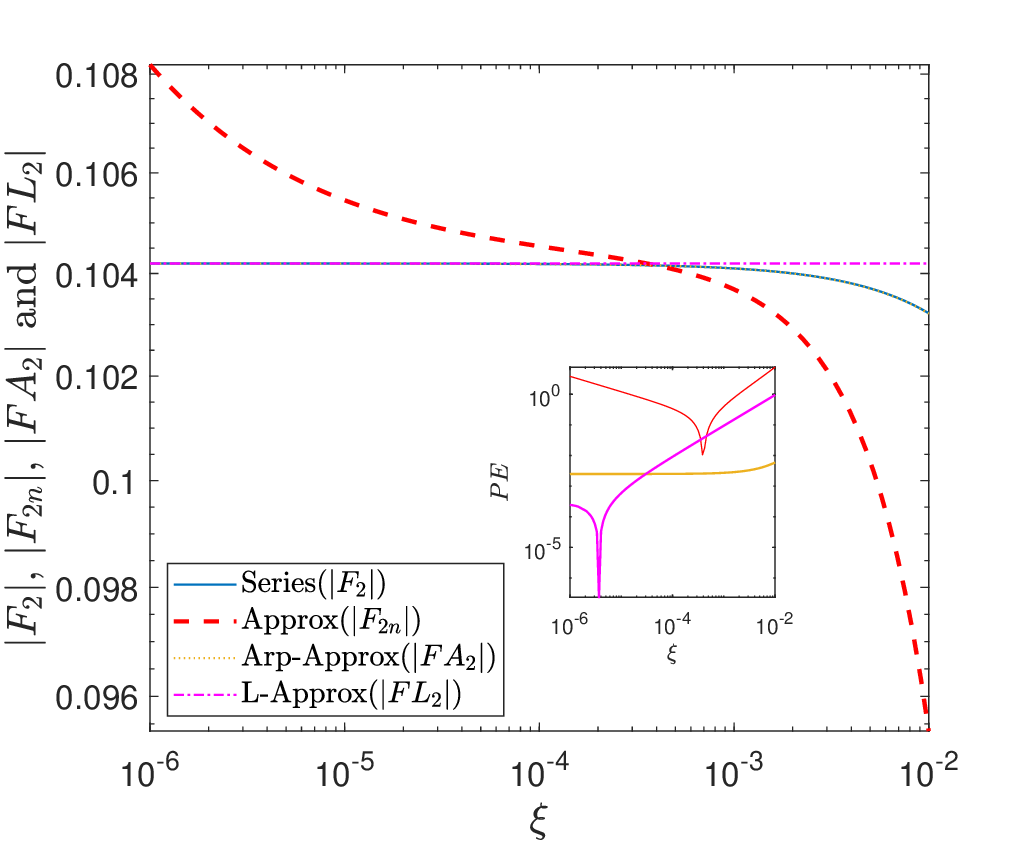}
	\caption{$F_2$-series and approximations }
	\label{F_2}
\end{figure}

\begin{figure}[!htbp]
	\includegraphics[scale=0.6,angle=0]{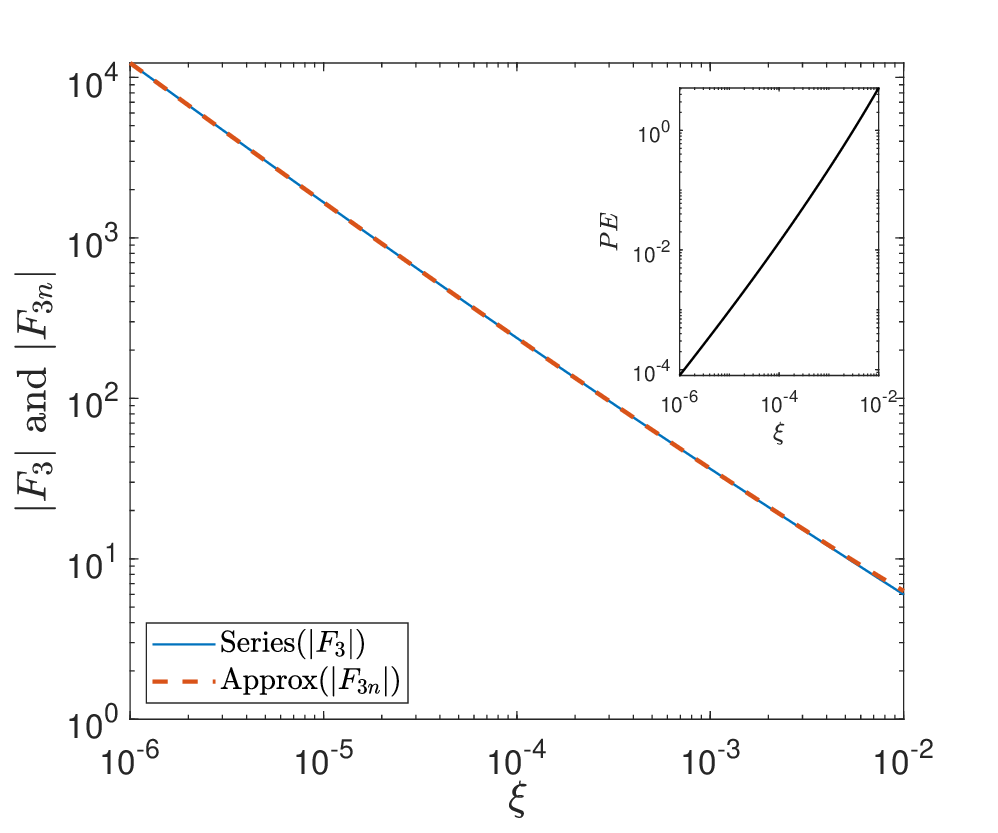}
	\caption{$F_3$-series and approximation }
	\label{F_3}
\end{figure}

\begin{figure}[!htbp]
	\includegraphics[scale=0.6,angle=0]{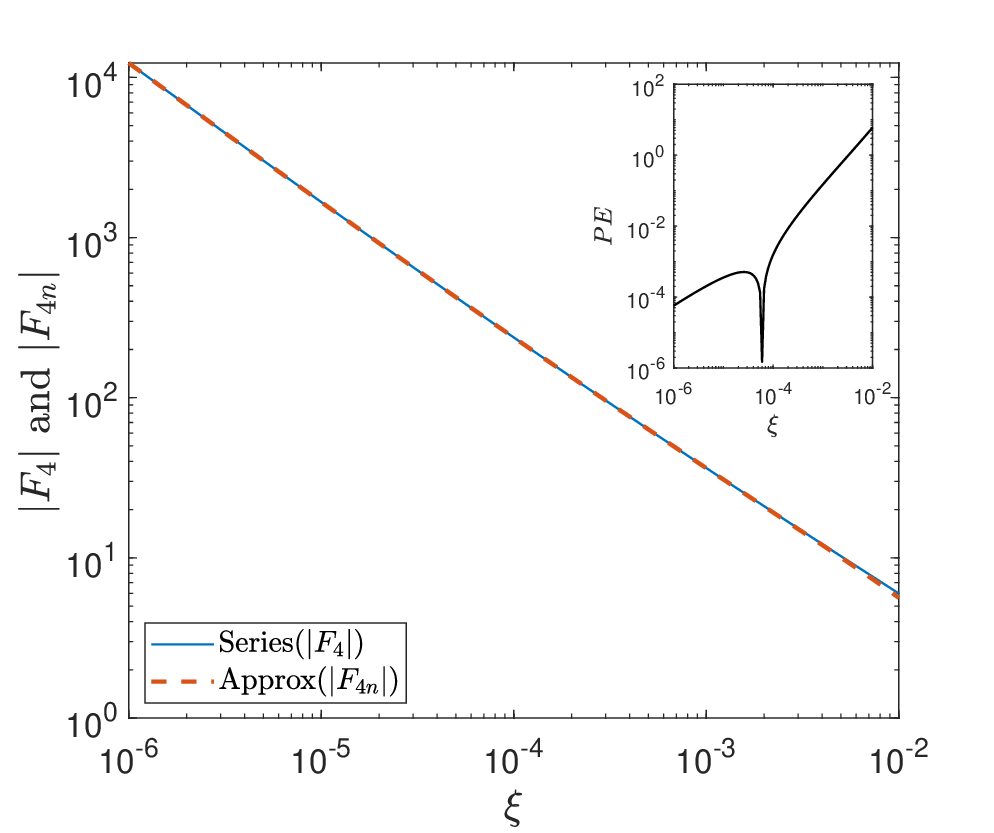}
	\caption{$F_4$-series and approximation}
	\label{F_4}
\end{figure}

\begin{figure}[!htbp]
	\includegraphics[scale=0.6,angle=0]{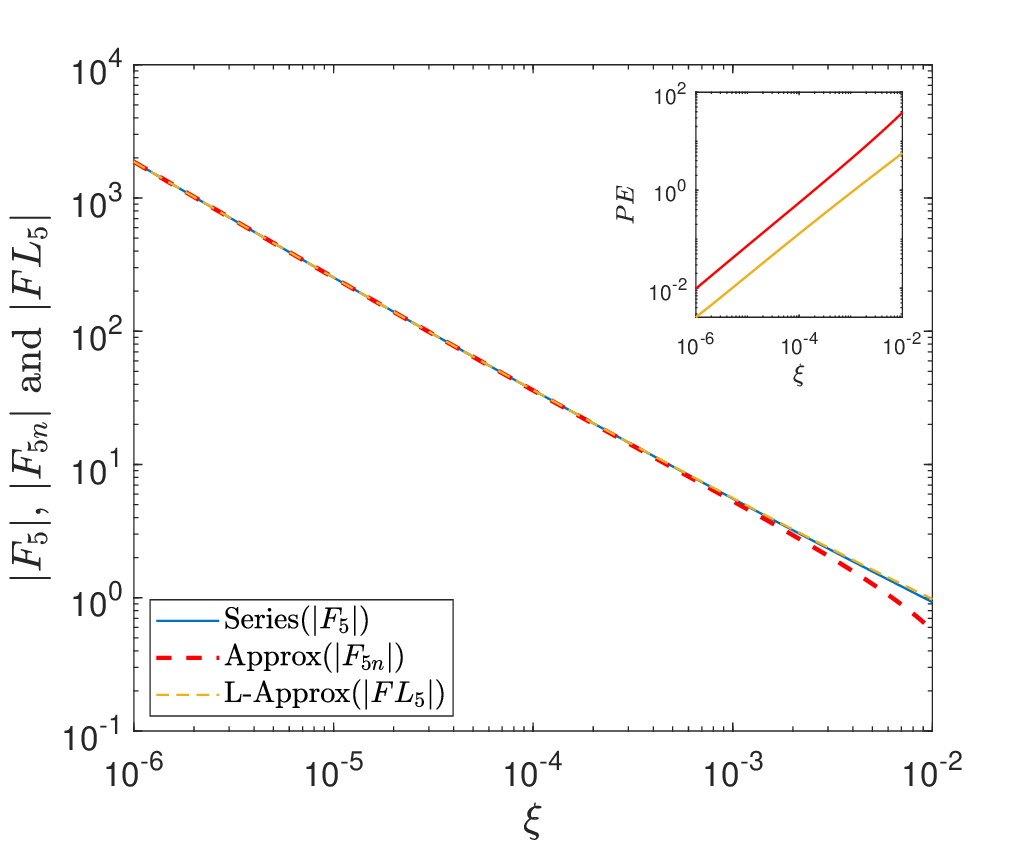}
	\caption{$F_5$-series and approximations }
	\label{F_5}
\end{figure}

\begin{figure}[!htpb]
	\includegraphics[scale=0.6,angle=0]{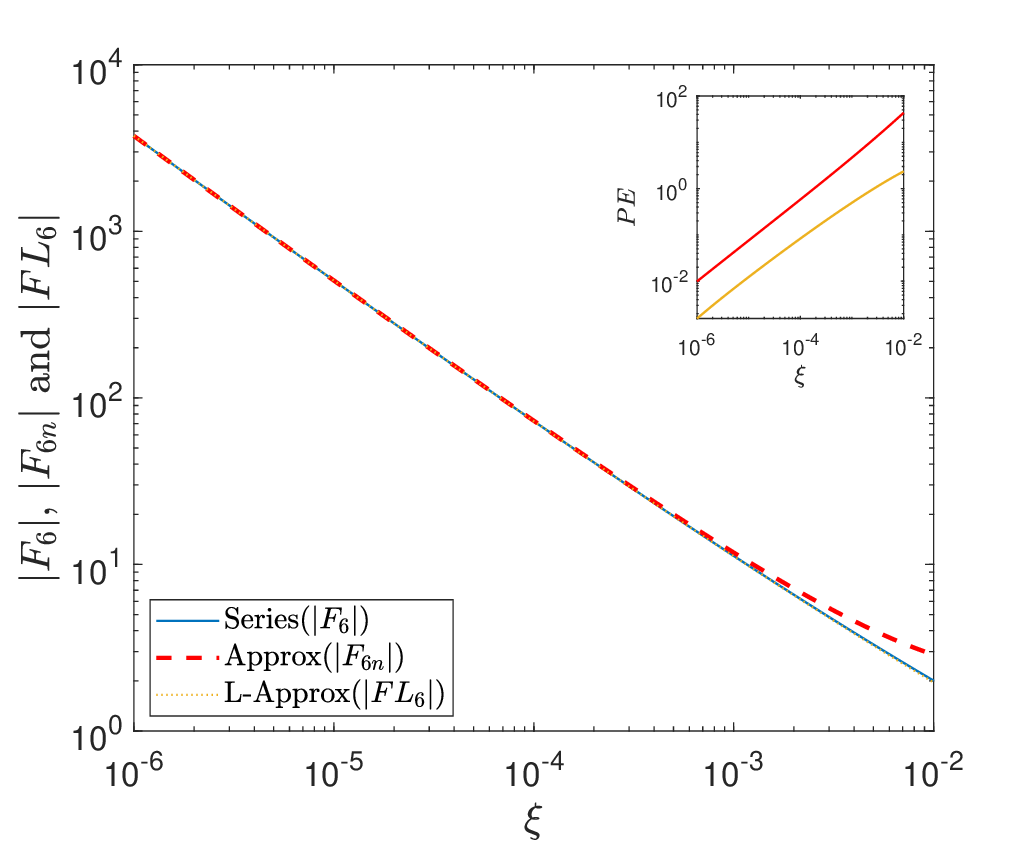}
	\caption{$F_6$-series and approximations}
	\label{F_6}
\end{figure}

\begin{figure}[!htpb]
	\includegraphics[scale=0.6,angle=0]{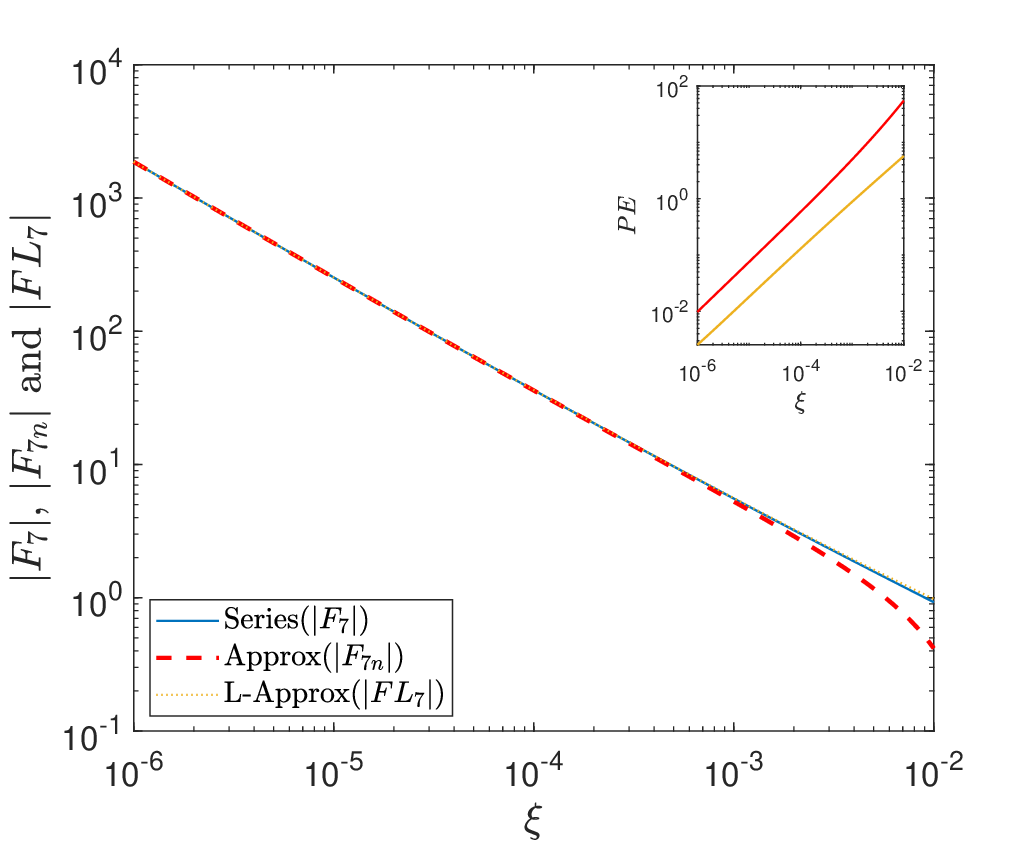}
	\caption{$F_7$-series and approximations}
	\label{F_7}
\end{figure}

\begin{figure}[!htbp]
	\includegraphics[scale=0.6,angle=0]{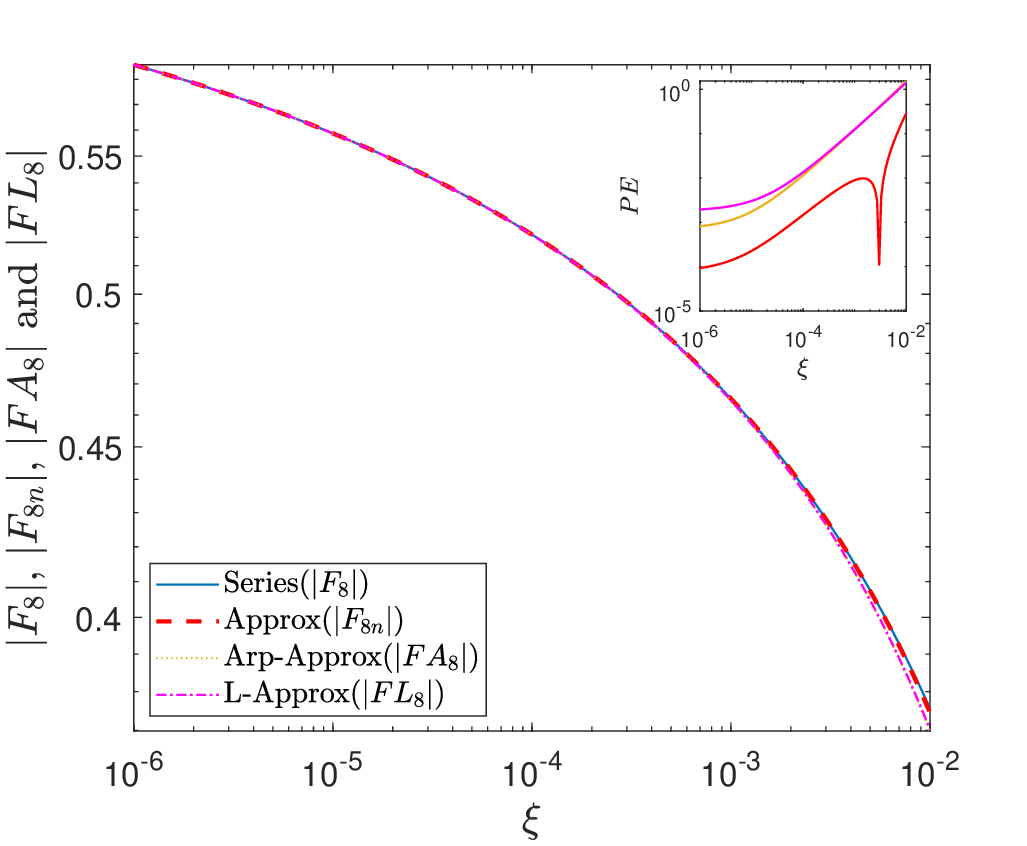}
	\caption{$F_8$-series and approximation}
	\label{F_8}
\end{figure}

\begin{figure}[!htbp]
	\includegraphics[scale=0.6,angle=0]{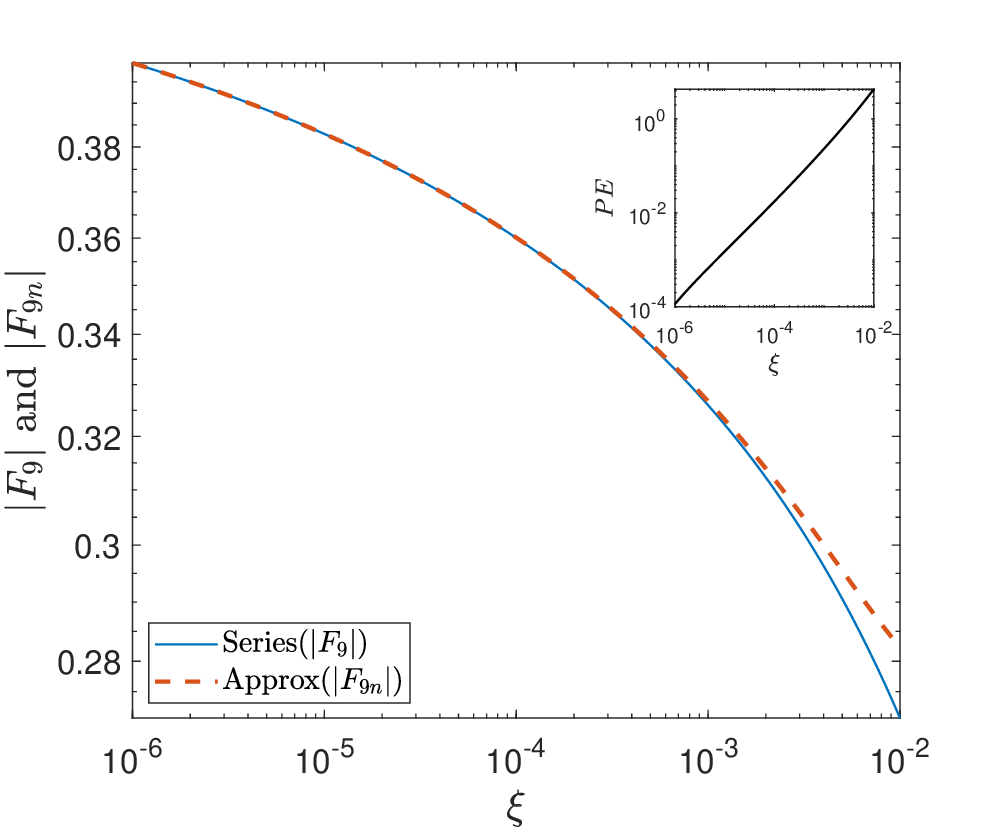}
	\caption{$F_9$-series and approximation}
	\label{F_9}
\end{figure}

\begin{enumerate}
	\item In the first category, when uncharged spheres are subjected to an external electric field, the force acting on them depends on three force coefficients: $F_1$, $F_2$, and $F_8$.
	\cite{arp1977particle} also provided an estimate for these force coefficients using the same approach. In order to assess the accuracy of our approximation (denoted as Approx in the plots), we compare the values obtained in the current work with that of values from  \cite{davis1964two} (which is denoted as Series $F_{n}$ in the plots) and the Arp and Mason's approximation [\cite{arp1977particle}] (denoted as Arp-Approx $|FA|$ in the plots) and the Lekner's approximation [\cite{lekner2013forces}] (denoted as L-Approx $|FL|$ in the plots) and presented the results in figures \ref{F_1}, \ref{F_2}, and \ref{F_8}. The plots of the percentage error for all the three approximations(our approximation, Arp and Mason's approximation, and Lekner's approximation) are given in red, yellow, and magenta colours, respectively. The approximations for $F_1$ and $F_8$ match well with the series and better than other approximations in the entire region. But the approximation for $F_2$ does not align with the other approximations and series, which can be attributed to the extremely small value of $F_2$ and the multiple (to be precise 18) infinite series involved in finding the value of $F_2$(\cite{davis1964two}). Nonetheless, the percentage error is less than 1.\\\\
	
	\item In the second category, when charged conducting spheres without an external electric field, the force acting on them depends on three force coefficients: $F_5$, $F_6$, and $F_7$. \cite{lekner2012electrostatics} proposed an alternative method for approximating the force coefficients of arbitrary spheres. This method differs from previous approaches and provides an accurate representation of the coefficients. In order to assess the accuracy of our approximation (denoted as Approx in the plots), We compare the values obtained in the current work with that of values from \cite{davis1964two} (which is denoted by Series $F_{n}$ in the plots) and the Lekner approximation \cite{lekner2012electrostatics} (denoted as L-Approx $|FL|$ in the plots) and presented the results in figures \ref{F_5}, \ref{F_6} and \ref{F_7}. The percentage error graph for all two approximations(our approximation and Lekner's approximation) are given in red and yellow colours, respectively. The approximations for $F_5$, $F_6$, and $F_7$ match well with the series and \cite{lekner2012electrostatics} approximations in the region of $10^{-6}$ to $10^{-3}$. In the range between $10^{-3}$ and $10^{-2}$, the percentage error surpasses 10.\\
	
	\item Finally, the other force coefficients $F_3$, $F_4$, $F_9$, $F_{10}$ depend on both the charges on the spheres and external electric field. The results are plotted in the figures \ref{F_3}, \ref{F_4}, \ref{F_9} and  \ref{F_{10}}.The approximations for $F_3$, $F_4$, $F_9$ and $F_{10}$ match well with the series in the region of $10^{-6}$ to $10^{-3}$, and the  deviations in the region $10^{-3}$ to $10^{-2}$ with percentage error of 1 to 6.
\end{enumerate}
	


\begin{figure}[!htbp]
	\includegraphics[scale=0.6,angle=0]{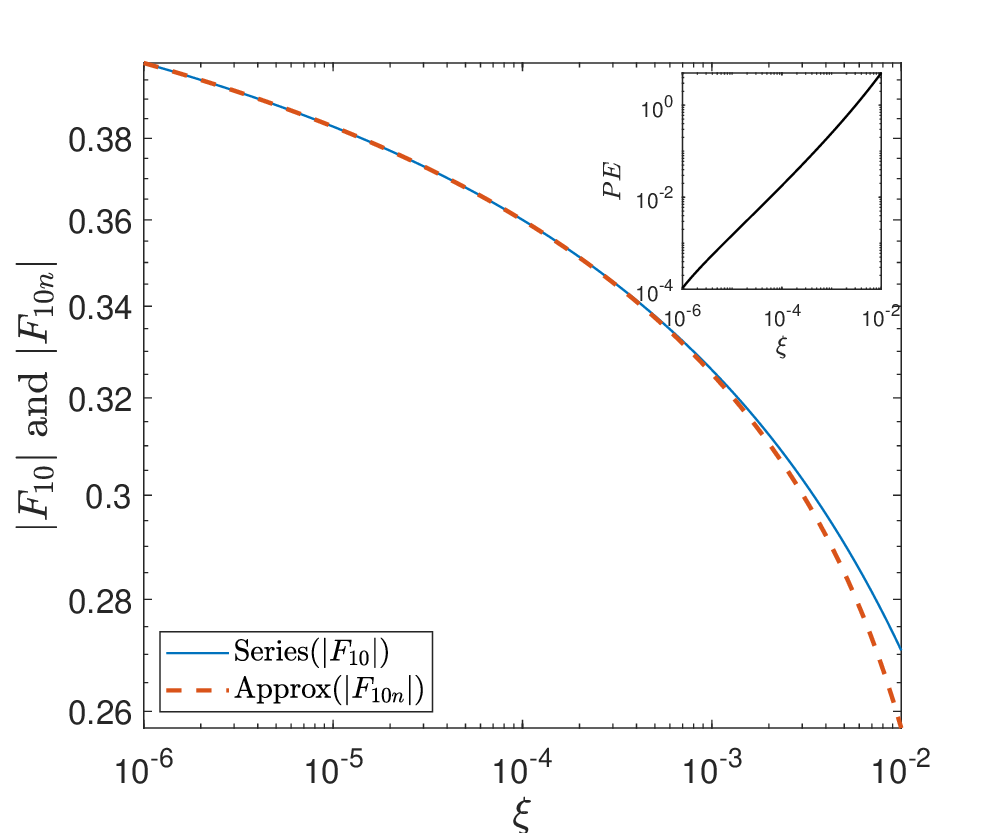}
	\caption{$F_{10}$-series and approximation}
	\label{F_{10}}
\end{figure}

\newpage

\subsection{Forces on charged conducting spheres in the presence of an external electric field}\label{subsec:6.4}
\begin{figure}[!htbp]
	\includegraphics[scale=0.8,angle=0]{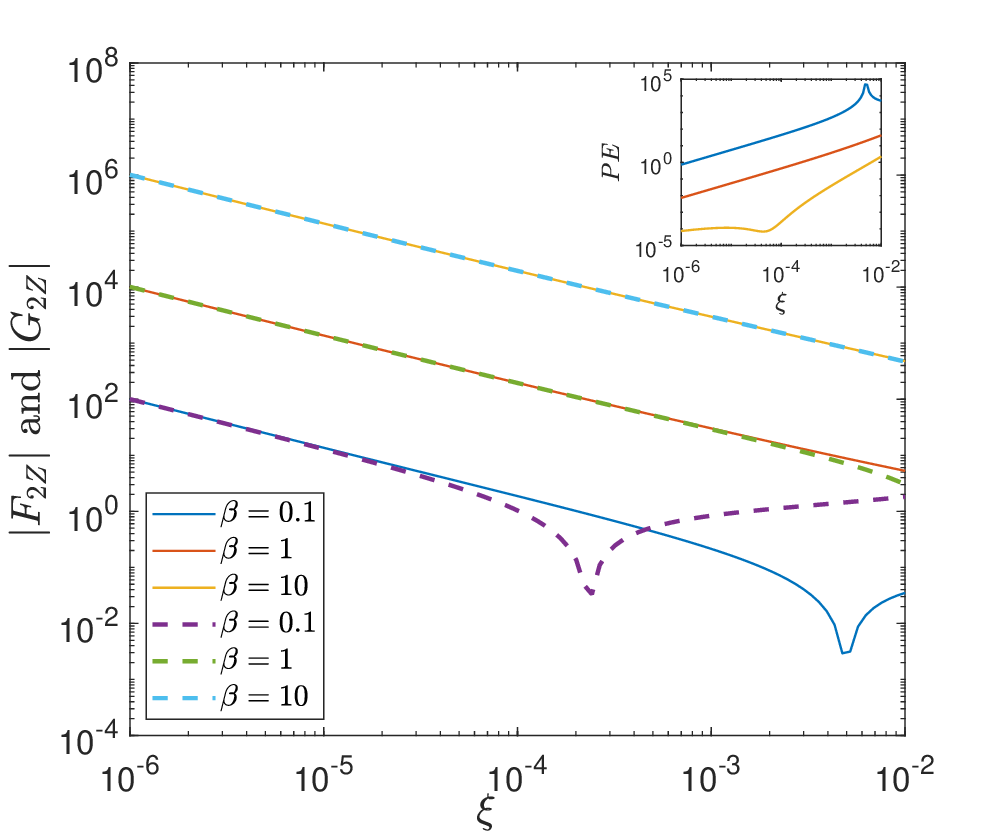}
	\caption{Force on $Z$ direction for charged conducting spheres with the external electric field of angle $\theta=\pi/4$,$\kappa=1$ and $\alpha=1$. $F_{2Z}$ is the force component in the $Z$ direction for sphere 2 from \cite{davis1964two}. $G_{2Z}$ is the approximation of the force component in the $Z$ direction }
	\label{F_z}
\end{figure}

\begin{figure}[!htbp]
	\includegraphics[scale=0.8,angle=0]{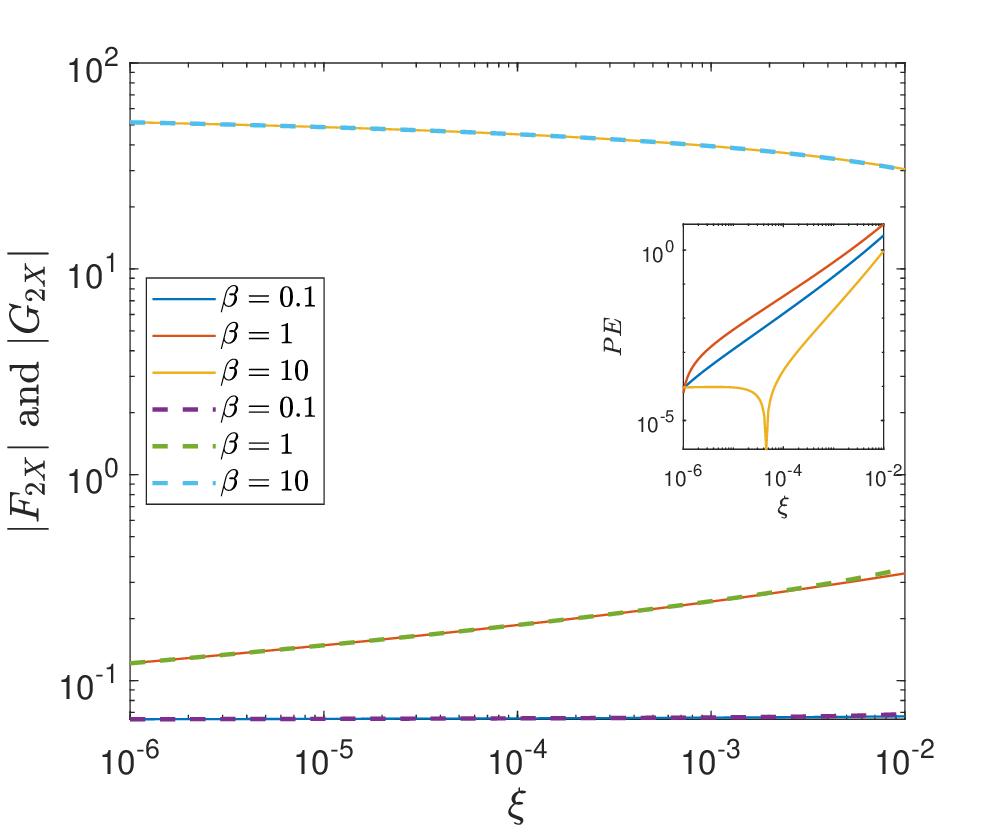}
	\caption{Force on $x$ direction for equally charged conducting spheres with the external electric field of angle $\theta=\pi/4$,$\kappa=1$ and $\alpha=1$. $F_{2X}$ is the force component in the $X$ direction for sphere 2 from \cite{davis1964two}. $G_{2X}$ is the approximation of the force component in the $X$ direction}
	\label{F_x}
\end{figure}

In the previous subsection \ref{subsec:6.3}, we obtained the values of force coefficients $F_1$, $F_2$, ..., $F_{10}$. We can now use these values to determine the components of the force on sphere 2 in the $Z$ and $X$ directions. The plots of the force components $F_{2Z}$ and $F_{2X}$ along with their corresponding approximations $G_{2Z}$ and $G_{2X}$ respectively, as a function of separation $\xi$ maintaining size ratio $\kappa = 1$, charge ratio $\alpha=1$ and the angle between external electric field and the line joining centres of the spheres ($\theta=45^{\circ}$) are as shown in the figures (\ref{F_z}) and (\ref{F_x}) respectively. We consider  absolute values of both the force components(continuous line) and its approximations(dotted line) as plots are on a $\log-\log$ scale with separation range from $10^{-6}$ to $10^{-2}$.  The approximations to the components of the force have been compared with those of \cite{davis1964two} for varying values of $\beta$. The findings indicate that the approximations for $\beta=1$ and $\beta=10$ align with Davis's approximations. However, a greater approximation error has been observed in the $Z$ direction for $\beta=0.1$. These results suggest the need for further investigation into the accuracy and reliability of approximations for $\beta=0.1$. The series expression by \cite{davis1964two} for all $\beta$ values matches our approximation of force in the $X$ direction. The separation($\xi$) between the spheres is inversely proportional to the forces in the $Z$ direction and directly proportional to the forces in the $X$ direction.



\clearpage
\section{Conclusion}\label{sec:7}
In this work, we deal with the problem of two charged conducting spheres whose separation tends to zero, and hence the expressions are functions of both the size ratio($\kappa$) and the separation($\xi$). There are a total of 24 series in the expressions for electrostatic force. 
\begin{itemize}
	\item As a result, we have developed asymptotic approximations for each of these 24 series.
	\item We have compared the asymptotic approximations with the its actual series and created percentage error plots for both. Here, these approximations have been derived for equal-sized conducting spheres ($\kappa=1$).
  \item  Hence, compute the 10 force coefficients, namely $F_1$, $F_2$,..., $F_{10}$ and finally apply them to compute the electrostatic force for small separations. 
\end{itemize}  


A similar approach can also be used to derive the expressions for various values of the size ratio ($\kappa$). 
\bmhead{Acknowledgements}
The authors would like to express their gratitude to Dr. Anubhab Roy, Department of Applied Mechanics and Biomedical Engineering, IIT Madras, India, for valuable discussions. They would also like to acknowledge the National Institute of Technology Karnataka, Surathkal, for providing financial support.

\section*{Declarations}
The authors declare no conflict of interest.
\begin{appendices}

\vspace{-1.5cm}
	
 \section{Asymptotic expression for the rest of the 22 infinite series}	\label{4Aapp}
\vspace{-0.5cm}
\paragraph*{\\}
In this Appendix, we are going to derive the asymptotic expression for the remaining 22 series. Recall Equations.\ref{E1} and \ref{E2}
\begin{eqnarray*}
	\displaystyle T_m(p)&=&\sum_{n=0}^{\infty} \dfrac{ (2n+1)^m e^{(2n+1)p}}{(e^{(2n+1)(\eta_1+\eta_2)}-1)^2}
	\\
	U_m(p)&=&\sum_{n=0}^{\infty} \dfrac{(2n+1)^m e^{(2n+1)p}}{(e^{(2n+1)(\eta_1+\eta_2)}-1) (e^{(2n+3)(\eta_1+\eta_2)}-1)}
\end{eqnarray*}
where, $m$=0, 1, 2, 3 and $p$=$\eta_1$, $2 \eta_1$,  $3 \eta_1$, which are 24 in number.

\paragraph*{Asymptotic expression for the series $T_1(\eta_1)$: }
Substituting $m=0$ and $p=\eta_1$ in Equation(\ref{E1}), we get $T_1(\eta_1)$ as follows,
\begin{eqnarray}
	T_1(\eta_1)&=&\sum_{n=0}^{\infty}\dfrac{ (2n+1) e^{(2n+1)\eta_1}}{(e^{(4n+2)\eta_1}-1)^2}
	\label{2_E1}
\end{eqnarray}
As outlined in the previous section, the series $T_1(\eta_1)$ is decomposed to an ``inner expansion'' ($f_{i}$) and an ``outer expansion'' ($f_{o}$) as follows:  
\begin{eqnarray}
	T_1(\eta_1)&=& f_{i}(\eta_1,N)+f_{o}(\eta_1,N)
	\label{2_E2}
\end{eqnarray}

where, 
\begin{eqnarray}
	f_{i}(\eta_1,N)&=&\sum_{n=0}^{N} \dfrac{ (2n+1) e^{(2n+1)\eta_1}}{(e^{(4n+2)\eta_1}-1)^2}
	\label{2_E3}
	\\
	f_{o}(\eta_1,N)&=&\sum_{n=N+1}^{\infty} \dfrac{ (2n+1) e^{(2n+1)\eta_1}}{(e^{(4n+2)\eta_1}-1)^2} 
	\label{2_E4}
\end{eqnarray}

Following similar steps as shown in  Section\ref{sec:3}, we find the inner ($f_i$) and outer ($f_o$) expansions for $T_1(\eta_1)$ in the next two subsections.
\subparagraph*{Inner expansion}\label{subsec:A2.1}
In this subsection, we derive the asymptotic expression for the inner expansion of $T_1(\eta_1)$.
The inner expansion can be written as follows,
\begin{eqnarray}
	f_{i}(\eta_1,N)&\approx&\frac{1}{4}(I_1 {\eta_1}^{-2}-I_2 {\eta_1}^{-1}+\frac{1}{6}I_3)
	\label{2_E9}
\end{eqnarray}
where
\begin{eqnarray}
	I_1&=&\sum_{n=0}^{N}\frac{1}{(2n+1)}
	\label{2_E10}\\
	I_2&=&\sum_{n=0}^{N}1
	\label{2_E11}\\
	I_3&=&\sum_{n=0}^{N}2n+1
	\label{2_E12}
\end{eqnarray}
Hence, as $N\to\infty$,
\begin{eqnarray}
	I_1&\sim&\frac{1}{2}\big(\gamma+\log{4N}+\frac{1}{N}\big)
	\label{2_E13}\\
	I_2&=& N+1
	\label{2_E14}\\
	I_3&=&(N+1)^2
	\label{2_E15}
\end{eqnarray}
Here $\gamma$ is Euler's constant, and it is approximately equal to 0.57721.  \\
Substitute Eqns. (\ref{2_E13}),(\ref{2_E14}),(\ref{2_E15}) and in Equation (\ref{2_E9}). Then we get,
\begin{eqnarray}
	f_{i}(\eta_1,N)&\sim&\bigg(\frac{1}{8\eta_1^2}\big(\gamma+\log{4N}+\frac{1}{N}\big)\bigg) -\bigg(\frac{ N+1}{4\eta_1}\bigg)+\frac{(N+1)^2}{24}
	\label{2_E16}
\end{eqnarray}

After introducing intermediate vanish $\displaystyle N=\frac{X}{\eta_1}$ and simplifying then Equation (\ref{2_E16}) becomes
\begin{eqnarray}
	f_{i}(\eta_1,X)&\sim&\frac{\gamma+\log{\frac{4}{\eta_1}}}{8\eta_1^2}-\frac{1}{4 \eta_1}+\frac{1}{24}+\frac{1}{8 X \eta_1}+\frac{\log{X}}{8\eta_1^2}-\frac{X}{4\eta_1^2}
	\label{2_E17}
\end{eqnarray}
Equation (\ref{2_E17}) is the asymptotic inner expansion which involves  $\eta_1$ and $X$ (which should vanish after summing with the outer expansion). 
In the next subsection (\ref{subsec:A2.2}), we derive the outer expansion for $T_1(\eta_1)$.
\subparagraph*{Outer expansion }\label{subsec:A2.2}
In order to derive the outer expansion $f_{o}$, let $m=n\eta_1$ in Equation (\ref{2_E4}) and examine the scenario where $\eta_1\to0$ with $m$ held fixed. After substituting the $n$ value in Equation (\ref{2_E4}), we get outer expansion in terms of $m$ and $\eta_1$ as follows,  
\begin{eqnarray}
	f_{o}(\eta_1,N)&=&\sum_{\substack{m=n\eta_1 \\ n=N+1}}^{\infty} \dfrac{ (2m+\eta_1)e^{2m+\eta_1}}{\eta_1(e^{4m+2\eta_1}-1)^2}  
	\label{2_E18}
\end{eqnarray}
For fixed $m$,  the function expansion for $\eta_1\to 0$ is as follows:
\begin{eqnarray}
	f_{o}(\eta_1,N)&=& \frac{1}{\eta_1} \sum_{\substack{m=n\eta_1 \\ n=N+1}}^{\infty}\Bigg(\frac{2m e^{2m}}{(-1+e^{4m})^2}-\frac{e^{2m}\bigg(1+2m+(6m-1)e^{4m}\bigg)}{(-1+e^{4m})^3}\eta_1 \nonumber\\
	&&+\frac{e^{2m}\bigg(1+m+(14m+2)e^{4m}-(9m-3)e^{8m}\bigg)}{(-1+e^{4m})^4}\eta_1^2\Bigg)
	\label{2_E19}
\end{eqnarray}

\begin{eqnarray}
	f_{o}(\eta_1,N)&=& \frac{1}{\eta_1^2}\sum_{\substack{m=n\eta_1 \\ n=N+1}}^{\infty}f_{1}(m)\Delta m-\frac{1}{\eta_1} \sum_{\substack{m=n\eta_1 \\ n=N+1}}^{\infty}f_{2}(m)\Delta m \nonumber\\
	&&+\sum_{\substack{m=n\eta_1 \\ n=N+1}}^{\infty}f_{3}(m)\Delta m
	\label{2_E20}
\end{eqnarray}

where, 
\begin{eqnarray}
	f_{1}(m)&=&\frac{2m e^{2m}}{(-1+e^{4m})^2}
	\label{2_E21}\\
	f_{2}(m)&=&\frac{e^{2m}\bigg(1+2m+(6m-1)e^{4m}\bigg)}{(-1+e^{4m})^3}
	\label{2_E22}\\
	f_{3}(m)&=&\frac{e^{2m}\bigg(1+m+(14m+2)e^{4m}-(9m-3)e^{8m}\bigg)}{(-1+e^{4m})^4}
	\label{2_E23}
\end{eqnarray}
Define $\Delta m =m_{n+1}-m_n $. This makes $\Delta m=(n+1)\eta_1-n\eta_1=\eta_1$.
Certainly, the outer expansion can be expressed in the following manner: 
\begin{eqnarray}
	f_{o}(\eta_1,X)&=&\frac{1}{\eta_1^2}\sum_{m=X}^{\infty}f_{1}(m)\Delta m -\frac{1}{\eta_1}\sum_{m=X}^{\infty}f_{2}(m)\Delta m \nonumber\\
	&&+\sum_{\substack{m=X}}^{\infty}f_{3}(m)\Delta m
	\label{2_E24}
\end{eqnarray}

As per the definition of $X$, it is apparent that $m=X$ at the lower bound of summation. As $\Delta m\to 0$, this expression remains asymptotically precise. Now, from Eqns. (\ref{2_E21}),(\ref{2_E22}) we get $f_1(\infty)=0$,$f_2(\infty)=0$. Furthermore, as $X\to 0$, 
\begin{eqnarray}
	f_{1}(X)\sim \dfrac{1}{8 X}-\dfrac{1}{4}
	\label{2_E25}
	\\
	f_{2}(X)\sim \dfrac{1}{16 X^2}-\dfrac{1}{24}
	\label{2_E26}\\
	f_{3}(X)\sim\dfrac{1}{32 X^3}
	\label{2_E27}
\end{eqnarray}
Thus, since $\Delta m=\eta_1$, we have from Eqns. (\ref{1_E24}),(\ref{2_E24})-(\ref{2_E27})  that as $\eta_1\to0$,
\begin{eqnarray} 
	f_{o}(\eta_1,X) &=& \frac{1}{\eta_1^2}\int_{X}^{\infty}f_{1}(m)dm-\frac{1}{16X\eta_1}+\frac{1}{8\eta_1}\nonumber\\
	&& -\frac{1}{\eta_1}\int_{X}^{\infty}f_{2}(m)dm+\frac{1}{48}+\int_{X}^{\infty}f_{3}(m)dm+\textit{o(1)}
	\label{2_E28}
\end{eqnarray}
The integral provided above converges for sufficiently large values of $m$. Now, as $\Delta m \to 0$, after simplification we get 
\begin{eqnarray}
	f_{o}(\eta_1,X)&\sim&-\frac{1}{8 X \eta_1}-\frac{\log{X}}{8\eta_1^2}+\frac{X}{4\eta_1^2}\nonumber\\
	&&-\frac{1}{4 \eta_1^2}+\frac{1}{6\eta_1}-\frac{1}{48}+\frac{K_{21}}{\eta_1^2}-\frac{K_{22}}{\eta_1}+C_{25}
	\label{2_E29}
\end{eqnarray}
\\ where,
\begin{eqnarray*}
	K_{21}&=&C_{21}+C_{22}=0.0665749 \\
	K_{22}&=&C_{23}+C_{24}=-0.0833333\\
	C_{21}&=&\int_{1}^{\infty} {f_1(m)}dm = 0.000984324\nonumber\\
	C_{22}&=&\int_{0}^{1} \Bigg({f_1(m)}-\frac{1}{8m}
	+\frac{1}{4}\Bigg)dm =0.06559053\nonumber\\
	C_{23}&=&\int_{1}^{\infty} \Bigg({f_2(m)}-\frac{1}{16m^2}\Bigg)dm =-0.0599279\nonumber\\
	C_{24}&=&\int_{0}^{1} \Bigg({f_2(m)}-\frac{1}{16m^2} +\frac{1}{24}\Bigg)dm=-0.0234054\\
	C_{25}&=&\int_{0}^{\infty} \Bigg({f_3(m)}-\frac{1}{32m^3}\Bigg)dm=-0.0104167
\end{eqnarray*}
Equation (\ref{2_E29}) is the asymptotic outer expansion, which also involves  $\eta_1$ and $X$.

Adding Equation (\ref{2_E17}) and Equation (\ref{2_E29}), we obtain the asymptotic expression for $T_1(\eta_1)$ in the the small separation region as follows:
\begin{eqnarray}
	T_1(\eta_1)&\sim&\frac{\gamma+\log{\frac{4}{\eta_1}}}{8\eta_1^2}-\frac{1}{4 \eta_1^2}-\frac{1}{12\eta_1}+\frac{1}{48}+\frac{K_{21}}{\eta_1^2}-\frac{K_{22}}{\eta_1}+C_{25}
	\label{2_E30}
\end{eqnarray}
At this stage, it can be observed that the final asymptotic expression is independent of $X$.
\paragraph*{Asymptotic expression for the series $T_2(\eta_1)$: } 
Substituting $m=0$ and $p=\eta_1$ in Equation(\ref{E1}), we get $T_2(\eta_1)$ as follows,
\begin{eqnarray}
	T_2(\eta_1)&=&\sum_{n=0}^{\infty}\dfrac{ (2n+1)^2 e^{(2n+1)\eta_1}}{(e^{(4n+2)\eta_1}-1)^2}
	\label{3_E1}
\end{eqnarray}
As outlined in the previous section, the series $T_2(\eta_1)$ is decomposed to an ``inner expansion'' ($f_{i}$) and an ``outer expansion'' ($f_{o}$) as follows:  
\begin{eqnarray}
	T_2(\eta_1)&=& f_{i}(\eta_1,N)+f_{o}(\eta_1,N)
	\label{3_E2}
\end{eqnarray}

where, 
\begin{eqnarray}
	f_{i}(\eta_1,N)&=&\sum_{n=0}^{N} \dfrac{ (2n+1)^2 e^{(2n+1)\eta_1}}{(e^{(4n+2)\eta_1}-1)^2}
	\label{3_E3}
	\\
	f_{o}(\eta_1,N)&=&\sum_{n=N+1}^{\infty} \dfrac{ (2n+1)^2 e^{(2n+1)\eta_1}}{(e^{(4n+2)\eta_1}-1)^2} 
	\label{3_E4}
\end{eqnarray}

Following similar steps as shown in  Section\ref{sec:3}, we find the inner ($f_i$) and outer ($f_o$) expansions for $T_2(\eta_1)$ in the next two subsections.
\subparagraph*{Inner expansion}\label{subsec:A3.1}
In this subsection, we derive the asymptotic expression for the inner expansion of $T_2(\eta_1)$.
The inner expansion can be written as follows,
\begin{eqnarray}
	f_{i}(\eta_1,N)&\approx&\frac{1}{4}(I_1 {\eta_1}^{-2}-I_2 {\eta_1}^{-1}+\frac{1}{6}I_3)
	\label{3_E9}
\end{eqnarray}
where
\begin{eqnarray}
	I_1&=&\sum_{n=0}^{N}1
	\label{3_E10}\\
	I_2&=&\sum_{n=0}^{N}2n+1
	\label{3_E11}\\
	I_3&=&\sum_{n=0}^{N}(2n+1)^2
	\label{3_E12}
\end{eqnarray}
Hence, as $N\to\infty$,
\begin{eqnarray}
	I_1&=&1+\textit{O(N)}
	\label{3_E13}\\
	I_2&=& 1+\textit{O(N)}
	\label{3_E14}\\
	I_3&=&1+\textit{O(N)}
	\label{3_E15}
\end{eqnarray}
Here $\gamma$ is Euler's constant, and it is approximately equal to 0.57721.  \\
Substitute Eqns. (\ref{3_E13}),(\ref{3_E14}),(\ref{3_E15}) and in Equation (\ref{3_E9}). Then we get,
\begin{eqnarray}
	f_{i}(\eta_1,N)&\sim&\frac{1}{4\eta_1^2} -\frac{ 1}{4\eta_1}+\frac{1}{24}
	\label{3_E16}
\end{eqnarray}

Equation (\ref{3_E16}) is the asymptotic inner expansion which involves  $\eta_1$. 
In the next subsection (\ref{subsec:A3.2}), we derive the outer expansion for $T_2(\eta_1)$.
\subparagraph*{Outer expansion }\label{subsec:A3.2}
In order to derive the outer expansion $f_{o}$, let $m=n\eta_1$ in Equation (\ref{3_E4}) and examine the scenario where $\eta_1\to0$ with $m$ held fixed. After substituting the $n$ value in Equation (\ref{3_E4}), we get outer expansion in terms of $m$ and $\eta_1$ as follows,  
\begin{eqnarray}
	f_{o}(\eta_1,N)&=&\sum_{\substack{m=n\eta_1 \\ n=N+1}}^{\infty} \dfrac{ (2m+\eta_1)^2e^{2m+\eta_1}}{\eta_1^2(e^{4m+2\eta_1}-1)^2}  
	\label{3_E18}
\end{eqnarray}
For fixed $m$,  the function expansion for $\eta_1\to 0$ is as follows:
\begin{eqnarray}
	f_{o}(\eta_1,N)&=& \frac{1}{\eta_1^2} \sum_{\substack{m=n\eta_1 \\ n=N+1}}^{\infty}\Bigg(\frac{4m^2 e^{2m}}{(-1+e^{4m})^2}-\frac{4me^{2m}\bigg(1+m+(3m-1)e^{4m}\bigg)}{(-1+e^{4m})^3}\eta_1 \nonumber\\
	&&+\frac{e^{2m}\bigg(1+4m+2m^2+(28m^2+8m-2)e^{4m}+(18m^2-12m+1)e^{8m}\bigg)\eta_1^2}{(-1+e^{4m})^4}\nonumber\\
	&&-\frac{e^{2m}}{3(-1+e^{4m})^5}\bigg(3+6m+2m^2+(94m^2+78m+3)e^{4m}+(230m^2-30m-15)e^{8m} \nonumber\\
	&&-(54m^2-54m+9)e^{12m}\bigg)\eta_1^3\Bigg)
	\label{3_E19}
\end{eqnarray}

\begin{eqnarray}
	f_{o}(\eta_1,N)&=& \frac{1}{\eta_1^3}\sum_{\substack{m=n\eta_1 \\ n=N+1}}^{\infty}f_{1}(m)\Delta m-\frac{1}{\eta_1^2} \sum_{\substack{m=n\eta_1 \\ n=N+1}}^{\infty}f_{2}(m)\Delta m \nonumber\\
	&&+\frac{1}{\eta_1} \sum_{\substack{m=n\eta_1 \\ n=N+1}}^{\infty}f_{3}(m)\Delta m- \sum_{\substack{m=n\eta_1 \\ n=N+1}}^{\infty}f_{4}(m)\Delta m
	\label{3_E20}
\end{eqnarray}

where, 
\begin{eqnarray}
	f_{1}(m)&=&\frac{4m^2 e^{2m}}{(-1+e^{4m})^2}
	\label{3_E21}\\
	f_{2}(m)&=&\frac{4me^{2m}\bigg(1+m+(3m-1)e^{4m}\bigg)}{(-1+e^{4m})^3}
	\label{3_E22}\\
	f_{3}(m)&=&\frac{e^{2m}\bigg(1+4m+2m^2+(28m^2+8m-2)e^{4m}+(18m^2-12m+1)e^{8m}\bigg)}{(-1+e^{4m})^4}
	\label{3_E23}\\
	f_{4}(m)&=&\frac{e^{2m}}{3(-1+e^{4m})^5}\bigg(3+6m+2m^2+(94m^2+78m+3)e^{4m}+(230m^2-30m-15)e^{8m} \nonumber\\
	&&-(54m^2-54m+9)e^{12m}\bigg)
	\label{3_E24}
\end{eqnarray}
Define $\Delta m =m_{n+1}-m_n $. This makes $\Delta m=(n+1)\eta_1-n\eta_1=\eta_1$.
Certainly, the outer expansion can be expressed in the following manner: 
\begin{eqnarray}
	f_{o}(\eta_1,X)&=&\frac{1}{\eta_1^3}\sum_{m=X}^{\infty}f_{1}(m)\Delta m -\frac{1}{\eta_1^2}\sum_{m=X}^{\infty}f_{2}(m)\Delta m \nonumber\\
	&&+\frac{1}{\eta_1}\sum_{\substack{m=X}}^{\infty}f_{3}(m)\Delta m-\sum_{\substack{m=X}}^{\infty}f_{4}(m)\Delta m
	\label{3_E25}
\end{eqnarray}

As per the definition of $X$, it is apparent that $m=X$ at the lower bound of summation. As $\Delta m\to 0$, this expression remains asymptotically precise. Now, from Eqns. (\ref{3_E21})-(\ref{3_E24}) we get $f_1(\infty)=0$,$f_2(\infty)=0$,$f_3(\infty)=0$,$f_4(\infty)=0$. Furthermore, as $X\to 0$, 
\begin{eqnarray}
	f_{1}(X)\sim \dfrac{1}{4}
	\label{3_E26}\\
	f_{2}(X)\sim \dfrac{1}{4}
	\label{3_E27}\\
	f_{3}(X)\sim\dfrac{1}{24}
	\label{3_E28}\\
	f_{4}(X)\sim \textit{O(1)}
	\label{3_E29}
\end{eqnarray}
Thus, since $\Delta m=\eta_1$, we have from Eqns. (\ref{1_E24}),(\ref{3_E25})-(\ref{3_E29})  that as $\eta_1\to0$,
\begin{eqnarray} 
	f_{o}(\eta_1,X) &=& \frac{1}{\eta_1^3}\int_{X}^{\infty}f_{1}(m)dm-\frac{1}{8\eta_1^2} -\frac{1}{\eta_1^2}\int_{X}^{\infty}f_{2}(m)dm-\frac{1}{8\eta_1}\nonumber\\
	&&+\frac{1}{\eta_1}\int_{X}^{\infty}f_{3}(m)dm-\frac{1}{48}-\int_{X}^{\infty}f_{4}(m)dm+\textit{o(1)}
	\label{3_E30}
\end{eqnarray}
The integral provided above converges for sufficiently large values of $m$. Now, as $\Delta m \to 0$, after simplification we get 
\begin{eqnarray}
	f_{o}(\eta_1,X)&\sim&\frac{1}{4\eta_1^3}-\frac{3}{8 \eta_1^2}+\frac{1}{6\eta_1}-\frac{1}{48}+\frac{K_{31}}{\eta_1^3}-\frac{K_{32}}{\eta_1^2}+\frac{K_{33}}{\eta_1}-C_{37}
	\label{3_E31}
\end{eqnarray}
\\ where,
\begin{eqnarray*}
	K_{31}&=&C_{31}+C_{32}=-0.15905 \\
	K_{32}&=&C_{33}+C_{34}=-0.125\\
	K_{33}&=&C_{35}+C_{36}=0.0208333\\
	C_{31}&=&\int_{1}^{\infty} {f_1(m)}dm = 0.00234029\nonumber\\
	C_{32}&=&\int_{0}^{1} \Bigg({f_1(m)}-\frac{1}{4}\Bigg)dm =-0.16139\nonumber\\
	C_{33}&=&\int_{1}^{\infty} {f_2(m)}dm =0.005144\nonumber\\
	C_{34}&=&\int_{0}^{1} \Bigg({f_2(m)}-\frac{1}{4}\Bigg)dm=-0.130144\\
	C_{35}&=&\int_{1}^{\infty} {f_3(m)}dm =0.0053361\nonumber\\
	C_{36}&=&\int_{0}^{1} \Bigg({f_3(m)}-\frac{1}{24}\Bigg)dm=0.0154972\\
	C_{37}&=&\int_{0}^{\infty} {f_4(m)}dm=0.00698606
\end{eqnarray*}
Equation (\ref{3_E31}) is the asymptotic outer expansion, which also involves  $\eta_1$.

Adding Equation (\ref{3_E16}) and Equation (\ref{3_E31}), we obtain the asymptotic expression for $T_2(\eta_1)$ in the the small separation region as follows:
\begin{eqnarray}
	T_2(\eta_1)&\sim&\frac{1}{4\eta_1^3}-\frac{1}{8 \eta_1^2}-\frac{1}{12\eta_1}+\frac{1}{48}+\frac{K_{31}}{\eta_1^3}-\frac{K_{32}}{\eta_1^2}+\frac{K_{33}}{\eta_1}-C_{37}
	\label{3_E32}
\end{eqnarray}
At this stage, it can be observed that the final asymptotic expression is independent of $X$.
\paragraph*{Asymptotic expression for the series $T_3(\eta_1)$: }
Substituting $m=0$ and $p=\eta_1$ in Equation(\ref{E1}), we get $T_3(\eta_1)$ as follows,
\begin{eqnarray}
	T_3(\eta_1)&=&\sum_{n=0}^{\infty}\dfrac{ (2n+1)^3 e^{(2n+1)\eta_1}}{(e^{(4n+2)\eta_1}-1)^2}
	\label{4_E1}
\end{eqnarray}
As outlined in the previous section, the series $T_3(\eta_1)$ is decomposed to an ``inner expansion'' ($f_{i}$) and an ``outer expansion'' ($f_{o}$) as follows:  
\begin{eqnarray}
	T_3(\eta_1)&=& f_{i}(\eta_1,N)+f_{o}(\eta_1,N)
	\label{4_E2}
\end{eqnarray}

where, 
\begin{eqnarray}
	f_{i}(\eta_1,N)&=&\sum_{n=0}^{N} \dfrac{ (2n+1)^3 e^{(2n+1)\eta_1}}{(e^{(4n+2)\eta_1}-1)^2}
	\label{4_E3}
	\\
	f_{o}(\eta_1,N)&=&\sum_{n=N+1}^{\infty} \dfrac{ (2n+1)^3 e^{(2n+1)\eta_1}}{(e^{(4n+2)\eta_1}-1)^2} 
	\label{4_E4}
\end{eqnarray}

Following similar steps as shown in  Section\ref{sec:3}, we find the inner ($f_i$) and outer ($f_o$) expansions for $T_3(\eta_1)$ in the next two subsections.
\subparagraph*{Inner expansion}\label{subsec:A4.1}
In this subsection, we derive the asymptotic expression for the inner expansion of $T_3(\eta_1)$.
The inner expansion can be written as follows,
\begin{eqnarray}
	f_{i}(\eta_1,N)&\approx&\frac{1}{4}(I_1 {\eta_1}^{-2}-I_2 {\eta_1}^{-1}+\frac{1}{6}I_3)
	\label{4_E9}
\end{eqnarray}
where
\begin{eqnarray}
	I_1&=&\sum_{n=0}^{N}2n+1
	\label{4_E10}\\
	I_2&=&\sum_{n=0}^{N}(2n+1)^2
	\label{4_E11}\\
	I_3&=&\sum_{n=0}^{N}(2n+1)^3
	\label{4_E12}
\end{eqnarray}
Hence, as $N\to\infty$,
\begin{eqnarray}
	I_1&=&1+\textit{O(N)}
	\label{4_E13}\\
	I_2&=& 1+\textit{O(N)}
	\label{4_E14}\\
	I_3&=&1+\textit{O(N)}
	\label{4_E15}
\end{eqnarray}
Here $\gamma$ is Euler's constant, and it is approximately equal to 0.57721.  \\
Substitute Eqns. (\ref{4_E13}),(\ref{4_E14}),(\ref{4_E15}) and in Equation (\ref{4_E9}). Then we get,
\begin{eqnarray}
	f_{i}(\eta_1,N)&\sim&\frac{1}{4\eta_1^2} -\frac{ 1}{4\eta_1}+\frac{1}{24}
	\label{4_E16}
\end{eqnarray}

Equation (\ref{4_E16}) is the asymptotic inner expansion which involves  $\eta_1$. 
In the next subsection (\ref{subsec:A4.2}), we derive the outer expansion for $T_3(\eta_1)$.
\subparagraph*{Outer expansion }\label{subsec:A4.2}
In order to derive the outer expansion $f_{o}$, let $m=n\eta_1$ in Equation (\ref{4_E4}) and examine the scenario where $\eta_1\to0$ with $m$ held fixed. After substituting the $n$ value in Equation (\ref{4_E4}), we get outer expansion in terms of $m$ and $\eta_1$ as follows,  
\begin{eqnarray}
	f_{o}(\eta_1,N)&=&\sum_{\substack{m=n\eta_1 \\ n=N+1}}^{\infty} \dfrac{ (2m+\eta_1)^3e^{2m+\eta_1}}{\eta_1^3(e^{4m+2\eta_1}-1)^2}  
	\label{4_E18}
\end{eqnarray}
For fixed $m$,  the function expansion for $\eta_1\to 0$ is as follows:
\begin{eqnarray}
	f_{o}(\eta_1,N)&=& \frac{1}{\eta_1^3} \sum_{\substack{m=n\eta_1 \\ n=N+1}}^{\infty}\Bigg(\frac{8m^3 e^{2m}}{(-1+e^{4m})^2}-\frac{4m^2e^{2m}\bigg(3+2m+(6m-3)e^{4m}\bigg)}{(-1+e^{4m})^3}\eta_1 \nonumber\\
	&&+\frac{2me^{2m}\bigg(3+6m+2m^2+(28m^2+12m-6)e^{4m}+(18m^2-18m+3)e^{8m}\bigg)\eta_1^2}{(-1+e^{4m})^4}\nonumber\\
	&&-\frac{\eta_1^3 e^{2m}}{3(-1+e^{4m})^5}\bigg(3+18m+18m^2+4m^3+(196m^3+234m^2+18m-9)e^{4m} \nonumber\\
	&&+(460m^3-90m^2-90m+9)e^{8m}+(108m^3-162m^2+54m-3)e^{12m}\bigg)\nonumber\\
	&&+\frac{\eta_1^4 e^{2m}}{3(-1+e^{4m})^6}\bigg(3+9m+6m^2+m^3+(156m^3+288m^2+108m)e^{4m} \nonumber\\
	&&+(918m^3+396m^2-162m-18)e^{8m}+(764m^3-528m^2-36m+24)e^{12m}\nonumber\\
	&&+(81m^3-162m^2+81m-9)e^{16m}\bigg)\Bigg)
	\label{4_E19}
\end{eqnarray}

\begin{eqnarray}
	f_{o}(\eta_1,N)&=& \frac{1}{\eta_1^4}\sum_{\substack{m=n\eta_1 \\ n=N+1}}^{\infty}f_{1}(m)\Delta m-\frac{1}{\eta_1^3} \sum_{\substack{m=n\eta_1 \\ n=N+1}}^{\infty}f_{2}(m)\Delta m +\frac{1}{\eta_1^2} \sum_{\substack{m=n\eta_1 \\ n=N+1}}^{\infty}f_{3}(m)\Delta m \nonumber\\
	&&- \frac{1}{\eta_1} \sum_{\substack{m=n\eta_1 \\ n=N+1}}^{\infty}f_{4}(m)\Delta m+\sum_{\substack{m=n\eta_1 \\ n=N+1}}^{\infty}f_{5}(m)\Delta m
	\label{4_E20}
\end{eqnarray}

where, 
\begin{eqnarray}
	f_{1}(m)&=&\frac{8m^3 e^{2m}}{(-1+e^{4m})^2}
	\label{4_E21}\\
	f_{2}(m)&=&\frac{4m^2e^{2m}\bigg(3+2m+(6m-3)e^{4m}\bigg)}{(-1+e^{4m})^3}
	\label{4_E22}\\
	f_{3}(m)&=&\frac{2me^{2m}}{(-1+e^{4m})^4}\bigg(3+6m+2m^2+(28m^2+12m-6)e^{4m} \nonumber\\
	&&+(18m^2-18m+3)e^{8m}\bigg)
	\label{4_E23}\\
	f_{4}(m)&=&\frac{e^{2m}}{3(-1+e^{4m})^5}\bigg(3+18m+18m^2+4m^3+(196m^3+234m^2+18m-9)e^{4m} \nonumber\\
	&&+(460m^3-90m^2-90m+9)e^{8m}+(108m^3-162m^2+54m-3)e^{12m}\bigg)
	\label{4_E24}\\
	f_{5}(m)&=&\frac{e^{2m}}{3(-1+e^{4m})^6}\bigg(3+9m+6m^2+m^3+(156m^3+288m^2+108m)e^{4m} \nonumber\\
	&&+(918m^3+396m^2-162m-18)e^{8m}+(764m^3-528m^2-36m+24)e^{12m}\nonumber\\
	&&+(81m^3-162m^2+81m-9)e^{16m}\bigg)
	\label{4_E25}
\end{eqnarray}
Define $\Delta m =m_{n+1}-m_n $. This makes $\Delta m=(n+1)\eta_1-n\eta_1=\eta_1$.
Certainly, the outer expansion can be expressed in the following manner: 
\begin{eqnarray}
	f_{o}(\eta_1,X)&=&\frac{1}{\eta_1^4}\sum_{m=X}^{\infty}f_{1}(m)\Delta m -\frac{1}{\eta_1^3}\sum_{m=X}^{\infty}f_{2}(m)\Delta m +\frac{1}{\eta_1^2}\sum_{\substack{m=X}}^{\infty}f_{3}(m)\Delta m
	\nonumber\\
	&&-\frac{1}{\eta_1}\sum_{\substack{m=X}}^{\infty}f_{4}(m)\Delta m+\sum_{\substack{m=X}}^{\infty}f_{5}(m)\Delta m
	\label{4_E26}
\end{eqnarray}

As per the definition of $X$, it is apparent that $m=X$ at the lower bound of summation. As $\Delta m\to 0$, this expression remains asymptotically precise. Now, from Eqns. (\ref{4_E21})-(\ref{4_E25}) we get $f_1(\infty)=0$,$f_2(\infty)=0$,$f_3(\infty)=0$,$f_4(\infty)=0$,$f_5(\infty)=0$. Furthermore, as $X\to 0$, 
\begin{eqnarray}
	f_{1}(X)\sim  \textit{O(X)}
	\label{4_E27}\\
	f_{2}(X)\sim \dfrac{-1}{4}
	\label{4_E28}\\
	f_{3}(X)\sim\dfrac{-1}{4}
	\label{4_E29}\\
	f_{4}(X)\sim\dfrac{-1}{24}
	\label{4_E30}\\
	f_{5}(X)\sim \textit{O(1)}
	\label{4_E31}
\end{eqnarray}
Thus, since $\Delta m=\eta_1$, we have from Eqns. (\ref{1_E24}),(\ref{4_E26})-(\ref{4_E31})  that as $\eta_1\to0$,
\begin{eqnarray} 
	f_{o}(\eta_1,X) &=& \frac{1}{\eta_1^4}\int_{X}^{\infty}f_{1}(m)dm -\frac{1}{\eta_1^3}\int_{X}^{\infty}f_{2}(m)dm+\frac{1}{8\eta_1^2}+\frac{1}{\eta_1^2}\int_{X}^{\infty}f_{3}(m)dm+\frac{1}{8\eta_1}\nonumber\\
	&&-\frac{1}{\eta_1}\int_{X}^{\infty}f_{4}(m)dm+\frac{1}{48}+\int_{X}^{\infty}f_{5}(m)dm+\textit{o(1)}
	\label{4_E32}
\end{eqnarray}
The integral provided above converges for sufficiently large values of $m$. Now, as $\Delta m \to 0$, after simplification we get 
\begin{eqnarray}
	f_{o}(\eta_1,X)&\sim&\frac{1}{4\eta_1^3}-\frac{3}{8 \eta_1^2}+\frac{1}{6\eta_1}-\frac{1}{48}+\frac{C_{41}}{\eta_1^4}\nonumber\\&&
	-\frac{K_{41}}{\eta_1^3}+\frac{K_{42}}{\eta_1^2}-\frac{K_{43}}{\eta_1}+C_{48}
	\label{4_E33}
\end{eqnarray}
\\ where,
\begin{eqnarray*}
	K_{41}&=&C_{42}+C_{43}=0.25 \\
	K_{42}&=&C_{44}+C_{45}=0.1875\\
	K_{43}&=&C_{46}+C_{47}=1.9893\times 10^{-9}\\
	C_{41}&=&\int_{0}^{\infty} {f_1(m)}dm = 0.0556826\nonumber\\
	C_{42}&=&\int_{1}^{\infty} {f_2(m)}dm =0.0102884\nonumber\\
	C_{43}&=&\int_{0}^{1} \Bigg({f_2(m)}+\frac{1}{4}\Bigg)dm=0.239712\\
	C_{44}&=&\int_{1}^{\infty} {f_3(m)}dm =0.00810024\nonumber\\
	C_{45}&=&\int_{0}^{1} \Bigg({f_3(m)}+\frac{1}{4}\Bigg)dm=0.1794\\
	C_{46}&=&\int_{1}^{\infty} {f_4(m)}dm=0.00322628\\
	C_{47}&=&\int_{0}^{1} \Bigg({f_4(m)}+\frac{1}{24}\Bigg)dm =-0.00322628\nonumber\\
	C_{48}&=&\int_{0}^{\infty} {f_5(m)}dm =-0.005243\nonumber\\
\end{eqnarray*}
Equation (\ref{4_E33}) is the asymptotic outer expansion, which also involves  $\eta_1$.

Adding Equation (\ref{4_E16}) and Equation (\ref{4_E33}), we obtain the asymptotic expression for $T_3(\eta_1)$ in the the small separation region as follows:
\begin{eqnarray}
	T_3(\eta_1)&\sim&\frac{1}{4\eta_1^3}-\frac{1}{8 \eta_1^2}-\frac{1}{12\eta_1}+\frac{1}{48}+\frac{C_{41}}{\eta_1^4}
	\nonumber\\&&-\frac{K_{41}}{\eta_1^3}+\frac{K_{42}}{\eta_1^2}-\frac{K_{43}}{\eta_1}+C_{48}
	\label{4_E34}
\end{eqnarray}
At this stage, it can be observed that the final asymptotic expression is independent of $X$.
\paragraph*{Asymptotic expression for the series $T_0(2 \eta_1)$: } 
Substituting $m=0$ and $p=\eta_1$ in Equation(\ref{E1}), we get $T_0(2 \eta_1)$ as follows,
\begin{eqnarray}
	T_0(2 \eta_1)&=&\sum_{n=0}^{\infty} \dfrac{ e^{(4n+2)\eta_1}}{(e^{(4n+2)\eta_1}-1)^2}
	\label{5_E1}
\end{eqnarray}
As outlined in the previous section, the series $T_0(2 \eta_1)$ is decomposed to an ``inner expansion'' ($f_{i}$) and an ``outer expansion'' ($f_{o}$) as follows:  
\begin{eqnarray}
	T_0(2 \eta_1)&=& f_{i}(\eta_1,N)+f_{o}(\eta_1,N)
	\label{5_E2}
\end{eqnarray}

where, 
\begin{eqnarray}
	f_{i}(\eta_1,N)&=&\sum_{n=0}^{N} \dfrac{ e^{(4n+2)\eta_1}}{(e^{(4n+2)\eta_1}-1)^2}
	\label{5_E3}
	\\
	f_{o}(\eta_1,N)&=&\sum_{n=N+1}^{\infty} \dfrac{ e^{(4n+2)\eta_1}}{(e^{(4n+2)\eta_1}-1)^2} 
	\label{5_E4}
\end{eqnarray}
Following similar steps as shown in  Section\ref{sec:3}, we find the inner ($f_i$) and outer ($f_o$) expansions for $T_0(2 \eta_1)$ in the next two subsections.

\subparagraph*{Inner expansion}\label{subsec:A5.1}
In this subsection, we derive the asymptotic expression for the inner expansion of $T_0(2 \eta_1)$.
The inner expansion can be written as follows,
\begin{eqnarray}
	f_{i}(\eta_1,N)&\approx&\frac{1}{4}(I_1 {\eta_1}^{-2}-\frac{1}{3} I_2)
	\label{5_E9}
\end{eqnarray}
where
\begin{eqnarray}
	I_1&=&\sum_{n=0}^{N}\frac{1}{(2n+1)^2}
	\label{5_E10}\\
	I_2&=&\sum_{n=0}^{N}1
	\label{5_E11}
\end{eqnarray}
Hence, as $N\to\infty$,
\begin{eqnarray}
	I_1&\sim&\frac{\pi^2}{8}-\frac{1}{4N}+\frac{1}{4N^2}
	\label{5_E13}\\
	I_2&=&N+1
	\label{5_E14}
\end{eqnarray}
Here $\gamma$ is Euler's constant, and it is approximately equal to 0.57721.  \\
Substitute Eqns. (\ref{5_E13}),(\ref{5_E14}) and in Equation (\ref{5_E9}). Then we get,
\begin{eqnarray}
	f_{i}(\eta_1,N)&\sim&\bigg(\frac{\pi^2}{32}-\frac{1}{16N}+\frac{1}{16N^2}\bigg){\eta_1}^{-2}-\frac{N+1}{12}
	\label{5_E16}
\end{eqnarray}

After introducing intermediate vanish $\displaystyle N=\frac{X}{\eta_1}$ and simplifying then Equation (\ref{5_E16}) becomes
\begin{eqnarray}
	f_{i}(\eta_1,X)&\sim&\frac{\pi^2}{32{\eta_1}^2}-\frac{1}{12} +\frac{1}{16X^2}-\frac{1}{16X \eta_1}-\frac{X\eta_1^{-1}}{12}
	\label{5_E17}
\end{eqnarray}
Equation (\ref{5_E17}) is the asymptotic inner expansion which involves  $\eta_1$ and $X$ (which should vanish after summing with the outer expansion). 
In the next subsection (\ref{subsec:A5.2}), we derive the outer expansion for $T_0(2 \eta_1)$.
\subparagraph*{Outer expansion }\label{subsec:A5.2}
In order to derive the outer expansion $f_{o}$, let $m=n\eta_1$ in Equation (\ref{5_E4}) and examine the scenario where $\eta_1\to0$ with $m$ held fixed. After substituting the $n$ value in Equation (\ref{5_E4}), we get outer expansion in terms of $m$ and $\eta_1$ as follows,  
\begin{eqnarray}
	f_{o}(\eta_1,N)&=&\sum_{\substack{m=n\eta_1 \\ n=N+1}}^{\infty} \dfrac{ e^{4m+2\eta_1}}{(e^{4m+2\eta_1}-1)^2}  
	\label{5_E18}
\end{eqnarray}
For fixed $m$,  the function expansion for $\eta_1\to 0$ is as follows:
\begin{eqnarray}
	f_{o}(\eta_1,N)&=& \sum_{\substack{m=n\eta_1 \\ n=N+1}}^{\infty}\Bigg(\frac{e^{4m}}{(-1+e^{4m})^2}-\frac{2 e^{4m}(1+e^{4m})}{(-1+e^{4m})^3}\eta_1 \Bigg)
	\label{5_E19}
\end{eqnarray}

\begin{eqnarray}
	f_{o}(\eta_1,N)&=& \frac{1}{\eta_1}\sum_{\substack{m=n\eta_1 \\ n=N+1}}^{\infty}f_{1}(m)\Delta m-\sum_{\substack{m=n\eta_1 \\ n=N+1}}^{\infty}f_{2}(m)\Delta m
	\label{5_E20}
\end{eqnarray}

where, 
\begin{eqnarray}
	f_{1}(m)&=&\frac{e^{4m}}{(-1+e^{4m})^2} 
	\label{5_E21}\\
	f_{2}(m)&=&\frac{2e^{4m}(1+e^{4m})}{(-1+e^{4m})^3}
	\label{5_E22}
\end{eqnarray}
Define $\Delta m =m_{n+1}-m_n $. This makes $\Delta m=(n+1)\eta_1-n\eta_1=\eta_1$.
Certainly, the outer expansion can be expressed in the following manner: 
\begin{eqnarray}
	f_{o}(\eta_1,X)&=&\frac{1}{\eta_1}\sum_{m=X}^{\infty}f_{1}(m)\Delta m -\sum_{m=X}^{\infty}f_{2}(m)\Delta m
	\label{5_E23}
\end{eqnarray}

As per the definition of $X$, it is apparent that $m=X$ at the lower bound of summation. As $\Delta m\to 0$, this expression remains asymptotically precise. Now, from Eqns. (\ref{5_E21}),(\ref{5_E22}) we get $f_1(\infty)=0$,$f_2(\infty)=0$. Furthermore, as $X\to 0$, 
\begin{eqnarray}
	f_{1}(X)\sim\dfrac{1}{16 X^2}-\dfrac{1}{12}
	\label{5_E25}
	\\
	f_{2}(X)\sim\dfrac{1}{16 X^3}
	\label{5_E26}
\end{eqnarray}
Thus, since $\Delta m=\eta_1$, we have from Eqns. (\ref{1_E24}),(\ref{5_E23}),(\ref{5_E25}),(\ref{5_E26})  that as $\eta_1\to0$,
\begin{eqnarray} 
	f_{o}(\eta_1,X) &=& \frac{1}{\eta_1}\int_{X}^{\infty}f_{1}(m)dm-\frac{1}{32X^2}\nonumber\\
	&&+\frac{1}{24} -\int_{X}^{\infty}f_{2}(m)dm+\textit{o(1)}
	\label{5_E27}
\end{eqnarray}
The integral provided above converges for sufficiently large values of $m$. Now, as $\Delta m \to 0$, after simplification we get 
\begin{eqnarray}
	f_{o}(\eta_1,X)&\sim&-\frac{1}{16X^2}+\frac{1}{16X \eta_1}+\frac{X\eta_1^{-1}}{12}-\frac{\eta_1^{-1}}{12}+\frac{1}{24}+\frac{K_{51}}{\eta_1}-C_{53}
	\label{5_E28}
\end{eqnarray}
\\ where,
\begin{eqnarray*}
	K_{51}&=&C_{51}+C_{52}=-0.0416667  \\
	C_{51}&=&\int_{1}^{\infty} \Bigg({f_1(m)}-\frac{1}{16m^2}\Bigg)dm = -0.0578357 \nonumber\\
	C_{52}&=&\int_{0}^{1} \Bigg({f_1(m)}-\frac{1}{16m^2}+\frac{1}{12}\Bigg)dm =0.016169\nonumber\\
	C_{53}&=&\int_{0}^{\infty} \Bigg({f_2(m)}-\frac{1}{16m^3}\Bigg)dm =-0.0416667
\end{eqnarray*}
Equation(\ref{5_E28}) is the asymptotic outer expansion, which also involves  $\eta_1$ and $X$.

Adding Equation (\ref{5_E17}) and Equation (\ref{5_E28}), we obtain the asymptotic expression for $T_0(2 \eta_1)$ in the the small separation region as follows:
\begin{eqnarray}
	T_0(2 \eta_1)&\sim&\frac{\pi^2}{32{\eta_1}^2}-\frac{1}{24}
	-\frac{1}{12 \eta_1}+\frac{K_{51}}{\eta_1}-C_{53}
	\label{5_E29}
\end{eqnarray}
At this stage, it can be observed that the final asymptotic expression is independent of $X$.
\paragraph*{Asymptotic expression for the series $T_1(2\eta_1)$: } 
Substituting $m=0$ and $p=\eta_1$ in Equation(\ref{E1}), we get $ T_1(2 \eta_1)$ as follows,
\begin{eqnarray}
	T_1(2 \eta_1)&=&\sum_{n=0}^{\infty}\dfrac{ (2n+1) e^{(4n+2)\eta_1}}{(e^{(4n+2)\eta_1}-1)^2}
	\label{6_E1}
\end{eqnarray}
As outlined in the previous section, the series $ T_1(2 \eta_1)$ is decomposed to an ``inner expansion'' ($f_{i}$) and an ``outer expansion'' ($f_{o}$) as follows:  
\begin{eqnarray}
	T_1(2 \eta_1)&=& f_{i}(\eta_1,N)+f_{o}(\eta_1,N)
	\label{6_E2}
\end{eqnarray}

where, 
\begin{eqnarray}
	f_{i}(\eta_1,N)&=&\sum_{n=0}^{N} \dfrac{ (2n+1) e^{(4n+2)\eta_1}}{(e^{(4n+2)\eta_1}-1)^2}
	\label{6_E3}
	\\
	f_{o}(\eta_1,N)&=&\sum_{n=N+1}^{\infty} \dfrac{ (2n+1) e^{(4n+2)\eta_1}}{(e^{(4n+2)\eta_1}-1)^2} 
	\label{6_E4}
\end{eqnarray}

Following similar steps as shown in  Section\ref{sec:3}, we find the inner ($f_i$) and outer ($f_o$) expansions for $ T_1(2 \eta_1)$ in the next two subsections.
\subparagraph*{Inner expansion}\label{subsec:A6.1}
In this subsection, we derive the asymptotic expression for the inner expansion of $ T_1(2 \eta_1)$.
The inner expansion can be written as follows,
\begin{eqnarray}
	f_{i}(\eta_1,N)&\approx&\frac{1}{4}(I_1 {\eta_1}^{-2}-\frac{1}{3}I_2)
	\label{6_E9}
\end{eqnarray}
where
\begin{eqnarray}
	I_1&=&\sum_{n=0}^{N}\frac{1}{(2n+1)}
	\label{6_E10}\\
	I_2&=&\sum_{n=0}^{N}2n+1
	\label{6_E11}
\end{eqnarray}
Hence, as $N\to\infty$,
\begin{eqnarray}
	I_1&\sim&\frac{1}{2}\big(\gamma+\log{4N}+\frac{1}{N}\big)
	\label{6_E13}\\
	I_2&=& 1+\textit{O(N)}
	\label{6_E14}
\end{eqnarray}
Here $\gamma$ is Euler's constant, and it is approximately equal to 0.57721.  \\
Substitute Eqns. (\ref{6_E13}),(\ref{6_E14}) and in Equation (\ref{6_E9}). Then we get,
\begin{eqnarray}
	f_{i}(\eta_1,N)&\sim&\bigg(\frac{1}{8\eta_1^2}\big(\gamma+\log{4N}+\frac{1}{N}\big)\bigg) -\frac{1}{12}
	\label{6_E16}
\end{eqnarray}

After introducing intermediate vanish $\displaystyle N=\frac{X}{\eta_1}$ and simplifying then Equation (\ref{6_E16}) becomes
\begin{eqnarray}
	f_{i}(\eta_1,X)&\sim&\frac{\gamma+\log{\frac{4}{\eta_1}}}{8\eta_1^2}-\frac{1}{12}+\frac{1}{8 X \eta_1}+\frac{\log{X}}{8\eta_1^2}
	\label{6_E17}
\end{eqnarray}
Equation (\ref{6_E17}) is the asymptotic inner expansion which involves  $\eta_1$ and $X$ (which should vanish after summing with the outer expansion). 
In the next subsection (\ref{subsec:A6.2}), we derive the outer expansion for $ T_1(2 \eta_1)$.
\subparagraph*{Outer expansion }\label{subsec:A6.2}
In order to derive the outer expansion $f_{o}$, let $m=n\eta_1$ in Equation (\ref{6_E4}) and examine the scenario where $\eta_1\to0$ with $m$ held fixed. After substituting the $n$ value in Equation (\ref{6_E4}), we get outer expansion in terms of $m$ and $\eta_1$ as follows,  
\begin{eqnarray}
	f_{o}(\eta_1,N)&=&\sum_{\substack{m=n\eta_1 \\ n=N+1}}^{\infty} \dfrac{ (2m+\eta_1)e^{4m+2\eta_1}}{\eta_1(e^{4m+2\eta_1}-1)^2}  
	\label{6_E18}
\end{eqnarray}
For fixed $m$,  the function expansion for $\eta_1\to 0$ is as follows:
\begin{eqnarray}
	f_{o}(\eta_1,N)&=& \frac{1}{\eta_1} \sum_{\substack{m=n\eta_1 \\ n=N+1}}^{\infty}\Bigg(\frac{2m e^{4m}}{(-1+e^{4m})^2}-\frac{e^{4m}\bigg(1+4m+(4m-1)e^{4m}\bigg)}{(-1+e^{4m})^3}\eta_1 \nonumber\\
	&&+\frac{2e^{4m}\bigg(1+2m+8m e^{4m}+(2m-1)e^{8m}\bigg)}{(-1+e^{4m})^4}\eta_1^2\Bigg)
	\label{6_E19}
\end{eqnarray}

\begin{eqnarray}
	f_{o}(\eta_1,N)&=& \frac{1}{\eta_1^2}\sum_{\substack{m=n\eta_1 \\ n=N+1}}^{\infty}f_{1}(m)\Delta m-\frac{1}{\eta_1} \sum_{\substack{m=n\eta_1 \\ n=N+1}}^{\infty}f_{2}(m)\Delta m \nonumber\\
	&&+\sum_{\substack{m=n\eta_1 \\ n=N+1}}^{\infty}f_{3}(m)\Delta m
	\label{6_E20}
\end{eqnarray}

where, 
\begin{eqnarray}
	f_{1}(m)&=&\frac{2m e^{4m}}{(-1+e^{4m})^2}
	\label{6_E21}\\
	f_{2}(m)&=&\frac{e^{4m}\bigg(1+4m+(4m-1)e^{4m}\bigg)}{(-1+e^{4m})^3}
	\label{6_E22}\\
	f_{3}(m)&=&\frac{2e^{4m}\bigg(1+2m+8m e^{4m}+(2m-1)e^{8m}\bigg)}{(-1+e^{4m})^4}
	\label{6_E23}
\end{eqnarray}
Define $\Delta m =m_{n+1}-m_n $. This makes $\Delta m=(n+1)\eta_1-n\eta_1=\eta_1$.
Certainly, the outer expansion can be expressed in the following manner: 
\begin{eqnarray}
	f_{o}(\eta_1,X)&=&\frac{1}{\eta_1^2}\sum_{m=X}^{\infty}f_{1}(m)\Delta m -\frac{1}{\eta_1}\sum_{m=X}^{\infty}f_{2}(m)\Delta m \nonumber\\
	&&+\sum_{\substack{m=X}}^{\infty}f_{3}(m)\Delta m
	\label{6_E24}
\end{eqnarray}

As per the definition of $X$, it is apparent that $m=X$ at the lower bound of summation. As $\Delta m\to 0$, this expression remains asymptotically precise. Now, from Eqns. (\ref{6_E21}),(\ref{6_E22}) we get $f_1(\infty)=0$,$f_2(\infty)=0$. Furthermore, as $X\to 0$, 
\begin{eqnarray}
	f_{1}(X)\sim \dfrac{1}{8 X}
	\label{6_E25}
	\\
	f_{2}(X)\sim \dfrac{1}{16 X^2}+\dfrac{1}{12}
	\label{6_E26}\\
	f_{3}(X)\sim\dfrac{1}{32 X^3}
	\label{6_E27}
\end{eqnarray}
Thus, since $\Delta m=\eta_1$, we have from Eqns. (\ref{1_E24}),(\ref{6_E24})-(\ref{6_E27})  that as $\eta_1\to0$,
\begin{eqnarray} 
	f_{o}(\eta_1,X) &=& \frac{1}{\eta_1^2}\int_{X}^{\infty}f_{1}(m)dm-\frac{1}{16X\eta_1}
	-\frac{1}{\eta_1}\int_{X}^{\infty}f_{2}(m)dm-\frac{1}{24}\nonumber\\&&+\int_{X}^{\infty}f_{3}(m)dm+\textit{o(1)}
	\label{6_E28}
\end{eqnarray}
The integral provided above converges for sufficiently large values of $m$. Now, as $\Delta m \to 0$, after simplification we get 
\begin{eqnarray}
	f_{o}(\eta_1,X)&\sim&-\frac{1}{8 X \eta_1}-\frac{\log{X}}{8\eta_1^2}-\frac{1}{4 \eta_1^2}-\frac{1}{12\eta_1}\nonumber\\
	&&+\frac{1}{24}+\frac{K_{61}}{\eta_1^2}-\frac{K_{62}}{\eta_1}+C_{65}
	\label{6_E29}
\end{eqnarray}
\\ where,
\begin{eqnarray*}
	K_{61}&=&C_{61}+C_{62}=-0.0482868 \\
	K_{62}&=&C_{63}+C_{64}=-0.0833333\\
	C_{61}&=&\int_{1}^{\infty} {f_1(m)}dm =0.0116394\nonumber\\
	C_{62}&=&\int_{0}^{1} \Bigg({f_1(m)}-\frac{1}{8m}\Bigg)dm =-0.0599262\nonumber\\
	C_{63}&=&\int_{1}^{\infty} \Bigg({f_2(m)}-\frac{1}{16m^2}\Bigg)dm =-0.0434945\nonumber\\
	C_{64}&=&\int_{0}^{1} \Bigg({f_2(m)}-\frac{1}{16m^2} -\frac{1}{12}\Bigg)dm=-0.0398388\\
	C_{65}&=&\int_{0}^{\infty} \Bigg({f_3(m)}-\frac{1}{32m^3}\Bigg)dm=0.0208333
\end{eqnarray*}
Equation (\ref{6_E29}) is the asymptotic outer expansion, which also involves  $\eta_1$ and $X$.

Adding Equation (\ref{6_E17}) and Equation (\ref{6_E29}), we obtain the asymptotic expression for $ T_1(2 \eta_1)$ in the the small separation region as follows:
\begin{eqnarray}
	T_1(2 \eta_1)&\sim&\frac{\gamma+\log{\frac{4}{\eta_1}}}{8\eta_1^2}-\frac{1}{12\eta_1}-\frac{1}{24}+\frac{K_{61}}{\eta_1^2}-\frac{K_{62}}{\eta_1}+C_{65}
	\label{6_E30}
\end{eqnarray}
At this stage, it can be observed that the final asymptotic expression is independent of $X$.
\paragraph*{Asymptotic expression for  the series $T_2(2 \eta_1)$:} 
Substituting $m=0$ and $p=\eta_1$ in Equation(\ref{E1}), we get $T_2(2 \eta_1)$ as follows,
\begin{eqnarray}
	T_2(2 \eta_1)&=&\sum_{n=0}^{\infty}\dfrac{ (2n+1)^2 e^{(4n+2)\eta_1}}{(e^{(4n+2)\eta_1}-1)^2}
	\label{7_E1}
\end{eqnarray}
As outlined in the previous section, the series $T_2(2 \eta_1)$ is decomposed to an ``inner expansion'' ($f_{i}$) and an ``outer expansion'' ($f_{o}$) as follows:  
\begin{eqnarray}
	T_2(2 \eta_1)&=& f_{i}(\eta_1,N)+f_{o}(\eta_1,N)
	\label{7_E2}
\end{eqnarray}

where, 
\begin{eqnarray}
	f_{i}(\eta_1,N)&=&\sum_{n=0}^{N} \dfrac{ (2n+1)^2 e^{(4n+2)\eta_1}}{(e^{(4n+2)\eta_1}-1)^2}
	\label{7_E3}
	\\
	f_{o}(\eta_1,N)&=&\sum_{n=N+1}^{\infty} \dfrac{ (2n+1)^2 e^{(4n+2)\eta_1}}{(e^{(4n+2)\eta_1}-1)^2} 
	\label{7_E4}
\end{eqnarray}

Following similar steps as shown in  Section\ref{sec:3}, we find the inner ($f_i$) and outer ($f_o$) expansions for $T_2(2 \eta_1)$ in the next two subsections.
\subparagraph*{Inner expansion}\label{subsec:A7.1}
In this subsection, we derive the asymptotic expression for the inner expansion of $T_2(2 \eta_1)$.
The inner expansion can be written as follows,
\begin{eqnarray}
	f_{i}(\eta_1,N)&\approx&\frac{1}{4}(I_1 {\eta_1}^{-2}-\frac{1}{3}I_2)
	\label{7_E9}
\end{eqnarray}
where
\begin{eqnarray}
	I_1&=&\sum_{n=0}^{N}1
	\label{7_E10}\\
	I_2&=&\sum_{n=0}^{N}(2n+1)^2
	\label{7_E11}
\end{eqnarray}
Hence, as $N\to\infty$,
\begin{eqnarray}
	I_1&=&1+\textit{O(N)}
	\label{7_E13}\\
	I_2&=& 1+\textit{O(N)}
	\label{7_E14}
\end{eqnarray}
Here $\gamma$ is Euler's constant, and it is approximately equal to 0.57721.  \\
Substitute Eqns. (\ref{7_E13}),(\ref{7_E14}) and in Equation (\ref{7_E9}). Then we get,
\begin{eqnarray}
	f_{i}(\eta_1,N)&\sim&\frac{1}{4\eta_1^2} -\frac{1}{12}
	\label{7_E16}
\end{eqnarray}

Equation (\ref{7_E16}) is the asymptotic inner expansion which involves  $\eta_1$. 
In the next subsection (\ref{subsec:A7.2}), we derive the outer expansion for $T_2(2 \eta_1)$.
\subparagraph*{Outer expansion }\label{subsec:A7.2}
In order to derive the outer expansion $f_{o}$, let $m=n\eta_1$ in Equation (\ref{7_E4}) and examine the scenario where $\eta_1\to0$ with $m$ held fixed. After substituting the $n$ value in Equation (\ref{7_E4}), we get outer expansion in terms of $m$ and $\eta_1$ as follows,  
\begin{eqnarray}
	f_{o}(\eta_1,N)&=&\sum_{\substack{m=n\eta_1 \\ n=N+1}}^{\infty} \dfrac{ (2m+\eta_1)^2e^{4m+2 \eta_1}}{\eta_1^2(e^{4m+2\eta_1}-1)^2}  
	\label{7_E18}
\end{eqnarray}
For fixed $m$,  the function expansion for $\eta_1\to 0$ is as follows:
\begin{eqnarray}
	f_{o}(\eta_1,N)&=& \frac{1}{\eta_1^2} \sum_{\substack{m=n\eta_1 \\ n=N+1}}^{\infty}\Bigg(\frac{4m^2 e^{4m}}{(-1+e^{4m})^2}-\frac{4me^{4m}\bigg(1+2m+(2m-1)e^{4m}\bigg)}{(-1+e^{4m})^3}\eta_1 \nonumber\\
	&&+\frac{e^{4m}\bigg(1+8m+8m^2+(32m^2-1)e^{4m}+(8m^2-8m+1)e^{8m}\bigg)\eta_1^2}{(-1+e^{4m})^4}\nonumber\\
	&&-\frac{2 e^{4m}}{3(-1+e^{4m})^5}\bigg(3+12m+8m^2+(88m^2+36m-3)e^{4m}+(88m^2-36m-3)e^{8m} \nonumber\\
	&&+(8m^2-12m+3)e^{12m}\bigg)\eta_1^3\Bigg)
	\label{7_E19}
\end{eqnarray}

\begin{eqnarray}
	f_{o}(\eta_1,N)&=& \frac{1}{\eta_1^3}\sum_{\substack{m=n\eta_1 \\ n=N+1}}^{\infty}f_{1}(m)\Delta m-\frac{1}{\eta_1^2} \sum_{\substack{m=n\eta_1 \\ n=N+1}}^{\infty}f_{2}(m)\Delta m \nonumber\\
	&&+\frac{1}{\eta_1} \sum_{\substack{m=n\eta_1 \\ n=N+1}}^{\infty}f_{3}(m)\Delta m- \sum_{\substack{m=n\eta_1 \\ n=N+1}}^{\infty}f_{4}(m)\Delta m
	\label{7_E20}
\end{eqnarray}

where, 
\begin{eqnarray}
	f_{1}(m)&=&\frac{4m^2 e^{4m}}{(-1+e^{4m})^2}
	\label{7_E21}\\
	f_{2}(m)&=&\frac{4me^{4m}\bigg(1+2m+(2m-1)e^{4m}\bigg)}{(-1+e^{4m})^3}
	\label{7_E22}\\
	f_{3}(m)&=&\frac{e^{4m}\bigg(1+8m+8m^2+(32m^2-1)e^{4m}+(8m^2-8m+1)e^{8m}\bigg)}{(-1+e^{4m})^4}
	\label{7_E23}\\
	f_{4}(m)&=&\frac{2 e^{4m}}{3(-1+e^{4m})^5}\bigg(3+12m+8m^2+(88m^2+36m-3)e^{4m}+(88m^2-36m-3)e^{8m} \nonumber\\
	&&+(8m^2-12m+3)e^{12m}\bigg)
	\label{7_E24}
\end{eqnarray}
Define $\Delta m =m_{n+1}-m_n $. This makes $\Delta m=(n+1)\eta_1-n\eta_1=\eta_1$.
Certainly, the outer expansion can be expressed in the following manner: 
\begin{eqnarray}
	f_{o}(\eta_1,X)&=&\frac{1}{\eta_1^3}\sum_{m=X}^{\infty}f_{1}(m)\Delta m -\frac{1}{\eta_1^2}\sum_{m=X}^{\infty}f_{2}(m)\Delta m \nonumber\\
	&&+\frac{1}{\eta_1}\sum_{\substack{m=X}}^{\infty}f_{3}(m)\Delta m-\sum_{\substack{m=X}}^{\infty}f_{4}(m)\Delta m
	\label{7_E25}
\end{eqnarray}

As per the definition of $X$, it is apparent that $m=X$ at the lower bound of summation. As $\Delta m\to 0$, this expression remains asymptotically precise. Now, from Eqns. (\ref{7_E21})-(\ref{7_E24}) we get $f_1(\infty)=0$,$f_2(\infty)=0$,$f_3(\infty)=0$,$f_4(\infty)=0$. Furthermore, as $X\to 0$, 
\begin{eqnarray}
	f_{1}(X)\sim \dfrac{1}{4}
	\label{7_E26}\\
	f_{2}(X)\sim\textit{O(X)}
	\label{7_E27}\\
	f_{3}(X)\sim\dfrac{-1}{12}
	\label{7_E28}\\
	f_{4}(X)\sim \textit{O(X)}
	\label{7_E29}
\end{eqnarray}
Thus, since $\Delta m=\eta_1$, we have from Eqns. (\ref{1_E24}),(\ref{7_E25})-(\ref{7_E29})  that as $\eta_1\to0$,
\begin{eqnarray} 
	f_{o}(\eta_1,X) &=& \frac{1}{\eta_1^3}\int_{X}^{\infty}f_{1}(m)dm-\frac{1}{8\eta_1^2} -\frac{1}{\eta_1^2}\int_{X}^{\infty}f_{2}(m)dm\nonumber\\
	&&+\frac{1}{\eta_1}\int_{X}^{\infty}f_{3}(m)dm
	+\frac{1}{24}-\int_{X}^{\infty}f_{4}(m)dm+\textit{o(1)}
	\label{7_E30}
\end{eqnarray}
The integral provided above converges for sufficiently large values of $m$. Now, as $\Delta m \to 0$, after simplification we get 
\begin{eqnarray}
	f_{o}(\eta_1,X)&\sim&\frac{1}{4\eta_1^3}-\frac{1}{8 \eta_1^2}-\frac{1}{12\eta_1}+\frac{1}{24}+\frac{K_{71}}{\eta_1^3}-\frac{C_{73}}{\eta_1^2}+\frac{K_{72}}{\eta_1}-C_{76}
	\label{7_E31}
\end{eqnarray}
\\ where,
\begin{eqnarray*}
	K_{71}&=&C_{71}+C_{72}=-0.0443832 \\
	K_{72}&=&C_{74}+C_{75}=0.0833333\\
	C_{71}&=&\int_{1}^{\infty} {f_1(m)}dm = 0.0302001\nonumber\\
	C_{72}&=&\int_{0}^{1} \Bigg({f_1(m)}-\frac{1}{4}\Bigg)dm =-0.0745833\nonumber\\
	C_{73}&=&\int_{0}^{\infty} {f_2(m)}dm =0.125\nonumber\\
	C_{74}&=&\int_{1}^{\infty} {f_3(m)}dm =0.0204238\nonumber\\
	C_{75}&=&\int_{0}^{1} \Bigg({f_3(m)}+\frac{1}{12}\Bigg)dm=0.0629095\\
	C_{76}&=&\int_{0}^{\infty} {f_4(m)}dm=-0.0138888
\end{eqnarray*}
Equation (\ref{7_E31}) is the asymptotic outer expansion, which also involves  $\eta_1$.

Adding Equation (\ref{7_E16}) and Equation (\ref{7_E31}), we obtain the asymptotic expression for $T_2(2 \eta_1)$ in the the small separation region as follows:
\begin{eqnarray}
	T_2(2 \eta_1)&\sim&\frac{1}{4\eta_1^3}+\frac{1}{8 \eta_1^2}-\frac{1}{12\eta_1}-\frac{1}{24}+\frac{K_{71}}{\eta_1^3}-\frac{C_{73}}{\eta_1^2}+\frac{K_{72}}{\eta_1}-C_{76}
	\label{7_E32}
\end{eqnarray}
At this stage, it can be observed that the final asymptotic expression is independent of $X$.
\paragraph*{Asymptotic expression for  the series $T_3(2 \eta_1)$:}
Substituting $m=0$ and $p=\eta_1$ in Equation(\ref{E1}), we get $T_3(2 \eta_1)$ as follows,
\begin{eqnarray}
	T_3(2 \eta_1)&=&\sum_{n=0}^{\infty}\dfrac{ (2n+1)^3 e^{(4n+2)\eta_1}}{(e^{(4n+2)\eta_1}-1)^2}
	\label{8_E1}
\end{eqnarray}
As outlined in the previous section, the series $T_3(2 \eta_1)$ is decomposed to an ``inner expansion'' ($f_{i}$) and an ``outer expansion'' ($f_{o}$) as follows:  
\begin{eqnarray}
	T_3(2 \eta_1)&=& f_{i}(\eta_1,N)+f_{o}(\eta_1,N)
	\label{8_E2}
\end{eqnarray}

where, 
\begin{eqnarray}
	f_{i}(\eta_1,N)&=&\sum_{n=0}^{N} \dfrac{ (2n+1)^3 e^{(4n+2)\eta_1}}{(e^{(4n+2)\eta_1}-1)^2}
	\label{8_E3}
	\\
	f_{o}(\eta_1,N)&=&\sum_{n=N+1}^{\infty} \dfrac{ (2n+1)^3 e^{(4n+2)\eta_1}}{(e^{(4n+2)\eta_1}-1)^2} 
	\label{8_E4}
\end{eqnarray}

Following similar steps as shown in  Section\ref{sec:3}, we find the inner ($f_i$) and outer ($f_o$) expansions for $T_3(2 \eta_1)$ in the next two subsections.
\subparagraph*{Inner expansion}\label{subsec:A8.1}
In this subsection, we derive the asymptotic expression for the inner expansion of $T_3(2 \eta_1)$.
The inner expansion can be written as follows,
\begin{eqnarray}
	f_{i}(\eta_1,N)&\approx&\frac{1}{4}(I_1 {\eta_1}^{-2}-\frac{1}{3}I_2)
	\label{8_E9}
\end{eqnarray}
where
\begin{eqnarray}
	I_1&=&\sum_{n=0}^{N}2n+1
	\label{8_E10}\\
	I_2&=&\sum_{n=0}^{N}(2n+1)^3
	\label{8_E11}
\end{eqnarray}
Hence, as $N\to\infty$,
\begin{eqnarray}
	I_1&=&1+\textit{O(N)}
	\label{8_E13}\\
	I_2&=& 1+\textit{O(N)}
	\label{8_E14}
\end{eqnarray}
Here $\gamma$ is Euler's constant, and it is approximately equal to 0.57721.  \\
Substitute Eqns. (\ref{8_E13}),(\ref{8_E14})and in Equation (\ref{8_E9}). Then we get,
\begin{eqnarray}
	f_{i}(\eta_1,N)&\sim&\frac{1}{4\eta_1^2} -\frac{1}{12}
	\label{8_E16}
\end{eqnarray}

Equation (\ref{8_E16}) is the asymptotic inner expansion which involves  $\eta_1$. 
In the next subsection (\ref{subsec:A8.2}), we derive the outer expansion for $T_3(2 \eta_1)$.
\subparagraph*{Outer expansion }\label{subsec:A8.2}
In order to derive the outer expansion $f_{o}$, let $m=n\eta_1$ in Equation (\ref{8_E4}) and examine the scenario where $\eta_1\to0$ with $m$ held fixed. After substituting the $n$ value in Equation (\ref{8_E4}), we get outer expansion in terms of $m$ and $\eta_1$ as follows,  
\begin{eqnarray}
	f_{o}(\eta_1,N)&=&\sum_{\substack{m=n\eta_1 \\ n=N+1}}^{\infty} \dfrac{ (2m+\eta_1)^3e^{4m+2\eta_1}}{\eta_1^3(e^{4m+2\eta_1}-1)^2}  
	\label{8_E18}
\end{eqnarray}
For fixed $m$,  the function expansion for $\eta_1\to 0$ is as follows:
\begin{eqnarray}
	f_{o}(\eta_1,N)&=& \frac{1}{\eta_1^3} \sum_{\substack{m=n\eta_1 \\ n=N+1}}^{\infty}\Bigg(\frac{8m^3 e^{4m}}{(-1+e^{4m})^2}-\frac{4m^2e^{4m}\bigg(3+4m+(4m-3)e^{4m}\bigg)}{(-1+e^{4m})^3}\eta_1 \nonumber\\
	&&+\frac{2me^{4m}\bigg(3+12m+8m^2+(32m^2-6)e^{4m}+(8m^2-12m+3)e^{8m}\bigg)\eta_1^2}{(-1+e^{4m})^4}\nonumber\\
	&&-\frac{\eta_1^3 e^{4m}}{3(-1+e^{4m})^5}\bigg(3+36m+72m^2+32m^3+(352m^3+216m^2-36m-9)e^{4m} \nonumber\\
	&&+(352m^3-216m^2-36m+9)e^{8m}+(32m^3-72m^2+36m-3)e^{12m}\bigg)\nonumber\\
	&&+\frac{\eta_1^4 2e^{4m}}{3(-1+e^{4m})^6}\bigg(3+18m+24m^2+8m^3+(208m^3+240m^2+36m-6)e^{4m} \nonumber\\
	&&+(528m^3-108m)e^{8m}+(208m^3-240m^2-36m+6)e^{12m}\nonumber\\
	&&+(8m^3-24m^2+18m-3)e^{16m}\bigg)\Bigg)
	\label{8_E19}
\end{eqnarray}

\begin{eqnarray}
	f_{o}(\eta_1,N)&=& \frac{1}{\eta_1^4}\sum_{\substack{m=n\eta_1 \\ n=N+1}}^{\infty}f_{1}(m)\Delta m-\frac{1}{\eta_1^3} \sum_{\substack{m=n\eta_1 \\ n=N+1}}^{\infty}f_{2}(m)\Delta m +\frac{1}{\eta_1^2} \sum_{\substack{m=n\eta_1 \\ n=N+1}}^{\infty}f_{3}(m)\Delta m \nonumber\\
	&&- \frac{1}{\eta_1} \sum_{\substack{m=n\eta_1 \\ n=N+1}}^{\infty}f_{4}(m)\Delta m+\sum_{\substack{m=n\eta_1 \\ n=N+1}}^{\infty}f_{5}(m)\Delta m
	\label{8_E20}
\end{eqnarray}

where, 
\begin{eqnarray}
	f_{1}(m)&=&\frac{8m^3 e^{4m}}{(-1+e^{4m})^2}
	\label{8_E21}\\
	f_{2}(m)&=&\frac{4m^2e^{4m}\bigg(3+4m+(4m-3)e^{4m}\bigg)}{(-1+e^{4m})^3}
	\label{8_E22}\\
	f_{3}(m)&=&\frac{2me^{4m}\bigg(3+12m+8m^2+(32m^2-6)e^{4m}+(8m^2-12m+3)e^{8m}\bigg)}{(-1+e^{4m})^4}
	\label{8_E23}\\
	f_{4}(m)&=&\frac{e^{4m}}{3(-1+e^{4m})^5}\bigg(3+36m+72m^2+32m^3+(352m^3+216m^2-36m-9)e^{4m} \nonumber\\
	&&+(352m^3-216m^2-36m+9)e^{8m}+(32m^3-72m^2+36m-3)e^{12m}
	\label{8_E24}\\
	f_{5}(m)&=&\frac{2e^{4m}}{3(-1+e^{4m})^6}\bigg(3+18m+24m^2+8m^3+(208m^3+240m^2+36m-6)e^{4m} \nonumber\\
	&&+(528m^3-108m)e^{8m}+(208m^3-240m^2-36m+6)e^{12m}\nonumber\\
	&&+(8m^3-24m^2+18m-3)e^{16m}\bigg)
	\label{8_E25}
\end{eqnarray}
Define $\Delta m =m_{n+1}-m_n $. This makes $\Delta m=(n+1)\eta_1-n\eta_1=\eta_1$.
Certainly, the outer expansion can be expressed in the following manner: 
\begin{eqnarray}
	f_{o}(\eta_1,X)&=&\frac{1}{\eta_1^4}\sum_{m=X}^{\infty}f_{1}(m)\Delta m -\frac{1}{\eta_1^3}\sum_{m=X}^{\infty}f_{2}(m)\Delta m +\frac{1}{\eta_1^2}\sum_{\substack{m=X}}^{\infty}f_{3}(m)\Delta m
	\nonumber\\
	&&-\frac{1}{\eta_1}\sum_{\substack{m=X}}^{\infty}f_{4}(m)\Delta m+\sum_{\substack{m=X}}^{\infty}f_{5}(m)\Delta m
	\label{8_E26}
\end{eqnarray}

As per the definition of $X$, it is apparent that $m=X$ at the lower bound of summation. As $\Delta m\to 0$, this expression remains asymptotically precise. Now, from Eqns. (\ref{8_E21})-(\ref{8_E25}) we get $f_1(\infty)=0$,$f_2(\infty)=0$,$f_3(\infty)=0$,$f_4(\infty)=0$,$f_5(\infty)=0$. Furthermore, as $X\to 0$, 
\begin{eqnarray}
	f_{1}(X)\sim  \textit{O(X)}
	\label{8_E27}\\
	f_{2}(X)\sim \dfrac{-1}{4}
	\label{8_E28}\\
	f_{3}(X)\sim \textit{O(X)}
	\label{8_E29}\\
	f_{4}(X)\sim \dfrac{1}{12}
	\label{8_E30}\\
	f_{5}(X)\sim \textit{O(1)}
	\label{8_E31}
\end{eqnarray}
Thus, since $\Delta m=\eta_1$, we have from Eqns. (\ref{1_E24}),(\ref{8_E26})-(\ref{8_E31})  that as $\eta_1\to0$,
\begin{eqnarray} 
	f_{o}(\eta_1,X) &=& \frac{1}{\eta_1^4}\int_{X}^{\infty}f_{1}(m)dm -\frac{1}{\eta_1^3}\int_{X}^{\infty}f_{2}(m)dm+\frac{1}{8\eta_1^2}+\frac{1}{\eta_1^2}\int_{X}^{\infty}f_{3}(m)dm\nonumber\\
	&&-\frac{1}{\eta_1}\int_{X}^{\infty}f_{4}(m)dm-\frac{1}{24}+\int_{X}^{\infty}f_{5}(m)dm+\textit{o(1)}
	\label{8_E32}
\end{eqnarray}
The integral provided above converges for sufficiently large values of $m$. Now, as $\Delta m \to 0$, after simplification we get 
\begin{eqnarray}
	f_{o}(\eta_1,X)&\sim&\frac{1}{4\eta_1^3}-\frac{1}{8 \eta_1^2}-\frac{1}{12\eta_1}+\frac{1}{24}+\frac{C_{81}}{\eta_1^4}\nonumber\\&&
	-\frac{K_{81}}{\eta_1^3}+\frac{C_{84}}{\eta_1^2}-\frac{K_{82}}{\eta_1}+C_{87}
	\label{8_E33}
\end{eqnarray}
\\ where,
\begin{eqnarray*}
	K_{81}&=&C_{82}+C_{83}=0.25 \\
	K_{82}&=&C_{85}+C_{86}=-0.0833333\\
	C_{81}&=&\int_{0}^{\infty} {f_1(m)}dm = 0.225386\nonumber\\
	C_{82}&=&\int_{1}^{\infty} {f_2(m)}dm =0.0760218\nonumber\\
	C_{83}&=&\int_{0}^{1} \Bigg({f_2(m)}+\frac{1}{4}\Bigg)dm=0.173978\\
	C_{84}&=&\int_{0}^{\infty} {f_3(m)}dm =-0.0625\nonumber\\
	C_{85}&=&\int_{1}^{\infty} {f_4(m)}dm=-0.00339257\\
	C_{86}&=&\int_{0}^{1} \Bigg({f_4(m)}-\frac{1}{12}\Bigg)dm =-0.0799408\nonumber\\
	C_{87}&=&\int_{0}^{\infty} {f_5(m)}dm =0.010415\nonumber\\
\end{eqnarray*}
Equation (\ref{8_E33}) is the asymptotic outer expansion, which also involves  $\eta_1$.

Adding Equation (\ref{8_E16}) and Equation (\ref{8_E33}), we obtain the asymptotic expression for $T_3(2 \eta_1)$ in the the small separation region as follows:
\begin{eqnarray}
	T_3(2 \eta_1)&\sim&\frac{1}{4\eta_1^3}+\frac{1}{8 \eta_1^2}-\frac{1}{12\eta_1}-\frac{1}{24}+\frac{C_{81}}{\eta_1^4}
	\nonumber\\&&-\frac{K_{81}}{\eta_1^3}+\frac{C_{84}}{\eta_1^2}-\frac{K_{82}}{\eta_1}+C_{87}
	\label{8_E34}
\end{eqnarray}
At this stage, it can be observed that the final asymptotic expression is independent of $X$.
\paragraph*{Asymptotic expression for the series $T_0(3 \eta_1)$: } \label{sec:9}
Substituting $m=0$ and $p=\eta_1$ in Equation(\ref{E1}), we get $T_0(3 \eta_1)$ as follows,
\begin{eqnarray}
	T_0(3 \eta_1)&=&\sum_{n=0}^{\infty} \dfrac{ e^{(6n+3)\eta_1}}{(e^{(4n+2)\eta_1}-1)^2}
	\label{9_E1}
\end{eqnarray}
As outlined in the previous section, the series $T_0(3 \eta_1)$ is decomposed to an ``inner expansion'' ($f_{i}$) and an ``outer expansion'' ($f_{o}$) as follows:  
\begin{eqnarray}
	T_0(3 \eta_1)&=& f_{i}(\eta_1,N)+f_{o}(\eta_1,N)
	\label{9_E2}
\end{eqnarray}

where, 
\begin{eqnarray}
	f_{i}(\eta_1,N)&=&\sum_{n=0}^{N} \dfrac{ e^{(6n+3)\eta_1}}{(e^{(4n+2)\eta_1}-1)^2}
	\label{9_E3}
	\\
	f_{o}(\eta_1,N)&=&\sum_{n=N+1}^{\infty} \dfrac{ e^{(6n+3)\eta_1}}{(e^{(4n+2)\eta_1}-1)^2} 
	\label{9_E4}
\end{eqnarray}

Following similar steps as shown in  Section\ref{sec:3}, we find the inner ($f_i$) and outer ($f_o$) expansions for $T_0(3 \eta_1)$ in the next two subsections.

\subparagraph*{Inner expansion}\label{subsec:A9.1}
In this subsection, we derive the asymptotic expression for the inner expansion of $T_0(3 \eta_1)$.
The inner expansion can be written as follows,
\begin{eqnarray}
	f_{i}(\eta_1,N)&\approx&\frac{1}{4}(I_1 {\eta_1}^{-2}+I_2 {\eta_1}^{-1}+\frac{1}{6}I_3)
	\label{9_E9}
\end{eqnarray}
where
\begin{eqnarray}
	I_1&=&\sum_{n=0}^{N}\frac{1}{(2n+1)^2}
	\label{9_E10}\\
	I_2&=&\sum_{n=0}^{N}\frac{1}{2n+1}
	\label{9_E11}\\
	I_3&=&\sum_{n=0}^{N}1
	\label{9_E12}
\end{eqnarray}
Hence, as $N\to\infty$,
\begin{eqnarray}
	I_1&\sim&\frac{\pi^2}{8}-\frac{1}{4N}+\frac{1}{4N^2}
	\label{9_E13}\\
	I_2&\sim&\frac{1}{2}\big(\gamma+\log{4N}+\frac{1}{N}\big)
	\label{9_E14}\\
	I_3&=&N+1
	\label{9_E15}
\end{eqnarray}
Here $\gamma$ is Euler's constant, and it is approximately equal to 0.57721.  \\
Substitute Eqns. (\ref{9_E13}),(\ref{9_E14}),(\ref{9_E15}) and in Equation (\ref{9_E9}). Then we get,
\begin{eqnarray}
	f_{i}(\eta_1,N)&\sim&\bigg(\frac{\pi^2}{32}-\frac{1}{16N}+\frac{1}{16N^2}\bigg){\eta_1}^{-2}\nonumber\\ 
	&&   +\bigg(\frac{ 1}{8}\big(\gamma+\log{4N}+\frac{1}{N}\big)\bigg){\eta_1}^{-1}+\frac{N+1}{24}
	\label{9_E16}
\end{eqnarray}

After introducing intermediate vanish $\displaystyle N=\frac{X}{\eta_1}$ and simplifying then Equation (\ref{9_E16}) becomes
\begin{eqnarray}
	f_{i}(\eta_1,X)&\sim&\frac{\pi^2}{32{\eta_1}^2}+\frac{\gamma+\log{\frac{4}{\eta_1}}}{8\eta_1}+\frac{1}{24}\nonumber\\
	&& +\frac{1}{16X^2}+\frac{1}{8X}-\frac{1}{16X \eta_1}+\frac{\log{X}}{8\eta_1}+\frac{X\eta_1^{-1}}{24}
	\label{9_E17}
\end{eqnarray}
Equation (\ref{9_E17}) is the asymptotic inner expansion which involves  $\eta_1$ and $X$ (which should vanish after summing with the outer expansion). 
In the next subsection (\ref{subsec:A9.2}), we derive the outer expansion for $T_0(3 \eta_1)$.
\subparagraph*{Outer expansion }\label{subsec:A9.2}
In order to derive the outer expansion $f_{o}$, let $m=n\eta_1$ in Equation (\ref{9_E4}) and examine the scenario where $\eta_1\to0$ with $m$ held fixed. After substituting the $n$ value in Equation (\ref{9_E4}), we get outer expansion in terms of $m$ and $\eta_1$ as follows,  
\begin{eqnarray}
	f_{o}(\eta_1,N)&=&\sum_{\substack{m=n\eta_1 \\ n=N+1}}^{\infty} \dfrac{ e^{6m+3\eta_1}}{(e^{4m+2\eta_1}-1)^2}  
	\label{9_E18}
\end{eqnarray}
For fixed $m$,  the function expansion for $\eta_1\to 0$ is as follows:
\begin{eqnarray}
	f_{o}(\eta_1,N)&=& \sum_{\substack{m=n\eta_1 \\ n=N+1}}^{\infty}\Bigg(\frac{e^{6m}}{(-1+e^{4m})^2}-\frac{e^{6m}(3+e^{4m})}{(-1+e^{4m})^3}\eta_1 \Bigg)
	\label{9_E19}
\end{eqnarray}

\begin{eqnarray}
	f_{o}(\eta_1,N)&=& \frac{1}{\eta_1}\sum_{\substack{m=n\eta_1 \\ n=N+1}}^{\infty}f_{1}(m)\Delta m-\sum_{\substack{m=n\eta_1 \\ n=N+1}}^{\infty}f_{2}(m)\Delta m
	\label{9_E20}
\end{eqnarray}

where, 
\begin{eqnarray}
	f_{1}(m)&=&\frac{e^{6m}}{(-1+e^{4m})^2} 
	\label{9_E21}\\
	f_{2}(m)&=&\frac{e^{6m}(3+e^{4m})}{(-1+e^{4m})^3}
	\label{9_E22}
\end{eqnarray}
Define $\Delta m =m_{n+1}-m_n $. This makes $\Delta m=(n+1)\eta_1-n\eta_1=\eta_1$.
Certainly, the outer expansion can be expressed in the following manner: 
\begin{eqnarray}
	f_{o}(\eta_1,X)&=&\frac{1}{\eta_1}\sum_{m=X}^{\infty}f_{1}(m)\Delta m -\sum_{m=X}^{\infty}f_{2}(m)\Delta m
	\label{9_E23}
\end{eqnarray}

As per the definition of $X$, it is apparent that $m=X$ at the lower bound of summation.
 As $\Delta m\to 0$, this expression remains asymptotically precise. Now, from Eqns. (\ref{9_E21}),(\ref{9_E22}) we get $f_1(\infty)=0$,$f_2(\infty)=0$. Furthermore, as $X\to 0$, 
\begin{eqnarray}
	f_{1}(X)\sim\dfrac{1}{16 X^2}+\dfrac{1}{8 X}+\dfrac{1}{24}
	\label{9_E25}
	\\
	f_{2}(X)\sim\dfrac{1}{16 X^3}+\dfrac{1}{16 X^2}
	\label{9_E26}
\end{eqnarray}
Thus, since $\Delta m=\eta_1$, we have from Eqns. (\ref{1_E24}),(\ref{9_E23})-(\ref{9_E26}) that as $\eta_1\to0$,
\begin{eqnarray} 
	f_{o}(\eta_1,X) &=& \frac{1}{\eta_1}\int_{X}^{\infty}f_{1}(m)dm-\frac{1}{32X^2}-\frac{1}{16X}-\frac{1}{48}\nonumber\\
	&& -\int_{X}^{\infty}f_{2}(m)dm+\textit{o(1)}
	\label{9_E27}
\end{eqnarray}
The integral provided above converges for sufficiently large values of $m$. Now, as $\Delta m \to 0$, after simplification we get 
\begin{eqnarray}
	f_{o}(\eta_1,X)&\sim&-\frac{1}{16X^2}-\frac{1}{8X}+\frac{1}{16X \eta_1}-\frac{\log{X}}{8\eta_1}-\frac{X\eta_1^{-1}}{24}\nonumber\\
	&&+\frac{\eta_1^{-1}}{24}-\frac{1}{48}+\frac{K_{91}}{\eta_1}-C_{93}
	\label{9_E28}
\end{eqnarray}
\\ where,
\begin{eqnarray*}
	K_{91}&=&C_{91}+C_{92}=-0.0416667  \\
	C_{91}&=&\int_{1}^{\infty} \Bigg({f_1(m)}-\frac{1}{16m^2}\Bigg)dm =0.00600775 \nonumber\\
	C_{92}&=&\int_{0}^{1} \Bigg({f_1(m)}-\frac{1}{16m^2}-\frac{1}{8m}-\frac{1}{24}\Bigg)dm =-0.0476744\nonumber\\
	C_{93}&=&\int_{0}^{\infty} \Bigg({f_2(m)}-\frac{1}{16m^3}-
	\frac{1}{16m^2}\Bigg)dm =-0.0235338
\end{eqnarray*}
Equation(\ref{9_E28}) is the asymptotic outer expansion, which also involves  $\eta_1$ and $X$.

Adding Equation (\ref{9_E17}) and Equation (\ref{9_E28}), we obtain the asymptotic expression for $T_0(3 \eta_1)$ in the the small separation region as follows:
\begin{eqnarray}
	T_0(3 \eta_1)&\sim&\frac{\pi^2}{32{\eta_1}^2}+\frac{\gamma+\log{\frac{4}{\eta_1}}}{8\eta_1}+\frac{1}{48}
	+\frac{\eta_1^{-1}}{24}+\frac{K_{91}}{\eta_1}-C_{3}
	\label{9_E29}
\end{eqnarray}
At this stage, it can be observed that the final asymptotic expression is independent of $X$.
\paragraph*{Asymptotic expression for the series $T_1(3 \eta_1)$: } \label{sec:10}
Substituting $m=0$ and $p=\eta_1$ in Equation(\ref{E1}), we get $T_1(3 \eta_1)$ as follows,
\begin{eqnarray}
	T_1(3 \eta_1)&=&\sum_{n=0}^{\infty}\dfrac{ (2n+1) e^{(6n+3)\eta_1}}{(e^{(4n+2)\eta_1}-1)^2}
	\label{10_E1}
\end{eqnarray}
As outlined in the previous section, the series $T_1(3 \eta_1)$ is decomposed to an ``inner expansion'' ($f_{i}$) and an ``outer expansion'' ($f_{o}$) as follows:  
\begin{eqnarray}
	T_1(3 \eta_1)&=& f_{i}(\eta_1,N)+f_{o}(\eta_1,N)
	\label{10_E2}
\end{eqnarray}

where, 
\begin{eqnarray}
	f_{i}(\eta_1,N)&=&\sum_{n=0}^{N} \dfrac{ (2n+1) e^{(6n+3)\eta_1}}{(e^{(4n+2)\eta_1}-1)^2}
	\label{10_E3}
	\\
	f_{o}(\eta_1,N)&=&\sum_{n=N+1}^{\infty} \dfrac{ (2n+1) e^{(6n+3)\eta_1}}{(e^{(4n+2)\eta_1}-1)^2} 
	\label{10_E4}
\end{eqnarray}

Following similar steps as shown in  Section\ref{sec:3}, we find the inner ($f_i$) and outer ($f_o$) expansions for $T_1(3 \eta_1)$ in the next two subsections.
\subparagraph*{Inner expansion}\label{subsec:A10.1}
In this subsection, we derive the asymptotic expression for the inner expansion of $T_1(3 \eta_1)$.
The inner expansion can be written as follows,
\begin{eqnarray}
	f_{i}(\eta_1,N)&\approx&\frac{1}{4}(I_1 {\eta_1}^{-2}+I_2 {\eta_1}^{-1}+\frac{1}{6}I_3)
	\label{10_E9}
\end{eqnarray}
where
\begin{eqnarray}
	I_1&=&\sum_{n=0}^{N}\frac{1}{(2n+1)}
	\label{10_E10}\\
	I_2&=&\sum_{n=0}^{N}1
	\label{10_E11}\\
	I_3&=&\sum_{n=0}^{N}2n+1
	\label{10_E12}
\end{eqnarray}
Hence, as $N\to\infty$,
\begin{eqnarray}
	I_1&\sim&\frac{1}{2}\big(\gamma+\log{4N}+\frac{1}{N}\big)
	\label{10_E13}\\
	I_2&=& N+1
	\label{10_E14}\\
	I_3&=&(N+1)^2
	\label{10_E15}
\end{eqnarray}
Here $\gamma$ is Euler's constant, and it is approximately equal to 0.57721.  \\
Substitute Eqns. (\ref{10_E13}),(\ref{10_E14}),(\ref{10_E15}) and in Equation (\ref{10_E9}). Then we get,
\begin{eqnarray}
	f_{i}(\eta_1,N)&\sim&\bigg(\frac{1}{8\eta_1^2}\big(\gamma+\log{4N}+\frac{1}{N}\big)\bigg) +\bigg(\frac{ N+1}{4\eta_1}\bigg)+\frac{(N+1)^2}{24}
	\label{10_E16}
\end{eqnarray}

After introducing intermediate vanish $\displaystyle N=\frac{X}{\eta_1}$ and simplifying then Equation (\ref{10_E16}) becomes
\begin{eqnarray}
	f_{i}(\eta_1,X)&\sim&\frac{\gamma+\log{\frac{4}{\eta_1}}}{8\eta_1^2}+\frac{1}{4 \eta_1}+\frac{1}{24}+\frac{1}{8 X \eta_1}+\frac{\log{X}}{8\eta_1^2}+\frac{X}{4\eta_1^2}
	\label{10_E17}
\end{eqnarray}
Equation (\ref{10_E17}) is the asymptotic inner expansion which involves  $\eta_1$ and $X$ (which should vanish after summing with the outer expansion). 
In the next subsection (\ref{subsec:A10.2}), we derive the outer expansion for $T_1(3 \eta_1)$.
\subparagraph*{Outer expansion }\label{subsec:A10.2}
In order to derive the outer expansion $f_{o}$, let $m=n\eta_1$ in Equation (\ref{10_E4}) and examine the scenario where $\eta_1\to0$ with $m$ held fixed. After substituting the $n$ value in Equation (\ref{10_E4}), we get outer expansion in terms of $m$ and $\eta_1$ as follows,  
\begin{eqnarray}
	f_{o}(\eta_1,N)&=&\sum_{\substack{m=n\eta_1 \\ n=N+1}}^{\infty} \dfrac{ (2m+\eta_1)e^{6m+3 \eta_1}}{\eta_1(e^{4m+2\eta_1}-1)^2}  
	\label{10_E18}
\end{eqnarray}
For fixed $m$,  the function expansion for $\eta_1\to 0$ is as follows:
\begin{eqnarray}
	f_{o}(\eta_1,N)&=& \frac{1}{\eta_1} \sum_{\substack{m=n\eta_1 \\ n=N+1}}^{\infty}\Bigg(\frac{2m e^{6m}}{(-1+e^{4m})^2}-\frac{e^{6m}\bigg(1+6m+(2m-1)e^{4m}\bigg)}{(-1+e^{4m})^3}\eta_1 \nonumber\\
	&&+\frac{e^{6m}\bigg(3+9m+(14m-2)e^{4m}+(m-1)e^{8m}\bigg)}{(-1+e^{4m})^4}\eta_1^2\Bigg)
	\label{10_E19}
\end{eqnarray}

\begin{eqnarray}
	f_{o}(\eta_1,N)&=& \frac{1}{\eta_1^2}\sum_{\substack{m=n\eta_1 \\ n=N+1}}^{\infty}f_{1}(m)\Delta m-\frac{1}{\eta_1} \sum_{\substack{m=n\eta_1 \\ n=N+1}}^{\infty}f_{2}(m)\Delta m \nonumber\\
	&&+\sum_{\substack{m=n\eta_1 \\ n=N+1}}^{\infty}f_{3}(m)\Delta m
	\label{10_E20}
\end{eqnarray}

where, 
\begin{eqnarray}
	f_{1}(m)&=&\frac{2m e^{6m}}{(-1+e^{4m})^2}
	\label{10_E21}\\
	f_{2}(m)&=&\frac{e^{6m}\bigg(1+6m+(2m-1)e^{4m}\bigg)}{(-1+e^{4m})^3}
	\label{10_E22}\\
	f_{3}(m)&=&\frac{e^{6m}\bigg(3+9m+(14m-2)e^{4m}+(m-1)e^{8m}\bigg)}{(-1+e^{4m})^4}
	\label{10_E23}
\end{eqnarray}
Define $\Delta m =m_{n+1}-m_n $. This makes $\Delta m=(n+1)\eta_1-n\eta_1=\eta_1$.
Certainly, the outer expansion can be expressed in the following manner: 
\begin{eqnarray}
	f_{o}(\eta_1,X)&=&\frac{1}{\eta_1^2}\sum_{m=X}^{\infty}f_{1}(m)\Delta m -\frac{1}{\eta_1}\sum_{m=X}^{\infty}f_{2}(m)\Delta m \nonumber\\
	&&+\sum_{\substack{m=X}}^{\infty}f_{3}(m)\Delta m
	\label{10_E24}
\end{eqnarray}

As per the definition of $X$, it is apparent that $m=X$ at the lower bound of summation. As $\Delta m\to 0$, this expression remains asymptotically precise. Now, from Eqns. (\ref{10_E21}),(\ref{10_E22}) we get $f_1(\infty)=0$,$f_2(\infty)=0$. Furthermore, as $X\to 0$, 
\begin{eqnarray}
	f_{1}(X)\sim \dfrac{1}{8 X}+\dfrac{1}{4}
	\label{10_E25}
	\\
	f_{2}(X)\sim \dfrac{1}{16 X^2}-\dfrac{1}{24}
	\label{10_E26}\\
	f_{3}(X)\sim\dfrac{1}{32 X^3}
	\label{10_E27}
\end{eqnarray}
Thus, since $\Delta m=\eta_1$, we have from Eqns. (\ref{1_E24}),(\ref{10_E24})-(\ref{10_E27})  that as $\eta_1\to0$,
\begin{eqnarray} 
	f_{o}(\eta_1,X) &=& \frac{1}{\eta_1^2}\int_{X}^{\infty}f_{1}(m)dm-\frac{1}{16X\eta_1}-\frac{1}{8\eta_1}\nonumber\\
	&& -\frac{1}{\eta_1}\int_{X}^{\infty}f_{2}(m)dm+\frac{1}{48}+\int_{X}^{\infty}f_{3}(m)dm+\textit{o(1)}
	\label{10_E28}
\end{eqnarray}
The integral provided above converges for sufficiently large values of $m$. Now, as $\Delta m \to 0$, after simplification we get 
\begin{eqnarray}
	f_{o}(\eta_1,X)&\sim&-\frac{1}{8 X \eta_1}-\frac{\log{X}}{8\eta_1^2}-\frac{X}{4\eta_1^2}\nonumber\\
	&&+\frac{1}{4 \eta_1^2}-\frac{1}{12\eta_1}-\frac{1}{48}+\frac{K_{101}}{\eta_1^2}-\frac{K_{102}}{\eta_1}+C_{105}
	\label{10_E29}
\end{eqnarray}
\\ where,
\begin{eqnarray*}
	K_{101}&=&C_{101}+C_{102}=0.183425 \\
	K_{102}&=&C_{103}+C_{104}=0.166667\\
	C_{101}&=&\int_{1}^{\infty} {f_1(m)}dm = 0.204961\nonumber\\
	C_{102}&=&\int_{0}^{1} \Bigg({f_1(m)}-\frac{1}{8m}
	-\frac{1}{4}\Bigg)dm =-0.0215362\nonumber\\
	C_{103}&=&\int_{1}^{\infty} \Bigg({f_2(m)}-\frac{1}{16m^2}\Bigg)dm =0.0779324\nonumber\\
	C_{104}&=&\int_{0}^{1} \Bigg({f_2(m)}-\frac{1}{16m^2} +\frac{1}{24}\Bigg)dm=0.0887343\\
	C_{105}&=&\int_{0}^{\infty} \Bigg({f_3(m)}-\frac{1}{32m^3}\Bigg)dm=0.03125
\end{eqnarray*}
Equation (\ref{10_E29}) is the asymptotic outer expansion, which also involves  $\eta_1$ and $X$.

Adding Equation (\ref{10_E17}) and Equation (\ref{10_E29}), we obtain the asymptotic expression for $T_1(3 \eta_1)$ in the the small separation region as follows:
\begin{eqnarray}
	T_1(3 \eta_1)&\sim&\frac{\gamma+\log{\frac{4}{\eta_1}}}{8\eta_1^2}+\frac{1}{4 \eta_1^2}+\frac{1}{6\eta_1}+\frac{1}{48}+\frac{K_{101}}{\eta_1^2}-\frac{K_{102}}{\eta_1}+C_{105}
	\label{10_E30}
\end{eqnarray}
At this stage, it can be observed that the final asymptotic expression is independent of $X$.
\paragraph*{Asymptotic expression for the series $T_2(3 \eta_1)$: } \label{sec:11}
Substituting $m=0$ and $p=\eta_1$ in Equation(\ref{E1}), we get $T_2(3 \eta_1)$ as follows,
\begin{eqnarray}
	T_2(3 \eta_1)&=&\sum_{n=0}^{\infty}\dfrac{ (2n+1)^2 e^{(6n+3)\eta_1}}{(e^{(4n+2)\eta_1}-1)^2}
	\label{11_E1}
\end{eqnarray}
As outlined in the previous section, the series $T_2(3 \eta_1)$ is decomposed to an ``inner expansion'' ($f_{i}$) and an ``outer expansion'' ($f_{o}$) as follows:  
\begin{eqnarray}
	T_2(3 \eta_1)&=& f_{i}(\eta_1,N)+f_{o}(\eta_1,N)
	\label{11_E2}
\end{eqnarray}

where, 
\begin{eqnarray}
	f_{i}(\eta_1,N)&=&\sum_{n=0}^{N} \dfrac{ (2n+1)^2 e^{(6n+3)\eta_1}}{(e^{(4n+2)\eta_1}-1)^2}
	\label{11_E3}
	\\
	f_{o}(\eta_1,N)&=&\sum_{n=N+1}^{\infty} \dfrac{ (2n+1)^2 e^{(6n+3)\eta_1}}{(e^{(4n+2)\eta_1}-1)^2} 
	\label{11_E4}
\end{eqnarray}

Following similar steps as shown in  Section\ref{sec:3}, we find the inner ($f_i$) and outer ($f_o$) expansions for $T_2(3 \eta_1)$ in the next two subsections.
\subparagraph*{Inner expansion}\label{subsec:A11.1}
In this subsection, we derive the asymptotic expression for the inner expansion of $T_2(3 \eta_1)$.
The inner expansion can be written as follows,
\begin{eqnarray}
	f_{i}(\eta_1,N)&\approx&\frac{1}{4}(I_1 {\eta_1}^{-2}+I_2 {\eta_1}^{-1}+\frac{1}{6}I_3)
	\label{11_E9}
\end{eqnarray}
where
\begin{eqnarray}
	I_1&=&\sum_{n=0}^{N}1
	\label{11_E10}\\
	I_2&=&\sum_{n=0}^{N}2n+1
	\label{11_E11}\\
	I_3&=&\sum_{n=0}^{N}(2n+1)^2
	\label{11_E12}
\end{eqnarray}
Hence, as $N\to\infty$,
\begin{eqnarray}
	I_1&=&1+\textit{O(N)}
	\label{11_E13}\\
	I_2&=& 1+\textit{O(N)}
	\label{11_E14}\\
	I_3&=&1+\textit{O(N)}
	\label{11_E15}
\end{eqnarray}
Here $\gamma$ is Euler's constant, and it is approximately equal to 0.57721.  \\
Substitute Eqns. (\ref{11_E13}),(\ref{11_E14}),(\ref{11_E15}) and in Equation (\ref{11_E9}). Then we get,
\begin{eqnarray}
	f_{i}(\eta_1,N)&\sim&\frac{1}{4\eta_1^2} +\frac{ 1}{4\eta_1}+\frac{1}{24}
	\label{11_E16}
\end{eqnarray}

Equation (\ref{11_E16}) is the asymptotic inner expansion which involves  $\eta_1$. 
In the next subsection (\ref{subsec:A11.2}), we derive the outer expansion for $T_2(3 \eta_1)$.
\subparagraph*{Outer expansion }\label{subsec:A11.2}
In order to derive the outer expansion $f_{o}$, let $m=n\eta_1$ in Equation (\ref{11_E4}) and examine the scenario where $\eta_1\to0$ with $m$ held fixed. After substituting the $n$ value in Equation (\ref{11_E4}), we get outer expansion in terms of $m$ and $\eta_1$ as follows,  
\begin{eqnarray}
	f_{o}(\eta_1,N)&=&\sum_{\substack{m=n\eta_1 \\ n=N+1}}^{\infty} \dfrac{ (2m+\eta_1)^2e^{6m+3 \eta_1}}{\eta_1^2(e^{4m+2\eta_1}-1)^2}  
	\label{11_E18}
\end{eqnarray}
For fixed $m$,  the function expansion for $\eta_1\to 0$ is as follows:
\begin{eqnarray}
	f_{o}(\eta_1,N)&=& \frac{1}{\eta_1^2} \sum_{\substack{m=n\eta_1 \\ n=N+1}}^{\infty}\Bigg(\frac{4m^2 e^{6m}}{(-1+e^{4m})^2}-\frac{4me^{6m}\bigg(1+3m+(m-1)e^{4m}\bigg)}{(-1+e^{4m})^3}\eta_1 \nonumber\\
	&&+\frac{e^{6m}\bigg(1+12m+18m^2+(28m^2-8m-2)e^{4m}+(2m^2-4m+1)e^{8m}\bigg)\eta_1^2}{(-1+e^{4m})^4}\nonumber\\
	&&-\frac{e^{6m}}{3(-1+e^{4m})^5}\bigg(9+54m+54m^2+(230m^2+30m-15)e^{4m}+(98m^2-78m+3)e^{8m} \nonumber\\
	&&+(2m^2-6m+3)e^{12m}\bigg)\eta_1^3\Bigg)
	\label{11_E19}
\end{eqnarray}

\begin{eqnarray}
	f_{o}(\eta_1,N)&=& \frac{1}{\eta_1^3}\sum_{\substack{m=n\eta_1 \\ n=N+1}}^{\infty}f_{1}(m)\Delta m-\frac{1}{\eta_1^2} \sum_{\substack{m=n\eta_1 \\ n=N+1}}^{\infty}f_{2}(m)\Delta m \nonumber\\
	&&+\frac{1}{\eta_1} \sum_{\substack{m=n\eta_1 \\ n=N+1}}^{\infty}f_{3}(m)\Delta m- \sum_{\substack{m=n\eta_1 \\ n=N+1}}^{\infty}f_{4}(m)\Delta m
	\label{11_E20}
\end{eqnarray}

where, 
\begin{eqnarray}
	f_{1}(m)&=&\frac{4m^2 e^{6m}}{(-1+e^{4m})^2}
	\label{11_E21}\\
	f_{2}(m)&=&\frac{4me^{6m}\bigg(1+3m+(m-1)e^{4m}\bigg)}{(-1+e^{4m})^3}
	\label{11_E22}\\
	f_{3}(m)&=&\frac{e^{6m}\bigg(1+12m+18m^2+(28m^2-8m-2)e^{4m}+(2m^2-4m+1)e^{8m}\bigg)}{(-1+e^{4m})^4}
	\label{11_E23}\\
	f_{4}(m)&=&\frac{e^{6m}}{3(-1+e^{4m})^5}\bigg(9+54m+54m^2+(230m^2+30m-15)e^{4m}+(98m^2-78m+3)e^{8m} \nonumber\\
	&&+(2m^2-6m+3)e^{12m}\bigg)
	\label{11_E24}
\end{eqnarray}
Define $\Delta m =m_{n+1}-m_n $. This makes $\Delta m=(n+1)\eta_1-n\eta_1=\eta_1$.
Certainly, the outer expansion can be expressed in the following manner: 
\begin{eqnarray}
	f_{o}(\eta_1,X)&=&\frac{1}{\eta_1^3}\sum_{m=X}^{\infty}f_{1}(m)\Delta m -\frac{1}{\eta_1^2}\sum_{m=X}^{\infty}f_{2}(m)\Delta m \nonumber\\
	&&+\frac{1}{\eta_1}\sum_{\substack{m=X}}^{\infty}f_{3}(m)\Delta m-\sum_{\substack{m=X}}^{\infty}f_{4}(m)\Delta m
	\label{11_E25}
\end{eqnarray}

As per the definition of $X$, it is apparent that $m=X$ at the lower bound of summation. As $\Delta m\to 0$, this expression remains asymptotically precise. Now, from Eqns. (\ref{11_E21})-(\ref{11_E24}) we get $f_1(\infty)=0$,$f_2(\infty)=0$,$f_3(\infty)=0$,$f_4(\infty)=0$. Furthermore, as $X\to 0$, 
\begin{eqnarray}
	f_{1}(X)\sim \dfrac{1}{4}
	\label{11_E26}\\
	f_{2}(X)\sim \dfrac{-1}{4}
	\label{11_E27}\\
	f_{3}(X)\sim\dfrac{1}{24}
	\label{11_E28}\\
	f_{4}(X)\sim \textit{O(1)}
	\label{11_E29}
\end{eqnarray}
Thus, since $\Delta m=\eta_1$, we have from Eqns. (\ref{1_E24}),(\ref{11_E25})-(\ref{11_E29})  that as $\eta_1\to0$,
\begin{eqnarray} 
	f_{o}(\eta_1,X) &=& \frac{1}{\eta_1^3}\int_{X}^{\infty}f_{1}(m)dm-\frac{1}{8\eta_1^2} -\frac{1}{\eta_1^2}\int_{X}^{\infty}f_{2}(m)dm+\frac{1}{8\eta_1}\nonumber\\
	&&+\frac{1}{\eta_1}\int_{X}^{\infty}f_{3}(m)dm-\frac{1}{48}-\int_{X}^{\infty}f_{4}(m)dm+\textit{o(1)}
	\label{11_E30}
\end{eqnarray}
The integral provided above converges for sufficiently large values of $m$. Now, as $\Delta m \to 0$, after simplification we get 
\begin{eqnarray}
	f_{o}(\eta_1,X)&\sim&\frac{1}{4\eta_1^3}+\frac{1}{8 \eta_1^2}-\frac{1}{12\eta_1}-\frac{1}{48}
	\nonumber\\
	&&+\frac{K_{111}}{\eta_1^3}-\frac{K_{112}}{\eta_1^2}+\frac{K_{113}}{\eta_1}-C_{117}
	\label{11_E31}
\end{eqnarray}
\\ where,
\begin{eqnarray*}
	K_{111}&=&C_{111}+C_{112}=0.89275 \\
	K_{112}&=&C_{113}+C_{114}=0.375\\
	K_{113}&=&C_{115}+C_{116}=-0.104167\\
	C_{111}&=&\int_{1}^{\infty} {f_1(m)}dm = 0.681334\nonumber\\
	C_{112}&=&\int_{0}^{1} \Bigg({f_1(m)}-\frac{1}{4}\Bigg)dm =0.211416\nonumber\\
	C_{113}&=&\int_{1}^{\infty} {f_2(m)}dm =0.280865\nonumber\\
	C_{114}&=&\int_{0}^{1} \Bigg({f_2(m)}+\frac{1}{4}\Bigg)dm=0.0941352\\
	C_{115}&=&\int_{1}^{\infty} {f_3(m)}dm =0.0104804\nonumber\\
	C_{116}&=&\int_{0}^{1} \Bigg({f_3(m)}-\frac{1}{24}\Bigg)dm=-0.114647\\
	C_{117}&=&\int_{0}^{\infty} {f_4(m)}dm=-0.0160274
\end{eqnarray*}
Equation (\ref{11_E31}) is the asymptotic outer expansion, which also involves  $\eta_1$.

Adding Equation (\ref{11_E16}) and Equation (\ref{11_E31}), we obtain the asymptotic expression for $T_2(3 \eta_1)$ in the the small separation region as follows:
\begin{eqnarray}
	T_2(3 \eta_1)&\sim&\frac{1}{4\eta_1^3}+\frac{3}{8 \eta_1^2}+\frac{1}{6\eta_1}+\frac{1}{48}+\frac{K_{111}}{\eta_1^3}-\frac{K_{112}}{\eta_1^2}+\frac{K_{113}}{\eta_1}-C_{117}
	\label{11_E32}
\end{eqnarray}
At this stage, it can be observed that the final asymptotic expression is independent of $X$.
\paragraph*{Asymptotic expression for the series $T_3(3 \eta_1)$: } \label{sec:12}
Substituting $m=0$ and $p=\eta_1$ in Equation(\ref{E1}), we get $T_3(3 \eta_1)$ as follows,
\begin{eqnarray}
	T_3(3 \eta_1)&=&\sum_{n=0}^{\infty}\dfrac{ (2n+1)^3 e^{(6n+3)\eta_1}}{(e^{(4n+2)\eta_1}-1)^2}
	\label{12_E1}
\end{eqnarray}
As outlined in the previous section, the series $T_3(3 \eta_1)$ is decomposed to an ``inner expansion'' ($f_{i}$) and an ``outer expansion'' ($f_{o}$) as follows:  
\begin{eqnarray}
	T_3(3 \eta_1)&=& f_{i}(\eta_1,N)+f_{o}(\eta_1,N)
	\label{12_E2}
\end{eqnarray}

where, 
\begin{eqnarray}
	f_{i}(\eta_1,N)&=&\sum_{n=0}^{N} \dfrac{ (2n+1)^3 e^{(6n+3)\eta_1}}{(e^{(4n+2)\eta_1}-1)^2}
	\label{12_E3}
	\\
	f_{o}(\eta_1,N)&=&\sum_{n=N+1}^{\infty} \dfrac{ (2n+1)^3 e^{(6n+3)\eta_1}}{(e^{(4n+2)\eta_1}-1)^2} 
	\label{12_E4}
\end{eqnarray}

Following similar steps as shown in  Section\ref{sec:3}, we find the inner ($f_i$) and outer ($f_o$) expansions for $T_3(3 \eta_1)$ in the next two subsections.
\subparagraph*{Inner expansion}\label{subsec:A12.1}
In this subsection, we derive the asymptotic expression for the inner expansion of $T_3(3 \eta_1)$.
The inner expansion can be written as follows,
\begin{eqnarray}
	f_{i}(\eta_1,N)&\approx&\frac{1}{4}(I_1 {\eta_1}^{-2}+I_2 {\eta_1}^{-1}+\frac{1}{6}I_3)
	\label{12_E9}
\end{eqnarray}
where
\begin{eqnarray}
	I_1&=&\sum_{n=0}^{N}2n+1
	\label{12_E10}\\
	I_2&=&\sum_{n=0}^{N}(2n+1)^2
	\label{12_E11}\\
	I_3&=&\sum_{n=0}^{N}(2n+1)^3
	\label{12_E12}
\end{eqnarray}
Hence, as $N\to\infty$,
\begin{eqnarray}
	I_1&=&1+\textit{O(N)}
	\label{12_E13}\\
	I_2&=& 1+\textit{O(N)}
	\label{12_E14}\\
	I_3&=&1+\textit{O(N)}
	\label{12_E15}
\end{eqnarray}
Here $\gamma$ is Euler's constant, and it is approximately equal to 0.57721.  \\
Substitute Eqns. (\ref{12_E13}),(\ref{12_E14}),(\ref{12_E15}) and in Equation (\ref{12_E9}). Then we get,
\begin{eqnarray}
	f_{i}(\eta_1,N)&\sim&\frac{1}{4\eta_1^2}+\frac{ 1}{4\eta_1}+\frac{1}{24}
	\label{12_E16}
\end{eqnarray}

Equation (\ref{12_E16}) is the asymptotic inner expansion which involves  $\eta_1$. 
In the next subsection (\ref{subsec:A12.2}), we derive the outer expansion for $T_3(3 \eta_1)$.
\subparagraph*{Outer expansion }\label{subsec:A12.2}
In order to derive the outer expansion $f_{o}$, let $m=n\eta_1$ in Equation (\ref{12_E4}) and examine the scenario where $\eta_1\to0$ with $m$ held fixed. After substituting the $n$ value in Equation (\ref{12_E4}), we get outer expansion in terms of $m$ and $\eta_1$ as follows,  
\begin{eqnarray}
	f_{o}(\eta_1,N)&=&\sum_{\substack{m=n\eta_1 \\ n=N+1}}^{\infty} \dfrac{ (2m+\eta_1)^3e^{6m+3 \eta_1}}{\eta_1^3(e^{4m+2\eta_1}-1)^2}  
	\label{12_E18}
\end{eqnarray}
For fixed $m$,  the function expansion for $\eta_1\to 0$ is as follows:
\begin{eqnarray}
	f_{o}(\eta_1,N)&=& \frac{1}{\eta_1^3} \sum_{\substack{m=n\eta_1 \\ n=N+1}}^{\infty}\Bigg(\frac{8m^3 e^{6m}}{(-1+e^{4m})^2}-\frac{4m^2e^{6m}\bigg(3+6m+(2m-3)e^{4m}\bigg)}{(-1+e^{4m})^3}\eta_1 \nonumber\\
	&&+\frac{2me^{6m}\bigg(3+18m+18m^2+(28m^2-12m-6)e^{4m}+(2m^2-6m+3)e^{8m}\bigg)\eta_1^2}{(-1+e^{4m})^4}\nonumber\\
	&&-\frac{\eta_1^3 e^{6m}}{3(-1+e^{4m})^5}\bigg(3+54m+162m^2+108m^3+(460m^3+90m^2-90m-9)e^{4m} \nonumber\\
	&&+(196m^3-234m^2+18m+9)e^{8m}+(4m^3-18m^2+18m-3)e^{12m}\bigg)\nonumber\\
	&&+\frac{\eta_1^4 e^{6m}}{3(-1+e^{4m})^6}\bigg(9+81m+162m^2+81m^3+(764m^3+528m^2-36m-24)e^{4m} \nonumber\\
	&&+(918m^3-396m^2-162m-18)e^{8m}+(156m^3-288m^2+108m)e^{12m}\nonumber\\
	&&+(m^3-6m^2+9m-3)e^{16m}\bigg)\Bigg)
	\label{12_E19}
\end{eqnarray}

\begin{eqnarray}
	f_{o}(\eta_1,N)&=& \frac{1}{\eta_1^4}\sum_{\substack{m=n\eta_1 \\ n=N+1}}^{\infty}f_{1}(m)\Delta m-\frac{1}{\eta_1^3} \sum_{\substack{m=n\eta_1 \\ n=N+1}}^{\infty}f_{2}(m)\Delta m +\frac{1}{\eta_1^2} \sum_{\substack{m=n\eta_1 \\ n=N+1}}^{\infty}f_{3}(m)\Delta m \nonumber\\
	&&- \frac{1}{\eta_1} \sum_{\substack{m=n\eta_1 \\ n=N+1}}^{\infty}f_{4}(m)\Delta m+\sum_{\substack{m=n\eta_1 \\ n=N+1}}^{\infty}f_{5}(m)\Delta m
	\label{12_E20}
\end{eqnarray}

where, 
\begin{eqnarray}
	f_{1}(m)&=&\frac{8m^3 e^{6m}}{(-1+e^{4m})^2}
	\label{12_E21}\\
	f_{2}(m)&=&\frac{4m^2e^{6m}\bigg(3+6m+(2m-3)e^{4m}\bigg)}{(-1+e^{4m})^3}
	\label{12_E22}\\
	f_{3}(m)&=&\frac{2me^{6m}\bigg(3+18m+18m^2+(28m^2-12m-6)e^{4m}+(2m^2-6m+3)e^{8m}\bigg)}{(-1+e^{4m})^4}
	\label{12_E23}\\
	f_{4}(m)&=&\frac{e^{6m}}{3(-1+e^{4m})^5}\bigg(3+54m+162m^2+108m^3+(460m^3+90m^2-90m-9)e^{4m} \nonumber\\
	&&+(196m^3-234m^2+18m+9)e^{8m}+(4m^3-18m^2+18m-3)e^{12m}\bigg)
	\label{12_E24}\\
	f_{5}(m)&=&\frac{e^{6m}}{3(-1+e^{4m})^6}\bigg(9+81m+162m^2+81m^3+(764m^3+528m^2-36m-24)e^{4m} \nonumber\\
	&&+(918m^3-396m^2-162m-18)e^{8m}+(156m^3-288m^2+108m)e^{12m}\nonumber\\
	&&+(m^3-6m^2+9m-3)e^{16m}\bigg)
	\label{12_E25}
\end{eqnarray}
Define $\Delta m =m_{n+1}-m_n $. This makes $\Delta m=(n+1)\eta_1-n\eta_1=\eta_1$.
Certainly, the outer expansion can be expressed in the following manner: 
\begin{eqnarray}
	f_{o}(\eta_1,X)&=&\frac{1}{\eta_1^4}\sum_{m=X}^{\infty}f_{1}(m)\Delta m -\frac{1}{\eta_1^3}\sum_{m=X}^{\infty}f_{2}(m)\Delta m +\frac{1}{\eta_1^2}\sum_{\substack{m=X}}^{\infty}f_{3}(m)\Delta m
	\nonumber\\
	&&-\frac{1}{\eta_1}\sum_{\substack{m=X}}^{\infty}f_{4}(m)\Delta m+\sum_{\substack{m=X}}^{\infty}f_{5}(m)\Delta m
	\label{12_E26}
\end{eqnarray}

As per the definition of $X$, it is apparent that $m=X$ at the lower bound of summation.
 
As $\Delta m\to 0$, this expression remains asymptotically precise. Now, from Eqns. (\ref{12_E21})-(\ref{12_E25}) we get $f_1(\infty)=0$,$f_2(\infty)=0$,$f_3(\infty)=0$,$f_4(\infty)=0$,$f_5(\infty)=0$. Furthermore, as $X\to 0$, 
\begin{eqnarray}
	f_{1}(X)\sim  \textit{O(X)}
	\label{12_E27}\\
	f_{2}(X)\sim \dfrac{-1}{4}
	\label{12_E28}\\
	f_{3}(X)\sim\dfrac{1}{4}
	\label{12_E29}\\
	f_{4}(X)\sim\dfrac{-1}{24}
	\label{12_E30}\\
	f_{5}(X)\sim \textit{O(1)}
	\label{12_E31}
\end{eqnarray}
Thus, since $\Delta m=\eta_1$, we have from Eqns. (\ref{1_E24}),(\ref{12_E26})-(\ref{12_E31})  that as $\eta_1\to0$,
\begin{eqnarray} 
	f_{o}(\eta_1,X) &=& \frac{1}{\eta_1^4}\int_{X}^{\infty}f_{1}(m)dm -\frac{1}{\eta_1^3}\int_{X}^{\infty}f_{2}(m)dm+\frac{1}{8\eta_1^2}+\frac{1}{\eta_1^2}\int_{X}^{\infty}f_{3}(m)dm-\frac{1}{8\eta_1}\nonumber\\
	&&-\frac{1}{\eta_1}\int_{X}^{\infty}f_{4}(m)dm+\frac{1}{48}+\int_{X}^{\infty}f_{5}(m)dm+\textit{o(1)}
	\label{12_E32}
\end{eqnarray}
The integral provided above converges for sufficiently large values of $m$. Now, as $\Delta m \to 0$, after simplification we get 
\begin{eqnarray}
	f_{o}(\eta_1,X)&\sim&\frac{1}{4\eta_1^3}+\frac{1}{8 \eta_1^2}-\frac{1}{12\eta_1}-\frac{1}{48}+\frac{C_{121}}{\eta_1^4}\nonumber\\&&
	-\frac{K_{121}}{\eta_1^3}+\frac{K_{122}}{\eta_1^2}-\frac{K_{123}}{\eta_1}+C_{128}
	\label{12_E33}
\end{eqnarray}
\\ where,
\begin{eqnarray*}
	K_{121}&=&C_{122}+C_{123}=0.25 \\
	K_{122}&=&C_{124}+C_{125}=-0.3125\\
	K_{123}&=&C_{126}+C_{127}=0.0833333\\
	C_{121}&=&\int_{0}^{\infty} {f_1(m)}dm = 3.09972\nonumber\\
	C_{122}&=&\int_{1}^{\infty} {f_2(m)}dm =0.56173\nonumber\\
	C_{123}&=&\int_{0}^{1} \Bigg({f_2(m)}+\frac{1}{4}\Bigg)dm=-0.31173\\
	C_{124}&=&\int_{1}^{\infty} {f_3(m)}dm =-0.119472\nonumber\\
	C_{125}&=&\int_{0}^{1} \Bigg({f_3(m)}-\frac{1}{4}\Bigg)dm=-0.193028\\
	C_{126}&=&\int_{1}^{\infty} {f_4(m)}dm=-0.0390417\\
	C_{127}&=&\int_{0}^{1} \Bigg({f_4(m)}+\frac{1}{24}\Bigg)dm =0.122375\nonumber\\
	C_{128}&=&\int_{0}^{\infty} {f_5(m)}dm =0.00588708\nonumber\\
\end{eqnarray*}
Equation (\ref{12_E33}) is the asymptotic outer expansion, which also involves  $\eta_1$.

Adding Equation (\ref{12_E16}) and Equation (\ref{12_E33}), we obtain the asymptotic expression for $T_3(3 \eta_1)$ in the the small separation region as follows:
\begin{eqnarray}
	T_3(3 \eta_1)&\sim&\frac{1}{4\eta_1^3}+\frac{3}{8 \eta_1^2}+\frac{1}{6\eta_1}+\frac{1}{48}+\frac{C_{121}}{\eta_1^4}
	\nonumber\\&&-\frac{K_{121}}{\eta_1^3}+\frac{K_{122}}{\eta_1^2}-\frac{K_{123}}{\eta_1}+C_{128}
	\label{12_E34}
\end{eqnarray}
At this stage, it can be observed that the final asymptotic expression is independent of $X$.
\paragraph*{Asymptotic expression  for the series $U_1(\eta_1)$: }\label{sec:14}
Substituting $m=0$ and $p=\eta_1$ in Equation(\ref{E2}), we get $U_1(\eta_1)$ as follows,
\begin{eqnarray}
	U_1(\eta_1)&=&\sum_{n=0}^{\infty}(2n+1) \dfrac{ e^{(2n+1)\eta_1}}{(e^{(4n+2)\eta_1}-1) (e^{(4n+6)\eta_1}-1)}
	\label{14_E1}
\end{eqnarray}
In this case, $U_1(\eta_1)$ is decomposed into an ``inner expansion'' ($g_{i}$) and an ``outer expansion'' ($g_{o}$) as follows:
\begin{eqnarray}
	U_1(\eta_1)&=& g_{i}(\eta_1,N)+g_{o}(\eta_1,N)
	\label{14_E2}
\end{eqnarray}
where, 
\begin{eqnarray}
	g_{i}(\eta_1,N)&=&\sum_{n=0}^{N} (2n+1)\dfrac{ e^{(2n+1)\eta_1}}{(e^{(4n+2)\eta_1}-1) (e^{(4n+6)\eta_1}-1)}
	\label{14_E3}
	\\
	g_{o}(\eta_1,N)&=&\sum_{n=N+1}^{\infty}(2n+1) \dfrac{ e^{(2n+1)\eta_1}}{(e^{(4n+2)\eta_1}-1) (e^{(4n+6)\eta_1}-1)} 
	\label{14_E4}
\end{eqnarray}

Following similar steps as shown in  Section\ref{sec:3}, we find the inner ($g_i$) and outer ($g_o$) expansions in the next two subsections.

\subparagraph*{Inner expansion}\label{subsec:A14.1}
In this subsection, we derive the inner expansion $g_{i}$ for $U_1(\eta_1)$.

\begin{eqnarray}
	g_{i}(\eta_1,N)&\approx&\frac{1}{4}\Bigg(I_1 ({\eta_1}^{-2}-\frac{2}{3})-I_2 ({\eta_1}^{-1}-\frac{2}{3})+\frac{1}{6}I_3\Bigg)
	\label{14_E9}
\end{eqnarray}
\\
where
\begin{eqnarray}
	I_1&=&\sum_{n=0}^{N}\frac{1}{(2n+3)}
	\label{14_E10}\\
	I_2&=&\sum_{n=0}^{N}1
	\label{14_E11}\\
	I_3&=&\sum_{n=0}^{N}2n+3
	\label{14_E12}
\end{eqnarray}
As $N\to\infty$, $I_1$,$I_2$ and $I_3$ is given by,
\begin{eqnarray}
	I_1&\sim&\frac{1}{2}\big(\gamma+\log{4N}-2\big)+\frac{1}{N}
	\label{14_E13}\\
	I_2&=&N+1
	\label{14_E14}\\
	I_3&=&3+\textit{O(N)}
	\label{14_E15}
\end{eqnarray}
Substituting Eqns. (\ref{14_E13}),(\ref{14_E14}),(\ref{14_E15}) in Equation (\ref{14_E9}), the asymptotic expression for $g_{i}$ in terms of  $N$ and $\eta_1$ is given by,
\begin{eqnarray}
	g_{i}(\eta_1,N)&\sim&\frac{1}{4}\Bigg(\bigg(\frac{1}{2}\big(\gamma+\log{4N}-2\big)+\frac{1}{N}\bigg)\bigg({\eta_1}^{-2}-\frac{2}{3}\bigg) \nonumber\\&&-\bigg({\eta_1}^{-1}-\frac{2}{3}\bigg)(N+1)+\frac{3}{6} \Bigg) 
	\label{14_E16}
\end{eqnarray}

After introducing intermediate vanish $N=\displaystyle\frac{X}{\eta_1}$ and simplifying then Equation (\ref{14_E16}) becomes
\begin{eqnarray}
	g_{i}(\eta_1,X)&\sim&\frac{-1}{4{\eta_1}}+\frac{\big({\eta_1}^{-2}-\frac{2}{3}\big)\big(\gamma+\log{\frac{4}{\eta_1}}-2\big)}{8}+\frac{7}{24}\nonumber\\
	&&+\frac{1}{4X \eta_1}+\frac{\big({\eta_1}^{-2}-\frac{2}{3}\big)\log{X}}{8}-\frac{X\eta_1^{-2}}{4}
	\label{14_E17}
\end{eqnarray}
Equation (\ref{14_E17}) can be used to calculate the inner expansion $g_{i}$ for $U_1(\eta_1)$. In the next subsection (\ref{subsec:A14.2}), we derive the outer expansion for $U_1(\eta_1)$.
\subparagraph*{Outer expansion }\label{subsec:A14.2}
In order to derive the outer expansion, let $m=n\eta_1$ in Equation (\ref{14_E4}) and examine the scenario where $\eta_1\to0$ with $m$ held fixed. After substituting the $n$ value in Equation (\ref{14_E4}), we get outer expansion in terms of $m$ and $\eta_1$ as follows,  
\begin{eqnarray}
	g_{o}(\eta_1,N)&=&\sum_{\substack{m=n\eta_1 \\ n=N+1}}^{\infty} \dfrac{ (2m+\eta_1)e^{(2m+\eta_1)}}{\eta_1(e^{(4m+2\eta_1)}-1) (e^{(4m+6\eta_1)}-1)}  
	\label{14_E18}
\end{eqnarray}
For fixed $m$,  the function expansion for $\eta_1\to 0$ is as follows:
\begin{eqnarray}
	g_{o}(\eta_1,N)&=&  \frac{1}{\eta_1}\sum_{\substack{m=n\eta_1 \\ n=N+1}}^{\infty}\Bigg(\frac{2m e^{2m}}{(-1+e^{4m})^2}-\frac{\eta_1e^{2m}}{(-1+e^{4m})^3}(1+2m+(14m-1)e^{4m})
	\nonumber
	\\&&+\frac{e^{2m}(1+m+(54m+6)e^{4m}+(49m-7)e^{8m})}{(-1+e^{4m})^4}\eta_1^2\Bigg)
	\label{14_E19}
\end{eqnarray}

\begin{eqnarray}
	g_{o}(\eta_1,N)&=& \frac{1}{\eta_1^2}\sum_{\substack{m=n\eta_1 \\ n=N+1}}^{\infty}g_{1}(m)\Delta m- \frac{1}{\eta_1}\sum_{\substack{m=n\eta_1 \\ n=N+1}}^{\infty}g_{2}(m)\Delta m \nonumber
	\\&& +\sum_{\substack{m=n\eta_1 \\ n=N+1}}^{\infty}g_{3}(m)\Delta m
	\label{14_E20}
\end{eqnarray}

where,
\begin{eqnarray}
	g_{1}(m)&=&\frac{2m e^{2m}}{(-1+e^{4m})^2}
	\label{14_E21}\\
	g_{2}(m)&=&\frac{e^{2m}(1+2m+(14m-1)e^{4m})}{(-1+e^{4m})^3}
	\label{14_E22}\\
	g_{3}(m)&=& \frac{e^{2m}(1+m+(54m+6)e^{4m}+(49m-7)e^{8m})}{(-1+e^{4m})^4}
	\label{14_E23}
\end{eqnarray}
Define $\Delta m =m_{n+1}-m_n $. This makes 
$\Delta m=(n+1)\eta_1-n\eta_1=\eta_1$
Certainly, the outer expansion can be expressed in the following manner:
\begin{eqnarray}
	g_{o}(\eta_1,X)&=&\frac{1}{\eta_1^2}\sum_{m=X}^{\infty}g_{1}(m)\Delta m -\frac{1}{\eta_1}\sum_{m=X}^{\infty}g_{2}(m)\Delta m \nonumber
	\\&&
	+\sum_{m=X}^{\infty}g_{3}(m)\Delta m
	\label{14_E24}
\end{eqnarray}
As per the definition of $X$, it is apparent that $m=X$ at the lower bound of summation.
 As $\Delta m\to 0$, this expression remains asymptotically precise. Now, from Eqns. (\ref{14_E21}-\ref{14_E23}) we get $g_1(\infty)=0$,$g_2(\infty)=0$,$g_3(\infty)=0$. Furthermore, as $X\to 0$, 
\begin{eqnarray}
	g_{1}(X)\sim\dfrac{1}{8 X}-\dfrac{1}{4}
	\label{14_E25}
	\\
	g_{2}(X)\sim\dfrac{3}{16 X^2}-\dfrac{7}{24}
	\label{14_E26}\\
	g_{3}(X)\sim\dfrac{9}{32 X^3}-\dfrac{1}{12 X}
	\label{14_E27}
\end{eqnarray}

Thus, since $\Delta m=\eta_1$, we have from Eqns. (\ref{1_E24}),(\ref{14_E24}-\ref{14_E27})  that as $\eta_1\to0$,
\begin{eqnarray} 
	g_{o}(\eta_1,X) &=& \frac{1}{\eta_1^2}\int_{X}^{\infty}g_{1}(m)dm-\frac{1}{16X\eta_1}+\frac{1}{8\eta_1}\nonumber\\
	&& - \frac{1}{\eta_1}\int_{X}^{\infty}g_{2}(m)dm+\frac{7}{48}+\int_{X}^{\infty}g_{3}(m)dm+\textit{o(1)}
	\label{14_E28}
\end{eqnarray}
The integral appearing above is convergent for the large value of m. In the limit, $\Delta m \to 0$, after simplification we get the asymptotic expression for the outer expansion $g_{o}$ for $U_1(\eta_1)$ as follows:
\begin{eqnarray}
	g_{o}(\eta_1,X)&\sim&\frac{-1}{4X \eta_1}-\frac{\big({\eta_1}^{-2}-\frac{2}{3}\big)\log{X}}{8}+\frac{X\eta_1^{-2}}{4}\nonumber\\
	&&-\frac{1}{4\eta_1^2}+\frac{5}{12 \eta_1}-\frac{7}{48}+\frac{K_{141}}{\eta_1^2}-\frac{K_{142}}{\eta_1}+K_{143}
	\label{14_E29}
\end{eqnarray}
\\ where,
\begin{eqnarray*}
	K_{141} &=& C_{141}+C_{142}=0.0665749  \nonumber \\ 
	K_{142} &=& C_{143}+C_{144}=-0.141758\\
	K_{143} &=& C_{145}+C_{146}=-0.204861
	\nonumber \\
	C_{141} &=& \int_{1}^{\infty} \Bigg({g_1(m)}\Bigg)dm = 0.000984324\nonumber \\
	C_{142} &=&\int_{0}^{1} \Bigg({g_1(m)}-\frac{1}{8m}+\frac{1}{4}\Bigg)dm =0.0655905\nonumber \\
	C_{143} &=&\int_{1}^{\infty} \Bigg({g_2(m)}-\frac{3}{16m^2}\Bigg)dm= -0.180949\nonumber \\
	C_{144} &=&\int_{0}^{1} \Bigg({g_2(m)}-\frac{3}{16m^2}+\frac{7}{24}\Bigg)dm=0.0391906 \nonumber \\
	C_{145} &=&\int_{1}^{\infty} \Bigg({g_3(m)}-\frac{9}{32m^3}\Bigg)dm= -0.118708\nonumber \\
	C_{146} &=&\int_{0}^{1} \Bigg({g_3(m)}-\frac{9}{32m^3}+\frac{1}{12m}\Bigg)dm=-0.0861535 \nonumber \\
\end{eqnarray*}

Adding Equation (\ref{14_E17}) and Equation (\ref{14_E29}), the asymptotic expression for $U_1(\eta_1)$ in the the small separation region is given by,
\begin{eqnarray}
	U_1(\eta_1)&\sim& \frac{\big({\eta_1}^{-2}-\frac{2}{3}\big)\big(\gamma+\log{\frac{4}{\eta_1}}-2\big)}{8}\nonumber\\
	&&-\frac{1}{4\eta_1^2}+\frac{1}{6 \eta_1}+\frac{7}{48}+\frac{K_{141}}{\eta_1^2}-\frac{K_{142}}{\eta_1}+K_{143}
	\label{14_E30}
\end{eqnarray}
At this stage, it can be observed that the final asymptotic expression is independent of $X$.
\paragraph*{Asymptotic expression  for the series $U_2(\eta_1)$: }\label{sec:15}
Substituting $m=0$ and $p=\eta_1$ in Equation(\ref{E2}), we get $U_2(\eta_1)$ as follows,
\begin{eqnarray}
	U_2(\eta_1)&=&\sum_{n=0}^{\infty}(2n+1)^2 \dfrac{ e^{(2n+1)\eta_1}}{(e^{(4n+2)\eta_1}-1) (e^{(4n+6)\eta_1}-1)}
	\label{15_E1}
\end{eqnarray}
In this case, $U_2(\eta_1)$ is decomposed into an ``inner expansion'' ($g_{i}$) and an ``outer expansion'' ($g_{o}$) as follows:
\begin{eqnarray}
	U_2(\eta_1)&=& g_{i}(\eta_1,N)+g_{o}(\eta_1,N)
	\label{15_E2}
\end{eqnarray}
where, 
\begin{eqnarray}
	g_{i}(\eta_1,N)&=&\sum_{n=0}^{N} (2n+1)^2\dfrac{ e^{(2n+1)\eta_1}}{(e^{(4n+2)\eta_1}-1) (e^{(4n+6)\eta_1}-1)}
	\label{15_E3}
	\\
	g_{o}(\eta_1,N)&=&\sum_{n=N+1}^{\infty}(2n+1)^2 \dfrac{ e^{(2n+1)\eta_1}}{(e^{(4n+2)\eta_1}-1) (e^{(4n+6)\eta_1}-1)} 
	\label{15_E4}
\end{eqnarray}

Following similar steps as shown in  Section\ref{sec:3}, we find the inner ($g_i$) and outer ($g_o$) expansions in the next two subsections.

\subparagraph*{Inner expansion}\label{subsec:A15.1}
In this subsection, we derive the inner expansion $g_{i}$ for $U_2(\eta_1)$.

\begin{eqnarray}
	g_{i}(\eta_1,N)&\approx&\Bigg(I_1 (\frac{1}{3}-\frac{{\eta_1}^{-2}}{2})+I_2 (\frac{{\eta_1}^{-2}}{4}+\frac{5}{24})-(\frac{{\eta_1}^{-1}}{4}+\frac{1}{12})I_3\Bigg)
	\label{15_E9}
\end{eqnarray}
\\
where
\begin{eqnarray}
	I_1&=&\sum_{n=0}^{N}\frac{1}{(2n+3)}
	\label{15_E10}\\
	I_2&=&\sum_{n=0}^{N}1
	\label{15_E11}\\
	I_3&=&\sum_{n=0}^{N}2n+1
	\label{15_E12}
\end{eqnarray}
As $N\to\infty$, $I_1$,$I_2$ and $I_3$ is given by,
\begin{eqnarray}
	I_1&\sim&\frac{1}{2}\big(\gamma+\log{4N}-2\big)+\frac{1}{N}
	\label{15_E13}\\
	I_2&=&N+1
	\label{15_E14}\\
	I_3&=&1+\textit{O(N)}
	\label{15_E15}
\end{eqnarray}
Substituting Eqns. (\ref{15_E13}),(\ref{15_E14}),(\ref{15_E15}) in Equation (\ref{15_E9}), the asymptotic expression for $g_{i}$ in terms of  $N$ and $\eta_1$ is given by,
\begin{eqnarray}
	g_{i}(\eta_1,N)&\sim&\Bigg(\bigg(\frac{1}{2}\big(\gamma+\log{4N}-2\big)+\frac{1}{N}\bigg)\bigg(\frac{1}{3}-\frac{{\eta_1}^{-2}}{2}\bigg) \nonumber\\&&\bigg(\frac{{\eta_1}^{-2}}{4}+\frac{5}{24}\bigg)\bigg(N+1\bigg)-\bigg(\frac{{\eta_1}^{-1}}{4}+\frac{1}{12}\bigg) \Bigg) 
	\label{15_E16}
\end{eqnarray}

After introducing intermediate vanish $N=\displaystyle\frac{X}{\eta_1}$ and simplifying then Equation (\ref{15_E16}) becomes
\begin{eqnarray}
	g_{i}(\eta_1,X)&\sim&\frac{\eta_1^{-2}}{4}-\frac{1}{4{\eta_1}}+\big(\frac{1}{6}-\frac{{\eta_1}^{-2}}{4}\big)\big(\gamma+\log{\frac{4}{\eta_1}}-2\big)+\frac{1}{8}\nonumber\\
	&&-\frac{1}{2X \eta_1}+\big(\frac{1}{6}-\frac{{\eta_1}^{-2}}{4}\big)\log{X}
	\label{15_E17}
\end{eqnarray}
Equation (\ref{15_E17}) can be used to calculate the inner expansion $g_{i}$ for $U_2(\eta_1)$. In the next subsection (\ref{subsec:A15.2}), we derive the outer expansion for $U_2(\eta_1)$.
\subparagraph*{Outer expansion }\label{subsec:A15.2}
In order to derive the outer expansion, let $m=n\eta_1$ in Equation (\ref{15_E4}) and examine the scenario where $\eta_1\to0$ with $m$ held fixed. After substituting the $n$ value in Equation (\ref{15_E4}), we get outer expansion in terms of $m$ and $\eta_1$ as follows,  
\begin{eqnarray}
	g_{o}(\eta_1,N)&=&\sum_{\substack{m=n\eta_1 \\ n=N+1}}^{\infty} \dfrac{ (2m+\eta_1)^2e^{(2m+\eta_1)}}{\eta_1^2(e^{(4m+2\eta_1)}-1) (e^{(4m+6\eta_1)}-1)}  
	\label{15_E18}
\end{eqnarray}
For fixed $m$,  the function expansion for $\eta_1\to 0$ is as follows:
\begin{eqnarray}
	g_{o}(\eta_1,N)&=&  \frac{1}{\eta_1^2}\sum_{\substack{m=n\eta_1 \\ n=N+1}}^{\infty}\Bigg(\frac{4m^2 e^{2m}}{(-1+e^{4m})^2}-\frac{4m\eta_1e^{2m}}{(-1+e^{4m})^3}(1+m+(7m-1)e^{4m})
	\nonumber
	\\&&+\frac{e^{2m}(1+4m+2m^2+(108m^2+24m-2)e^{4m}+(98m^2-28m+1)e^{8m})}{(-1+e^{4m})^4}\eta_1^2
	\nonumber\\
	&&-\frac{e^{2m}}{3(-1+e^{4m})^5}\bigg(3+6m+2m^2+(730m^2+318m+15)e^{4m}+(2422m^2-30m-39)e^{8m} \nonumber\\
	&&+(686m^2-249m+21)e^{12m}\bigg)\eta_1^3\Bigg)
	\label{15_E19}
\end{eqnarray}

\begin{eqnarray}
	g_{o}(\eta_1,N)&=& \frac{1}{\eta_1^3}\sum_{\substack{m=n\eta_1 \\ n=N+1}}^{\infty}g_{1}(m)\Delta m- \frac{1}{\eta_1^2}\sum_{\substack{m=n\eta_1 \\ n=N+1}}^{\infty}g_{2}(m)\Delta m \nonumber
	\\&& +\frac{1}{\eta_1}\sum_{\substack{m=n\eta_1 \\ n=N+1}}^{\infty}g_{3}(m)\Delta m-\sum_{\substack{m=n\eta_1 \\ n=N+1}}^{\infty}g_{4}(m)\Delta m
	\label{15_E20}
\end{eqnarray}

where,
\begin{eqnarray}
	g_{1}(m)&=&\frac{4m^2 e^{2m}}{(-1+e^{4m})^2}
	\label{15_E21}\\
	g_{2}(m)&=&\frac{4m e^{2m}}{(-1+e^{4m})^3}(1+m+(7m-1)e^{4m})
	\label{15_E22}\\
	g_{3}(m)&=& \frac{e^{2m}(1+4m+2m^2+(108m^2+24m-2)e^{4m}+(98m^2-28m+1)e^{8m})}{(-1+e^{4m})^4}
	\label{15_E23}\\
	g_{4}(m)&=& \frac{e^{2m}}{3(-1+e^{4m})^5}\bigg(3+6m+2m^2+(730m^2+318m+15)e^{4m}+(2422m^2-30m-39)e^{8m} \nonumber\\
	&&+(686m^2-249m+21)e^{12m}\bigg)
	\label{15_E24}
\end{eqnarray}
Define $\Delta m =m_{n+1}-m_n $. This makes 
$\Delta m=(n+1)\eta_1-n\eta_1=\eta_1$
Certainly, the outer expansion can be expressed in the following manner:
\begin{eqnarray}
	g_{o}(\eta_1,X)&=&\frac{1}{\eta_1^3}\sum_{m=X}^{\infty}g_{1}(m)\Delta m -\frac{1}{\eta_1^2}\sum_{m=X}^{\infty}g_{2}(m)\Delta m \nonumber
	\\&&
	+\frac{1}{\eta_1}\sum_{m=X}^{\infty}g_{3}(m)\Delta m
	-\sum_{m=X}^{\infty}g_{4}(m)\Delta m
	\label{15_E25}
\end{eqnarray}
As per the definition of $X$, it is apparent that $m=X$ at the lower bound of summation.
 As $\Delta m\to 0$, this expression remains asymptotically precise. Now, from Eqns. (\ref{15_E21}-\ref{15_E24}) we get $g_1(\infty)=0$,$g_2(\infty)=0$,$g_3(\infty)=0$,$g_4(\infty)=0$. Furthermore, as $X\to 0$, 
\begin{eqnarray}
	g_{1}(X)\sim\dfrac{1}{4}
	\label{15_E26}
	\\
	g_{2}(X)\sim\dfrac{1}{4 X}+\dfrac{1}{4}
	\label{15_E27}\\
	g_{3}(X)\sim\dfrac{3}{8 X^2}+\dfrac{1}{8}
	\label{15_E28}\\
	g_{4}(X)\sim\dfrac{9}{16 X^3}-\dfrac{1}{6 X}
	\label{15_E29}
\end{eqnarray}

Thus, since $\Delta m=\eta_1$, we have from Eqns. (\ref{1_E24}),(\ref{15_E25}-\ref{15_E29})  that as $\eta_1\to0$,
\begin{eqnarray} 
	g_{o}(\eta_1,X) &=& \frac{1}{\eta_1^3}\int_{X}^{\infty}g_{1}(m)dm-\frac{1}{8\eta_1^2} - \frac{1}{\eta_1^2}\int_{X}^{\infty}g_{2}(m)dm+\frac{1}{8X\eta_1}+\frac{1}{8\eta_1}\nonumber\\&&+ \frac{1}{\eta_1}\int_{X}^{\infty}g_{3}(m)dm-\frac{1}{16}-\int_{X}^{\infty}g_{4}(m)dm+\textit{o(1)}
	\label{15_E30}
\end{eqnarray}

The integral appearing above is convergent for the large value of m. In the limit, $\Delta m \to 0$, after simplification we get the asymptotic expression for the outer expansion $g_{o}$ for $U_2(\eta_1)$ as follows:
\begin{eqnarray}
	g_{o}(\eta_1,X)&\sim&\frac{1}{2X \eta_1}-\big(\frac{1}{6}-\frac{{\eta_1}^{-2}}{4}\big)\log{X}+\frac{1}{4\eta_1^3}-\frac{3}{8\eta_1^2}\nonumber\\
	&&+\frac{1}{4 \eta_1}-\frac{1}{16}+\frac{K_{151}}{\eta_1^3}-\frac{K_{152}}{\eta_1^2}+\frac{K_{153}}{\eta_1}-K_{154}
	\label{15_E31}
\end{eqnarray}
\\ where,
\begin{eqnarray*}
	K_{151} &=& C_{151}+C_{152}=-0.15905 \nonumber \\ 
	K_{152} &=& C_{153}+C_{154}=-0.2759\\
	K_{153} &=& C_{155}+C_{156}=-0.473734\\
	K_{154} &=& C_{157}+C_{158}=-0.0747
	\nonumber \\
	C_{151} &=& \int_{1}^{\infty} {g_1(m)} dm = 0.00234029\nonumber \\
	C_{152} &=&\int_{0}^{1} \Bigg({g_1(m)}-\frac{1}{4}\Bigg)dm =-0.16139\nonumber \\
	C_{153} &=&\int_{1}^{\infty}{g_2(m)} dm= 0.0145974\nonumber \\
	C_{154} &=&\int_{0}^{1} \Bigg({g_2(m)}-\frac{1}{4m}-\frac{1}{4}\Bigg)dm=-0.290497 \nonumber \\
	C_{155} &=&\int_{1}^{\infty} \Bigg({g_3(m)}-\frac{3}{8m^2}\Bigg)dm= -0.329424\nonumber \\
	C_{156} &=&\int_{0}^{1} \Bigg({g_3(m)}-\frac{3}{8m^2}-\frac{1}{8}\Bigg)dm=-0.144309\nonumber \\
	C_{157} &=&\int_{1}^{\infty} \Bigg({g_4(m)}-\frac{9}{16m^3}\Bigg)dm= -0.185521\nonumber \\
	C_{158} &=&\int_{0}^{1} \Bigg({g_4(m)}-\frac{9}{16m^3}+\frac{1}{6m}\Bigg)dm=0.11082 \nonumber \\
\end{eqnarray*}
Adding Equation (\ref{15_E17}) and Equation (\ref{15_E31}), the asymptotic expression for $U_2(\eta_1)$ in the the small separation region is given by,
\begin{eqnarray}
	U_2(\eta_1)&\sim& \frac{\eta_1^{-3}}{4}-\frac{1}{8{\eta_1}^2}+\big(\frac{1}{6}-\frac{{\eta_1}^{-2}}{4}\big)\big(\gamma+\log{\frac{4}{\eta_1}}-2\big)+\frac{1}{16}\nonumber\\
	&&+\frac{K_{151}}{\eta_1^3}-\frac{K_{152}}{\eta_1^2}+\frac{K_{153}}{\eta_1}-K_{154}
	\label{15_E32}
\end{eqnarray}
At this stage, it can be observed that the final asymptotic expression is independent of $X$.
\paragraph*{Asymptotic expression  for the series $U_3(\eta_1)$: }\label{sec:16}
Substituting $m=0$ and $p=\eta_1$ in Equation(\ref{E2}), we get $U_3(\eta_1)$ as follows,
\begin{eqnarray}
	U_3(\eta_1)&=&\sum_{n=0}^{\infty}(2n+1)^3 \dfrac{ e^{(2n+1)\eta_1}}{(e^{(4n+2)\eta_1}-1) (e^{(4n+6)\eta_1}-1)}
	\label{16_E1}
\end{eqnarray}
In this case, $U_3(\eta_1)$ is decomposed into an ``inner expansion'' ($g_{i}$) and an ``outer expansion'' ($g_{o}$) as follows:
\begin{eqnarray}
	U_3(\eta_1)&=& g_{i}(\eta_1,N)+g_{o}(\eta_1,N)
	\label{16_E2}
\end{eqnarray}
where, 
\begin{eqnarray}
	g_{i}(\eta_1,N)&=&\sum_{n=0}^{N} (2n+1)^3\dfrac{ e^{(2n+1)\eta_1}}{(e^{(4n+2)\eta_1}-1) (e^{(4n+6)\eta_1}-1)}
	\label{16_E3}
	\\
	g_{o}(\eta_1,N)&=&\sum_{n=N+1}^{\infty}(2n+1)^3 \dfrac{ e^{(2n+1)\eta_1}}{(e^{(4n+2)\eta_1}-1) (e^{(4n+6)\eta_1}-1)} 
	\label{16_E4}
\end{eqnarray}

Following similar steps as shown in  Section\ref{sec:3}, we find the inner ($g_i$) and outer ($g_o$) expansions in the next two subsections.

\subparagraph*{Inner expansion}\label{subsec:A16.1}
In this subsection, we derive the inner expansion $g_{i}$ for $U_3(\eta_1)$.

\begin{eqnarray}
	g_{i}(\eta_1,N)&\approx&\Bigg(I_1 ({\eta_1}^{-2}-\frac{2}{3})-I_2 (\frac{{\eta_1}^{-2}}{2}+\frac{{\eta_1}^{-1}}{4}-\frac{1}{12})\nonumber\\&&+(\frac{{\eta_1}^{-2}}{4}+\frac{3}{8})I_3\Bigg)
	\label{16_E9}
\end{eqnarray}
\\
where
\begin{eqnarray}
	I_1&=&\sum_{n=0}^{N}\frac{1}{(2n+3)}
	\label{16_E10}\\
	I_2&=&\sum_{n=0}^{N}1
	\label{16_E11}\\
	I_3&=&\sum_{n=0}^{N}2n+1
	\label{16_E12}
\end{eqnarray}
As $N\to\infty$, $I_1$,$I_2$ and $I_3$ is given by,
\begin{eqnarray}
	I_1&\sim&\frac{1}{2}\big(\gamma+\log{4N}-2\big)+\frac{1}{N}
	\label{16_E13}\\
	I_2&=&N+1
	\label{16_E14}\\
	I_3&=&1+\textit{O(N)}
	\label{16_E15}
\end{eqnarray}
Substituting Eqns. (\ref{16_E13}),(\ref{16_E14}),(\ref{16_E15}) in Equation (\ref{16_E9}), the asymptotic expression for $g_{i}$ in terms of  $N$ and $\eta_1$ is given by,
\begin{eqnarray}
	g_{i}(\eta_1,N)&\sim&\Bigg(\bigg(\frac{1}{2}\big(\gamma+\log{4N}-2\big)+\frac{1}{N}\bigg)\bigg(\frac{1}{3}-\frac{{\eta_1}^{-2}}{2}\bigg) \nonumber\\&&\bigg(\frac{{\eta_1}^{-2}}{4}+\frac{5}{24}\bigg)\bigg(N+1\bigg)-\bigg(\frac{{\eta_1}^{-1}}{4}+\frac{1}{12}\bigg) \Bigg) 
	\label{16_E16}
\end{eqnarray}

After introducing intermediate vanish $N=\displaystyle\frac{X}{\eta_1}$ and simplifying then Equation (\ref{16_E16}) becomes
\begin{eqnarray}
	g_{i}(\eta_1,X)&\sim&-\frac{\eta_1^{-2}}{4}-\frac{1}{4{\eta_1}}+\big(\frac{{\eta_1}^{-2}}{2}-\frac{1}{3}\big)\big(\gamma+\log{\frac{4}{\eta_1}}-2\big)+\frac{11}{24}\nonumber\\
	&&+\frac{1}{X \eta_1}+\big(\frac{{\eta_1}^{-2}}{2}-\frac{1}{3}\big)\log{X}
	\label{16_E17}
\end{eqnarray}
Equation (\ref{16_E17}) can be used to calculate the inner expansion $g_{i}$ for $U_3(\eta_1)$. In the next subsection (\ref{subsec:A16.2}), we derive the outer expansion for $U_3(\eta_1)$.
\subparagraph*{Outer expansion }\label{subsec:A16.2}
In order to derive the outer expansion, let $m=n\eta_1$ in Equation (\ref{16_E4}) and examine the scenario where $\eta_1\to0$ with $m$ held fixed. After substituting the $n$ value in Equation (\ref{16_E4}), we get outer expansion in terms of $m$ and $\eta_1$ as follows,  
\begin{eqnarray}
	g_{o}(\eta_1,N)&=&\sum_{\substack{m=n\eta_1 \\ n=N+1}}^{\infty} \dfrac{ (2m+\eta_1)^3e^{(2m+\eta_1)}}{\eta_1^3(e^{(4m+2\eta_1)}-1) (e^{(4m+6\eta_1)}-1)}  
	\label{16_E18}
\end{eqnarray}
For fixed $m$,  the function expansion for $\eta_1\to 0$ is as follows:
\begin{eqnarray}
	g_{o}(\eta_1,N)&=&  \frac{1}{\eta_1^3}\sum_{\substack{m=n\eta_1 \\ n=N+1}}^{\infty}\Bigg(\frac{8m^3 e^{2m}}{(-1+e^{4m})^2}-\frac{4m^2\eta_1e^{2m}}{(-1+e^{4m})^3}(3+2m+(14m-3)e^{4m})
	\nonumber
	\\&&+\frac{2me^{2m}(3+6m+2m^2+(108m^2+36m-6)e^{4m}+(98m^2-42m+3)e^{8m})}{(-1+e^{4m})^4}\eta_1^2
	\nonumber\\
	&&-\frac{e^{2m}}{3(-1+e^{4m})^5}\bigg(3+18m+18m^2+4m^3+(1460m^3+954m^2+90m-9)e^{4m}\nonumber\\
	&&+(4844m^3-90m^2-234m+9)e^{8m}+(1372m^3-882m^2+126m-3)e^{12m}\bigg)\eta_1^3\nonumber\\
	&&+\frac{\eta_1^4 e^{2m}}{3(-1+e^{4m})^6}\bigg(3+9m+6m^2+m^3+(2476m^3+2184m^2+468m+12)e^{4m} \nonumber\\
	&&+(20870m^3+5076m^2-522m-54)e^{8m}+(20716m^3-5208m^2-396m+60)e^{12m}\nonumber\\
	&&+(2401m^3-2058m^2+441m-21)e^{16m}\bigg)\Bigg)
	\label{16_E19}
\end{eqnarray}

\begin{eqnarray}
	g_{o}(\eta_1,N)&=& \frac{1}{\eta_1^4}\sum_{\substack{m=n\eta_1 \\ n=N+1}}^{\infty}g_{1}(m)\Delta m- \frac{1}{\eta_1^3}\sum_{\substack{m=n\eta_1 \\ n=N+1}}^{\infty}g_{2}(m)\Delta m  +\frac{1}{\eta_1^2}\sum_{\substack{m=n\eta_1 \\ n=N+1}}^{\infty}g_{3}(m)\Delta m\nonumber
	\\&&-\frac{1}{\eta_1}\sum_{\substack{m=n\eta_1 \\ n=N+1}}^{\infty}g_{4}(m)\Delta m+\sum_{\substack{m=n\eta_1 \\ n=N+1}}^{\infty}g_{5}(m)\Delta m
	\label{16_E20}
\end{eqnarray}

where,
\begin{eqnarray}
	g_{1}(m)&=&\frac{8m^3 e^{2m}}{(-1+e^{4m})^2}
	\label{16_E21}\\
	g_{2}(m)&=&\frac{4m^2e^{2m}}{(-1+e^{4m})^3}(3+2m+(14m-3)e^{4m})
	\label{16_E22}\\
	g_{3}(m)&=& \frac{2me^{2m}(3+6m+2m^2+(108m^2+36m-6)e^{4m}+(98m^2-42m+3)e^{8m})}{(-1+e^{4m})^4}
	\label{16_E23}\\
	g_{4}(m)&=& \frac{e^{2m}}{3(-1+e^{4m})^5}\bigg(3+18m+18m^2+4m^3+(1460m^3+954m^2+90m-9)e^{4m}\nonumber\\
	&&+(4844m^3-90m^2-234m+9)e^{8m}+(1372m^3-882m^2+126m-3)e^{12m}\bigg)
	\label{16_E24}\\
	g_{5}(m)&=& \frac{e^{2m}}{3(-1+e^{4m})^6}\bigg(3+9m+6m^2+m^3+(2476m^3+2184m^2+468m+12)e^{4m} \nonumber\\
	&&+(20870m^3+5076m^2-522m-54)e^{8m}+(20716m^3-5208m^2-396m+60)e^{12m}\nonumber\\
	&&+(2401m^3-2058m^2+441m-21)e^{16m}\bigg)
	\label{16_E25}
\end{eqnarray}
Define $\Delta m =m_{n+1}-m_n $. This makes 
$\Delta m=(n+1)\eta_1-n\eta_1=\eta_1$
Certainly, the outer expansion can be expressed in the following manner:
\begin{eqnarray}
	g_{o}(\eta_1,X)&=&\frac{1}{\eta_1^4}\sum_{m=X}^{\infty}g_{1}(m)\Delta m -\frac{1}{\eta_1^3}\sum_{m=X}^{\infty}g_{2}(m)\Delta m+\frac{1}{\eta_1^2}\sum_{m=X}^{\infty}g_{3}(m)\Delta m\nonumber
	\\&&
	-\frac{1}{\eta_1}\sum_{m=X}^{\infty}g_{4}(m)\Delta m+\sum_{m=X}^{\infty}g_{5}(m)\Delta m
	\label{16_E26}
\end{eqnarray}
As per the definition of $X$, it is apparent that $m=X$ at the lower bound of summation.
 As $\Delta m\to 0$, this expression remains asymptotically precise. Now, from Eqns. (\ref{16_E21}-\ref{16_E25}) we get $g_1(\infty)=0$,$g_2(\infty)=0$,$g_3(\infty)=0$,$g_4(\infty)=0$,$g_5(\infty)=0$. Furthermore, as $X\to 0$, 
\begin{eqnarray}
	g_{1}(X)\sim\textit{O(X)}
	\label{16_E27} \\
	g_{2}(X)\sim\dfrac{1}{4}
	\label{16_E28}\\
	g_{3}(X)\sim\dfrac{1}{2 X}-\dfrac{1}{4}
	\label{16_E29}\\
	g_{4}(X)\sim\dfrac{3}{4 X^2}-\dfrac{11}{24}
	\label{16_E30}\\
	g_{5}(X)\sim\dfrac{9}{8 X^3}-\dfrac{1}{3 X}
	\label{16_E31}
\end{eqnarray}

Thus, since $\Delta m=\eta_1$, we have from Eqns. (\ref{1_E24}),(\ref{16_E26}-\ref{16_E31})  that as $\eta_1\to0$,
\begin{eqnarray} 
	g_{o}(\eta_1,X) &=& \frac{1}{\eta_1^4}\int_{X}^{\infty}g_{1}(m)dm - \frac{1}{\eta_1^3}\int_{X}^{\infty}g_{2}(m)dm+\frac{1}{8\eta_1^2}+ \frac{1}{\eta_1^2}\int_{X}^{\infty}g_{3}(m)dm\nonumber\\&&-\frac{1}{4X\eta_1}+\frac{1}{8\eta_1}-\frac{1}{\eta_1}\int_{X}^{\infty}g_{4}(m)dm-\frac{11}{48}+\int_{X}^{\infty}g_{5}(m)dm+\textit{o(1)}
	\label{16_E32}
\end{eqnarray}

The integral appearing above is convergent for the large value of m. In the limit, $\Delta m \to 0$, after simplification we get the asymptotic expression for the outer expansion $g_{o}$ for $U_3(\eta_1)$ as follows:
\begin{eqnarray}
	g_{o}(\eta_1,X)&\sim&-\frac{1}{X \eta_1}-\big(\frac{{\eta_1}^{-2}}{2}-\frac{1}{3}\big)\log{X}-\frac{1}{4\eta_1^3}-\frac{1}{8\eta_1^2}+\frac{7}{12\eta_1}\nonumber\\
	&&-\frac{11}{48}+\frac{C_{161}}{\eta_1^4}-\frac{K_{161}}{\eta_1^3}+\frac{K_{162}}{\eta_1^2}-\frac{K_{163}}{\eta_1}+K_{164}
	\label{16_E33}
\end{eqnarray}
\\ where,
\begin{eqnarray*}
	K_{161} &=& C_{162}+C_{163}=0.0785338 \nonumber \\ 
	K_{162} &=& C_{164}+C_{165}=0.302367\\
	K_{163} &=& C_{166}+C_{167}=-0.309695\\
	K_{164} &=& C_{168}+C_{169}=-0.275837
	\nonumber \\
	C_{161} &=& \int_{0}^{\infty} {g_1(m)} dm = 0.0556826\nonumber \\
	C_{162} &=&\int_{1}^{\infty}{g_2(m)} dm= 0.0332989\nonumber \\
	C_{163} &=&\int_{0}^{1} \Bigg({g_2(m)}-\frac{1}{4}\Bigg)dm=0.0452349 \nonumber \\
	C_{164} &=&\int_{1}^{\infty} {g_3(m)}dm= 0.0968758\nonumber \\
	C_{165} &=&\int_{0}^{1} \Bigg({g_3(m)}-\frac{1}{2m}+\frac{1}{4}\Bigg)dm=0.205491\nonumber \\
	C_{166} &=&\int_{1}^{\infty} \Bigg({g_4(m)}-\frac{3}{4m^2}\Bigg)dm=-0.56128\nonumber \\
	C_{167} &=&\int_{0}^{1} \Bigg({g_4(m)}-\frac{3}{4m^2}+\frac{11}{24}\Bigg)dm=0.251585 \nonumber \\
	C_{168} &=& \int_{1}^{\infty} \Bigg({g_5(m)}-\frac{9}{8m^3}\Bigg) dm = -0.280511\nonumber \\
	C_{169} &=&\int_{0}^{1} \Bigg({g_5(m)}-\frac{9}{8m^3}-\frac{1}{3m}\Bigg)dm =0.00467345\nonumber \\
\end{eqnarray*}
Adding Equation (\ref{16_E17}) and Equation (\ref{16_E33}), the asymptotic expression for $U_3(\eta_1)$ in the the small separation region is given by,
\begin{eqnarray}
	U_3(\eta_1)&\sim&-\frac{1}{4\eta_1^{3}} -\frac{3}{8\eta_1^{2}}+\frac{1}{3{\eta_1}}+\big(\frac{{\eta_1}^{-2}}{2}-\frac{1}{3}\big)\big(\gamma+\log{\frac{4}{\eta_1}}-2\big)+\frac{11}{48}\nonumber\\
	&&+\frac{C_{161}}{\eta_1^4}-\frac{K_{161}}{\eta_1^3}+\frac{K_{162}}{\eta_1^2}-\frac{K_{163}}{\eta_1}+K_{164}
	\label{16_E34}
\end{eqnarray}
At this stage, it can be observed that the final asymptotic expression is independent of $X$.
\paragraph*{Asymptotic expression  for the series $U_0(2 \eta_1)$: }\label{sec:17}
Substituting $m=0$ and $p=\eta_1$ in Equation(\ref{E2}), we get $U_0(2 \eta_1)$ as follows,
\begin{eqnarray}
	U_0(2 \eta_1)&=&\sum_{n=0}^{\infty} \dfrac{ e^{(4n+2)\eta_1}}{(e^{(4n+2)\eta_1}-1) (e^{(4n+6)\eta_1}-1)}
	\label{17_E1}
\end{eqnarray}
In this case, $U_0(2 \eta_1)$ is decomposed into an ``inner expansion'' ($g_{i}$) and an ``outer expansion'' ($g_{o}$) as follows:
\begin{eqnarray}
	U_0(2 \eta_1)&=& g_{i}(\eta_1,N)+g_{o}(\eta_1,N)
	\label{17_E2}
\end{eqnarray}
where, 
\begin{eqnarray}
	g_{i}(\eta_1,N)&=&\sum_{n=0}^{N} \dfrac{ e^{(4n+2)\eta_1}}{(e^{(4n+2)\eta_1}-1) (e^{(4n+6)\eta_1}-1)}
	\label{17_E3}
	\\
	g_{o}(\eta_1,N)&=&\sum_{n=N+1}^{\infty} \dfrac{ e^{(4n+2)\eta_1}}{(e^{(4n+2)\eta_1}-1) (e^{(4n+6)\eta_1}-1)} 
	\label{17_E4}
\end{eqnarray}

Following similar steps as shown in  Section\ref{sec:3}, we find the inner ($g_i$) and outer ($g_o$) expansions in the next two subsections.

\subparagraph*{Inner expansion}\label{subsec:A17.1}
In this subsection, we derive the inner expansion $g_{i}$ for $U_0(2 \eta_1)$.

\begin{eqnarray}
	g_{i}(\eta_1,N)&\approx&\Bigg(I_1 (\frac{{\eta_1}^{-2}}{4}-\frac{\eta_1^{-1}}{2}+\frac{1}{3})+\frac{1}{12}I_2\Bigg)
	\label{17_E9}
\end{eqnarray}
\\
where
\begin{eqnarray}
	I_1&=&\sum_{n=0}^{N}\frac{1}{(2n+1)(2n+3)}
	\label{17_E10}\\
	I_2&=&\sum_{n=0}^{N}1
	\label{17_E11}
\end{eqnarray}
As $N\to\infty$, $I_1$and $I_2$ is given by,
\begin{eqnarray}
	I_1&\sim&\frac{1}{2}-\frac{1}{4N}+\frac{3}{8N^2}
	\label{17_E13}\\
	I_2&=&N+1
	\label{17_E14}
\end{eqnarray}
Substituting Eqns. (\ref{17_E13}),(\ref{17_E14}) in Equation (\ref{17_E9}), the asymptotic expression for $g_{i}$ in terms of  $N$ and $\eta_1$ is given by,
\begin{eqnarray}
	g_{i}(\eta_1,N)&\sim&\Bigg(\bigg(\frac{1}{2}-\frac{1}{4N}+\frac{3}{8N^2}\bigg)\bigg(\frac{{\eta_1}^{-2}}{4}-\frac{\eta_1^{-1}}{2}+\frac{1}{3}\bigg) \nonumber\\&&+\frac{N+1}{12}\Bigg) 
	\label{17_E16}
\end{eqnarray}

After introducing intermediate vanish $N=\displaystyle\frac{X}{\eta_1}$ and simplifying then Equation (\ref{17_E16}) becomes
\begin{eqnarray}
	g_{i}(\eta_1,X)&\sim&\frac{{\eta_1}^{-2}}{8}-\frac{\eta_1^{-1}}{4}+\frac{1}{12}+\frac{3}{32X^2}+\frac{1}{8X}-\frac{1}{16X \eta_1}-\frac{X\eta_1^{-1}}{12}
	\label{17_E17}
\end{eqnarray}
Equation (\ref{17_E17}) can be used to calculate the inner expansion $g_{i}$ for $U_0(2 \eta_1)$. In the next subsection (\ref{subsec:A17.2}), we derive the outer expansion for $U_0(2 \eta_1)$.
\subparagraph*{Outer expansion }\label{subsec:A17.2}
In order to derive the outer expansion, let $m=n\eta_1$ in Equation (\ref{17_E4}) and examine the scenario where $\eta_1\to0$ with $m$ held fixed. After substituting the $n$ value in Equation (\ref{17_E4}), we get outer expansion in terms of $m$ and $\eta_1$ as follows,  
\begin{eqnarray}
	g_{o}(\eta_1,N)&=&\sum_{\substack{m=n\eta_1 \\ n=N+1}}^{\infty} \dfrac{ e^{(4m+2 \eta_1)}}{(e^{(4m+2\eta_1)}-1) (e^{(4m+6\eta_1)}-1)}  
	\label{17_E18}
\end{eqnarray}
For fixed $m$,  the function expansion for $\eta_1\to 0$ is as follows:
\begin{eqnarray}
	g_{o}(\eta_1,N)&=&  \sum_{\substack{m=n\eta_1 \\ n=N+1}}^{\infty}\Bigg(\frac{e^{4m}}{(-1+e^{4m})^2}-\frac{2e^{4m}(1+3e^{4m})}{(-1+e^{4m})^3}\eta_1 \Bigg)
	\label{17_E19}
\end{eqnarray}

\begin{eqnarray}
	g_{o}(\eta_1,N)&=& \frac{1}{\eta_1}\sum_{\substack{m=n\eta_1 \\ n=N+1}}^{\infty}g_{1}(m)\Delta m-\sum_{\substack{m=n\eta_1 \\ n=N+1}}^{\infty}g_{2}(m)\Delta m
	\label{17_E20}
\end{eqnarray}

where,
\begin{eqnarray}
	g_{1}(m)&=&\frac{e^{4m}}{(-1+e^{4m})^2}
	\label{17_E21}\\
	g_{2}(m)&=&\frac{2e^{4m}(1+3e^{4m})}{(-1+e^{4m})^3}
	\label{17_E22}
\end{eqnarray}
Define $\Delta m =m_{n+1}-m_n $. This makes 
$\Delta m=(n+1)\eta_1-n\eta_1=\eta_1$
Certainly, the outer expansion can be expressed in the following manner:
\begin{eqnarray}
	g_{o}(\eta_1,X)&=&\frac{1}{\eta_1}\sum_{m=X}^{\infty}g_{1}(m)\Delta m -\sum_{m=X}^{\infty}g_{2}(m)\Delta m
	\label{17_E23}
\end{eqnarray}
As per the definition of $X$, it is apparent that $m=X$ at the lower bound of summation.
 
As $\Delta m\to 0$, this expression remains asymptotically precise. Now, from Eqns. (\ref{17_E21}),(\ref{17_E22}) we get $g_1(\infty)=0$,$g_2(\infty)=0$. Furthermore, as $X\to 0$, 
\begin{eqnarray}
	g_{1}(X)\sim\dfrac{1}{16 X^2}-\dfrac{1}{12}
	\label{17_E25}
	\\
	g_{2}(X)\sim\dfrac{1}{8 X^3}+\dfrac{1}{8 X^2}
	\label{17_E26}
\end{eqnarray}

Thus, since $\Delta m=\eta_1$, we have from Eqns. (\ref{1_E24}),(\ref{17_E23}-\ref{17_E26})  that as $\eta_1\to0$,
\begin{eqnarray} 
	g_{o}(\eta_1,X) &=& \frac{1}{\eta_1}\int_{X}^{\infty}g_{1}(m)dm-\frac{1}{32X^2}+\frac{1}{24}\nonumber\\
	&& -\int_{X}^{\infty}g_{2}(m)dm+\textit{o(1)}
	\label{17_E27}
\end{eqnarray}
The integral appearing above is convergent for the large value of m. In the limit, $\Delta m \to 0$, after simplification we get the asymptotic expression for the outer expansion $g_{o}$ for $U_0(2 \eta_1)$ as follows:
\begin{eqnarray}
	g_{o}(\eta_1,X)&\sim&-\frac{3}{32X^2}-\frac{1}{8X}+\frac{1}{16X \eta_1}+\frac{X\eta_1^{-1}}{12}\nonumber\\
	&&-\frac{1}{12\eta_1}+\frac{1}{24}+\frac{K_{171}}{\eta_1}-C_{173}
	\label{17_E28}
\end{eqnarray}
\\ where,
\begin{eqnarray*}
	K_{171} &=& C_{171}+C_{172}=-0.0416667  
	\nonumber \\
	C_{171} &=& \int_{1}^{\infty} \Bigg({g_1(m)}-\frac{1}{16m^2}\Bigg)dm = -0.0578357\nonumber \\
	C_{172} &=&\int_{0}^{1} \Bigg({g_1(m)}-\frac{1}{16m^2}+\frac{1}{12}\Bigg)dm =0.016169\nonumber \\
	C_{173} &=&\int_{0}^{\infty} \Bigg({g_2(m)}-\frac{1}{8m^3}-\frac{1}{8m^2}\Bigg)dm= -0.159166\nonumber \\
\end{eqnarray*}

Adding Equation (\ref{17_E17}) and Equation (\ref{17_E28}), the asymptotic expression for $U_0(2 \eta_1)$ in the the small separation region is given by,
\begin{eqnarray}
	U_0(2 \eta_1)&\sim&\frac{{\eta_1}^{-2}}{8}-\frac{\eta_1^{-1}}{3}+\frac{1}{8}+\frac{K_{171}}{\eta_1}-C_{173}
	\label{17_E29}
\end{eqnarray}
At this stage, it can be observed that the final asymptotic expression is independent of $X$.
\paragraph*{Asymptotic expression  for the series $U_1(2 \eta_1)$: }\label{sec:18}
Substituting $m=0$ and $p=\eta_1$ in Equation(\ref{E2}), we get $U_1(2 \eta_1)$ as follows,
\begin{eqnarray}
	U_1(2 \eta_1)&=&\sum_{n=0}^{\infty}(2n+1) \dfrac{ e^{(4n+2)\eta_1}}{(e^{(4n+2)\eta_1}-1) (e^{(4n+6)\eta_1}-1)}
	\label{18_E1}
\end{eqnarray}
In this case, $U_1(2 \eta_1)$ is decomposed into an ``inner expansion'' ($g_{i}$) and an ``outer expansion'' ($g_{o}$) as follows:
\begin{eqnarray}
	U_1(2 \eta_1)&=& g_{i}(\eta_1,N)+g_{o}(\eta_1,N)
	\label{18_E2}
\end{eqnarray}
where, 
\begin{eqnarray}
	g_{i}(\eta_1,N)&=&\sum_{n=0}^{N} (2n+1)\dfrac{ e^{(4n+2)\eta_1}}{(e^{(4n+2)\eta_1}-1) (e^{(4n+6)\eta_1}-1)}
	\label{18_E3}
	\\
	g_{o}(\eta_1,N)&=&\sum_{n=N+1}^{\infty}(2n+1) \dfrac{ e^{(4n+2)\eta_1}}{(e^{(4n+2)\eta_1}-1) (e^{(4n+6)\eta_1}-1)} 
	\label{18_E4}
\end{eqnarray}

Following similar steps as shown in  Section\ref{sec:3}, we find the inner ($g_i$) and outer ($g_o$) expansions in the next two subsections.

\subparagraph*{Inner expansion}\label{subsec:A18.1}
In this subsection, we derive the inner expansion $g_{i}$ for $U_1(2 \eta_1)$.

\begin{eqnarray}
	g_{i}(\eta_1,N)&\approx&\Bigg(I_1 \bigg(\frac{{\eta_1}^{-2}}{4}-\frac{\eta_1^{-1}}{2}+\frac{1}{3}\bigg)+ \frac{1}{6}I_2-\frac{1}{12}I_3\Bigg)
	\label{18_E9}
\end{eqnarray}
\\
where
\begin{eqnarray}
	I_1&=&\sum_{n=0}^{N}\frac{1}{(2n+3)}
	\label{18_E10}\\
	I_2&=&\sum_{n=0}^{N}1
	\label{18_E11}\\
	I_3&=&\sum_{n=0}^{N}2n+3
	\label{18_E12}
\end{eqnarray}
As $N\to\infty$, $I_1$,$I_2$ and $I_3$ is given by,
\begin{eqnarray}
	I_1&\sim&\frac{1}{2}\big(\gamma+\log{4N}-2\big)+\frac{1}{N}
	\label{18_E13}\\
	I_2&=&1+\textit{O(N)}
	\label{18_E14}\\
	I_3&=&3+\textit{O(N)}
	\label{18_E15}
\end{eqnarray}
Substituting Eqns. (\ref{18_E13}),(\ref{18_E14}),(\ref{18_E15}) in Equation (\ref{18_E9}), the asymptotic expression for $g_{i}$ in terms of  $N$ and $\eta_1$ is given by,
\begin{eqnarray}
	g_{i}(\eta_1,N)&\sim&\Bigg(\bigg(\frac{1}{2}\big(\gamma+\log{4N}-2\big)+\frac{1}{N}\bigg)\bigg(\frac{{\eta_1}^{-2}}{4}-\frac{\eta_1^{-1}}{2}+\frac{1}{3}\bigg) \nonumber\\&&+\frac{1}{6}-\frac{3}{12} \Bigg) 
	\label{18_E16}
\end{eqnarray}

After introducing intermediate vanish $N=\displaystyle\frac{X}{\eta_1}$ and simplifying then Equation (\ref{18_E16}) becomes
\begin{eqnarray}
	g_{i}(\eta_1,X)&\sim&\big(\frac{{\eta_1}^{-2}}{8}-\frac{\eta_1^{-1}}{4}+\frac{1}{6}\big)\big(\gamma+\log{\frac{4}{\eta_1}}-2\big)-\frac{1}{12}\nonumber\\
	&&+\frac{1}{4X \eta_1}-\frac{1}{2X}+\big(\frac{{\eta_1}^{-2}}{8}-\frac{\eta_1^{-1}}{4}+\frac{1}{6}\big)\log{X}
	\label{18_E17}
\end{eqnarray}
Equation (\ref{18_E17}) can be used to calculate the inner expansion $g_{i}$ for $U_1(2 \eta_1)$. In the next subsection (\ref{subsec:A18.2}), we derive the outer expansion for $U_1(2 \eta_1)$.
\subparagraph*{Outer expansion }\label{subsec:A18.2}
In order to derive the outer expansion, let $m=n\eta_1$ in Equation (\ref{18_E4}) and examine the scenario where $\eta_1\to0$ with $m$ held fixed. After substituting the $n$ value in Equation (\ref{18_E4}), we get outer expansion in terms of $m$ and $\eta_1$ as follows,  
\begin{eqnarray}
	g_{o}(\eta_1,N)&=&\sum_{\substack{m=n\eta_1 \\ n=N+1}}^{\infty} \dfrac{ (2m+\eta_1)e^{(4m+2 \eta_1)}}{\eta_1(e^{(4m+2\eta_1)}-1) (e^{(4m+6\eta_1)}-1)}  
	\label{18_E18}
\end{eqnarray}
For fixed $m$,  the function expansion for $\eta_1\to 0$ is as follows:
\begin{eqnarray}
	g_{o}(\eta_1,N)&=&  \frac{1}{\eta_1}\sum_{\substack{m=n\eta_1 \\ n=N+1}}^{\infty}\Bigg(\frac{2m e^{4m}}{(-1+e^{4m})^2}-\frac{\eta_1e^{4m}}{(-1+e^{4m})^3}(1+4m+(12m-1)e^{4m})
	\nonumber
	\\&&+\frac{2e^{4m}(1+2m+(32m+2)e^{4m}+(18m-3)e^{8m})}{(-1+e^{4m})^4}\eta_1^2\Bigg)
	\label{18_E19}
\end{eqnarray}

\begin{eqnarray}
	g_{o}(\eta_1,N)&=& \frac{1}{\eta_1^2}\sum_{\substack{m=n\eta_1 \\ n=N+1}}^{\infty}g_{1}(m)\Delta m- \frac{1}{\eta_1}\sum_{\substack{m=n\eta_1 \\ n=N+1}}^{\infty}g_{2}(m)\Delta m \nonumber
	\\&& +\sum_{\substack{m=n\eta_1 \\ n=N+1}}^{\infty}g_{3}(m)\Delta m
	\label{18_E20}
\end{eqnarray}

where,
\begin{eqnarray}
	g_{1}(m)&=&\frac{2m e^{4m}}{(-1+e^{4m})^2}
	\label{18_E21}\\
	g_{2}(m)&=&\frac{e^{4m}(1+4m+(12m-1)e^{4m})}{(-1+e^{4m})^3}
	\label{18_E22}\\
	g_{3}(m)&=&\frac{2e^{4m}(1+2m+(32m+2)e^{4m}+(18m-3)e^{8m})}{(-1+e^{4m})^4}
	\label{18_E23}
\end{eqnarray}
Define $\Delta m =m_{n+1}-m_n $. This makes 
$\Delta m=(n+1)\eta_1-n\eta_1=\eta_1$
Certainly, the outer expansion can be expressed in the following manner:
\begin{eqnarray}
	g_{o}(\eta_1,X)&=&\frac{1}{\eta_1^2}\sum_{m=X}^{\infty}g_{1}(m)\Delta m -\frac{1}{\eta_1}\sum_{m=X}^{\infty}g_{2}(m)\Delta m \nonumber
	\\&&
	+\sum_{m=X}^{\infty}g_{3}(m)\Delta m
	\label{18_E24}
\end{eqnarray}
As per the definition of $X$, it is apparent that $m=X$ at the lower bound of summation.
 As $\Delta m\to 0$, this expression remains asymptotically precise. Now, from Eqns. (\ref{18_E21}-\ref{18_E23}) we get $g_1(\infty)=0$,$g_2(\infty)=0$,$g_3(\infty)=0$. Furthermore, as $X\to 0$, 
\begin{eqnarray}
	g_{1}(X)\sim\dfrac{1}{8 X}
	\label{18_E25}
	\\
	g_{2}(X)\sim\dfrac{3}{16 X^2}+\dfrac{1}{4X}+\dfrac{1}{12}
	\label{18_E26}\\
	g_{3}(X)\sim\dfrac{9}{32 X^3}+\dfrac{3}{8X^2}+\dfrac{1}{6X}
	\label{18_E27}
\end{eqnarray}

Thus, since $\Delta m=\eta_1$, we have from Eqns. (\ref{1_E24}),(\ref{18_E24}-\ref{18_E27})  that as $\eta_1\to0$,
\begin{eqnarray} 
	g_{o}(\eta_1,X) &=& \frac{1}{\eta_1^2}\int_{X}^{\infty}g_{1}(m)dm-\frac{1}{16X\eta_1} - \frac{1}{\eta_1}\int_{X}^{\infty}g_{2}(m)dm+\frac{1}{24}\nonumber\\
	&&+\int_{X}^{\infty}g_{3}(m)dm+\textit{o(1)}
	\label{18_E28}
\end{eqnarray}
The integral appearing above is convergent for the large value of m. In the limit, $\Delta m \to 0$, after simplification we get the asymptotic expression for the outer expansion $g_{o}$ for $U_1(2 \eta_1)$ as follows:
\begin{eqnarray}
	g_{o}(\eta_1,X)&\sim&-\frac{1}{4X \eta_1}+\frac{1}{2X}-\big(\frac{{\eta_1}^{-2}}{8}-\frac{\eta_1^{-1}}{4}+\frac{1}{6}\big)\log{X}\nonumber\\
	&&-\frac{1}{12 \eta_1}+\frac{1}{24}+\frac{K_{181}}{\eta_1^2}-\frac{K_{182}}{\eta_1}+K_{183}
	\label{18_E29}
\end{eqnarray}
\\ where,
\begin{eqnarray*}
	K_{181} &=& C_{181}+C_{182}=-0.0482868 \nonumber \\ 
	K_{182} &=& C_{183}+C_{184}=-0.304907\\
	K_{183} &=& C_{185}+C_{186}=-0.321327
	\nonumber \\
	C_{181} &=& \int_{1}^{\infty} {g_1(m)}dm = 0.0116394\nonumber \\
	C_{182} &=&\int_{0}^{1} \Bigg({g_1(m)}-\frac{1}{8m}\Bigg)dm =-0.0599262\nonumber \\
	C_{183} &=&\int_{1}^{\infty} \Bigg({g_2(m)}-\frac{3}{16m^2}\Bigg)dm= -0.121546\nonumber \\
	C_{184} &=&\int_{0}^{1} \Bigg({g_2(m)}-\frac{3}{16m^2}-\frac{1}{4m}-\frac{1}{12}\Bigg)dm=-0.183361 \nonumber \\
	C_{185} &=&\int_{1}^{\infty} \Bigg({g_3(m)}-\frac{9}{32m^3}-\frac{3}{8m^2}\Bigg)dm= -0.327741\nonumber \\
	C_{186} &=&\int_{0}^{1} \Bigg({g_3(m)}-\frac{9}{32m^3}-\frac{3}{8m^2}-\frac{1}{6m}\Bigg)dm=0.00641461\nonumber \\
\end{eqnarray*}

Adding Equation (\ref{18_E17}) and Equation (\ref{18_E29}), the asymptotic expression for $U_1(2 \eta_1)$ in the the small separation region is given by,
\begin{eqnarray}
	U_1(2 \eta_1)&\sim& \big(\frac{{\eta_1}^{-2}}{8}-\frac{\eta_1^{-1}}{4}+\frac{1}{6}\big)\big(\gamma+\log{\frac{4}{\eta_1}}-2\big)-\frac{1}{24}\nonumber\\
	&&-\frac{1}{12 \eta_1}+\frac{K_{181}}{\eta_1^2}-\frac{K_{182}}{\eta_1}+K_{183}
	\label{18_E30}
\end{eqnarray}
At this stage, it can be observed that the final asymptotic expression is independent of $X$.
\paragraph*{Asymptotic expression  for the series $U_2(2 \eta_1)$: }\label{sec:19}
Substituting $m=0$ and $p=\eta_1$ in Equation(\ref{E2}), we get $U_2(2 \eta_1)$ as follows,
\begin{eqnarray}
	U_2(2 \eta_1)&=&\sum_{n=0}^{\infty}(2n+1)^2 \dfrac{ e^{(4n+2)\eta_1}}{(e^{(4n+2)\eta_1}-1) (e^{(4n+6)\eta_1}-1)}
	\label{19_E1}
\end{eqnarray}
In this case, $U_2(2 \eta_1)$ is decomposed into an ``inner expansion'' ($g_{i}$) and an ``outer expansion'' ($g_{o}$) as follows:
\begin{eqnarray}
	U_2(2 \eta_1)&=& g_{i}(\eta_1,N)+g_{o}(\eta_1,N)
	\label{19_E2}
\end{eqnarray}
where, 
\begin{eqnarray}
	g_{i}(\eta_1,N)&=&\sum_{n=0}^{N} (2n+1)^2\dfrac{ e^{(4n+2)\eta_1}}{(e^{(4n+2)\eta_1}-1) (e^{(4n+6)\eta_1}-1)}
	\label{19_E3}
	\\
	g_{o}(\eta_1,N)&=&\sum_{n=N+1}^{\infty}(2n+1)^2 \dfrac{ e^{(4n+2)\eta_1}}{(e^{(4n+2)\eta_1}-1) (e^{(4n+6)\eta_1}-1)} 
	\label{19_E4}
\end{eqnarray}

Following similar steps as shown in  Section\ref{sec:3}, we find the inner ($g_i$) and outer ($g_o$) expansions in the next two subsections.

\subparagraph*{Inner expansion}\label{subsec:A19.1}
In this subsection, we derive the inner expansion $g_{i}$ for $U_2(2 \eta_1)$.

\begin{eqnarray}
	g_{i}(\eta_1,N)&\approx&\Bigg(I_1 \bigg(\frac{{-\eta_1}^{-2}}{2}+\eta_1^{-1}-\frac{2}{3}\bigg)\nonumber\\&&+I_2 \bigg(\frac{{\eta_1}^{-2}}{4}-\frac{\eta_1^{-1}}{2}+\frac{1}{12}\bigg)+\frac{1}{6}I_3\Bigg)
	\label{19_E9}
\end{eqnarray}
\\
where
\begin{eqnarray}
	I_1&=&\sum_{n=0}^{N}\frac{1}{(2n+3)}
	\label{19_E10}\\
	I_2&=&\sum_{n=0}^{N}1
	\label{19_E11}\\
	I_3&=&\sum_{n=0}^{N}2n+1
	\label{19_E12}
\end{eqnarray}
As $N\to\infty$, $I_1$,$I_2$ and $I_3$ is given by,
\begin{eqnarray}
	I_1&\sim&\frac{1}{2}\big(\gamma+\log{4N}-2\big)+\frac{1}{N}
	\label{19_E13}\\
	I_2&=&1+\textit{O(N)}
	\label{19_E14}\\
	I_3&=&1+\textit{O(N)}
	\label{19_E15}
\end{eqnarray}
Substituting Eqns. (\ref{19_E13}),(\ref{19_E14}),(\ref{19_E15}) in Equation (\ref{19_E9}), the asymptotic expression for $g_{i}$ in terms of  $N$ and $\eta_1$ is given by,
\begin{eqnarray}
	g_{i}(\eta_1,N)&\sim&\Bigg(\bigg(\frac{1}{2}\big(\gamma+\log{4N}-2\big)+\frac{1}{N}\bigg)\bigg(\frac{{-\eta_1}^{-2}}{2}+\eta_1^{-1}-\frac{2}{3}\bigg)\nonumber\\&&\bigg(\frac{{\eta_1}^{-2}}{4}-\frac{\eta_1^{-1}}{2}+\frac{1}{12}\bigg)+\frac{1}{6} \Bigg) 
	\label{19_E16}
\end{eqnarray}

After introducing intermediate vanish $N=\displaystyle\frac{X}{\eta_1}$ and simplifying then Equation (\ref{19_E16}) becomes
\begin{eqnarray}
	g_{i}(\eta_1,X)&\sim&\frac{\eta_1^{-2}}{4}-\frac{1}{2{\eta_1}}+\bigg(\frac{-{\eta_1}^{-2}}{4}+\frac{\eta_1^{-1}}{2}-\frac{1}{3}\bigg)\bigg(\gamma+\log{\frac{4}{\eta_1}}-2\bigg)+\frac{1}{4}\nonumber\\
	&&+\frac{1}{X}-\frac{1}{2X \eta_1}+\bigg(\frac{-{\eta_1}^{-2}}{4}+\frac{\eta_1^{-1}}{2}-\frac{1}{3}\bigg)\log{X}
	\label{19_E17}
\end{eqnarray}
Equation (\ref{19_E17}) can be used to calculate the inner expansion $g_{i}$ for $U_2(2 \eta_1)$. In the next subsection (\ref{subsec:A19.2}), we derive the outer expansion for $U_2(2 \eta_1)$.
\subparagraph*{Outer expansion }\label{subsec:A19.2}
In order to derive the outer expansion, let $m=n\eta_1$ in Equation (\ref{19_E4}) and examine the scenario where $\eta_1\to0$ with $m$ held fixed. After substituting the $n$ value in Equation (\ref{19_E4}), we get outer expansion in terms of $m$ and $\eta_1$ as follows,  
\begin{eqnarray}
	g_{o}(\eta_1,N)&=&\sum_{\substack{m=n\eta_1 \\ n=N+1}}^{\infty} \dfrac{ (2m+\eta_1)^2e^{(4m+2\eta_1)}}{\eta_1^2(e^{(4m+2\eta_1)}-1) (e^{(4m+6\eta_1)}-1)}  
	\label{19_E18}
\end{eqnarray}
For fixed $m$,  the function expansion for $\eta_1\to 0$ is as follows:
\begin{eqnarray}
	g_{o}(\eta_1,N)&=&  \frac{1}{\eta_1^2}\sum_{\substack{m=n\eta_1 \\ n=N+1}}^{\infty}\Bigg(\frac{4m^2 e^{4m}}{(-1+e^{4m})^2}-\frac{4m\eta_1e^{4m}}{(-1+e^{4m})^3}(1+2m+(6m-1)e^{4m})
	\nonumber
	\\&&+\frac{e^{4m}(1+8m+8m^2+(128m^2+16m-2)e^{4m}+(72m^2-24m+1)e^{8m})}{(-1+e^{4m})^4}\eta_1^2
	\nonumber\\
	&&-\frac{e^{4m}}{3(-1+e^{4m})^5}\bigg(3+12m+8m^2+(536m^2+180m+3)e^{4m}+(1160m^2-84m-15)e^{8m} \nonumber\\
	&&+(216m^2-108m+9)e^{12m}\bigg)\eta_1^3\Bigg)
	\label{19_E19}
\end{eqnarray}

\begin{eqnarray}
	g_{o}(\eta_1,N)&=& \frac{1}{\eta_1^3}\sum_{\substack{m=n\eta_1 \\ n=N+1}}^{\infty}g_{1}(m)\Delta m- \frac{1}{\eta_1^2}\sum_{\substack{m=n\eta_1 \\ n=N+1}}^{\infty}g_{2}(m)\Delta m \nonumber
	\\&& +\frac{1}{\eta_1}\sum_{\substack{m=n\eta_1 \\ n=N+1}}^{\infty}g_{3}(m)\Delta m-\sum_{\substack{m=n\eta_1 \\ n=N+1}}^{\infty}g_{4}(m)\Delta m
	\label{19_E20}
\end{eqnarray}

where,
\begin{eqnarray}
	g_{1}(m)&=&\frac{4m^2 e^{4m}}{(-1+e^{4m})^2}
	\label{19_E21}\\
	g_{2}(m)&=&\frac{4m e^{4m}}{(-1+e^{4m})^3}(1+2m+(6m-1)e^{4m})
	\label{19_E22}\\
	g_{3}(m)&=&\frac{e^{4m}(1+8m+8m^2+(128m^2+16m-2)e^{4m}+(72m^2-24m+1)e^{8m})}{(-1+e^{4m})^4}
	\label{19_E23}\\
	g_{4}(m)&=& \frac{e^{4m}}{3(-1+e^{4m})^5}\bigg(3+12m+8m^2+(536m^2+180m+3)e^{4m}+(1160m^2-84m-15)e^{8m} \nonumber\\
	&&+(216m^2-108m+9)e^{12m}\bigg)
	\label{19_E24}
\end{eqnarray}
Define $\Delta m =m_{n+1}-m_n $. This makes 
$\Delta m=(n+1)\eta_1-n\eta_1=\eta_1$
Certainly, the outer expansion can be expressed in the following manner:
\begin{eqnarray}
	g_{o}(\eta_1,X)&=&\frac{1}{\eta_1^3}\sum_{m=X}^{\infty}g_{1}(m)\Delta m -\frac{1}{\eta_1^2}\sum_{m=X}^{\infty}g_{2}(m)\Delta m \nonumber
	\\&&
	+\frac{1}{\eta_1}\sum_{m=X}^{\infty}g_{3}(m)\Delta m
	-\sum_{m=X}^{\infty}g_{4}(m)\Delta m
	\label{19_E25}
\end{eqnarray}
As per the definition of $X$, it is apparent that $m=X$ at the lower bound of summation. As $\Delta m\to 0$, this expression remains asymptotically precise. Now, from Eqns. (\ref{19_E21}-\ref{19_E24}) we get $g_1(\infty)=0$,$g_2(\infty)=0$,$g_3(\infty)=0$,$g_4(\infty)=0$. Furthermore, as $X\to 0$, 
\begin{eqnarray}
	g_{1}(X)\sim\dfrac{1}{4}
	\label{19_E26}
	\\
	g_{2}(X)\sim\dfrac{1}{4 X}+\dfrac{1}{2}
	\label{19_E27}\\
	g_{3}(X)\sim\dfrac{3}{8 X^2}+\dfrac{1}{2X}+\dfrac{1}{4}
	\label{19_E28}\\
	g_{4}(X)\sim\dfrac{9}{16 X^3}+\dfrac{3}{4 X^2}+\dfrac{1}{3 X}
	\label{19_E29}
\end{eqnarray}

Thus, since $\Delta m=\eta_1$, we have from Eqns. (\ref{1_E24}),(\ref{19_E25}-\ref{19_E29})  that as $\eta_1\to0$,
\begin{eqnarray} 
	g_{o}(\eta_1,X) &=& \frac{1}{\eta_1^3}\int_{X}^{\infty}g_{1}(m)dm-\frac{1}{8\eta_1^2} - \frac{1}{\eta_1^2}\int_{X}^{\infty}g_{2}(m)dm+\frac{1}{8X\eta_1}+\frac{1}{4\eta_1}\nonumber\\&&+ \frac{1}{\eta_1}\int_{X}^{\infty}g_{3}(m)dm-\frac{1}{4X}-\frac{1}{8}-\int_{X}^{\infty}g_{4}(m)dm+\textit{o(1)}
	\label{19_E30}
\end{eqnarray}

The integral appearing above is convergent for the large value of m. In the limit, $\Delta m \to 0$, after simplification we get the asymptotic expression for the outer expansion $g_{o}$ for $U_2(2 \eta_1)$ as follows:
\begin{eqnarray}
	g_{o}(\eta_1,X)&\sim&-\frac{1}{X}+\frac{1}{2X \eta_1}-\bigg(\frac{-{\eta_1}^{-2}}{4}+\frac{\eta_1^{-1}}{2}-\frac{1}{3}\bigg)\log{X}+\frac{1}{4\eta_1^3}-\frac{5}{8\eta_1^2}\nonumber\\
	&&+\frac{1}{2 \eta_1}-\frac{1}{8}+\frac{K_{191}}{\eta_1^3}-\frac{K_{192}}{\eta_1^2}+\frac{K_{193}}{\eta_1}-K_{194}
	\label{19_E31}
\end{eqnarray}
\\ where,
\begin{eqnarray*}
	K_{191} &=& C_{191}+C_{192}=-0.0443832 \nonumber \\ 
	K_{192} &=& C_{193}+C_{194}=0.0646599\\
	K_{193} &=& C_{195}+C_{196}=0.164342\\
	K_{194} &=& C_{197}+C_{198}=-0.170431
	\nonumber \\
	C_{191} &=& \int_{1}^{\infty} {g_1(m)} dm = 0.0302001\nonumber \\
	C_{192} &=&\int_{0}^{1} \Bigg({g_1(m)}-\frac{1}{4}\Bigg)dm =-0.0745833\nonumber \\
	C_{193} &=&\int_{1}^{\infty}{g_2(m)} dm= 0.159701\nonumber \\
	C_{194} &=&\int_{0}^{1} \Bigg({g_2(m)}-\frac{1}{4m}-\frac{1}{2}\Bigg)dm=-0.0950408 \nonumber \\
	C_{195} &=&\int_{1}^{\infty} \Bigg({g_3(m)}-\frac{3}{8m^2}\Bigg)dm= 0.0472824\nonumber \\
	C_{196} &=&\int_{0}^{1} \Bigg({g_3(m)}-\frac{3}{8m^2}-\frac{1}{2m}-\frac{1}{4}\Bigg)dm=0.117059\nonumber \\
	C_{197} &=&\int_{1}^{\infty} \Bigg({g_4(m)}-\frac{9}{16m^3}-\frac{3}{4m^2}\Bigg)dm= -0.279644\nonumber \\
	C_{198} &=&\int_{0}^{1} \Bigg({g_4(m)}-\frac{9}{16m^3}-\frac{3}{4m^2}-\frac{1}{3m}\Bigg)dm=0.109213 \nonumber \\
\end{eqnarray*}
Adding Equation (\ref{19_E17}) and Equation (\ref{19_E31}), the asymptotic expression for $U_2(2 \eta_1)$ in the the small separation region is given by,
\begin{eqnarray}
	U_2(2 \eta_1)&\sim& \bigg(\frac{-{\eta_1}^{-2}}{4}+\frac{\eta_1^{-1}}{2}-\frac{1}{3}\bigg)\bigg(\gamma+\log{\frac{4}{\eta_1}}-2\bigg)+\frac{1}{8}\nonumber\\
	&&+\frac{1}{4\eta_1^3}-\frac{3}{8\eta_1^2}+\frac{K_{191}}{\eta_1^3}-\frac{K_{192}}{\eta_1^2}+\frac{K_{193}}{\eta_1}-K_{194}
	\label{19_E32}
\end{eqnarray}
At this stage, it can be observed that the final asymptotic expression is independent of $X$.
\paragraph*{Asymptotic expression  for the series $U_3(2 \eta_1)$: }\label{sec:20}
Substituting $m=0$ and $p=\eta_1$ in Equation(\ref{E2}), we get $U_3(2 \eta_1)$ as follows,
\begin{eqnarray}
	U_3(2 \eta_1)&=&\sum_{n=0}^{\infty}(2n+1)^3 \dfrac{ e^{(4n+2)\eta_1}}{(e^{(4n+2)\eta_1}-1) (e^{(4n+6)\eta_1}-1)}
	\label{20_E1}
\end{eqnarray}
In this case, $U_3(2 \eta_1)$ is decomposed into an ``inner expansion'' ($g_{i}$) and an ``outer expansion'' ($g_{o}$) as follows:
\begin{eqnarray}
	U_3(2 \eta_1)&=& g_{i}(\eta_1,N)+g_{o}(\eta_1,N)
	\label{20_E2}
\end{eqnarray}
where, 
\begin{eqnarray}
	g_{i}(\eta_1,N)&=&\sum_{n=0}^{N} (2n+1)^3\dfrac{ e^{(4n+2)\eta_1}}{(e^{(4n+2)\eta_1}-1) (e^{(4n+6)\eta_1}-1)}
	\label{20_E3}
	\\
	g_{o}(\eta_1,N)&=&\sum_{n=N+1}^{\infty}(2n+1)^3 \dfrac{ e^{(4n+2)\eta_1}}{(e^{(4n+2)\eta_1}-1) (e^{(4n+6)\eta_1}-1)} 
	\label{20_E4}
\end{eqnarray}

Following similar steps as shown in  Section\ref{sec:3}, we find the inner ($g_i$) and outer ($g_o$) expansions in the next two subsections.

\subparagraph*{Inner expansion}\label{subsec:A20.1}
In this subsection, we derive the inner expansion $g_{i}$ for $U_3(2 \eta_1)$.

\begin{eqnarray}
	g_{i}(\eta_1,N)&\approx&\Bigg(I_1 \bigg({\eta_1}^{-2}-2 \eta_1^{-1}+\frac{4}{3}\bigg)-I_2 \bigg(\frac{{\eta_1}^{-2}}{2}-\eta_1^{-1}+\frac{2}{3}\bigg)\nonumber\\&&+\bigg(\frac{{\eta_1}^{-2}}{4}-\frac{\eta_1^{-1}}{2}+\frac{1}{4}\bigg)I_3\Bigg)
	\label{20_E9}
\end{eqnarray}
\\
where
\begin{eqnarray}
	I_1&=&\sum_{n=0}^{N}\frac{1}{(2n+3)}
	\label{20_E10}\\
	I_2&=&\sum_{n=0}^{N}1
	\label{20_E11}\\
	I_3&=&\sum_{n=0}^{N}2n+1
	\label{20_E12}
\end{eqnarray}
As $N\to\infty$, $I_1$,$I_2$ and $I_3$ is given by,
\begin{eqnarray}
	I_1&\sim&\frac{1}{2}\big(\gamma+\log{4N}-2\big)+\frac{1}{N}
	\label{20_E13}\\
	I_2&=&N+1
	\label{20_E14}\\
	I_3&=&1+\textit{O(N)}
	\label{20_E15}
\end{eqnarray}
Substituting Eqns. (\ref{20_E13}),(\ref{20_E14}),(\ref{20_E15}) in Equation (\ref{20_E9}), the asymptotic expression for $g_{i}$ in terms of  $N$ and $\eta_1$ is given by,
\begin{eqnarray}
	g_{i}(\eta_1,N)&\sim&\Bigg(\bigg(\frac{1}{2}\big(\gamma+\log{4N}-2\big)+\frac{1}{N}\bigg)\bigg({\eta_1}^{-2}-2 \eta_1^{-1}+\frac{4}{3}\bigg) \nonumber\\&&-\bigg(\frac{{\eta_1}^{-2}}{2}-\eta_1^{-1}+\frac{2}{3}\bigg)+\bigg(\frac{{\eta_1}^{-2}}{4}-\frac{\eta_1^{-1}}{2}+\frac{1}{4}\bigg) \Bigg) 
	\label{20_E16}
\end{eqnarray}

After introducing intermediate vanish $N=\displaystyle\frac{X}{\eta_1}$ and simplifying then Equation (\ref{20_E16}) becomes
\begin{eqnarray}
	g_{i}(\eta_1,X)&\sim&-\frac{\eta_1^{-2}}{4}+\frac{1}{2{\eta_1}}+\bigg(\frac{{\eta_1}^{-2}}{2}-\eta_1^{-1}+\frac{2}{3}\bigg)\bigg(\gamma+\log{\frac{4}{\eta_1}}-2\bigg)-\frac{5}{12}\nonumber\\
	&&+\frac{1}{X \eta_1}-\frac{2}{X}+\bigg(\frac{{\eta_1}^{-2}}{2}-\eta_1^{-1}+\frac{2}{3}\bigg)\log{X}
	\label{20_E17}
\end{eqnarray}
Equation (\ref{20_E17}) can be used to calculate the inner expansion $g_{i}$ for $U_3(2 \eta_1)$. In the next subsection (\ref{subsec:A20.2}), we derive the outer expansion for $U_3(2 \eta_1)$.
\subparagraph*{Outer expansion }\label{subsec:A20.2}
In order to derive the outer expansion, let $m=n\eta_1$ in Equation (\ref{20_E4}) and examine the scenario where $\eta_1\to0$ with $m$ held fixed. After substituting the $n$ value in Equation (\ref{20_E4}), we get outer expansion in terms of $m$ and $\eta_1$ as follows,  
\begin{eqnarray}
	g_{o}(\eta_1,N)&=&\sum_{\substack{m=n\eta_1 \\ n=N+1}}^{\infty} \dfrac{ (2m+\eta_1)^3e^{(4m+2\eta_1)}}{\eta_1^3(e^{(4m+2\eta_1)}-1) (e^{(4m+6\eta_1)}-1)}  
	\label{20_E18}
\end{eqnarray}
For fixed $m$,  the function expansion for $\eta_1\to 0$ is as follows:
\begin{eqnarray}
	g_{o}(\eta_1,N)&=&  \frac{1}{\eta_1^3}\sum_{\substack{m=n\eta_1 \\ n=N+1}}^{\infty}\Bigg(\frac{8m^3 e^{4m}}{(-1+e^{4m})^2}-\frac{4m^2\eta_1e^{4m}}{(-1+e^{4m})^3}(3+4m+(12m-3)e^{4m})
	\nonumber
	\\&&+\frac{2me^{4m}(3+12m+8m^2+(128m^2+24m-6)e^{4m}+(72m^2-36m+3)e^{8m})}{(-1+e^{4m})^4}\eta_1^2
	\nonumber\\
	&&-\frac{e^{4m}}{3(-1+e^{4m})^5}\bigg(3+36m+72m^2+32m^3+(2144m^3+1080m^2+36m-9)e^{4m}\nonumber\\
	&&+(4640m^3-504m^2-180m+9)e^{8m}+(864m^3-648m^2+108m-3)e^{12m}\bigg)\eta_1^3\nonumber\\
	&&+\frac{\eta_1^4 2 e^{4m}}{3(-1+e^{4m})^6}\bigg(3+18m+24m^2+8m^3+(2128m^3+1584m^2+252m)e^{4m} \nonumber\\
	&&+(11920m^3+1872m^2-396m-18)e^{8m}+(8528m^3-2832m^2-36m+24)e^{12m}\nonumber\\
	&&+(648m^3-648m^2+162m-9)e^{16m}\bigg)\Bigg)
	\label{20_E19}
\end{eqnarray}

\begin{eqnarray}
	g_{o}(\eta_1,N)&=& \frac{1}{\eta_1^4}\sum_{\substack{m=n\eta_1 \\ n=N+1}}^{\infty}g_{1}(m)\Delta m- \frac{1}{\eta_1^3}\sum_{\substack{m=n\eta_1 \\ n=N+1}}^{\infty}g_{2}(m)\Delta m  +\frac{1}{\eta_1^2}\sum_{\substack{m=n\eta_1 \\ n=N+1}}^{\infty}g_{3}(m)\Delta m\nonumber
	\\&&-\frac{1}{\eta_1}\sum_{\substack{m=n\eta_1 \\ n=N+1}}^{\infty}g_{4}(m)\Delta m+\sum_{\substack{m=n\eta_1 \\ n=N+1}}^{\infty}g_{5}(m)\Delta m
	\label{20_E20}
\end{eqnarray}

where,
\begin{eqnarray}
	g_{1}(m)&=&\frac{8m^3 e^{4m}}{(-1+e^{4m})^2}
	\label{20_E21}\\
	g_{2}(m)&=&\frac{4m^2e^{4m}}{(-1+e^{4m})^3}(3+4m+(12m-3)e^{4m})
	\label{20_E22}\\
	g_{3}(m)&=&\frac{2me^{4m}(3+12m+8m^2+(128m^2+24m-6)e^{4m}+(72m^2-36m+3)e^{8m})}{(-1+e^{4m})^4}
	\label{20_E23}\\
	g_{4}(m)&=&\frac{e^{4m}}{3(-1+e^{4m})^5}\bigg(3+36m+72m^2+32m^3+(2144m^3+1080m^2+36m-9)e^{4m}\nonumber\\
	&&+(4640m^3-504m^2-180m+9)e^{8m}+(864m^3-648m^2+108m-3)e^{12m}\bigg)
	\label{20_E24}\\
	g_{5}(m)&=& \frac{2 e^{4m}}{3(-1+e^{4m})^6}\bigg(3+18m+24m^2+8m^3+(2128m^3+1584m^2+252m)e^{4m} \nonumber\\
	&&+(11920m^3+1872m^2-396m-18)e^{8m}+(8528m^3-2832m^2-36m+24)e^{12m}\nonumber\\
	&&+(648m^3-648m^2+162m-9)e^{16m}\bigg)
	\label{20_E25}
\end{eqnarray}
Define $\Delta m =m_{n+1}-m_n $. This makes 
$\Delta m=(n+1)\eta_1-n\eta_1=\eta_1$
Certainly, the outer expansion can be expressed in the following manner:
\begin{eqnarray}
	g_{o}(\eta_1,X)&=&\frac{1}{\eta_1^4}\sum_{m=X}^{\infty}g_{1}(m)\Delta m -\frac{1}{\eta_1^3}\sum_{m=X}^{\infty}g_{2}(m)\Delta m+\frac{1}{\eta_1^2}\sum_{m=X}^{\infty}g_{3}(m)\Delta m\nonumber
	\\&&
	-\frac{1}{\eta_1}\sum_{m=X}^{\infty}g_{4}(m)\Delta m+\sum_{m=X}^{\infty}g_{5}(m)\Delta m
	\label{20_E26}
\end{eqnarray}
As per the definition of $X$, it is apparent that $m=X$ at the lower bound of summation. As $\Delta m\to 0$, this expression remains asymptotically precise. Now, from Eqns. (\ref{20_E21}-\ref{20_E25}) we get $g_1(\infty)=0$,$g_2(\infty)=0$,$g_3(\infty)=0$,$g_4(\infty)=0$,$g_5(\infty)=0$. Furthermore, as $X\to 0$, 
\begin{eqnarray}
	g_{1}(X)\sim\textit{O(X)}
	\label{20_E27} \\
	g_{2}(X)\sim\dfrac{1}{4}
	\label{20_E28}\\
	g_{3}(X)\sim\dfrac{1}{2 X}+\dfrac{1}{2}
	\label{20_E29}\\
	g_{4}(X)\sim\dfrac{3}{4 X^2}+\dfrac{1}{X}+\dfrac{5}{12}
	\label{20_E30}\\
	g_{5}(X)\sim\dfrac{9}{8 X^3}+\dfrac{3}{2 X^2}+\dfrac{2}{3 X}
	\label{20_E31}
\end{eqnarray}

Thus, since $\Delta m=\eta_1$, we have from Eqns. (\ref{1_E24}),(\ref{20_E26}-\ref{20_E31})  that as $\eta_1\to0$,
\begin{eqnarray} 
	g_{o}(\eta_1,X) &=& \frac{1}{\eta_1^4}\int_{X}^{\infty}g_{1}(m)dm - \frac{1}{\eta_1^3}\int_{X}^{\infty}g_{2}(m)dm+\frac{1}{8\eta_1^2}+ \frac{1}{\eta_1^2}\int_{X}^{\infty}g_{3}(m)dm-\frac{1}{4X\eta_1}\nonumber\\&&-\frac{1}{4\eta_1}-\frac{1}{\eta_1}\int_{X}^{\infty}g_{4}(m)dm+\frac{1}{2X}+\frac{5}{24}+\int_{X}^{\infty}g_{5}(m)dm+\textit{o(1)}
	\label{20_E32}
\end{eqnarray}

The integral appearing above is convergent for the large value of m. In the limit, $\Delta m \to 0$, after simplification we get the asymptotic expression for the outer expansion $g_{o}$ for $U_3(2 \eta_1)$ as follows:
\begin{eqnarray}
	g_{o}(\eta_1,X)&\sim&-\frac{1}{X \eta_1}+\frac{2}{X}-\bigg(\frac{{\eta_1}^{-2}}{2}-\eta_1^{-1}+\frac{2}{3}\bigg)\log{X}-\frac{1}{4\eta_1^3}+\frac{5}{8\eta_1^2}-\frac{2}{3\eta_1}\nonumber\\
	&&+\frac{5}{24}+\frac{C_{201}}{\eta_1^4}-\frac{K_{201}}{\eta_1^3}+\frac{K_{202}}{\eta_1^2}-\frac{K_{203}}{\eta_1}+K_{204}
	\label{20_E33}
\end{eqnarray}
\\ where,
\begin{eqnarray*}
	K_{201} &=& C_{202}+C_{203}=0.817622 \nonumber \\ 
	K_{202} &=& C_{204}+C_{205}=1.44523\\
	K_{203} &=& C_{206}+C_{207}=0.977839\\
	K_{204} &=& C_{208}+C_{209}=-0.0566794
	\nonumber \\
	C_{201} &=& \int_{0}^{\infty} {g_1(m)} dm =0.225386\nonumber \\
	C_{202} &=&\int_{1}^{\infty}{g_2(m)} dm= 0.407214\nonumber \\
	C_{203} &=&\int_{0}^{1} \Bigg({g_2(m)}-\frac{1}{4}\Bigg)dm=0.410407 \nonumber \\
	C_{204} &=&\int_{1}^{\infty} {g_3(m)}dm= 1.00228\nonumber \\
	C_{205} &=&\int_{0}^{1} \Bigg({g_3(m)}-\frac{1}{2m}-\frac{1}{2}\Bigg)dm=0.442949\nonumber \\
	C_{206} &=&\int_{1}^{\infty} \Bigg({g_4(m)}-\frac{3}{4m^2}\Bigg)dm=0.900755\nonumber \\
	C_{207} &=&\int_{0}^{1} \Bigg({g_4(m)}-\frac{3}{4m^2}-\frac{1}{m}-\frac{5}{12}\Bigg)dm=0.0770848 \nonumber \\
	C_{208} &=& \int_{1}^{\infty} \Bigg({g_5(m)}-\frac{9}{8m^3}-\frac{3}{2m^2}\Bigg) dm = 0.0290593\nonumber \\
	C_{209} &=&\int_{0}^{1} \Bigg({g_5(m)}-\frac{9}{8m^3}-\frac{3}{2m^2}-\frac{2}{3m}\Bigg)dm =-0.0857387\nonumber \\
\end{eqnarray*}
Adding Equation (\ref{20_E17}) and Equation (\ref{20_E33}), the asymptotic expression for $U_3(2 \eta_1)$ in the the small separation region is given by,
\begin{eqnarray}
	U_3(2 \eta_1)&\sim&\frac{1}{2{\eta_1}}+\bigg(\frac{{\eta_1}^{-2}}{2}-\eta_1^{-1}+\frac{2}{3}\bigg)\bigg(\gamma+\log{\frac{4}{\eta_1}}-2\bigg)-\frac{1}{4\eta_1^3}+\frac{3}{8\eta_1^2}-\frac{1}{6\eta_1}-\frac{5}{24}\nonumber\\
	&&+\frac{C_{201}}{\eta_1^4}-\frac{K_{201}}{\eta_1^3}+\frac{K_{202}}{\eta_1^2}-\frac{K_{203}}{\eta_1}+K_{204}
	\label{20_E34}
\end{eqnarray}
At this stage, it can be observed that the final asymptotic expression is independent of $X$.
\paragraph*{Asymptotic expression  for the series $U_0(3 \eta_1)$: }\label{sec:21}
Substituting $m=0$ and $p=\eta_1$ in Equation(\ref{E2}), we get $U_0(3 \eta_1)$ as follows,
\begin{eqnarray}
	U_0(3 \eta_1)&=&\sum_{n=0}^{\infty} \dfrac{ e^{(6n+3)\eta_1}}{(e^{(4n+2)\eta_1}-1) (e^{(4n+6)\eta_1}-1)}
	\label{21_E1}
\end{eqnarray}
In this case, $U_0(3 \eta_1)$ is decomposed into an ``inner expansion'' ($g_{i}$) and an ``outer expansion'' ($g_{o}$) as follows:
\begin{eqnarray}
	U_0(3 \eta_1)&=& g_{i}(\eta_1,N)+g_{o}(\eta_1,N)
	\label{21_E2}
\end{eqnarray}
where, 
\begin{eqnarray}
	g_{i}(\eta_1,N)&=&\sum_{n=0}^{N} \dfrac{ e^{(6n+3)\eta_1}}{(e^{(4n+2)\eta_1}-1) (e^{(4n+6)\eta_1}-1)}
	\label{21_E3}
	\\
	g_{o}(\eta_1,N)&=&\sum_{n=N+1}^{\infty} \dfrac{ e^{(6n+3)\eta_1}}{(e^{(4n+2)\eta_1}-1) (e^{(4n+6)\eta_1}-1)} 
	\label{21_E4}
\end{eqnarray}

Following similar steps as shown in  Section\ref{sec:3}, we find the inner ($g_i$) and outer ($g_o$) expansions in the next two subsections.

\subparagraph*{Inner expansion}\label{subsec:A21.1}
In this subsection, we derive the inner expansion $g_{i}$ for $U_0(3 \eta_1)$.

\begin{eqnarray}
	g_{i}(\eta_1,N)&\approx&\Bigg(I_1 \bigg(\frac{{\eta_1}^{-2}}{4}-\frac{\eta_1^{-1}}{2}+\frac{1}{3}\bigg)+I_2 (\frac{{\eta_1}^{-1}}{4}-\frac{3}{4})+\frac{1}{24}I_3\Bigg)
	\label{21_E9}
\end{eqnarray}
\\
where
\begin{eqnarray}
	I_1&=&\sum_{n=0}^{N}\frac{1}{(2n+1)(2n+3)}
	\label{21_E10}\\
	I_2&=&\sum_{n=0}^{N}\frac{1}{2n+3}
	\label{21_E11}\\
	I_3&=&\sum_{n=0}^{N}1
	\label{21_E12}
\end{eqnarray}
As $N\to\infty$, $I_1$,$I_2$ and $I_3$ is given by,
\begin{eqnarray}
	I_1&\sim&\frac{1}{2}-\frac{1}{4N}+\frac{3}{8N^2}
	\label{21_E13}\\
	I_2&\sim&\frac{1}{2}\big(\gamma+\log{4N}-2\big)+\frac{1}{N}
	\label{21_E14}\\
	I_3&=&N+1
	\label{21_E15}
\end{eqnarray}
Substituting Eqns. (\ref{21_E13}),(\ref{21_E14}),(\ref{21_E15}) in Equation (\ref{21_E9}), the asymptotic expression for $g_{i}$ in terms of  $N$ and $\eta_1$ is given by,
\begin{eqnarray}
	g_{i}(\eta_1,N)&\sim&\Bigg(\bigg(\frac{1}{2}-\frac{1}{4N}+\frac{3}{8N^2}\bigg)\bigg(\frac{{\eta_1}^{-2}}{4}-\frac{\eta_1^{-1}}{2}+\frac{1}{3}\bigg)\nonumber\\&&+\bigg(\frac{1}{2}\big(\gamma+\log{4N}-2\big)+\frac{1}{N}\bigg)\bigg(\frac{{\eta_1}^{-1}}{4}-\frac{3}{4}\bigg)+\frac{N+1}{24}\Bigg) 
	\label{21_E16}
\end{eqnarray}

After introducing intermediate vanish $N=\displaystyle\frac{X}{\eta_1}$ and simplifying then Equation (\ref{21_E16}) becomes
\begin{eqnarray}
	g_{i}(\eta_1,X)&\sim&\frac{1}{8{\eta_1}^2}-\frac{1}{4\eta_1}+\frac{(\eta_1^{-1}-3)(\gamma+\log{\frac{4}{\eta_1}}-2)}{8}+\frac{5}{24}\nonumber\\
	&& +\frac{3}{32X^2}+\frac{3}{8X}-\frac{1}{16X \eta_1}+\frac{(\eta_1^{-1}-3)\log{X}}{8}+\frac{X\eta_1^{-1}}{24}
	\label{21_E17}
\end{eqnarray}
Equation (\ref{21_E17}) can be used to calculate the inner expansion $g_{i}$ for $U_0(3 \eta_1)$. In the next subsection (\ref{subsec:A21.2}), we derive the outer expansion for $U_0(3 \eta_1)$.
\subparagraph*{Outer expansion }\label{subsec:A21.2}
In order to derive the outer expansion, let $m=n\eta_1$ in Equation (\ref{21_E4}) and examine the scenario where $\eta_1\to0$ with $m$ held fixed. After substituting the $n$ value in Equation (\ref{21_E4}), we get outer expansion in terms of $m$ and $\eta_1$ as follows,  
\begin{eqnarray}
	g_{o}(\eta_1,N)&=&\sum_{\substack{m=n\eta_1 \\ n=N+1}}^{\infty} \dfrac{ e^{(6m+3\eta_1)}}{(e^{(4m+2\eta_1)}-1) (e^{(4m+6\eta_1)}-1)}  
	\label{21_E18}
\end{eqnarray}
For fixed $m$,  the function expansion for $\eta_1\to 0$ is as follows:
\begin{eqnarray}
	g_{o}(\eta_1,N)&=&  \sum_{\substack{m=n\eta_1 \\ n=N+1}}^{\infty}\Bigg(\frac{e^{6m}}{(-1+e^{4m})^2}-\frac{e^{6m}(3+5e^{4m})}{(-1+e^{4m})^3}\eta_1 \Bigg)
	\label{21_E19}
\end{eqnarray}

\begin{eqnarray}
	g_{o}(\eta_1,N)&=& \frac{1}{\eta_1}\sum_{\substack{m=n\eta_1 \\ n=N+1}}^{\infty}g_{1}(m)\Delta m-\sum_{\substack{m=n\eta_1 \\ n=N+1}}^{\infty}g_{2}(m)\Delta m
	\label{21_E20}
\end{eqnarray}
where,
\begin{eqnarray}
	g_{1}(m)&=&\frac{e^{6m}}{(-1+e^{4m})^2}
	\label{21_E21}\\
	g_{2}(m)&=&\frac{e^{6m}(3+5e^{4m})}{(-1+e^{4m})^3}
	\label{21_E22}
\end{eqnarray}
Define $\Delta m =m_{n+1}-m_n $. This makes 
$\Delta m=(n+1)\eta_1-n\eta_1=\eta_1$
Certainly, the outer expansion can be expressed in the following manner:
\begin{eqnarray}
	g_{o}(\eta_1,X)&=&\frac{1}{\eta_1}\sum_{m=X}^{\infty}g_{1}(m)\Delta m -\sum_{m=X}^{\infty}g_{2}(m)\Delta m
	\label{21_E23}
\end{eqnarray}
As per the definition of $X$, it is apparent that $m=X$ at the lower bound of summation. As $\Delta m\to 0$, this expression remains asymptotically precise. Now, from Eqns. (\ref{21_E21}),(\ref{21_E22}) we get $g_1(\infty)=0$,$g_2(\infty)=0$. Furthermore, as $X\to 0$, 
\begin{eqnarray}
	g_{1}(X)\sim\dfrac{1}{16 X^2}+\dfrac{1}{8 X}+\dfrac{1}{24}
	\label{21_E25}
	\\
	g_{2}(X)\sim\dfrac{1}{8 X^3}+\dfrac{5}{16 X^2}+\dfrac{3}{8 X}
	\label{21_E26}
\end{eqnarray}

Thus, since $\Delta m=\eta_1$, we have from Eqns. (\ref{1_E24}),(\ref{21_E23}),(\ref{21_E25}),(\ref{21_E26})  that as $\eta_1\to0$,
\begin{eqnarray} 
	g_{o}(\eta_1,X) &=& \frac{1}{\eta_1}\int_{X}^{\infty}g_{1}(m)dm-\frac{1}{32X^2}-\frac{1}{16X}-\frac{1}{48}\nonumber\\
	&& -\int_{X}^{\infty}g_{2}(m)dm+\textit{o(1)}
	\label{21_E27}
\end{eqnarray}
The integral appearing above is convergent for the large value of m. In the limit, $\Delta m \to 0$, after simplification we get the asymptotic expression for the outer expansion $g_{o}$ for $U_0(3 \eta_1)$ as follows:
\begin{eqnarray}
	g_{o}(\eta_1,X)&\sim&-\frac{3}{32X^2}-\frac{3}{8X}+\frac{1}{16X \eta_1}-\frac{(\eta_1^{-1}-3)\log{X}}{8}-\frac{X\eta_1^{-1}}{24}\nonumber\\
	&&+\frac{\eta_1^{-1}}{24}-\frac{1}{48}+\frac{K_{211}}{\eta_1}-K_{212}
	\label{21_E28}
\end{eqnarray}
\\ where,
\begin{eqnarray*}
	K_{211} &=& C_{211}+C_{212}=-0.0416667  \nonumber \\ 
	K_{212} &=& C_{213}+C_{214}=0.0416667
	\nonumber \\
	C_{211} &=& \int_{1}^{\infty} \Bigg({g_1(m)}-\frac{1}{16m^2}\Bigg)dm = 0.00600775\nonumber \\
	C_{212} &=&\int_{0}^{1} \Bigg({g_1(m)}-\frac{1}{16m^2}-\frac{1}{8m}-\frac{1}{24}\Bigg)dm =-0.0476744\nonumber \\
	C_{213} &=&\int_{1}^{\infty} \Bigg({g_2(m)}-\frac{1}{8m^3}-\frac{5}{16m^2}\Bigg)dm= -0.0290443\nonumber \\
	C_{214} &=&\int_{0}^{1} \Bigg({g_2(m)}-\frac{1}{8m^3}-\frac{5}{16m^2}-\frac{3}{8m}\Bigg)dm=0.070711 \nonumber \\
\end{eqnarray*}

Adding Equation (\ref{21_E17}) and Equation (\ref{21_E28}), the asymptotic expression for $U_0(3 \eta_1)$ in the the small separation region is given by,
\begin{eqnarray}
	U_0(3 \eta_1)&\sim&\frac{(\eta_1^{-1}-3)(\gamma+\log{\frac{4}{\eta_1}}-2)}{8}+\frac{3}{16}\nonumber\\&&\frac{1}{8{\eta_1}^2}-\frac{5\eta_1^{-1}}{24}+\frac{K_{211}}{\eta_1}-K_{212}
	\label{21_E29}
\end{eqnarray}
At this stage, it can be observed that the final asymptotic expression is independent of $X$.
\paragraph*{Asymptotic expression  for the series $U_1(3 \eta_1)$: }\label{sec:22}
Substituting $m=0$ and $p=\eta_1$ in Equation(\ref{E2}), we get $U_1(3 \eta_1)$ as follows,
\begin{eqnarray}
	U_1(3 \eta_1)&=&\sum_{n=0}^{\infty}(2n+1) \dfrac{ e^{(6n+3)\eta_1}}{(e^{(4n+2)\eta_1}-1) (e^{(4n+6)\eta_1}-1)}
	\label{22_E1}
\end{eqnarray}
In this case, $U_1(3 \eta_1)$ is decomposed into an ``inner expansion'' ($g_{i}$) and an ``outer expansion'' ($g_{o}$) as follows:
\begin{eqnarray}
	U_1(3 \eta_1)&=& g_{i}(\eta_1,N)+g_{o}(\eta_1,N)
	\label{22_E2}
\end{eqnarray}
where, 
\begin{eqnarray}
	g_{i}(\eta_1,N)&=&\sum_{n=0}^{N} (2n+1)\dfrac{ e^{(6n+3)\eta_1}}{(e^{(4n+2)\eta_1}-1) (e^{(4n+6)\eta_1}-1)}
	\label{22_E3}
	\\
	g_{o}(\eta_1,N)&=&\sum_{n=N+1}^{\infty}(2n+1) \dfrac{ e^{(6n+3)\eta_1}}{(e^{(4n+2)\eta_1}-1) (e^{(4n+6)\eta_1}-1)} 
	\label{22_E4}
\end{eqnarray}

Following similar steps as shown in  Section\ref{sec:3}, we find the inner ($g_i$) and outer ($g_o$) expansions in the next two subsections.

\subparagraph*{Inner expansion}\label{subsec:A22.1}
In this subsection, we derive the inner expansion $g_{i}$ for $U_1(3 \eta_1)$.

\begin{eqnarray}
	g_{i}(\eta_1,N)&\approx&\Bigg(I_1\bigg(\frac{{\eta_1}^{-2}}{4}-\eta_1^{-1}+\frac{11}{6}\bigg)+I_2 (\frac{{\eta_1}^{-1}}{4}-\frac{5}{6})+\frac{1}{24}I_3\Bigg)
	\label{22_E9}
\end{eqnarray}
\\
where
\begin{eqnarray}
	I_1&=&\sum_{n=0}^{N}\frac{1}{(2n+3)}
	\label{22_E10}\\
	I_2&=&\sum_{n=0}^{N}1
	\label{22_E11}\\
	I_3&=&\sum_{n=0}^{N}2n+3
	\label{22_E12}
\end{eqnarray}
As $N\to\infty$, $I_1$,$I_2$ and $I_3$ is given by,
\begin{eqnarray}
	I_1&\sim&\frac{1}{2}\big(\gamma+\log{4N}-2\big)+\frac{1}{N}
	\label{22_E13}\\
	I_2&=&1+\textit{O(N)}
	\label{22_E14}\\
	I_3&=&3+\textit{O(N)}
	\label{22_E15}
\end{eqnarray}
Substituting Eqns. (\ref{22_E13}),(\ref{22_E14}),(\ref{22_E15}) in Equation (\ref{22_E9}), the asymptotic expression for $g_{i}$ in terms of  $N$ and $\eta_1$ is given by,
\begin{eqnarray}
	g_{i}(\eta_1,N)&\sim&\Bigg(\bigg(\frac{1}{2}\big(\gamma+\log{4N}-2\big)+\frac{1}{N}\bigg)\bigg(\frac{{\eta_1}^{-2}}{4}-\eta_1^{-1}+\frac{11}{6}\bigg)\nonumber\\&&+\bigg (\frac{{\eta_1}^{-1}}{4}-\frac{5}{6}\bigg)+\frac{3}{24} \Bigg) 
	\label{22_E16}
\end{eqnarray}

After introducing intermediate vanish $N=\displaystyle\frac{X}{\eta_1}$ and simplifying then Equation (\ref{22_E16}) becomes
\begin{eqnarray}
	g_{i}(\eta_1,X)&\sim&\bigg(\frac{{\eta_1}^{-2}}{8}-\frac{\eta_1^{-1}}{2}+\frac{11}{12}\bigg)\bigg(\gamma+\log{\frac{4}{\eta_1}}-2\bigg)+\frac{1}{4{\eta_1}}-\frac{17}{24}\nonumber\\
	&&+\frac{1}{4X \eta_1}-\frac{1}{X}+\bigg(\frac{{\eta_1}^{-2}}{8}-\frac{\eta_1^{-1}}{2}+\frac{11}{12}\bigg)\log{X}
	\label{22_E17}
\end{eqnarray}
Equation (\ref{22_E17}) can be used to calculate the inner expansion $g_{i}$ for $U_1(3 \eta_1)$. In the next subsection (\ref{subsec:A22.2}), we derive the outer expansion for $U_1(3 \eta_1)$.
\subparagraph*{Outer expansion }\label{subsec:A22.2}
In order to derive the outer expansion, let $m=n\eta_1$ in Equation (\ref{22_E4}) and examine the scenario where $\eta_1\to0$ with $m$ held fixed. After substituting the $n$ value in Equation (\ref{22_E4}), we get outer expansion in terms of $m$ and $\eta_1$ as follows,  
\begin{eqnarray}
	g_{o}(\eta_1,N)&=&\sum_{\substack{m=n\eta_1 \\ n=N+1}}^{\infty} \dfrac{ (2m+\eta_1)e^{(6m+3\eta_1)}}{\eta_1(e^{(4m+2\eta_1)}-1) (e^{(4m+6\eta_1)}-1)}  
	\label{22_E18}
\end{eqnarray}
For fixed $m$,  the function expansion for $\eta_1\to 0$ is as follows:
\begin{eqnarray}
	g_{o}(\eta_1,N)&=&  \frac{1}{\eta_1}\sum_{\substack{m=n\eta_1 \\ n=N+1}}^{\infty}\Bigg(\frac{2m e^{6m}}{(-1+e^{4m})^2}-\frac{\eta_1e^{6m}}{(-1+e^{4m})^3}(1+6m+(10m-1)e^{4m})
	\nonumber
	\\&&+\frac{e^{6m}(3+9m+(70m+2)e^{4m}+(25m-5)e^{8m})}{(-1+e^{4m})^4}\eta_1^2\Bigg)
	\label{22_E19}
\end{eqnarray}

\begin{eqnarray}
	g_{o}(\eta_1,N)&=& \frac{1}{\eta_1^2}\sum_{\substack{m=n\eta_1 \\ n=N+1}}^{\infty}g_{1}(m)\Delta m- \frac{1}{\eta_1}\sum_{\substack{m=n\eta_1 \\ n=N+1}}^{\infty}g_{2}(m)\Delta m \nonumber
	\\&& +\sum_{\substack{m=n\eta_1 \\ n=N+1}}^{\infty}g_{3}(m)\Delta m
	\label{22_E20}
\end{eqnarray}

where,
\begin{eqnarray}
	g_{1}(m)&=&\frac{2m e^{6m}}{(-1+e^{4m})^2}
	\label{22_E21}\\
	g_{2}(m)&=&\frac{e^{6m}(1+6m+(10m-1)e^{4m})}{(-1+e^{4m})^3}
	\label{22_E22}\\
	g_{3}(m)&=& \frac{e^{6m}(3+9m+(70m+2)e^{4m}+(25m-5)e^{8m})}{(-1+e^{4m})^4}
	\label{22_E23}
\end{eqnarray}
Define $\Delta m =m_{n+1}-m_n $. This makes 
$\Delta m=(n+1)\eta_1-n\eta_1=\eta_1$
Certainly, the outer expansion can be expressed in the following manner:
\begin{eqnarray}
	g_{o}(\eta_1,X)&=&\frac{1}{\eta_1^2}\sum_{m=X}^{\infty}g_{1}(m)\Delta m -\frac{1}{\eta_1}\sum_{m=X}^{\infty}g_{2}(m)\Delta m \nonumber
	\\&&
	+\sum_{m=X}^{\infty}g_{3}(m)\Delta m
	\label{22_E24}
\end{eqnarray}
As per the definition of $X$, it is apparent that $m=X$ at the lower bound of summation. As $\Delta m\to 0$, this expression remains asymptotically precise. Now, from Eqns. (\ref{22_E21}-\ref{22_E23}) we get $g_1(\infty)=0$,$g_2(\infty)=0$,$g_3(\infty)=0$. Furthermore, as $X\to 0$, 
\begin{eqnarray}
	g_{1}(X)\sim\dfrac{1}{8 X}+\dfrac{1}{4}
	\label{22_E25}
	\\
	g_{2}(X)\sim\dfrac{3}{16 X^2}+\dfrac{1}{2X}+\dfrac{17}{24}
	\label{22_E26}\\
	g_{3}(X)\sim\dfrac{9}{32 X^3}+\dfrac{3}{4 X^2}+\dfrac{11}{12 X}
	\label{22_E27}
\end{eqnarray}

Thus, since $\Delta m=\eta_1$, we have from Eqns. (\ref{1_E24}),(\ref{22_E24}-\ref{22_E27})  that as $\eta_1\to0$,
\begin{eqnarray} 
	g_{o}(\eta_1,X) &=& \frac{1}{\eta_1^2}\int_{X}^{\infty}g_{1}(m)dm-\frac{1}{16X\eta_1}-\frac{1}{8\eta_1}\nonumber\\
	&& - \frac{1}{\eta_1}\int_{X}^{\infty}g_{2}(m)dm+\frac{1}{4 X}+\frac{17}{48}+\int_{X}^{\infty}g_{3}(m)dm+\textit{o(1)}
	\label{22_E28}
\end{eqnarray}
The integral appearing above is convergent for the large value of m. In the limit, $\Delta m \to 0$, after simplification we get the asymptotic expression for the outer expansion $g_{o}$ for $U_1(3 \eta_1)$ as follows:
\begin{eqnarray}
	g_{o}(\eta_1,X)&\sim&-\frac{1}{4X \eta_1}+\frac{1}{X}-\bigg(\frac{{\eta_1}^{-2}}{8}-\frac{\eta_1^{-1}}{2}+\frac{11}{12}\bigg)\log{X}\nonumber\\
	&&+\frac{1}{4\eta_1^2}-\frac{5}{6 \eta_1}+\frac{17}{48}+\frac{K_{221}}{\eta_1^2}-\frac{K_{222}}{\eta_1}+K_{223}
	\label{22_E29}
\end{eqnarray}
\\ where,
\begin{eqnarray*}
	K_{221} &=& C_{221}+C_{222}=0.183425  \nonumber \\ 
	K_{222} &=& C_{223}+C_{224}=0.841942\\
	K_{223} &=& C_{225}+C_{226}=2.52884
	\nonumber \\
	C_{221} &=& \int_{1}^{\infty} \Bigg({g_1(m)}\Bigg)dm = 0.204961\nonumber \\
	C_{222} &=&\int_{0}^{1} \Bigg({g_1(m)}-\frac{1}{8m}-\frac{1}{4}\Bigg)dm =-0.0215362\nonumber \\
	C_{223} &=&\int_{1}^{\infty} \Bigg({g_2(m)}-\frac{3}{16m^2}\Bigg)dm=0.776757\nonumber \\
	C_{224} &=&\int_{0}^{1} \Bigg({g_2(m)}-\frac{3}{16m^2}-\frac{1}{2m}-\frac{17}{24}\Bigg)dm=0.0651856 \nonumber \\
	C_{225} &=&\int_{1}^{\infty} \Bigg({g_3(m)}-\frac{9}{32m^3}-\frac{3}{4 m^2}\Bigg)dm= 1.38567\nonumber \\
	C_{226} &=&\int_{0}^{1} \Bigg({g_3(m)}-\frac{9}{32m^3}-\frac{3}{4 m^2}-\frac{11}{12m}\Bigg)dm=1.14317 \nonumber \\
\end{eqnarray*}

Adding Equation (\ref{22_E17}) and Equation (\ref{22_E29}), the asymptotic expression for $U_1(3 \eta_1)$ in the the small separation region is given by,
\begin{eqnarray}
	U_1(3 \eta_1)&\sim& \bigg(\frac{{\eta_1}^{-2}}{8}-\frac{\eta_1^{-1}}{2}+\frac{11}{12}\bigg)\bigg(\gamma+\log{\frac{4}{\eta_1}}-2\bigg)-\frac{7}{12{\eta_1}}-\frac{17}{48}\nonumber\\
	&&+\frac{1}{4\eta_1^2}+\frac{K_{221}}{\eta_1^2}-\frac{K_{222}}{\eta_1}+K_{223}
	\label{22_E30}
\end{eqnarray}
At this stage, it can be observed that the final asymptotic expression is independent of $X$.
\paragraph*{Asymptotic expression  for the series $U_2(3 \eta_1)$: }\label{sec:23}
Substituting $m=0$ and $p=\eta_1$ in Equation(\ref{E2}), we get $U_2(3 \eta_1)$ as follows,
\begin{eqnarray}
	U_2(3 \eta_1)&=&\sum_{n=0}^{\infty}(2n+1)^2 \dfrac{ e^{(6n+3)\eta_1}}{(e^{(4n+2)\eta_1}-1) (e^{(4n+6)\eta_1}-1)}
	\label{23_E1}
\end{eqnarray}
In this case, $U_2(3 \eta_1)$ is decomposed into an ``inner expansion'' ($g_{i}$) and an ``outer expansion'' ($g_{o}$) as follows:
\begin{eqnarray}
	U_2(3 \eta_1)&=& g_{i}(\eta_1,N)+g_{o}(\eta_1,N)
	\label{23_E2}
\end{eqnarray}
where, 
\begin{eqnarray}
	g_{i}(\eta_1,N)&=&\sum_{n=0}^{N} (2n+1)^2\dfrac{ e^{(6n+3)\eta_1}}{(e^{(4n+2)\eta_1}-1) (e^{(4n+6)\eta_1}-1)}
	\label{23_E3}
	\\
	g_{o}(\eta_1,N)&=&\sum_{n=N+1}^{\infty}(2n+1)^2 \dfrac{ e^{(6n+3)\eta_1}}{(e^{(4n+2)\eta_1}-1) (e^{(4n+6)\eta_1}-1)} 
	\label{23_E4}
\end{eqnarray}

Following similar steps as shown in  Section\ref{sec:3}, we find the inner ($g_i$) and outer ($g_o$) expansions in the next two subsections.

\subparagraph*{Inner expansion}\label{subsec:A23.1}
In this subsection, we derive the inner expansion $g_{i}$ for $U_2(3 \eta_1)$.

\begin{eqnarray}
	g_{i}(\eta_1,N)&\approx&\Bigg(I_1\bigg(\frac{-{\eta_1}^{-2}}{2}+2\eta_1^{-1}-\frac{11}{3}\bigg)+I_2\bigg(\frac{{\eta_1}^{-2}}{4}-\eta_1^{-1}+\frac{47}{24}\bigg)\nonumber\\&& +I_3(\frac{{\eta_1}^{-1}}{4}-\frac{5}{6})\Bigg)
	\label{23_E9}
\end{eqnarray}
\\
where
\begin{eqnarray}
	I_1&=&\sum_{n=0}^{N}\frac{1}{(2n+3)}
	\label{23_E10}\\
	I_2&=&\sum_{n=0}^{N}1
	\label{23_E11}\\
	I_3&=&\sum_{n=0}^{N}2n+1
	\label{23_E12}
\end{eqnarray}
As $N\to\infty$, $I_1$,$I_2$ and $I_3$ is given by,
\begin{eqnarray}
	I_1&\sim&\frac{1}{2}\big(\gamma+\log{4N}-2\big)+\frac{1}{N}
	\label{23_E13}\\
	I_2&=&1+\textit{O(N)}
	\label{23_E14}\\
	I_3&=&1+\textit{O(N)}
	\label{23_E15}
\end{eqnarray}
Substituting Eqns. (\ref{23_E13}),(\ref{23_E14}),(\ref{23_E15}) in Equation (\ref{23_E9}), the asymptotic expression for $g_{i}$ in terms of  $N$ and $\eta_1$ is given by,
\begin{eqnarray}
	g_{i}(\eta_1,N)&\sim&\Bigg(\bigg(\frac{1}{2}\big(\gamma+\log{4N}-2\big)+\frac{1}{N}\bigg)\bigg(\frac{-{\eta_1}^{-2}}{2}+2\eta_1^{-1}-\frac{11}{3}\bigg) \nonumber\\&&\bigg(\frac{{\eta_1}^{-2}}{4}-\eta_1^{-1}+\frac{47}{24}\bigg)-\bigg(\frac{{\eta_1}^{-1}}{4}-\frac{5}{6}\bigg) \Bigg) 
	\label{23_E16}
\end{eqnarray}

After introducing intermediate vanish $N=\displaystyle\frac{X}{\eta_1}$ and simplifying then Equation (\ref{23_E16}) becomes
\begin{eqnarray}
	g_{i}(\eta_1,X)&\sim&\frac{1}{4\eta_1^2}-\frac{3}{4{\eta_1}}-\bigg(\frac{{\eta_1}^{-2}}{4}-\eta_1^{-1}+\frac{11}{6}\bigg) \bigg(\gamma+\log{\frac{4}{\eta_1}}-2\bigg)+\frac{27}{24}\nonumber\\
	&&-\frac{1}{2X \eta_1}+\frac{2}{X}-\bigg(\frac{{\eta_1}^{-2}}{4}+\eta_1^{-1}+\frac{11}{6}\bigg)\log{X}
	\label{23_E17}
\end{eqnarray}
Equation (\ref{23_E17}) can be used to calculate the inner expansion $g_{i}$ for $U_2(3 \eta_1)$. In the next subsection (\ref{subsec:A23.2}), we derive the outer expansion for $U_2(3 \eta_1)$.
\subparagraph*{Outer expansion }\label{subsec:A23.2}
In order to derive the outer expansion, let $m=n\eta_1$ in Equation (\ref{23_E4}) and examine the scenario where $\eta_1\to0$ with $m$ held fixed. After substituting the $n$ value in Equation (\ref{23_E4}), we get outer expansion in terms of $m$ and $\eta_1$ as follows,  
\begin{eqnarray}
	g_{o}(\eta_1,N)&=&\sum_{\substack{m=n\eta_1 \\ n=N+1}}^{\infty} \dfrac{ (2m+\eta_1)^2e^{(6m+3 \eta_1)}}{\eta_1^2(e^{(4m+2\eta_1)}-1) (e^{(4m+6\eta_1)}-1)}  
	\label{23_E18}
\end{eqnarray}
For fixed $m$,  the function expansion for $\eta_1\to 0$ is as follows:
\begin{eqnarray}
	g_{o}(\eta_1,N)&=&  \frac{1}{\eta_1^2}\sum_{\substack{m=n\eta_1 \\ n=N+1}}^{\infty}\Bigg(\frac{4m^2 e^{6m}}{(-1+e^{4m})^2}-\frac{4m\eta_1e^{6m}}{(-1+e^{4m})^3}(1+3m+(5m-1)e^{4m})
	\nonumber
	\\&&+\frac{e^{6m}(1+12m+18m^2+(140m^2+8m-2)e^{4m}+(50m^2-20m+1)e^{8m})}{(-1+e^{4m})^4}\eta_1^2
	\nonumber\\
	&&-\frac{e^{6m}}{3(-1+e^{4m})^5}\bigg(9+54m+54m^2+(1438m^2+366m-3)e^{4m}+(2098m^2-270m-21)e^{8m} \nonumber\\
	&&+(250m^2-150m+15)e^{12m}\bigg)\eta_1^3\Bigg)
	\label{23_E19}
\end{eqnarray}

\begin{eqnarray}
	g_{o}(\eta_1,N)&=& \frac{1}{\eta_1^3}\sum_{\substack{m=n\eta_1 \\ n=N+1}}^{\infty}g_{1}(m)\Delta m- \frac{1}{\eta_1^2}\sum_{\substack{m=n\eta_1 \\ n=N+1}}^{\infty}g_{2}(m)\Delta m \nonumber
	\\&& +\frac{1}{\eta_1}\sum_{\substack{m=n\eta_1 \\ n=N+1}}^{\infty}g_{3}(m)\Delta m-\sum_{\substack{m=n\eta_1 \\ n=N+1}}^{\infty}g_{4}(m)\Delta m
	\label{23_E20}
\end{eqnarray}

where,
\begin{eqnarray}
	g_{1}(m)&=&\frac{4m^2 e^{6m}}{(-1+e^{4m})^2}
	\label{23_E21}\\
	g_{2}(m)&=&\frac{4m e^{6m}}{(-1+e^{4m})^3}(1+3m+(5m-1)e^{4m})
	\label{23_E22}\\
	g_{3}(m)&=& \frac{e^{6m}(1+12m+18m^2+(140m^2+8m-2)e^{4m}+(50m^2-20m+1)e^{8m})}{(-1+e^{4m})^4}
	\label{23_E23}\\
	g_{4}(m)&=& \frac{e^{6m}}{3(-1+e^{4m})^5}\bigg(9+54m+54m^2+(1438m^2+366m-3)e^{4m}+(2098m^2-270m-21)e^{8m} \nonumber\\
	&&+(250m^2-150m+15)e^{12m}\bigg)
	\label{23_E24}
\end{eqnarray}
Define $\Delta m =m_{n+1}-m_n $. This makes 
$\Delta m=(n+1)\eta_1-n\eta_1=\eta_1$
Certainly, the outer expansion can be expressed in the following manner:
\begin{eqnarray}
	g_{o}(\eta_1,X)&=&\frac{1}{\eta_1^3}\sum_{m=X}^{\infty}g_{1}(m)\Delta m -\frac{1}{\eta_1^2}\sum_{m=X}^{\infty}g_{2}(m)\Delta m \nonumber
	\\&&
	+\frac{1}{\eta_1}\sum_{m=X}^{\infty}g_{3}(m)\Delta m
	-\sum_{m=X}^{\infty}g_{4}(m)\Delta m
	\label{23_E25}
\end{eqnarray}
As per the definition of $X$, it is apparent that $m=X$ at the lower bound of summation.
 As $\Delta m\to 0$, this expression remains asymptotically precise. Now, from Eqns. (\ref{23_E21}-\ref{23_E24}) we get $g_1(\infty)=0$,$g_2(\infty)=0$,$g_3(\infty)=0$,$g_4(\infty)=0$. Furthermore, as $X\to 0$, 
\begin{eqnarray}
	g_{1}(X)\sim\dfrac{1}{4}
	\label{23_E26}
	\\
	g_{2}(X)\sim\dfrac{1}{4 X}+\dfrac{3}{4}
	\label{23_E27}\\
	g_{3}(X)\sim\dfrac{3}{8 X^2}+\dfrac{1}{X}+\dfrac{9}{8}
	\label{23_E28}\\
	g_{4}(X)\sim\dfrac{9}{16 X^3}+\dfrac{3}{2 X^2}+\dfrac{11}{6 X}
	\label{23_E29}
\end{eqnarray}

Thus, since $\Delta m=\eta_1$, we have from Eqns. (\ref{1_E24}),(\ref{23_E25}-\ref{23_E29})  that as $\eta_1\to0$,
\begin{eqnarray} 
	g_{o}(\eta_1,X) &=& \frac{1}{\eta_1^3}\int_{X}^{\infty}g_{1}(m)dm-\frac{1}{8\eta_1^2} - \frac{1}{\eta_1^2}\int_{X}^{\infty}g_{2}(m)dm+\frac{1}{8X\eta_1}+\frac{3}{8\eta_1}\nonumber\\&&+ \frac{1}{\eta_1}\int_{X}^{\infty}g_{3}(m)dm-\frac{1}{2 X}-\frac{9}{16}-\int_{X}^{\infty}g_{4}(m)dm+\textit{o(1)}
	\label{23_E30}
\end{eqnarray}

The integral appearing above is convergent for the large value of m. In the limit, $\Delta m \to 0$, after simplification we get the asymptotic expression for the outer expansion $g_{o}$ for $U_2(3 \eta_1)$ as follows:
\begin{eqnarray}
	g_{o}(\eta_1,X)&\sim&\frac{1}{2X \eta_1}-\frac{2}{X}+\bigg(\frac{{\eta_1}^{-2}}{4}+\eta_1^{-1}+\frac{11}{6}\bigg)\log{X}+\frac{1}{4\eta_1^3}-\frac{7}{8\eta_1^2}\nonumber\\
	&&+\frac{3}{2 \eta_1}-\frac{9}{16}+\frac{K_{231}}{\eta_1^3}-\frac{K_{232}}{\eta_1^2}+\frac{K_{233}}{\eta_1}-K_{234}
	\label{23_E31}
\end{eqnarray}
\\ where,
\begin{eqnarray*}
	K_{231} &=& C_{231}+C_{232}=0.89275 \nonumber \\ 
	K_{232} &=& C_{233}+C_{234}=3.7951\\
	K_{233} &=& C_{235}+C_{236}=7.43967\\
	K_{234} &=& C_{237}+C_{238}=10.0125
	\nonumber \\
	C_{231} &=& \int_{1}^{\infty} {g_1(m)} dm = 0.681334\nonumber \\
	C_{232} &=&\int_{0}^{1} \Bigg({g_1(m)}-\frac{1}{4}\Bigg)dm =0.211416\nonumber \\
	C_{233} &=&\int_{1}^{\infty}{g_2(m)} dm= 3.01566\nonumber \\
	C_{234} &=&\int_{0}^{1} \Bigg({g_2(m)}-\frac{1}{4m}-\frac{3}{4}\Bigg)dm=0.779446\nonumber \\
	C_{235} &=&\int_{1}^{\infty} \Bigg({g_3(m)}-\frac{3}{8m^2}\Bigg)dm= 6.28767\nonumber \\
	C_{236} &=&\int_{0}^{1} \Bigg({g_3(m)}-\frac{3}{8m^2}-\frac{1}{m}-\frac{9}{8}\Bigg)dm=1.152\nonumber \\
	C_{237} &=&\int_{1}^{\infty} \Bigg({g_4(m)}-\frac{9}{16m^3}-\frac{3}{2 m^2}\Bigg)dm= 8.08805\nonumber \\
	C_{238} &=&\int_{0}^{1} \Bigg({g_4(m)}-\frac{9}{16m^3}-\frac{3}{2 m^2}-\frac{11}{6m}\Bigg)dm=1.92445 \nonumber \\
\end{eqnarray*}
Adding Equation (\ref{23_E17}) and Equation (\ref{23_E31}), the asymptotic expression for $U_2(3 \eta_1)$ in the the small separation region is given by,
\begin{eqnarray}
	U_2(3 \eta_1)&\sim&-\bigg(\frac{{\eta_1}^{-2}}{4}-\eta_1^{-1}+\frac{11}{6}\bigg) \bigg(\gamma+\log{\frac{4}{\eta_1}}-2\bigg)+\frac{27}{48}+\frac{1}{4\eta_1^3}-\frac{5}{8\eta_1^2}\nonumber\\
	&&+\frac{3}{4 \eta_1}+\frac{K_{231}}{\eta_1^3}-\frac{K_{232}}{\eta_1^2}+\frac{K_{233}}{\eta_1}-K_{234}
	\label{23_E32}
\end{eqnarray}
At this stage, it can be observed that the final asymptotic expression is independent of $X$.
\paragraph*{Asymptotic expression  for the series $U_3(3 \eta_1)$: }\label{sec:24}
Substituting $m=0$ and $p=\eta_1$ in Equation(\ref{E2}), we get $U_3(3 \eta_1)$ as follows,
\begin{eqnarray}
	U_3(3 \eta_1)&=&\sum_{n=0}^{\infty}(2n+1)^3 \dfrac{ e^{(6n+3)\eta_1}}{(e^{(4n+2)\eta_1}-1) (e^{(4n+6)\eta_1}-1)}
	\label{24_E1}
\end{eqnarray}
In this case, $U_3(3 \eta_1)$ is decomposed into an ``inner expansion'' ($g_{i}$) and an ``outer expansion'' ($g_{o}$) as follows:
\begin{eqnarray}
	U_3(3 \eta_1)&=& g_{i}(\eta_1,N)+g_{o}(\eta_1,N)
	\label{24_E2}
\end{eqnarray}
where, 
\begin{eqnarray}
	g_{i}(\eta_1,N)&=&\sum_{n=0}^{N} (2n+1)^3\dfrac{ e^{(6n+3)\eta_1}}{(e^{(4n+2)\eta_1}-1) (e^{(4n+6)\eta_1}-1)}
	\label{24_E3}
	\\
	g_{o}(\eta_1,N)&=&\sum_{n=N+1}^{\infty}(2n+1)^3 \dfrac{ e^{(6n+3)\eta_1}}{(e^{(4n+2)\eta_1}-1) (e^{(4n+6)\eta_1}-1)} 
	\label{24_E4}
\end{eqnarray}

Following similar steps as shown in  Section\ref{sec:3}, we find the inner ($g_i$) and outer ($g_o$) expansions in the next two subsections.

\subparagraph*{Inner expansion}\label{subsec:A24.1}
In this subsection, we derive the inner expansion $g_{i}$ for $U_3(3 \eta_1)$.

\begin{eqnarray}
	g_{i}(\eta_1,N)&\approx&\Bigg(I_1\bigg(\frac{2{\eta_1}^{-2}}{2}-4\eta_1^{-1}+\frac{22}{3}\bigg)+I_2\bigg(\frac{-1}{2\eta_1^2}+\frac{9}{4\eta_1}-\frac{9}{2}\bigg)\nonumber\\&&+I_3\bigg(\frac{{\eta_1}^{-2}}{4}-\eta_1^{-1}+\frac{47}{24}\bigg)\Bigg)
	\label{24_E9}
\end{eqnarray}
\\
where
\begin{eqnarray}
	I_1&=&\sum_{n=0}^{N}\frac{1}{(2n+3)}
	\label{24_E10}\\
	I_2&=&\sum_{n=0}^{N}1
	\label{24_E11}\\
	I_3&=&\sum_{n=0}^{N}2n+1
	\label{24_E12}
\end{eqnarray}
As $N\to\infty$, $I_1$,$I_2$ and $I_3$ is given by,
\begin{eqnarray}
	I_1&\sim&\frac{1}{2}\big(\gamma+\log{4N}-2\big)+\frac{1}{N}
	\label{24_E13}\\
	I_2&=&1+\textit{O(N)}
	\label{24_E14}\\
	I_3&=&1+\textit{O(N)}
	\label{24_E15}
\end{eqnarray}
Substituting Eqns. (\ref{24_E13}),(\ref{24_E14}),(\ref{24_E15}) in Equation (\ref{24_E9}), the asymptotic expression for $g_{i}$ in terms of  $N$ and $\eta_1$ is given by,
\begin{eqnarray}
	g_{i}(\eta_1,N)&\sim&\Bigg(\bigg(\frac{1}{2}\big(\gamma+\log{4N}-2\big)+\frac{1}{N}\bigg)\bigg(\frac{2{\eta_1}^{-2}}{2}-4\eta_1^{-1}+\frac{22}{3}\bigg)\nonumber\\&&\bigg(\frac{-1}{2\eta_1^2}+\frac{9}{4\eta_1}-\frac{9}{2}\bigg)+\bigg(\frac{{\eta_1}^{-2}}{4}-\eta_1^{-1}+\frac{47}{24}\bigg)\Bigg) \Bigg) 
	\label{24_E16}
\end{eqnarray}

After introducing intermediate vanish $N=\displaystyle\frac{X}{\eta_1}$ and simplifying then Equation (\ref{24_E16}) becomes
\begin{eqnarray}
	g_{i}(\eta_1,X)&\sim&-\frac{\eta_1^{-2}}{4}+\frac{5}{4{\eta_1}}+\bigg(\frac{{\eta_1}^{-2}}{2}-2\eta_1^{-1}+\frac{11}{3}\bigg)\bigg(\gamma+\log{\frac{4}{\eta_1}}-2\bigg)-\frac{61}{24}\nonumber\\
	&&+\frac{1}{X \eta_1}-\frac{4}{X}+\bigg(\frac{{\eta_1}^{-2}}{2}-2\eta_1^{-1}+\frac{11}{3}\bigg)\log{X}
	\label{24_E17}
\end{eqnarray}
Equation (\ref{24_E17}) can be used to calculate the inner expansion $g_{i}$ for $U_3(3 \eta_1)$. In the next subsection (\ref{subsec:A24.2}), we derive the outer expansion for $U_3(3 \eta_1)$.
\subparagraph*{Outer expansion }\label{subsec:A24.2}
In order to derive the outer expansion, let $m=n\eta_1$ in Equation (\ref{24_E4}) and examine the scenario where $\eta_1\to0$ with $m$ held fixed. After substituting the $n$ value in Equation (\ref{24_E4}), we get outer expansion in terms of $m$ and $\eta_1$ as follows,  
\begin{eqnarray}
	g_{o}(\eta_1,N)&=&\sum_{\substack{m=n\eta_1 \\ n=N+1}}^{\infty} \dfrac{ (2m+\eta_1)^3e^{(6m+3\eta_1)}}{\eta_1^3(e^{(4m+2\eta_1)}-1) (e^{(4m+6\eta_1)}-1)}  
	\label{24_E18}
\end{eqnarray}
For fixed $m$,  the function expansion for $\eta_1\to 0$ is as follows:
\begin{eqnarray}
	g_{o}(\eta_1,N)&=&  \frac{1}{\eta_1^3}\sum_{\substack{m=n\eta_1 \\ n=N+1}}^{\infty}\Bigg(\frac{8m^3 e^{6m}}{(-1+e^{4m})^2}-\frac{4m^2\eta_1e^{6m}}{(-1+e^{4m})^3}(3+6m+(10m-3)e^{4m})
	\nonumber
	\\&&+\frac{2me^{6m}(3+18m+18m^2+(140m^2+12m-6)e^{4m}+(50m^2-30m+3)e^{8m})}{(-1+e^{4m})^4}\eta_1^2
	\nonumber\\
	&&-\frac{e^{6m}}{3(-1+e^{4m})^5}\bigg(3+54m+162m^2+108m^3+(2876m^3+1098m^2-18m-9)e^{4m}\nonumber\\
	&&+(4196m^3-810m^2-126m+9)e^{8m}+(500m^3-450m^2+90m-3)e^{12m}\bigg)\eta_1^3\nonumber\\
	&&+\frac{\eta_1^4 e^{6m}}{3(-1+e^{4m})^6}\bigg(9+81m+162m^2+81m^3+(6700m^3+4152m^2+468m-12)e^{4m} \nonumber\\
	&&+(25766m^3+1980m^2-954m-18)e^{8m}+(13292m^3-5544m^2+180m+36)e^{12m}\nonumber\\
	&&+(625m^3-750m^2+225m-15)e^{16m}\bigg)\Bigg)
	\label{24_E19}
\end{eqnarray}

\begin{eqnarray}
	g_{o}(\eta_1,N)&=& \frac{1}{\eta_1^4}\sum_{\substack{m=n\eta_1 \\ n=N+1}}^{\infty}g_{1}(m)\Delta m- \frac{1}{\eta_1^3}\sum_{\substack{m=n\eta_1 \\ n=N+1}}^{\infty}g_{2}(m)\Delta m  +\frac{1}{\eta_1^2}\sum_{\substack{m=n\eta_1 \\ n=N+1}}^{\infty}g_{3}(m)\Delta m\nonumber
	\\&&-\frac{1}{\eta_1}\sum_{\substack{m=n\eta_1 \\ n=N+1}}^{\infty}g_{4}(m)\Delta m+\sum_{\substack{m=n\eta_1 \\ n=N+1}}^{\infty}g_{5}(m)\Delta m
	\label{24_E20}
\end{eqnarray}

where,
\begin{eqnarray}
	g_{1}(m)&=&\frac{8m^3 e^{6m}}{(-1+e^{4m})^2}
	\label{24_E21}\\
	g_{2}(m)&=&\frac{4m^2e^{6m}}{(-1+e^{4m})^3}(3+6m+(10m-3)e^{4m})
	\label{24_E22}\\
	g_{3}(m)&=& \frac{2me^{6m}(3+18m+18m^2+(140m^2+12m-6)e^{4m}+(50m^2-30m+3)e^{8m})}{(-1+e^{4m})^4}
	\label{24_E23}\\
	g_{4}(m)&=& \frac{e^{6m}}{3(-1+e^{4m})^5}\bigg(3+54m+162m^2+108m^3+(2876m^3+1098m^2-18m-9)e^{4m}\nonumber\\
	&&+(4196m^3-810m^2-126m+9)e^{8m}+(500m^3-450m^2+90m-3)e^{12m}\bigg)
	\label{24_E24}\\
	g_{5}(m)&=&\frac{e^{6m}}{3(-1+e^{4m})^6}\bigg(9+81m+162m^2+81m^3+(6700m^3+4152m^2+468m-12)e^{4m} \nonumber\\
	&&+(25766m^3+1980m^2-954m-18)e^{8m}+(13292m^3-5544m^2+180m+36)e^{12m}\nonumber\\
	&&+(625m^3-750m^2+225m-15)e^{16m}\bigg)
	\label{24_E25}
\end{eqnarray}
Define $\Delta m =m_{n+1}-m_n $. This makes 
$\Delta m=(n+1)\eta_1-n\eta_1=\eta_1$
Certainly, the outer expansion can be expressed in the following manner:
\begin{eqnarray}
	g_{o}(\eta_1,X)&=&\frac{1}{\eta_1^4}\sum_{m=X}^{\infty}g_{1}(m)\Delta m -\frac{1}{\eta_1^3}\sum_{m=X}^{\infty}g_{2}(m)\Delta m+\frac{1}{\eta_1^2}\sum_{m=X}^{\infty}g_{3}(m)\Delta m\nonumber
	\\&&
	-\frac{1}{\eta_1}\sum_{m=X}^{\infty}g_{4}(m)\Delta m+\sum_{m=X}^{\infty}g_{5}(m)\Delta m
	\label{24_E26}
\end{eqnarray}
As per the definition of $X$, it is apparent that $m=X$ at the lower bound of summation.
 As $\Delta m\to 0$, this expression remains asymptotically precise. Now, from Eqns. (\ref{24_E21}-\ref{24_E25}) we get $g_1(\infty)=0$,$g_2(\infty)=0$,$g_3(\infty)=0$,$g_4(\infty)=0$,$g_5(\infty)=0$. Furthermore, as $X\to 0$, 
\begin{eqnarray}
	g_{1}(X)\sim\textit{O(X)}
	\label{24_E27} \\
	g_{2}(X)\sim\dfrac{1}{4}
	\label{24_E28}\\
	g_{3}(X)\sim\dfrac{1}{2 X}+\dfrac{5}{4}
	\label{24_E29}\\
	g_{4}(X)\sim\dfrac{3}{4 X^2}+\dfrac{2}{X}+\dfrac{61}{24}
	\label{24_E30}\\
	g_{5}(X)\sim\dfrac{9}{8 X^3}+\dfrac{3}{X^2}+\dfrac{11}{3 X}
	\label{24_E31}
\end{eqnarray}

Thus, since $\Delta m=\eta_1$, we have from Eqns. (\ref{1_E24}),(\ref{24_E26}-\ref{24_E31})  that as $\eta_1\to0$,
\begin{eqnarray} 
	g_{o}(\eta_1,X) &=& \frac{1}{\eta_1^4}\int_{X}^{\infty}g_{1}(m)dm - \frac{1}{\eta_1^3}\int_{X}^{\infty}g_{2}(m)dm+\frac{1}{8\eta_1^2}+ \frac{1}{\eta_1^2}\int_{X}^{\infty}g_{3}(m)dm-\frac{1}{4X\eta_1}\nonumber\\&&-\frac{5}{8\eta_1}-\frac{1}{\eta_1}\int_{X}^{\infty}g_{4}(m)dm+\frac{61}{48}+\frac{1}{X}+\int_{X}^{\infty}g_{5}(m)dm+\textit{o(1)}
	\label{24_E32}
\end{eqnarray}

The integral appearing above is convergent for the large value of m. In the limit, $\Delta m \to 0$, after simplification we get the asymptotic expression for the outer expansion $g_{o}$ for $U_3(3 \eta_1)$ as follows:
\begin{eqnarray}
	g_{o}(\eta_1,X)&\sim&-\frac{1}{X \eta_1}+\frac{4}{X}-\bigg(\frac{{\eta_1}^{-2}}{2}-2\eta_1^{-1}+\frac{11}{3}\bigg)\log{X}-\frac{1}{4\eta_1^3}+\frac{11}{8\eta_1^2}-\frac{19}{6\eta_1}\nonumber\\
	&&+\frac{61}{48}+\frac{C_{241}}{\eta_1^4}-\frac{K_{241}}{\eta_1^3}+\frac{K_{242}}{\eta_1^2}-\frac{K_{243}}{\eta_1}+K_{244}
	\label{24_E33}
\end{eqnarray}
\\ where,
\begin{eqnarray*}
	K_{241} &=& C_{242}+C_{243}=12.4774 \nonumber \\ 
	K_{242} &=& C_{244}+C_{245}=24.9142\\
	K_{243} &=& C_{246}+C_{247}=32.675\\
	K_{244} &=& C_{248}+C_{249}=33.9727
	\nonumber \\
	C_{241} &=& \int_{0}^{\infty} {g_1(m)} dm =3.09972\nonumber \\
	C_{242} &=&\int_{1}^{\infty}{g_2(m)} dm=10.9156\nonumber \\
	C_{243} &=&\int_{0}^{1} \Bigg({g_2(m)}-\frac{1}{4}\Bigg)dm=1.56177 \nonumber \\
	C_{244} &=&\int_{1}^{\infty} {g_3(m)}dm= 22.97\nonumber \\
	C_{245} &=&\int_{0}^{1} \Bigg({g_3(m)}-\frac{1}{2m}-\frac{5}{4}\Bigg)dm=1.9442\nonumber \\
	C_{246} &=&\int_{1}^{\infty} \Bigg({g_4(m)}-\frac{3}{4m^2}\Bigg)dm=31.5006\nonumber \\
	C_{247} &=&\int_{0}^{1} \Bigg({g_4(m)}-\frac{3}{4m^2}-\frac{2}{m}-\frac{61}{24}\Bigg)dm=1.1744 \nonumber \\
	C_{248} &=& \int_{1}^{\infty} \Bigg({g_5(m)}-\frac{9}{8m^3}-\frac{3}{m^2}\Bigg) dm = 30.8516\nonumber \\
	C_{249} &=&\int_{0}^{1} \Bigg({g_5(m)}-\frac{9}{8m^3}-\frac{3}{m^2}-\frac{11}{3m}\Bigg)dm =3.12108\nonumber \\
\end{eqnarray*}
Adding Equation (\ref{24_E17}) and Equation (\ref{24_E33}), the asymptotic expression for $U_3(3 \eta_1)$ in the the small separation region is given by,
\begin{eqnarray}
	U_3(3 \eta_1)&\sim&\bigg(\frac{{\eta_1}^{-2}}{2}-2\eta_1^{-1}+\frac{11}{3}\bigg)\bigg(\gamma+\log{\frac{4}{\eta_1}}-2\bigg)+\frac{9\eta_1^{-2}}{8}-\frac{23}{12{\eta_1}}-\frac{1}{4 \eta_1^3}-\frac{61}{48}\nonumber\\
	&&+\frac{C_{241}}{\eta_1^4}-\frac{K_{241}}{\eta_1^3}+\frac{K_{242}}{\eta_1^2}-\frac{K_{243}}{\eta_1}+K_{244}
	\label{24_E34}
\end{eqnarray}
At this stage, it can be observed that the final asymptotic expression is independent of $X$.


\end{appendices}
\bibliography{sn-bibliography}

\end{document}